%% file: minerva_proposal_E938.tex
\documentclass[11pt,twoside]{article}
\usepackage{times,latexsym,epsfig,epsf,graphicx,wrapfig,url,psfig,times}
\newcommand{\AmS}{{\protect\the\textfont2 
A\kern-.1667em\lower.5ex\hbox{M}\kern-.125emS}}
\def\be{\begin{equation}}
\def\ee{\end{equation}}
\def\bea{\begin{eqnarray*}}
\def\eea{\end{eqnarray*}}

\newcommand{\bd}{\begin{displaymath} }
\newcommand{\ed}{\end{displaymath} }

\newcommand{\neut}{\mbox{$\nu$}}                         
\newcommand{\numu}{\mbox{$\nu_{\mu}$}}                   
\newcommand{\nue}{\mbox{$\nu_{e}$}}                      
\newcommand{\anu}{\mbox{$\overline{\nu}$}} 
\newcommand{\anumu}{\mbox{$\overline{\nu}_{\mu}$}}  

\newcommand{\pip}{\mbox{$\pi^{+}$}}       
\newcommand{\piz}{\mbox{$\pi^{0}$}} 
\newcommand{\g}{\mbox{$\gamma$}} 

\newcommand{\nubar}[0]{\overline{\nu}}
\newcommand{\qbar}[0]{\overline{q}}

\newcommand{\stw}{\mbox{$\sin^2\theta_W$}}

\newcommand{\ubar}{\overline{u}}
\newcommand{\dbar}{\overline{d}}
\newcommand{\cbar}{\overline{c}}
\newcommand{\sbar}{\overline{s}}

\newcommand{\qsq}{\mbox{$Q^2$}}

\newcommand{\minerva}{\mbox{\hbox{MINER}$\nu$\hbox{A}}}
\newcommand{\numi}{\mbox{NuMI}}

\newcommand{\D}{\displaystyle}

\newcommand{\nova}{\mbox{NO$\nu$A}}
\newcommand{\qdif}{($Q^2_{\mu}-2M{\nu})$/err}

\begin{document}
\ifx\href\undefined\else\hypersetup{linktocpage=true}\fi
\renewcommand{\thepage}{}

\input{title}
\cleardoublepage

\input{collaboration}

\cleardoublepage

\pagenumbering{roman}
\setcounter{page}{5}

\tableofcontents
\cleardoublepage

\addcontentsline{toc}{section}{List of Figures}
\listoffigures
\cleardoublepage

\addcontentsline{toc}{section}{List of Tables}
\listoftables
\cleardoublepage

\pagenumbering{arabic}
\setcounter{page}{1}

\part{Introduction}
\cleardoublepage

\input{execSummary}
\cleardoublepage

\input{overview}

\cleardoublepage

\input{phenomenology}
\cleardoublepage

\input{nudata}

\cleardoublepage

\input{numiBeam}
\cleardoublepage

\part{Physics Motivation and Goals}
\label{part:physics}
\cleardoublepage

\input{quasielastic}

\cleardoublepage

\input{resonant}
\cleardoublepage

\input{coherent}
\cleardoublepage

\input{strangeCharm}

\cleardoublepage

\input{duality}

\cleardoublepage

\input{gpd}
\cleardoublepage

 \input{nuclear}

 \cleardoublepage

\input{oscillation}

\cleardoublepage

\part{Project Description}
\cleardoublepage

\input{numi}

\cleardoublepage

\input{monteCarlo}

\cleardoublepage

\input{detector}

\cleardoublepage

\input{costSchedule}

\cleardoublepage

%

\part{Appendices}
\cleardoublepage

\appendix

\input{appendixCryogenic}
\cleardoublepage

\input{appendixOffAxis}
\cleardoublepage

\part{Addendum to the \minerva Proposal}

\input{summary.tex}

\pagebreak
\input{qe_analysis.tex}

\pagebreak

\input{coherent_addendum.tex}

\pagebreak

\input{nuclupdate.tex}

\pagebreak

\input{res_addendum.tex}

\pagebreak

\input{MINERVA-osc.tex}

\input{bibliography}
\end{document}

%% file: title.tex
\begin{titlepage}
\begin{center}
{\LARGE \bf Proposal to Perform a High-Statistics\\
\vspace*{5mm}
 Neutrino Scattering Experiment\\ 
\vspace*{5mm}
 Using a Fine-grained Detector\\
\vspace*{5mm}
 in the NuMI Beam}
\vskip 1cm
\vspace*{5mm}
{\it \large 2 February 2004} \\
\vspace*{5mm}
\vspace{5mm}
\end{center}
\begin{abstract}

\noindent The NuMI facility at Fermilab will provide an extremely intense beam of 
neutrinos for the MINOS neutrino-oscillation experiment.  The spacious and
fully-outfitted MINOS near detector hall will be the ideal venue for a high-statistics,
high-resolution $\nu$ and $\nubar$--nucleon/nucleus scattering
experiment.  The experiment described here will measure neutrino cross-sections
and probe nuclear effects essential to present and future neutrino-oscillation
experiments. Moreover, with the high NuMI beam intensity, the experiment
will either initially address or significantly improve our knowledge of a wide
variety of neutrino physics topics of interest and importance to  the
elementary-particle and nuclear-physics communities.

\vspace{1pc}
\end{abstract}
\end{titlepage}

%% file: collaboration.tex
\centerline{\LARGE \bf The \minerva\ Collaboration}
\medskip

\centerline{D.~Drakoulakos, P.~Stamoulis, G.~Tzanakos, M.~Zois}
\centerline{\it University of Athens; Athens, Greece}
\medskip

\centerline{D.~Casper}
\centerline{\it University of California, Irvine; Irvine, California, USA}
\medskip

\centerline{E.~Paschos}
\centerline{\it University of Dortmund, Dortmund, Germany}
\medskip

\centerline{D. Boehnlein, D.~Harris, M.~Kostin, J.G.~Morf\'{\i}n$^*$, P.~Shanahan}
\centerline{\it Fermi National Accelerator Laboratory; Batavia, Illinois, USA}
\medskip

\centerline{M.E.~Christy, W.~Hinton, C.E.~Keppel$^1$}
\centerline{\it Hampton University; Hampton, Virginia, USA}
\medskip

\centerline{R.~Burnstein, A.~Chakravorty$^2$, O.~Kamaev, N.~Solomey}
\centerline{\it Illinois Institute of Technology; Chicago, Illinois, USA}
\medskip

\centerline{S.~Kulagin}
\centerline{\it Institute for Nuclear Research, Moscow, Russia}
\medskip

\centerline{W.K.~Brooks, A.~Bruell, R.~Ent, D.~Gaskell, W.~Melnitchouk, S.A.~Wood}
\centerline{\it Thomas Jefferson National Accelerator Facility; Newport News, Virginia, USA}
\medskip

\centerline{I.~Niculescu, G.~Niculescu}
\centerline{\it James Madison University, Harrisonburg, Virginia, USA}
\medskip

\centerline{G.~Blazey, M.A.C.~Cummings, V.~Rykalin}
\centerline{\it Northern Illinois University; DeKalb, Illinois, USA}
\medskip

\centerline{S.~Boyd, D.~Naples, V.~Paolone}
\centerline{\it University of Pittsburgh; Pittsburgh, Pennsylvania, USA}
\medskip

\centerline{A.~Bodek, H.~Budd, J.~Chvojka, P.~deBarbaro, S.~Manly, K.~McFarland$^*$,
I.~Park, W.~Sakumoto, R.~Teng}
\centerline{\it University of Rochester; Rochester, New York, USA}
\medskip

\centerline{R.~Gilman, C. ~Glashausser, X.~Jiang, G.~Kumbartzki, K.~McCormick,
R.~Ransome}
\centerline{\it Rutgers, The State University of New Jersey; Piscataway, New Jersey, USA}
\medskip

\centerline{H.~Gallagher, T.~Kafka, W.A.~Mann, W.~Oliver}
\centerline{\it Tufts University; Boston, Massachusetts, USA}

\vspace*{0.3in}
\noindent
$^1$ Also at Thomas Jefferson National Accelerator Facility \\
$^2$ Also at Saint Xavier College, Chicago, Illinois, USA \\
$^*$ Co-spokespersons

%% file: execSummary.tex
\section{Executive Summary}

The imminent completion of the NuMI beamline, which will be the
highest intensity neutrino beamline in the world for many years after
its completion, offers the particle and nuclear physics community a
new opportunity.  By constructing a fully active neutrino detector to
run for the first time in a high rate neutrino beam, the \minerva\
experiment, a collaboration between the high energy physics community
already working at Fermilab and groups of new users from the medium
energy nuclear physics community, proposes to exploit this opportunity
to access a broad and rich program in neutrino scattering physics.

\minerva\ will be able to complete a physics program of high rate
studies of exclusive final states in neutrino scattering, as described in
Chapters~\ref{sect:quasielastic}--\ref{sect:coherent}, of elucidation
of the connection between pQCD and QCD in non-perturbative regime, as
described in Chapter~\ref{sect:duality}, and of studies of the axial
current in the elastic (Chapter~\ref{sect:quasielastic}), DIS
(Chapter~\ref{sect:duality}) and off-forward (Chapter~\ref{sect:gpd})
regimes, as well as inside the nucleus (Chapter~\ref{sect:nuclear}).
\minerva\ then seeks the application of its data to aid present and
future neutrino oscillation experiments
(Chapter~\ref{sect:oscillation}), where understanding the details of
neutrino cross-sections and final states is essential for separating
backgrounds to oscillation from signal.

\minerva\ can address all these topics, and can bring a new physics
focus to the Fermilab program with a simple, low-risk detector of
modest cost, as detailed in Chapters~\ref{sect:numihall} and
\ref{sect:detector}--\ref{sect:costs}.  The performance of this
detector is expected to be excellent for resolving individual final
states as well as measuring kinematics in inclusive reactions as
documented in Chapter~\ref{sect:MC}.

As we submit this proposal to Fermilab, we are also preparing to
request funding from sources outside Fermilab to pursue this
interdisciplinary experiment at the intersection of particle and
nuclear physics.  We request that the lab and its Physics Advisory
Committee support this physics and our efforts to seek outside funding
by granting stage one approval to \minerva\ in time for this approval
to enter into funding deliberations this spring and summer.

%% file: overview.tex
\section{Overview of the \minerva\ Experiment}

Upcoming neutrino oscillation experiments in the United States, Europe and
Japan are driving the construction of new, very intense neutrino beamlines
required to achieve reasonable event rates at detectors located hundreds of
kilometers away. These new beamlines will allow us to initiate a vigorous
research program at a detector, located close to the production target, where
event rates are much higher than at the previous generation of neutrino beam
facilities. In addition, it is neutrino oscillation experiments, with their
low-energy neutrinos and massive nuclear targets, which highlight the need for
much improved knowledge of low-energy neutrino--Nucleus interactions, the
overall goal of this experiment.

At Fermilab, the new neutrino facility \numi, designed for the MINOS
neutrino oscillation experiment, will be based on the Main Injector (MI)
accelerator. The neutrino beams from the MI yield several orders of magnitude
more events per kg of detector per year of exposure than the higher-energy
Tevatron neutrino beam.  This highlights the major improvement of this next
generation of neutrino experiments.  One can now perform
statistically-significant experiments with much lighter targets than the
massive iron, marble and other high-A detector materials used in the past. 
That these facilities are designed to study neutrino oscillations points out
the second advantage of these neutrino experiments: An excellent knowledge of
the neutrino beam will be required to reduce the beam-associated systematic
uncertainties of the oscillation result.  This knowledge of the neutrino
spectrum will also reduce the beam systematics in the measurement of
neutrino-scattering phenomena.

To take advantage of these major improvements in experimental neutrino physics
possible with the \numi\ beam and facility, a collaboration of elementary
particle and nuclear physics groups and institutions named ``\minerva" (Main
INjector ExpeRiment: $\nu$--A) has been formed.  This collaboration represents
the combined efforts of two earlier groups that submitted Expressions of
Interest (EOI)~\cite{EOI} to the Fermilab PAC in December, 2002.  The goal of
the \minerva\ experiment is to perform a high-statistics neutrino-nucleus
scattering experiment using a fine-grained detector located on-axis, upstream
of the MINOS near detector. 

\subsection{The Fermilab \numi\ Facility}

The Fermilab \numi\ on-site facility is made up of the beamline
components, the underground facilities to contain these components and a
large, on-site experimental detector hall to contain the MINOS near detector,
located just over 1~km downstream of the target and $\sim$~100~meters
underground.  

\subsubsection{The \numi\ near experimental hall} 

This experimental hall is being constructed and completely outfitted for the
MINOS near detector. The hall is 45~m long, 9.5~m wide and 9.6~m high.
There is a space upstream of the MINOS near detector amounting to, roughly, a
cylindrical volume 26~m long and 3~m in radius for additional detector(s) which, were it
desired, could use the MINOS near detector as an external muon-identifier and
spectrometer.

\subsubsection{The \numi\ neutrino beam}

The neutrino energy distribution of the \numi\ beam can be chosen by changing the
distance of the target and second horn from the first horn, as in a
zoom lens. The energy of the beamline can also be varied, essentially
continuously, by simply changing the target's distance from the first horn and
leaving the second horn in a fixed position. There is a loss of event
rate with this procedure compared to also moving the second horn, and the most
efficient energy tunes will always require moving the second horn. However,
moving the target and second horn involves considerably more time and expense
then simply moving the target. It is now expected that the Main Injector will
deliver $2.5 \times 10^{20}$~POT/year at the start of MINOS running, and will ramp up
to to higher proton intensities if the required funds can be obtained.  The charged-current
neutrino event rates per ton (of detector) per year at startup of MINOS would
then range from just under 200~K to over 1200~K depending on the position of
the target.

For the MINOS experiment the beamline will be operating mainly in its lowest
possible neutrino energy configuration to probe the desired low values
of $\Delta m^2$.  However, to minimize systematics, there will also be running
in higher-energy configurations that will significantly increase the event rates and
kinematic reach of \minerva.

The $\nue$ content of the low-energy beam is estimated at just over 1\% of the
flux.  An important function of \minerva\ will be to provide a far more
accurate measurement of the $\nue$ flux and energy spectrum within the \numi\
beam than is possible with the much coarser MINOS near detector.  This
important figure-of-merit is needed for the design of next-generation
neutrino-oscillation experiments using the \numi\ beam, as well as $\nue$
studies in the MINOS experiment.

\subsection{Neutrino Scattering Physics}

A neutrino scattering experiment in the \numi\ near experimental hall offers
a unique opportunity to study a broad spectrum of physics topics with
measurement precision heretofore unachievable.   Several of these topics have
not yet been studied in any systematic, dedicated way. For other topics, the few results that do
exist are compromised by large statistical and systematic errors. Topics
particularly open to rapid progress upon exposure of \minerva\ in the
\numi\ beam include:

\begin{itemize}

\item  Precision measurement of the quasi-elastic neutrino--nucleus
       cross-section, including its $E_{\nu}$ and $q^{2}$ dependence, and
       study of the nucleon axial form factors.

\item   Determination of single- and double-pion production cross-sections in
	the resonance production region for both neutral-current and
	charged-current interactions, including a study of isospin amplitudes,
	measurement of pion angular distributions, isolation of dominant form
	factors, and measurement of the effective axial-vector mass.

\item   Clarification of the W ($\equiv$ mass of the hadronic system)
	transition region wherein resonance production merges with
	deep-inelastic scattering, including tests of phenomenological
	characterizations of this transition such as quark/hadron duality.

\item   Precision measurement of coherent single-pion production
	cross-sections, with particular attention to target A dependence. 
	Coherent $\pi^{0}$ production, especially via neutral-currents, is a 
	significant background for next-generation neutrino oscillation
	experiments seeking to observe $\nu_{\mu} \rightarrow \nu_{e}$
	oscillation.

\item   Examination of nuclear effects in neutrino-induced interactions
	including  energy loss and  final-state modifications in heavy 
	nuclei. With sufficient $\nubar$ running, a study of quark 
	flavor-dependent nuclear effects can also be performed.

\item   Clarification of the role of nuclear effects as they influence 
        the determination of $\stw$ via measurement of the
        ratio of neutral-current to charged-current cross-sections off different nuclei.
	
\item   With sufficient $\nubar$ running, much-improved measurement of the
	parton distribution functions will be possible using a
	measurement of all six $\nu$ and $\nubar$ structure functions.

\item   Examination of the leading exponential contributions of perturbative
QCD. 

\item   Precision measurement of exclusive strange-particle production channels near threshold,
        thereby improving 
	knowledge of backgrounds in nucleon-decay searches, 
	determination of $V_{us}$, and enabling searches for 
	strangeness-changing neutral-currents and candidate pentaquark 
	resonances. Measurement of hyperon-production cross-sections, including 
	hyperon polarization, is feasible with exposure of \minerva\ to
	$\nubar$ beams.

\item   Improved determination of the effective charm-quark mass ($m_c$)
        near threshold, and new measurements of $V_{cd}$, $s(x)$ and, 
        independently, $\sbar(x)$.
	
\item   Studies of nuclear physics for which neutrino reactions provide
        information complementary to JLab
        studies in the same kinematic range. 
\end{itemize}

In addition to being significant fields of study in their own right, {\bf
improved knowledge of many of these topics is essential to minimizing
systematic uncertainties in neutrino-oscillation experiments}. 

\subsubsection{Low-energy neutrino cross-sections}

   This is a topic of considerable importance to both present and proposed
future (off-axis) neutrino oscillation experiments.  Available measurements of
both total and exclusive cross-sections from early experiments at ANL, BNL,
CERN and FNAL
 all have considerable uncertainties due to low statistics and
large systematic errors, including poor knowledge of the incoming neutrino
flux\cite{Vovenko}. A working group\cite{Gallagher} to assemble all
available data on $\nu$ and $\nubar$ cross-sections and to determine
quantitative requirements for new experiments has been established by members
of this collaboration.  \minerva\ will be able to measure these cross-sections
with negligible statistical errors and with the well-controlled beam systematic
errors needed for the MINOS experiment

\subsubsection{Quasi-elastic scattering}

    Charged-current quasi-elastic reactions play a crucial role in both 
non-accel\-er\-a\-tor and accelerator neutrino oscillation studies, and
cross-section uncertainties - often expressed as uncertainty in the value of
the axial-vector mass - are a significant component in error budgets of
these experiments. There have been recent advances in the measurement of the
vector component of elastic scattering from SLAC and Jefferson Lab. Measurement
of the neutrino quasi-elastic channel is the most direct way to improve our
knowledge of the axial-vector component to this channel.  \minerva's ability to
carefully measure $d\sigma/dQ^2$ to high $Q^2$ allows investigation of the
non-dipole component of the axial-vector form factor to an unprecedented
accuracy.  Combining these \minerva\ measurements with present and future
Jefferson Lab data will permit precision extraction of all form factors needed
to improve and test models of the nucleon.  In addition, due to the
well-constrained kinematics of this channel, a careful study of the muon and
proton momentum vectors allows an important probe of nuclear effects.

\subsubsection{Resonances and transition to deep-inelastic scattering}

    Existing data on neutrino resonance-production is insufficient for the task
of specifying the complex overlapping $\Delta$ and $N^{*}$ resonance amplitudes
and related form-factors which characterize the 1--5~GeV $E_{{\nu}}$ regime.
Neutrino Monte-Carlo programs trying to simulate this kinematic region have
used early theoretical predictions by Rein and Sehgal\cite{Rein:1980wg} or
results from electro-production experiments.  Recently Lee and
Sato\cite{LeeSato} have developed a new model for weak production of the
$\Delta$ resonance.  Paschos and collaborators\cite{Paschos} have also
contributed to this effort. It is noteworthy that the theoretical and
experimental picture of the resonance region is far more obscure than the
quasi-elastic and deep-inelastic scattering (DIS) regions which border it and
that much of the relevant MINOS event sample falls inside this poorly-understood resonance
region. 

Recent work at Jefferson Lab\cite{Wood} shows strong support for quark/hadron
duality, which relates the average resonance production cross-section to
the DIS $F_2$ structure function.  How to incorporate this new paradigm into neutrino Monte
Carlos is currently being studied. An analysis by Bodek and
Yang\cite{Bodek2002} offers a very promising procedure for fitting $F_2$ in
the low-$Q^2$/high-$x$ region.  Extrapolating their results through the
resonance region yields values of $F_2$  consistent with duality arguments and
the Jefferson Lab results mentioned above.  The resonance and transition region
will be carefully examined by \minerva.

\subsubsection{Coherent pion production}

    Both charged- and neutral-current coherent production of pions result in a
single forward-going pion with little energy transfer to the target nucleus. 
In the neutral-current case, the single forward-going $\pi^0$ can mimic an
electron and be misinterpreted as a $\nue$ event.  Existing cross-section
measurements for this reaction are only accurate to $\sim$ 35\% and are only 
available for a limited number of target nuclei.      

\subsubsection{Studying nuclear effects with neutrinos}

   The study of nuclear effects with neutrinos can be broadly divided into two
areas. The first area involves the kinematics of the initial interaction
(spectral function of the struck nucleon within the nucleus and Pauli-excluded
interactions) and the evolution of the hadronic cascade as it proceeds through
the nucleus.  This aspect has direct and important application to the MINOS
neutrino oscillation experiment since it can drastically distort the
initial neutrino energy by mixing final states to such an extent that the
visible energy observed in the detector is much different than the initial
energy.

   The second area involves modification of the structure functions, $F_{i}$ and,
consequently, the cross-section of $\nu$--A scattering compared to $\nu$--nucleon
scattering.  Nuclear effects in DIS have been studied extensively using
muon and electron beams, but only superficially for neutrinos (in
low-statistics bubble-chamber experiments). High-statistics neutrino experiments
have, to date, only been possible using heavy nuclear targets such as
iron-dominated target-calorimeters.  For these experiments, results from e/$\mu$--A
analyses have been applied to the results.  However, there are strong
indications that the nuclear corrections for e/$\mu$--A and $\nu$--A are
different.  Among these differences is growing evidence for quark-flavor
dependent nuclear effects.  A neutrino-scattering program at \numi\ would permit
a systematic, precision study of these effects,
by using a variety of heavy nuclear targets and both $\nu$
and $\nubar$ beams.

\subsubsection{Strangeness and charm production}

  \minerva\ will allow precise measurement of cross-sections for exclusive-channel
strangeness associated-production ($\Delta S = 0$) and 
Cabbibo-suppressed ($\Delta S = 1$) reactions.  Detailed studies of the
hadronic systems  will be carried out, including $q^{2}$ dependence, resonant
structure, and polarization states for produced lambda hyperons.  A detailed
study of coupling strengths and form-factors characterizing the $\Delta
S$ weak hadronic current is envisaged, which will hopefully reawaken efforts at
detailed modelling of these reactions\cite{Shrock}.  \minerva\ observations
of strangeness production near threshold will have ramifications in other areas
of particle physics, as for example with estimation of atmospheric-neutrino
$\Delta S$ backgrounds for nucleon decay searches with megaton-year
exposure.  Searches for new resonant states and new physics will of
course be possible:  we envisage a dedicated search for strangeness-changing
neutral-current reactions and investigation of unusual baryon resonances
such as the recently reported candidate pentaquark state (in $K^{+}n$ and
$K^{0}p$ systems).  Clean measurement of $V_{us}$ should be feasible; it may be
possible to address long-standing discrepancies between theory and experiment
concerning hyperon beta-decay by exploring the related inverse reactions
obtained via $\Delta S = 1$ single-hyperon production by
antineutrinos\cite{solomey}. The production of hyperons by neutrinos and
antineutrinos would provide new information in the form of hyperon polarization
which would reduce ambiguities which currently compromise the analysis of
hyperon beta-decay processes.

Although the neutrino energy spectrum is relatively low for a high-statistics
charm study, it does cover the important threshold region where production
rates are highly dependent on the mass of the charm quark.  Depending on the
value of $m_c$, the expected number of charm events could change by as much as
50\% in this sensitive region.

\subsubsection{Extracting parton distribution functions}

Neutrinos have long been a particularly sensitive probe of nucleon structure. 
One obvious reason is the neutrino's ability to directly resolve the flavor of
the nucleon's constituents: $\nu$ interacts with $d$, $s$, $\ubar$ and $\cbar$
while the $\overline{\nu}$ interacts with $u$, $c$, $\dbar$ and $\sbar$. This
unique ability of the neutrino to ``taste" only particular flavors of quarks
enhances any study of parton distribution functions.  Study of the partonic
structure of the nucleon, using the neutrino's weak probe, would complement the
on-going study of this subject with electromagnetic probes at Jefferson Lab as
well as earlier studies at SLAC, CERN and FNAL.  With the high statistics
foreseen for \minerva, as well as the special attention to minimizing neutrino
beam systematics, it should  be possible for the first time to determine the
separate structure functions $2xF_1^{\nu N}(x,Q^2)$, $ 2xF_1^{\bar \nu
N}(x,Q^2)$, $F_2^{\nu N}(x,Q^2)$, $F_2^{\bar \nu N}(x,Q^2)$, x$F_3^{\nu
N}(x,Q^2)$ and x$F_3^{\bar \nu N}(x,Q^2)$.  This in turn would allow
much-improved knowledge of the individual sea-quark distributions.

\subsection{The \minerva\ Detector}

    To perform the full spectrum of physics outlined in this proposal,
the \minerva\ target/detector must be able to:

\begin{itemize}
\item  Identify muons and measure their momentum with high precision,
\item  identify individual hadrons and $\pi^0$ and measure their momentum,
\item  measure the energy of both hadronic and electromagnetic showers 
	with reasonable precision,
\item  minimize confusion of neutral-current and charged-current 
	event classifications, and
\item  accommodate other nuclear targets.
\end{itemize}

\begin{figure}[b!]
\epsfxsize=\textwidth
\epsfbox{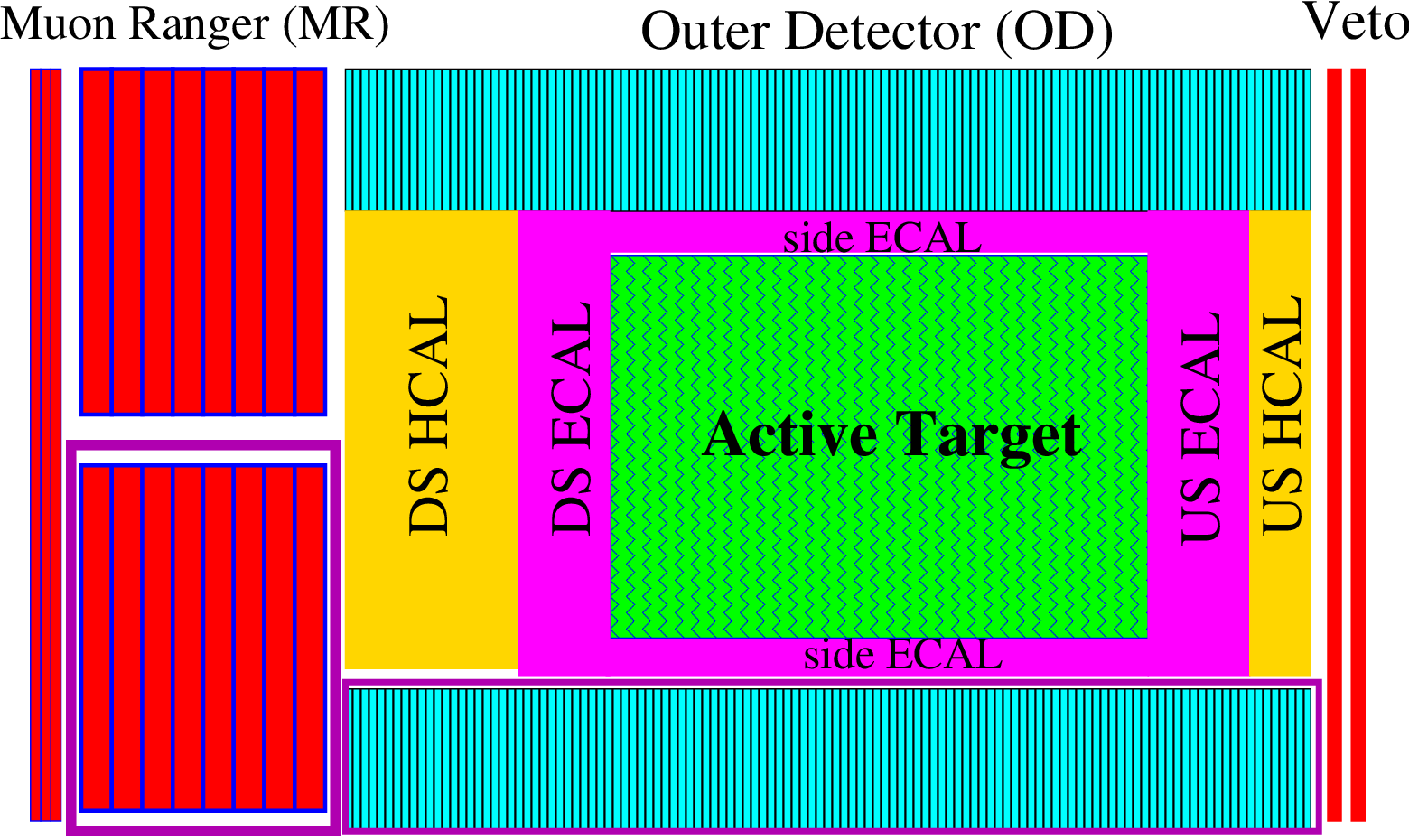}
\caption[Sub-detectors of the \minerva\ detector]{A schematic side
  view of the \minerva\ detector with sub-detectors labeled.  The
  neutrino beam enters from the right.}
\label{fig:sideSchematicOverview}
\end{figure}

These goals can be met by a relatively compact and active target/detector
consisting of a central section of essentially solid scintillator bars
(Figure~\ref{fig:sideSchematicOverview}).  This
central detector is surrounded on all sides by an electromagnetic calorimeter,
a hadronic calorimeter and a magnetized muon-identifier/spectrometer. The
detector has the approximate overall shape of a hexagon (to permit three stereo
views) with a cross-section of 3.55~m minor and 4.10~m major axis. The length
is up to 5.9~m depending on how close \minerva\ can be placed to the MINOS near
detector. The active plastic scintillator volume is 6.1 tons allowing variable
sized fiducial volumes depending on the channel being studied. At the upstream
end of the detector are nuclear targets consisting of 1~ton of Fe and Pb. 
Significant granularity and vertex-reconstruction accuracy can be achieved by
the use of triangular-shaped extruded plastic scintillator(CH) bars with 3.3~cm
base, 1.7~cm height and length up to 4.0~m, with an optical fiber placed in a
groove at the base of the bar for readout.  A second triangular shape with base
1.65~cm and height 1.7~cm (1/2 of the larger triangles) will be used in the
barrel and downstream calorimeter detectors.  Recent work at the Fermilab
Scintillator R\&D Facility has shown that using light division across
triangularly shaped scintillator strips of this size can yield coordinate
resolutions of a few millimeters.  The orientation of the scintillator strips
are alternated so that efficient pattern recognition and tracking can be
performed.

Following the downstream end of the central detector are  electromagnetic and
hadronic calorimeters.  \minerva\ should be placed as close as possible to the
upstream face of the MINOS near detector in order to use that detector's
magnetic field and steel as an external muon-identifier and spectrometer for
the forward-going muons, and as a calorimeter for any hadronic energy exiting
\minerva\ itself.  Moving the \minerva\ detector further upstream from the
MINOS detector will decrease the acceptance for muons in the MINOS detector. 
If necessary a "muon ranger", consisting of 1.2~m of segmented and magnetised
iron, will be added to help identify and measure the momentum of low-energy
muons.  For high-energy muons, the MINOS near detector will provide much better
momentum resolution than the muon ranger.

With this design, even at the lowest beam-energy setting, \minerva\ will
collect more than 580~K events per $2.5 \times 10^{20}$~POT in a 3~ton active
target fiducial volume and just under 200~K events in each of the nuclear
targets.

The statistics from a several-year MINOS run will suffice to study
all the physics topics listed above, although some measurements would be limited
in kinematic reach by the beam energies used for MINOS. 
In addition, all studies involving $\nubar$ channels would be somewhat
limited with the currently-planned MINOS exposure, which includes relatively little
$\nubar$ running.

%% file: phenomenology.tex
\section{Low-Energy Neutrino Scattering Overview}

\subsection{Form Factors and Structure Functions}

 Several formalisms are used to discuss electron-nucleon and
  neutrino-nucleon scattering, and the corresponding reactions on 
  nuclear targets. 
  
  Inclusive lepton scattering can be described in the language of structure
  functions or in terms of form factors for the production of resonant   final
  states. The two descriptions are equivalent and  there are expressions
  relating form-factors to structure functions. In electron scattering, the
  vector form factors can be related to the two structure functions $W_1$ and
  $W_2$ (which are different for neutrons and protons), or equivalently 
  $F_{2}$ and $R$.

   In neutrino scattering, there are three structure functions $W_1$, $W_2$ and
   $W_3$ (or $F_{2}$, $R$ and $xF_{3}$), different for neutrons and protons,
   and containing both vector and axial-vector components. There are also two
   other structure functions (important only at very low energies) whose
   contributions depend on the final-state lepton mass; these can be related to
   the  dominant structure functions within the framework of theoretical
   models.

\subsection{Electron versus Neutrino Scattering}

From the conservation of the vector current (CVC), the vector structure
functions (or form-factors) measured in  electron scattering can be related to
their counterparts in neutrino scattering for specific isospin final states. 
For elastic scattering from spin-$1 \over 2$ quarks or nucleons, these
relationships between vector form factors are simple. For production of higher
spin resonances, the relations are more complicated and involve Clebsch-Gordon
coefficients.

In contrast, the axial structure functions in neutrino scattering cannot be
related to those from electron scattering, except in certain limiting cases
(for example, within the quark/parton model at high energies with V=A). At low
$Q^2$, the axial and vector form factors are different, e.g. because of the
different interactions with the pion cloud around the nucleon.

Another difference arises from nuclear effects in inclusive neutrino vs.
electron scattering.  Nuclear effects on  the axial and vector components of
the cross-section can differ due to shadowing, and can also affect valence and
sea quarks differently.

\subsection {Sum Rules and Constraints}

Several theoretical constraints
and sum rules can be
tested in electron and neutrino reactions (or
applied in the analysis of data).
Some of the sum rules and constraints are valid at all
values of $Q^2$, and some are valid only in certain limits.

The Adler sum rules apply separately to the axial and vector parts of 
$W_{1}$,  $W_{2}$,  and $W_{3}$ and are valid for all values of $Q^2$ (since
they are based on current algebra considerations). At high $Q^2$, these sum 
rules are equivalent to the statement that the number of u valence quarks in
the proton minus the number of d  valence quarks is equal to 1.

Other sum rules, such as the
momentum sum rule (sum of the momentum
carried by quarks and gluons is 1) and
the Gross/Llewelyn-Smith sum rule (number of valence quarks
is equal to 3), have QCD corrections and break down
at very low $Q^2$.  

As $Q^2 \rightarrow 0$, 
the vector structure functions are further constrained by the
measured photoproduction cross-section. Conversely, as $Q^2 \rightarrow \infty$
it is expected that the structure functions
are described by QCD and satisfy QCD sum rules.

\subsection{Final States}

Quasi-elastic\footnote{We should clarify  that the neutrino community uses
the term `quasi-elastic' to describe a charged-current
process in which a neutrino interacts with a nucleon
to produce a charged lepton
in the final state. The nucleon can be a free nucleon
or a nucleon bound in the nucleus. The term `quasi-elastic' refers
to the fact that the initial state neutrino changes
into a different lepton, and there is a single recoil
nucleon in the final state (which changes its
charge state). In contrast, the electron scattering community
refers to electron-nucleon scattering with
a single recoil nucleon as `elastic' scattering.
The term `quasi-elastic' scattering is used by the electron
scattering community to describe elastic electron-nucleon scattering
from bound nucleons in a nucleus.  Here the term `quasi-elastic'
refers to the fact that the bound nucleon is quasi-free. Both
nomenclatures are used in the literature.}
reactions,
resonance production, and deep-inelastic scattering are all important components of 
neutrino scattering
at low energies.

To describe specific final states, one can use the 
language of structure functions, combined with fragmentation functions,
at high values of $Q^2$. At low values of $Q^2$, many experiments describe
the cross-sections for specific exclusive final states.   Both of
these pictures need to be modified when the scattering takes
place on a complex nucleus.

\begin{figure*}[htb]
\centerline{\psfig{figure=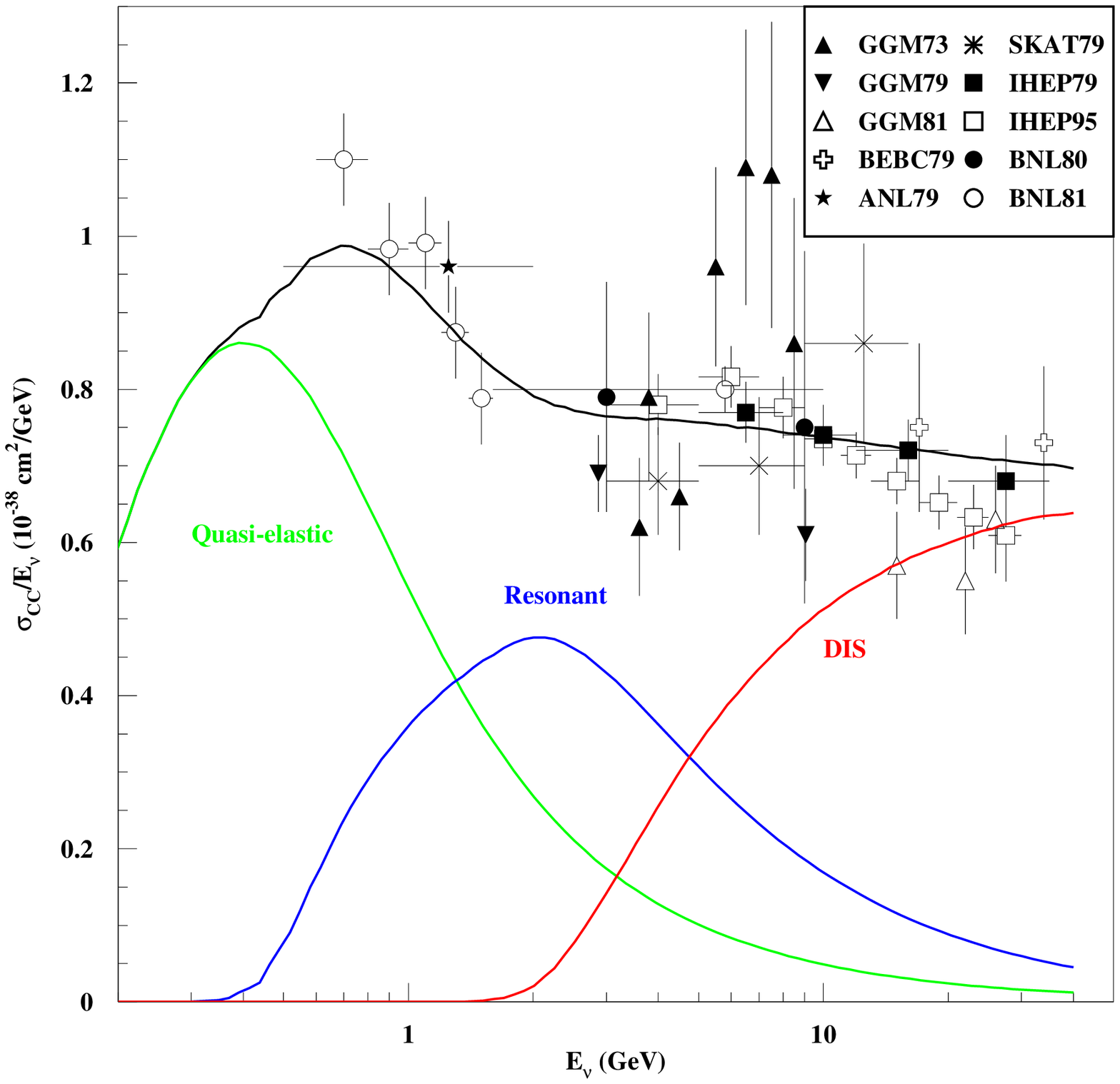,width=2.9in}}
\centerline{\psfig{figure=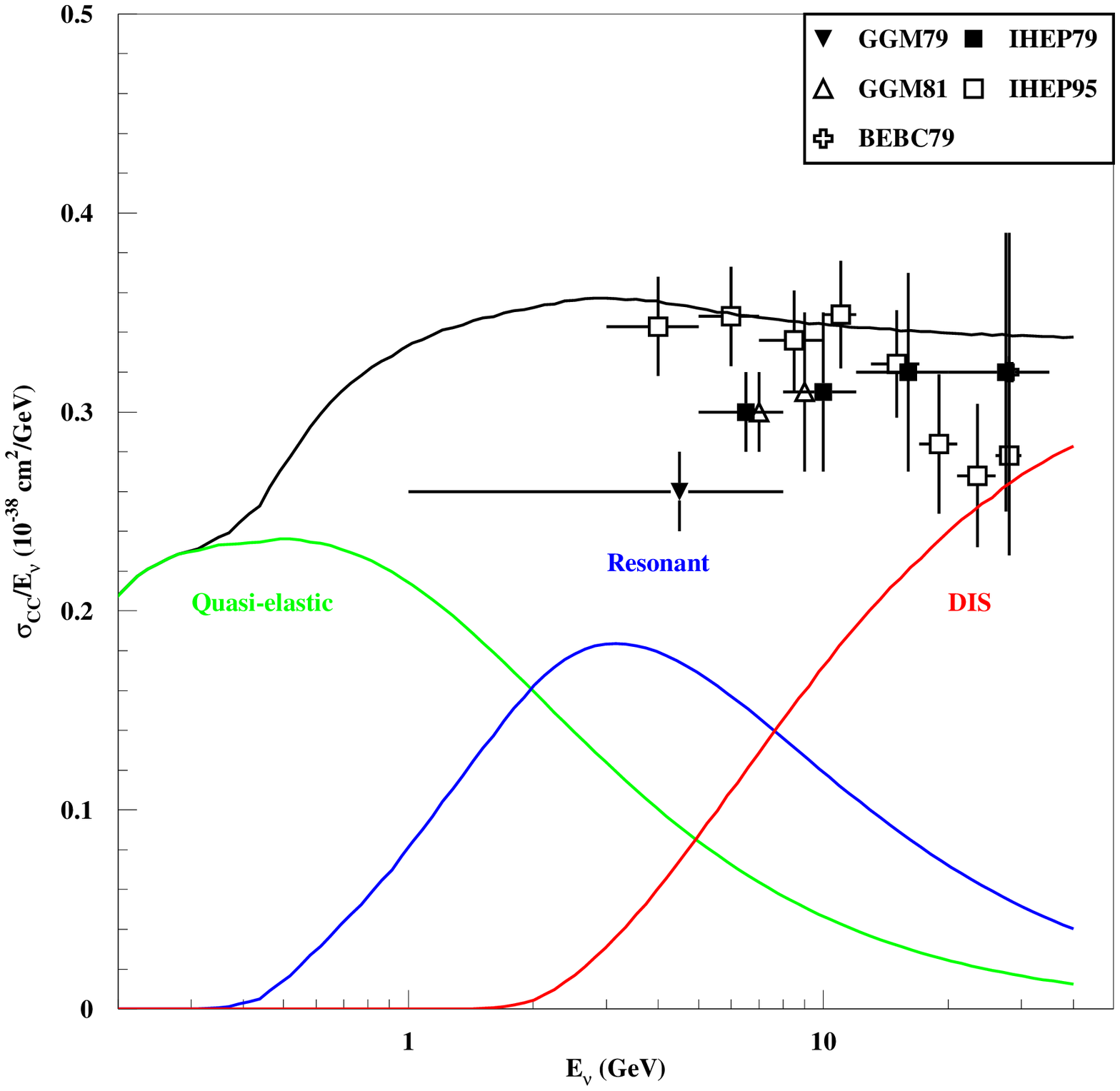,width=2.9in}}
    \caption[Total neutrino and anti-neutrino cross-sections] {Total neutrino
(top) and anti-neutrino (bottom)  cross-sections divided by energy versus
energy compared to the sum of quasi-elastic, resonant, and inelastic
contributions from the NUANCE model. The sum is constructed  to be continuous
in $W$ ($\equiv$ mass of the hadronic system) as follows. For $W>2~\hbox{GeV}$ 
the Bodek-Yang model is  used.  The Rein-Sehgal model is used for
$W<2~\hbox{GeV}$. In addition, a fraction of the Bodek-Yang cross-section is
added to the Rein-Sehgal cross-section between $W=1.7~\hbox{GeV}$ and
$W=2~\hbox{GeV}$. The fraction increases linearly with $W$ from 0 to 0.38
between $W=1.7$ and $W =2~\hbox{GeV}$.}
\label{fig:total}
\end{figure*}

Figure~\ref{fig:total} shows the total
neutrino and anti-neutrino cross-sections (per nucleon for an isoscalar target)
    versus energy (at low energies) compared
    to the sum of quasi-elastic, resonant,
    and inelastic contributions.
These two figures also show the various contributions
to the neutrino and anti-neutrino total
cross-sections that will be investigated
in this experiment.

%% file: nudata.tex
\section{Existing Neutrino Scattering Data}
\label{sec:nudata}

Neutrino experiments dating back to the 1960's have played an important role 
in particle physics, including discovery of neutral currents, electroweak measurements,
  determination of the flavor composition of the nucleon,  
measurements of the weak hadronic current, and QCD studies based on 
scaling violations in structure and fragmentation functions. 

In the 1--10~GeV energy range of interest to the current and future generation
of neutrino oscillation studies, relevant data comes from bubble-chamber
experiments that ran from the 1960's through the 1980's.  Gargamelle, the 
12-foot bubble chamber at the Argonne ZGS, the 7-foot bubble chamber at the AGS at Brookhaven, 
the Big European Bubble Chamber (BEBC) at CERN, the Serpukhov bubble chamber SKAT, and the FNAL 15-foot 
bubble chamber studied neutrino and anti-neutrino interactions
off free nucleons and heavy liquid targets.  
Spark-chamber and emulsion experiments from this era played a less prominent
role but did make crucial measurements in a number of areas.

Despite limited statistics, the excellent imaging capabilities of bubble chambers made
a wide range of physics topics accessible.  It is primarily this data that is used
to tune our Monte Carlo simulations and provides the basis for our present understanding of low-energy
neutrino cross-sections.   While adequate
for validating the models at some level, most of the bubble-chamber data-sets 
are limited in size and do not cover the full range of neutrino energy, 
nuclear targets and neutrino species ($\nu$/$\overline{\nu}$) required for a complete 
understanding of neutrino interactions.    
Some of the main topics of interest for experiments of this era are described below.  For each topic,
an approximate count of the number of SPIRES publication references 
is included.

\subsection{Quasi-elastic Scattering}

{\bf (8 pubs)}
Studies of quasi-elastic charged-current (CC) interactions were among the first results
from bubble-chamber neutrino exposures, and are the primary tool for studying 
the axial component of the weak nucleon current.  While data were taken on both 
light (H$_2$/D$_2$) and heavy (Neon/propane/freon) targets, no attempts were
made to extract measurements related to the nuclear system.  Rather, the
nuclear system was treated as
a complication requiring corrections.  In many instances even this 
correction was not done, and the published data are for interactions on nucleons 
in a particular nucleus.  This helps account for the large spread in data points
between different experiments in Figure~\ref{elas_JhaKJhaJ_nu}.

\subsection{Other Exclusive Charged-current Channels}

\begin{figure}[b]
\center
\epsfxsize=100mm\leavevmode
\epsffile{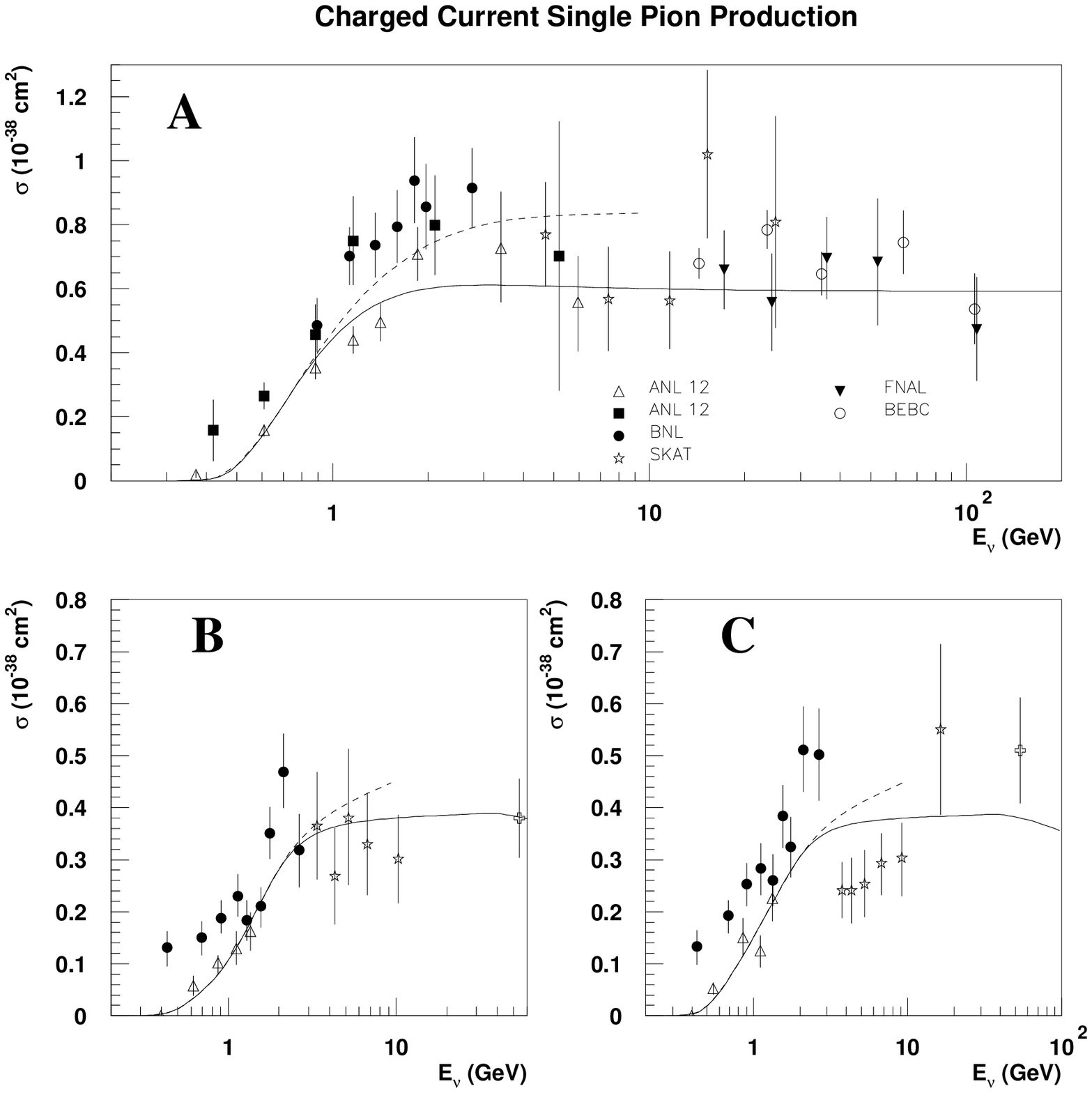}
\caption[Cross-sections for charged-current single-pion production]{Cross-sections for charged-current
single-pion production.  
Plot A:  $\nu_\mu + p \rightarrow
\mu^- + p + \pi^+$, Plot B:  $\nu_\mu + n \rightarrow \mu^- + n + \pi^+$, 
Plot C:  $\nu_\mu + n \rightarrow
\mu^- + p + \pi^0$.  Solid lines are the NEUGEN predictions
for W$<$1.4 GeV (plot A) and W$<$2.0 GeV (plots B and C).  The dashed curve 
is the NEUGEN prediction with no
invariant mass cut, for comparison with the BNL data.  Data are from:  ANL 
\cite{ANLpi1,Barish:1979pj}, BNL \cite{BNL1pi},
FNAL \cite{FNALpi}, BEBC \cite{BEBC1pi}}
\label{fig:modes}
\end{figure}

{\bf  (19 pubs)}
Total cross-section measurements and studies of differential distributions were
made for both light and heavy targets in each of the three charged-current single-pion channels. In nearly all cases cuts were placed on the hadronic invariant mass 
({\it e.g.} $W < 1.4~\hbox{GeV}/c^2$) 
to limit the analysis to the resonant region. 
The results are shown in Figure~\ref{fig:modes}. 
Fewer experiments published cross-sections for two- and three-pion channels. 

\subsection{Neutral-current Measurements}

{\bf  (22 pubs)}  
Neutral-current (NC) measurements fall into three categories:  elastic measurements in 
dedicated experiments, single-pion exclusive final-state measurements, or NC/CC 
ratio measurements in the deep-inelastic scattering (DIS) regime.  NC/CC ratio measurements were 
made at high energies and applied cuts on the energy transfer $\nu$ to isolate
the DIS regime.  Single-pion studies of the NC electroweak
couplings and the isospin characteristics of the hadronic current in the 
resonance region suffered from lack
of statistics.  Table~\ref{tab:nc} summarizes the published 
data.  These processes are of particular interest,
as they constitute one of the primary backgrounds
to $\nu_e$ appearance in oscillation experiments.   

\subsection{Hadronic Final States} 

{\bf (32 pubs)}
A number of publications were devoted to inclusive measurements of the hadronic 
system produced in neutrino interactions.  Multiplicity measurements, 
transverse momentum distributions, inclusive particle production, 
fragmentation functions, and evaluation of the universality of hadron
dynamics were studied.  In this area, hadronic mass cuts ({\it e.g.} $W > 2~\hbox{GeV}$)
were applied to limit the analysis to the DIS region. 

\subsection{Strange and Charmed Particle Production}

{\bf (27 pubs)}
Because of their clear signatures in photographic quality bubble 
chambers, exclusive and inclusive measurements of strange and charm
particle production were popular topics.  A survey of these
results is given in Section~\ref{sect:strangeCharmData}.

\subsection{Total Cross-sections} 

\begin{figure}[!htb]
\centerline{\epsfig{file=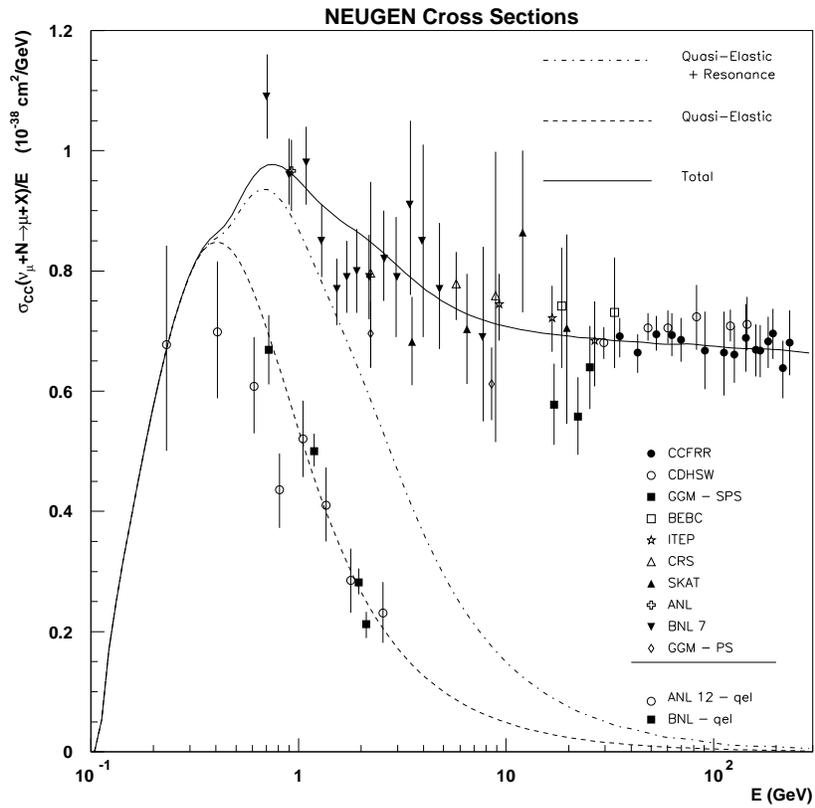,width=12.0cm}}
\caption[Existing charged-current $\nu_\mu$ cross-section data]{The NEUGEN prediction for the $\nu_\mu$  charged-current cross-section ($\sigma/E_\nu$) from an 
isoscalar target compared with data from a number of experiments.  Quasi-elastic and resonance  
contributions are also shown.
Data are from: CCFRR \cite{MacFarlane:1984ax}, CDHSW \cite{Berge:1987zw}, 
GGM-SPS \cite{Morfin:1981kg}, BEBC \cite{Colley:1979rt},
ITEP \cite{Mukhin:1979bd}, SKAT \cite{Baranov:1979sx}, CRS \cite{Baltay:1980pr}, 
ANL \cite{Barish:1979pj}, BNL \cite{Baker:1982ty}, GGM-PS 
\cite{Ciampolillo:1979wp},
ANL-QEL \cite{Barish:1977qk}, BNL-QEL \cite{Baker:1981su}.  }
\label{fig:neugen}
\end{figure}

{\bf (19 pubs)}
Total charged-current cross-section measurements were a staple of 
bubble-chamber experiments.  Their data is shown in Figure~\ref{fig:neugen}. The
large errors are due to a combination of low statistics and poor knowledge of
the $\nu$ beam.

\subsection{Structure Functions}

{\bf (18 pubs)}
Numerous experiments, particulary those at higher energies (and 
of course all the large calorimetric neutrino detectors like CDHS, CCFR,
NuTeV, etc that followed) measured structure functions. 
Neutrino experiments are complementary to studies
with electron and muon beams as they allow extraction of the 
valence quark distributions through measurement of $xF_3$ as well 
as independent analysis of the strange quark content via di-muon
production.  These experiments made possible precision electroweak 
and QCD measurements with the NC/CC ratio and scaling violation in 
the structure functions.  

\subsection{Summary}

Viewed from a historical perspective, the results 
from these experiments clearly reflected the topics of interest (and 
the theoretical tools available) {\em at the time they were 
performed}, and some general trends are clear.  
These experiments focused on two regimes.  First, low $Q^2$
scattering: the non-perturbative regime where the scattering takes
place from a single nucleon.  By measuring total and 
differential cross-sections for exclusive channels (like quasi-elastic and $\Delta$
production), these experiments studied in detail the weak hadronic current of the nucleon.
Parton-model studies form a second, complementary class of experiments, studying scaling phemonena like 
total cross-sections, structure functions, scaling-variable distributions, and 
inclusive final-state dynamics, and applying kinematic cuts to remove resonant and 
quasi-elastic contributions. 

This dichotomy reflects the fact that decent models only existed for the
extreme perturbative and non-perturbative limiting cases.
The resonant/DIS transition region, where perturbative QCD breaks down, was avoided because a clear
theoretical framework for it was not available.
With the current generation of duality studies at JLab and elsewhere, 
this complex but fundamental region is just now being fruitfully probed.

Another area of difficulty was treatment of nuclear effects.
While heavy targets gave bubble chambers increased
target mass, the confounding effects of the nuclear environment on the target kinematics and observed final states
were a topic which was largely ignored.  Very few nuclear physics studies were ever
carried out with neutrinos, and these had only the most na\"ive models available for
comparison.  These studies focused on nuclear rescattering
of produced pions, shadowing and EMC effects, formation-zone studies, and
inclusive production of slow particles.   Neither the small samples nor the models available
allowed neutrinos to probe the nuclear environment in detail.

These ``holes" in existing neutrino data and related phenomenology are now becoming increasingly evident. The MINOS experiment, for instance, will see a wide-band beam of 1--50~GeV neutrinos.  Since a significant fraction
of the interactions in MINOS are in the ``transition" region, and nearly all take place on
Iron nuclei, the areas of study neglected during the bubble-chamber era begin to loom large.  MINOS, and
the neutrino-oscillation experiments that will follow it, will be forced to confront them to achieve maximum sensitivity.  Part~\ref{part:physics} of this proposal explains in detail how \minerva\ will not only address fundamental topics in nuclear and neutrino physics, which are compelling in their own right, but also substantially improve the quality of results from future oscillation experiments.

\begin{table}[tbp]
\begin{center}
\begin{tabular}{|l|r|r|c|r|c|r|}
\hline
 Experiment     & Year   &       Reaction                        & Measurement                      &       Events   &           Ref         \\
\hline
Gargamelle      &  1977  &   $\nu/\overline{\nu}$ - propane/freon &  semi-inclusive                 &  $\nu$: 1061  & \cite{Kluttig:1977gb} \\  
                &  1977  &   $\nu/\overline{\nu}$ - propane/freon &  $\pi$ production               &  $\overline{\nu}$: 1200  &                \\  \hline  
Gargamelle	        &  1978  &   $\overline{\nu}$-propane/freon       &  $\overline{\nu}(\pi^o)$         &          139  & \cite{Erriquez:1978yc} \\ 
	        &  1978  &   $\overline{\nu}$-propane/freon       &  $\overline{\nu}(\pi^-)$         &           73  & \\  \hline

Gargamelle                &  1978  &   $\nu$-propane/freon                  &  $\nu$ p $\rightarrow \nu$ p $\pi^o$& 240        &  \cite{Krenz:1978sw}   \\
                &  1978  &   $\nu$-propane/freon                  &  $\nu$ p $\rightarrow \nu$ n $\pi^+$& 104        & \\
                &  1978  &   $\nu$-propane/freon                  &  $\nu$ n $\rightarrow \nu$ n $\pi^o$&  31        & \\
                &  1978  &   $\nu$-propane/freon                  &  $\nu$ n $\rightarrow \nu$ p $\pi^-$&  94        & \\  \hline
 
Gargamelle      &  1979  &   $\nu/\overline{\nu}$ - propane/freon & $\nu(1\pi^o)$                    &    178      & \cite{Pohl:1979ya}  \\
                &  1979  &   $\nu/\overline{\nu}$ - propane/freon & $\overline{\nu}(1\pi^o)$         &    139      &   \\ \hline

BNL - Counter   &  1977  &   $\nu/\overline{\nu}$ - Al/C          & $\nu(1\pi^o)$                   &    204      & \cite{Lee:1977wr} \\
                &  1977  &   $\nu/\overline{\nu}$ - Al/C          & $\overline{\nu}(1\pi^o)$                   &    22      &  \\ \hline

ANL - 12'       &  1974  &  $\nu$-D$_2$/$\nu$-H$_2$  & $\nu$ p $\rightarrow \nu$ n $\pi^+$           &   8        & \cite{Barish:1974fe} \\
                &  1974  &  $\nu$-D$_2$/$\nu$-H$_2$  & $\nu$ p $\rightarrow \nu$ p $\pi^o$           &   18       &  \\ \hline

ANL - 12'       &  1980  &  $\nu$-D$_2$              & $\nu$ n $\rightarrow \nu$ p $\pi^-$           &   ?        & \cite{Derrick:1980nr} \\ \hline

ANL - 12'       &  1981  &  $\nu$-D$_2$              & $\nu$ n $\rightarrow \nu$ p $\pi^-$           &   ?        & \cite{Derrick:1981xw} \\
                &  1981  &  $\nu$-D$_2$              & $\nu$ p $\rightarrow \nu$ p $\pi^o$           &   8        &  \\
                &  1981  &  $\nu$-D$_2$              & $\nu$ p $\rightarrow \nu$ n $\pi^+$           &  22        &  \\ \hline

BNL 7'          &  1981  &   $\nu$-D$_2$             & $\nu$ n $\rightarrow \nu$ p $\pi^-$                   &    200      & \cite{Baker:1981pj} \\ \hline

\end{tabular}
\end{center}
\caption{Neutral-current measurements
}
\label{tab:nc}
\end{table}

%% file: numiBeam.tex
\section{The \numi\ Beam and \mbox{\minerva\ Event Sample}}

\begin{figure}[h]
\epsfysize=2.0in
\centerline{\epsffile{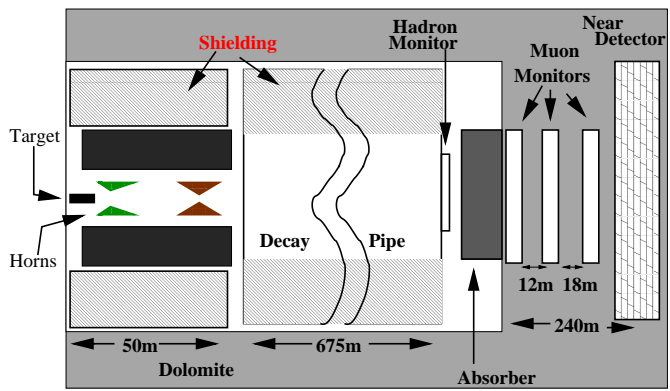}}
\caption{Layout of \numi\ beamline components and near detector hall.}
\label{Sideview}
\end{figure}

The \numi\ neutrino beam is produced from $\pi$- and $K$-decay in a  675~m
decay pipe beginning 50~m downstream of a double horn focusing  system. At the
end of the decay pipe a 10~m long hadron absorber stops the undecayed
secondaries and non-interacting primary protons. Just downstream of the
absorber,  240~m Dolomite is used to range out muons before the $\nu$ beam
enters the near detector hall. Figure~\ref{Sideview} shows the beam component
and near detector hall layout. 

\subsection{Energy Options}

\begin{figure}[hp]
\epsfysize=2.5in
\centerline{
{\includegraphics[width=0.75\linewidth,width=2.5in,angle=270]{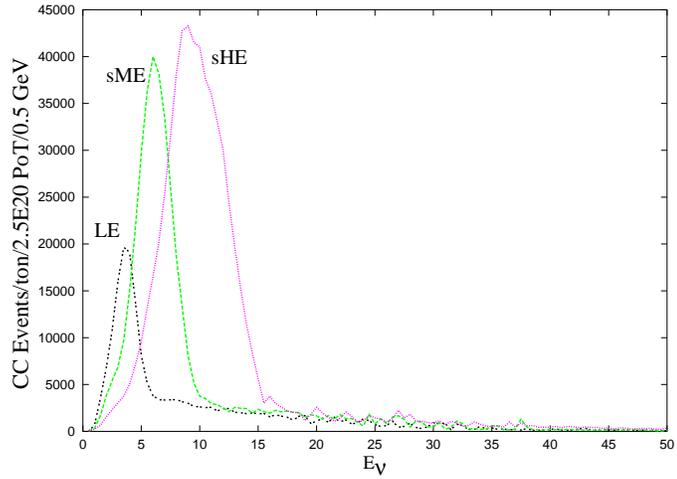}}}
\caption[Charged-current interaction spectra for the LE, sME and sHE beams]{The neutrino charged-current event energy distribution for the three
     configurations of the NuMI beam corresponding to low-energy (LE),
     medium-energy (sME) and high-energy (sHE).}
\label{lemehe}
\end{figure}

\begin{figure}[hp]
\epsfysize=2.5in
\centerline{
{\includegraphics[width=0.75\linewidth,width=2.5in,angle=270]{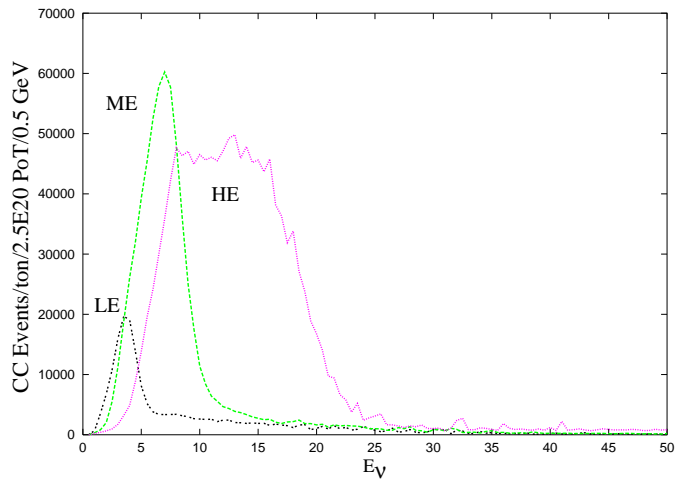}}}
\caption[Charged-current interaction spectra for the ME and HE beams]{The neutrino charged-current event energy distribution for
the high-rate medium and high-energy beam configurations (ME and HE)
which involve movement of the second horn as well as target position.}
\label{mehe}
\end{figure}

The neutrino energy distribution of the NuMI beam can be chosen by changing the
distance of the target and second horn with respect to the first horn, as in a
zoom lens. These three configurations result in three beam energy tunes for the
low (LE), medium (ME), and high (HE) energy ranges respectively. However, to
switch from one the beam mode into  an alternate configuration will require
down time to reconfigure the  target hall and a loss of beam time. An
alternative to this which allows the peak energy to be varied is to change  the
distance of target from the first horn and leave the second horn fixed in the
LE position.  This can be accomplished remotely with maximum transit of -2.5~m
motion of the target upstream of the first horn from its nominal low energy
position.  The configurations corresponding to target -1.0~m from nominal
results in a ``semi-medium'' energy beam tune (sME) and target -2.5~m from
nominal will produce ``semi-high'' energy beam (sHE). These semi-beam
configurations are less efficient and result in lower event rates than the ME
and HE beams.  A considerably more efficient sHE beam is possible with
three-day downtime to allow the target to be moved back to its nominal HE
position of -4.0 m. This more efficient sHE(-4.0) beam would yield over 50\%
more events than the sHE(-2.5) beam.  For the MINOS experiment the beamline
will be operating mainly in its lowest possible neutrino energy configuration
to be able to reach desired low values of $\Delta m^2$.  However, to minimize
systematics, there will also be running in the sME and sHE configurations
described above.  The neutrino energy distributions for the LE, sME, and sHE
running  modes are shown in Figure~\ref{lemehe}.  Figure~\ref{mehe} shows the
event energy distributions  for the ME and HE beam configurations for
comparison. 

\subsection{\minerva\ Event Rates}

Table~\ref{tab:NumiRates} shows the charged-current event rates  per $10^{20}$
protons on target (PoT) per ton for the three beam configurations discussed
above. In addition, the same configurations but with horn-current reversed
provide anti-neutrino beams. Event rates for  $\overline{\nu}_{\mu}$
charged-current events using  anti-neutrino beam configurations (LErev, MErev,
and HErev)  are also shown along with their $\nu_{\mu}$ background components.
Running in these modes would be highly desirable for \minerva\ physics. 

\begin{table}[hp]
\begin{center}
\begin{tabular}{@{}lll}
\hline
\multicolumn{3}{c}{CC Events/$10^{20}$ PoT/ton} \\ \hline \\
Beam  & CC $\nu_{\mu}$ & CC $\nu_{e}$\\ \cline{1-3}
LE & 78~K & 1.1~K \\
sME & 158~K  & 1.8~K \\
sHE & 257~K & 2~K \\ \cline{1-3}
  & CC $\overline{\nu}_{\mu}$ & CC $\nu_{\mu}$\\ \cline{1-3}
LErev & 26~K  & 34~K \\
MErev & 56~K  & 10~K \\
HErev & 75~K & 13~K \\ \cline{1-3}
\end{tabular}
\end{center}
\caption{\minerva\ event rates for different beam configurations.}
\label{tab:NumiRates}
\end{table}

\subsection{Baseline MINOS Run Plan}
\label{sect:runPlan}

Table~\ref{tab:run} shows a scenario  for predicted PoT over a conservative
hypothetical four-year MINOS run. From this table the total integrated
charged-current event samples for a four-year \minerva\ run would be 940~K
$\nu_{\mu}$ charged-current events per ton and 275~K $\overline{\nu}_{\mu}$
charged-current events per ton.

\begin{table}[hp]
\begin{center}
\begin{tabular}{@{}llllllll}
\hline
\multicolumn{8}{c}{Scenario for PoT per year ($\times 10^{20}$)} \\ \hline \\
year  & total PoT & LE & sME & sHE & LErev &  MErev & HErev \\ \cline{1-8}
2006 & 3.0 & 3.0 & 0.0 & 0.0 & 0.0 & 0.0 & 0.0  \\
2007 & 4.0 & 3.0 & 0.7 & 0.3 & 0.0 & 0.0 & 0.0  \\
2008 & 4.0 & 0.0 & 0.0 & 0.0 & 2.5 & 1.0 & 0.5  \\
2009 & 4.0 & 1.0 & 0.5 & 0.5 & 0.5 & 0.5 & 1.0  \\
Total & 15.0 & 7.0 & 1.2 & 0.8 & 3.0 & 1.5 & 1.5 \\ \cline{1-8}
\end{tabular}
\end{center}
\caption[Hypothetical four-year proton luminosity scenario]{Hypothetical proton luminosity scenario for a four-year run.}
\label{tab:run}
\end{table}

\subsection{\minerva\ Data Samples}
\label{sect:numiSample}

The event rates for physics processes of interest to \minerva\
for the four-year scenario discussed in the previous section are summarized 
in the Table~\ref{tab:channelRates}.

\begin{table}[hp]
\begin{center}
\begin{tabular}{@{}lll} 
\hline
\multicolumn{3}{c}{Event Rates per ton} \\ \hline \\
Process & CC  & NC \\ \cline{1-3}
Elastic  & 103~K & 42~K \\
Resonance & 196~K & 70~K \\
Transition & 210~K & 65~K \\
DIS  & 420~K & 125~K \\
Coherent & 8.3~K & 4.2~K \\
Total  & 940~K & 288~K     \\ \cline{1-3}
\end{tabular}
\end{center}
\caption[Total event rates for different reaction channels in four-year run]{Total event rates for different reaction types,
per ton, for the four-year scenario outlined in Table~\ref{tab:run}.}
\label{tab:channelRates}
\end{table}

\begin{figure}[p]
\centerline{{\includegraphics[width=5.0in,height=5.0in]{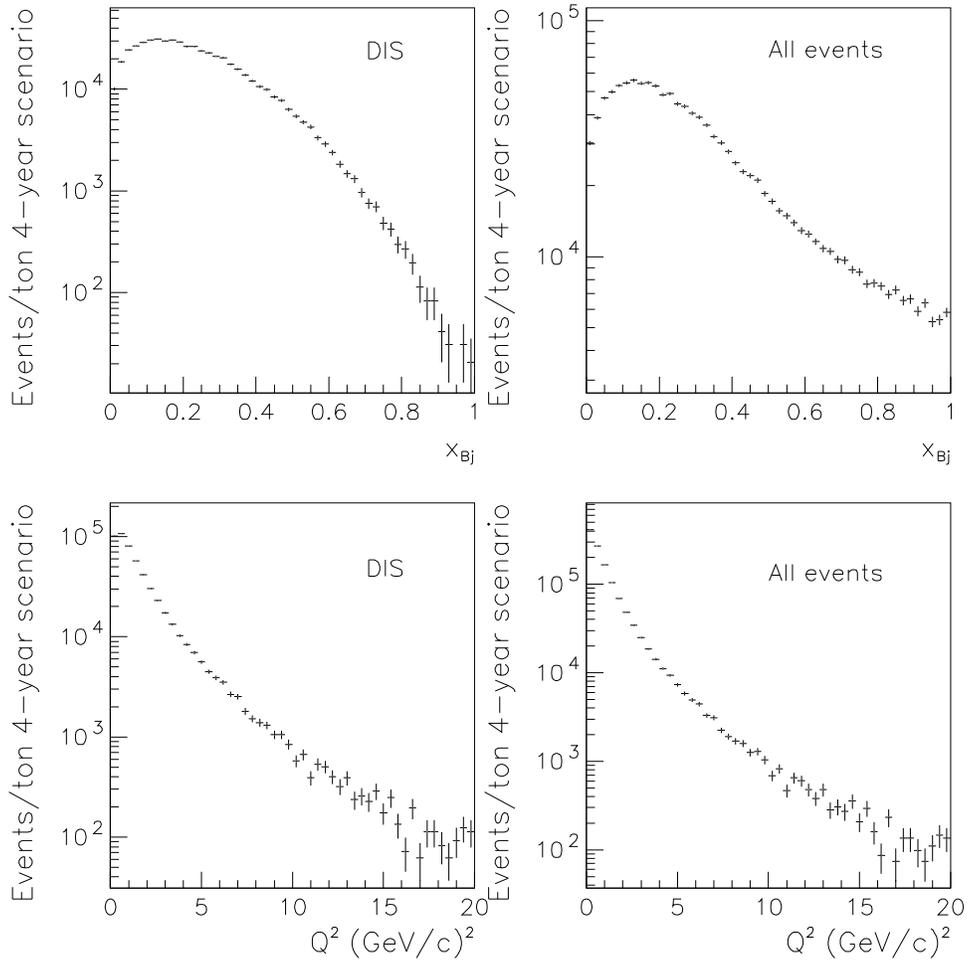}}}
\caption[Expected $x_{Bj}$ and $Q^{2}$ distributions in \minerva]{Kinematic distributions ($x_{Bj}$ and $Q^{2}$) expected for deep-inelastic (top left and bottom left) and all event types (top right and bottom right).}
\label{fig:xq2all}
\end{figure}

The distribution of the number of interactions expected for different $x_{Bj}$
and $Q^{2}$ values are shown for  the quasi-elastic, resonant and
deep-inelastic channels in Tables~\ref{tab:xq2qel},~\ref{tab:xq2res}
and~\ref{tab:xq2dis}, respectively. The spread of the quasi-elastic events in x
is due to the smearing from the Fermi motion of the target nucleon. For clarity
the $x_{Bj}$ and $Q^{2}$ distributions of the total and deep-inelastic event
samples are shown in Figure~\ref{fig:xq2all}. These tables are based on the
four-year scenario outlined in Table~\ref{tab:run}.

The number of interactions expected during
the full four-year exposure of the detector in the NuMI beam eclipses the number of events recorded in the bubble-chamber
experiments described in Section~\ref{sec:nudata} by several orders of magnitude. 
The implications of this unprecedented event sample for physics are described in later sections.
It would not be an exaggeration to observe that this large sample
of neutrino interactions will reduce many of the systematic errors currently limiting the sensitivity of 
neutrino oscillation experiments and allow detailed study of kinematic regions that are
presently rather poorly understood.

\begin{table}[p]
\begin{center}
\begin{tabular}{@{}llllllllll|l}
\hline
\multicolumn{11}{c}{Quasi-elastic events per ton vs. (x$_{Bj}$,$Q^{2}$) for four-year scenario} \\ \hline \\
         &  \multicolumn{10}{c}{$Q^2~\hbox{(GeV/c)}^2$} \\
$x_{Bj}$ & 0.0-0.5 & 0.5-1.0 & 1.0-1.5 & 1.5-2.0 & 2.0-2.5 & 2.5-3.0 & 3.0-3.5 & 3.5-4.0 & 4.0+ & Total \\ \cline{1-11}
0.0-0.1  &    1068 &        0 &        0 &        0 &        0 &        0 &        0 &        0 &        0 &     1068  \\
0.1-0.2  &    2372 &        0 &        0 &        0 &        0 &        0 &        0 &        0 &        0 &     2372  \\
0.2-0.3  &    3366 &       12 &        0 &        0 &        0 &        0 &        0 &        0 &        0 &     3378  \\
0.3-0.4  &    5552 &      199 &       49 &        0 &        0 &        0 &        0 &        0 &        0 &     5800  \\
0.4-0.5  &    7812 &      361 &       49 &        0 &        0 &        0 &        0 &        0 &        0 &     8222  \\
0.5-0.6  &    9974 &     1217 &      173 &       37 &       39 &        0 &        0 &        0 &       12 &    11452  \\
0.6-0.7  &   11377 &     3717 &      956 &      161 &      198 &       37 &        0 &       12 &       12 &    16470  \\
0.7-0.8  &    9663 &     4682 &     2061 &      906 &      427 &      173 &      136 &       12 &       62 &    18122  \\
0.8-0.9  &    8645 &     4744 &     2235 &     1296 &      608 &      322 &      161 &       99 &      248 &    18358  \\
0.9-1.0  &    6868 &     4956 &     2198 &     1246 &      509 &      360 &      223 &       86 &      248 &    16694  \\ \cline{1-11}
Total  &   66697 &    19888 &     7721 &     3646 &     1781 &      892 &      520 &      209 &      582 &   101936  \\ \cline{1-11}
\end{tabular}
\end{center}
\caption[Kinematic distribution of quasi-elastic sample.]{Quasi-elastic interactions 
expected per ton for the four-year scenario of Table~\ref{tab:run}.}
\label{tab:xq2qel}
\end{table}

\begin{table}[p]
\begin{center}
\begin{tabular}{@{}llllllllll|l}
\hline
\multicolumn{11}{c}{Resonant events per ton vs. (x$_{Bj}$,$Q^{2}$) for four-year scenario} \\ \hline \\
         &  \multicolumn{10}{c}{$Q^2~\hbox{(GeV/c)}^2$} \\
$x_{Bj}$ & 0.0-0.5 & 0.5-1.0 & 1.0-1.5 & 1.5-2.0 & 2.0-2.5 & 2.5-3.0 & 3.0-3.5 & 3.5-4.0 & 4.0+ & Total \\ \cline{1-11}
0.0-0.1  &   48169 &       49 &        0 &        0 &        0 &        0 &        0 &        0 &        0 &    48219  \\
0.1-0.2  &   46132 &     7763 &      173 &        0 &        0 &        0 &        0 &        0 &        0 &    54069  \\
0.2-0.3  &   27649 &    15104 &     2881 &      310 &       12 &        0 &        0 &        0 &        0 &    45958  \\
0.3-0.4  &   16135 &    15613 &     6508 &     1689 &      298 &       62 &       12 &       12 &        0 &    40331  \\
0.4-0.5  &    5974 &    13576 &     6359 &     2943 &     1416 &      521 &       86 &       86 &       24 &    30990  \\
0.5-0.6  &    1018 &     8968 &     5924 &     3279 &     1738 &     1018 &      496 &      211 &      236 &    22892  \\
0.6-0.7  &      74 &     4012 &     4571 &     2956 &     1577 &      993 &      434 &      459 &      571 &    15650  \\
0.7-0.8  &      12 &     1217 &     2583 &     1788 &     1217 &      919 &      558 &      496 &      770 &     9564  \\
0.8-0.9  &       0 &      260 &      844 &      745 &      757 &      732 &      472 &      347 &      633 &     4794  \\
0.9-1.0  &       0 &      111 &      347 &      285 &      397 &      310 &      124 &      136 &      534 &     2248  \\ \cline{1-11}
Total  &  145163 &    66673 &    30190 &    13995 &     7412 &     4555 &     2182 &     1747 &     2768 &   274715  \\ \cline{1-11}
\end{tabular}
\end{center}
\caption[Kinematic distribution of resonant sample.]{Resonant interactions
expected per ton for the four-year scenario of Table~\ref{tab:run}.}
\label{tab:xq2res}
\end{table}

\begin{table}[p]
\begin{center}
\begin{tabular}{@{}lllllllllll|l}
\hline
\multicolumn{12}{c}{Deep-inelastic events per ton vs. (x$_{Bj}$,$Q^{2}$) for four-year scenario} \\ \hline \\
         &  \multicolumn{11}{c}{$Q^2~\hbox{(GeV/c)}^2$} \\
$x_{Bj}$ & 0-2     & 2-4      & 4-6      &   6-8    &  8-10    &  10-12   &  12-14   &  14-16   & 16-20    &  20+     & Total \\
0.0-0.1  & 100276  & 1987     &     198  &      24  &       0  &       0  &       0  &       0  &   0      &       0  &   102485  \\
0.1-0.2  &  123988 &    13688 &     2670 &      832 &      310 &       86 &       12 &        0 &       12 &        0 &   141598  \\
0.2-0.3  &   79632 &    24954 &     5738 &     1676 &      956 &      360 &      223 &      111 &       37 &       86 &   113773  \\
0.3-0.4  &   39598 &    23028 &     8011 &     2633 &     1279 &      521 &      211 &      186 &      136 &      211 &    75814  \\
0.4-0.5  &   15091 &    15104 &     5614 &     2571 &     1291 &      658 &      322 &      248 &      173 &      322 &    41394  \\
0.5-0.6  &    4670 &     7154 &     3316 &     1726 &      894 &      645 &      161 &      111 &       99 &      223 &    18999  \\
0.6-0.7  &    1366 &     2620 &     1664 &     1043 &      397 &      236 &      186 &       86 &       24 &      136 &     7740  \\
0.7-0.8  &     472 &      670 &      509 &      273 &      223 &      111 &      149 &       12 &       37 &       74 &     2530  \\
0.8-0.9  &      99 &      173 &      149 &       24 &       99 &       37 &       12 &       12 &        0 &       12 &      617  \\
0.9-1.0  &      74 &       37 &       37 &        0 &       24 &        0 &        0 &        0 &       12 &        0 &      184  \\ \cline{1-12}
Total  &  365276 &    89423 &    27906 &    10802 &     5473 &     2654 &     1276 &      766 &      530 &   1064     &   505134  \\ \cline{1-12}
\end{tabular}
\end{center}
\caption[Kinematic distribution of deep-inelastic sample.]{Deep-inelastic 
interactions expected per ton for the four-year scenario of Table~\ref{tab:run}.}
\label{tab:xq2dis}
\end{table}

Were \minerva\ the prime user of \numi, the beamline would be run in the
high-energy configuration with energies in the 5--25~GeV range. This
configuration offers the ability to study neutrino interactions across an
appreciable fraction of the $x_{Bj}$ range at reasonable $Q^2$. In HE beam mode
expected event rates would be 580~K charged-current $\nu_{\mu}$ events per
$10^{20}$ PoT per ton, over twice as many as the sHE(-2.5) beam.

\subsection{Accuracy of Predicted Neutrino Flux}

As mentioned earlier, one of the significant advantages of \minerva\ over
previous wide-band neutrino experiments is the expected accuracy with which the
neutrino absolute and energy dependent flux is known.  Since the NuMI beamline
has been designed for the MINOS neutrino oscillation experiment, particular
attention has been paid to control and knowledge of the beam of neutrinos being
used in the experiment.

The biggest uncertainty in the predicted energy spectrum of the neutrinos comes
directly from the uncertainty of hadron prodution spectrum of the $\pi$ and K
parents of the neutrinos.  To help reduce this uncertainty, there is an
approved Fermilab experiment E-907\cite{MIPP,GNuMI} which has as it's main goal
the measurement of hadron production spectra off various nuclear targets.  One
of the measurements that will be made by E-907 is an exposure of the NuMI target
to the 120 GeV Main Injector proton beam.  By using the NuMI target material
and shape, E-907 will be able to provide the spectra coming off the target
including all of the secondary and tertiary interactions which can
significantly modify the produced spectra.  It is expected that with the input
from E-907, the absolute and energy dependent shape of neutrinos per POT will be
known to $\approx$ 3\%.

For the absolute flux of neutrinos there is a second uncertainty which must be
considered and that is the accuracy with which we know the number of protons on
target. With the planned NuMI primary proton beamline
instrumentation\cite{numiinst}, the number of protons on target will be
known to between (1 - 3)\%, the range being determined by the calibration
techniques used to control drift of the primary beam toroid devices.  

To summarize, the energy shape of the NuMI beam should be known to 3\% while
the absolute flux should be known to between (3 - 5)\%.

%% file: quasielastic.tex
\section{Quasi-Elastic Scattering}
\label{sect:quasielastic}

\subsection{Quasi-elastic Cross-sections}

\begin{figure*}[ph]
\centerline{\psfig{figure=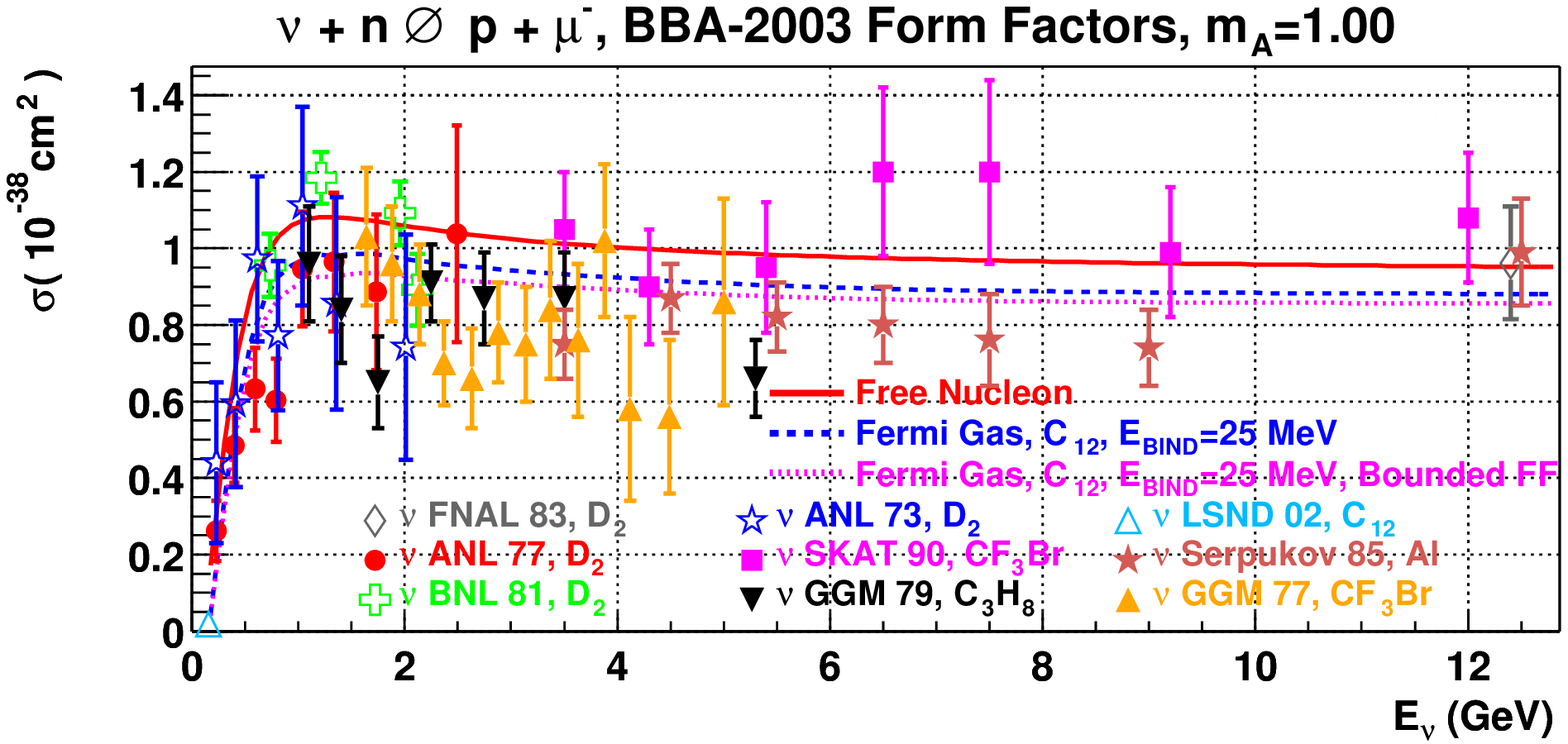,width=5.in}}
\centerline{\psfig{figure=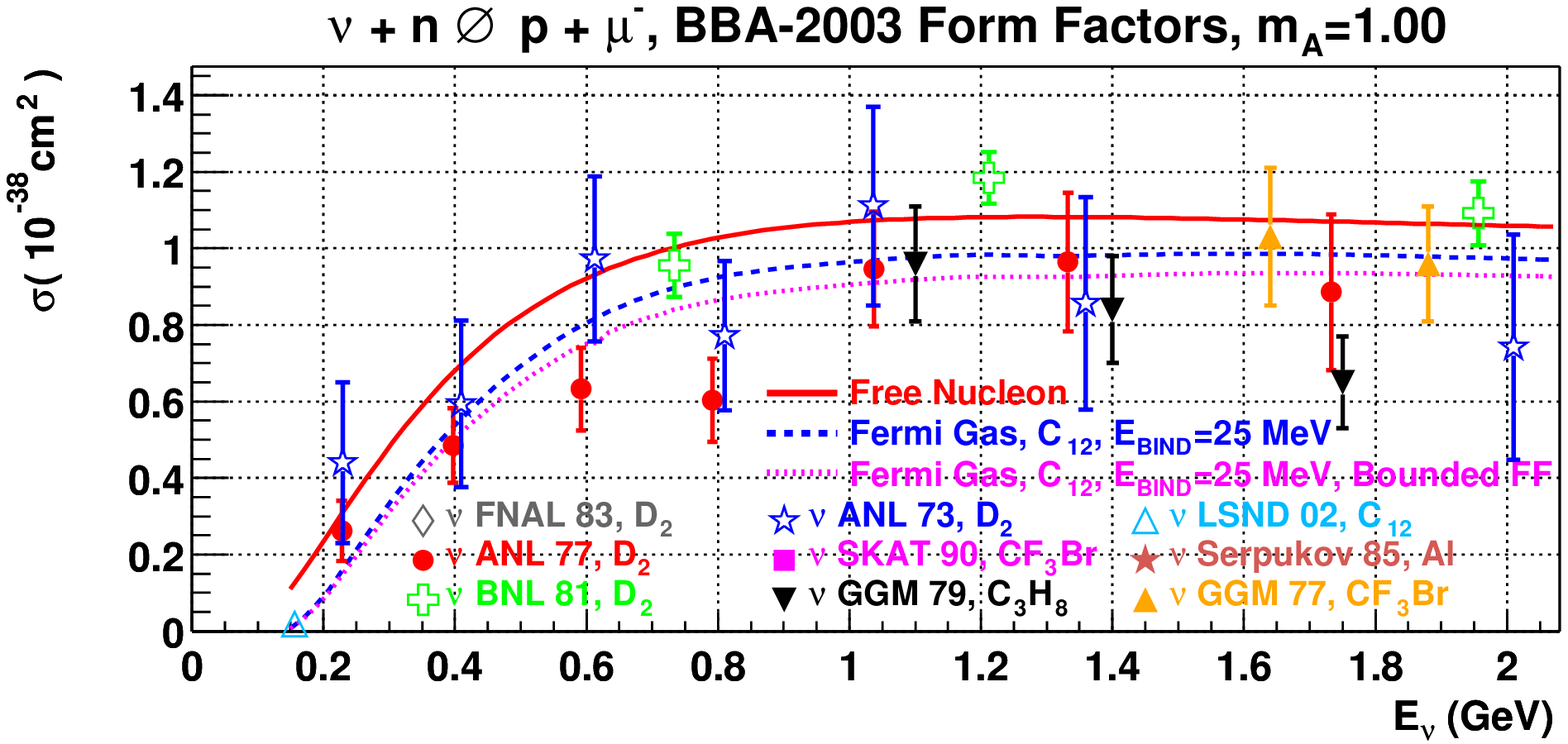,width=5.in}}
\caption[Quasi-elastic neutrino cross-section data]
{Compilation of neutrino
quasi-elastic cross-section data.
   The data have large errors and are only marginally consistent throughout
the $E_{\nu}$ range
   accessible to \minerva\ (upper plot), and particularly in the 
threshold region (lower plot).  Representative calculations are shown using BBA-2003
form factors with $M_A$=1.00 GeV.  The solid curve is without nuclear corrections, the
dashed curve includes a Fermi gas model~\cite{Zeller_03},
   and the dotted curve includes Pauli blocking and nuclear
   binding.  The data shown are from
   FNAL 1983~\cite{Kitagaki_83}, ANL 1977~\cite{Barish:1977qk},
   BNL 1981~\cite{Baker:1981su}, ANL 1973~\cite{Mann_73},
   SKAT 1990~\cite{Brunner_90}, GGM 1979~\cite{Pohl_79},
   LSND 2002~\cite{Auerbach_02}, Serpukov 1985~\cite{Belikov_85},
   and GGM 1977~\cite{Bonetti_77}.  }
\label{elas_JhaKJhaJ_nu}
\end{figure*}

\begin{figure*}[htb]
\centerline{\psfig{figure=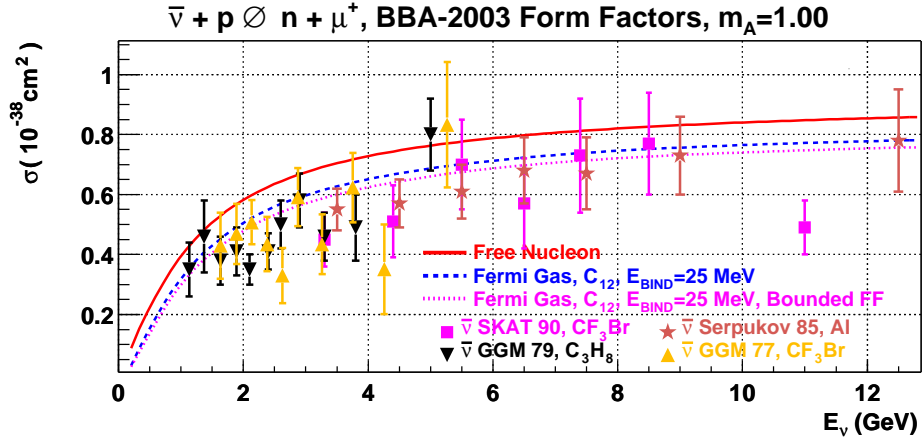,width=5.in}}
\caption[Quasi-elastic anti-neutrino cross section data]
{Compilation of anti-neutrino
quasi-elastic cross-section data.
   As in Figure~\ref{elas_JhaKJhaJ_nu}, the data have large errors, and considerable scatter among the
   different experiments.   Theoretical predictions without (solid curve) and including
   nuclear corrections (dashed, dotted curves) are shown for comparison.
The data shown are from SKAT 1990~\cite{Brunner_90},
GGM 1979~\cite{Armenise_79}, Serpukov 1985~\cite{Belikov_85},
and GGM 1977~\cite{Bonetti_77}.}
\label{elas_JhaKJhaJ_nub}
\end{figure*}

     Quasi-elastic scattering makes up
the largest single component of the total $\nu$--N interaction rate in the
threshold regime $E_{\nu} \le 2~\hbox{GeV}$.  Precision measurement of the
cross-section for this reaction, including its energy dependence and
variation with target nuclei, is essential to current and
future neutrino-oscillation experiments.
Figures~\ref{elas_JhaKJhaJ_nu} and~\ref{elas_JhaKJhaJ_nub} summarize current knowledge of
neutrino and anti-neutrino quasi-elastic cross-sections.
Among the results shown, there are typically 10--20\% normalization uncertainties
from knowledge of the fluxes.  These plots show that
existing measurements have large errors throughout the $E_{\nu}$
range accessible to \minerva\ (Figure~\ref{elas_JhaKJhaJ_nu}, upper plot), and
especially in the threshold regime which is crucial to future oscillation experiments
(Figure~\ref{elas_JhaKJhaJ_nu}, lower plot).
Figure~\ref{elas_JhaKJhaJ_nub} shows these large uncertainties extend to anti-neutrino
measurements as well.

     \begin{figure*}[htb]
\centerline{\psfig{figure=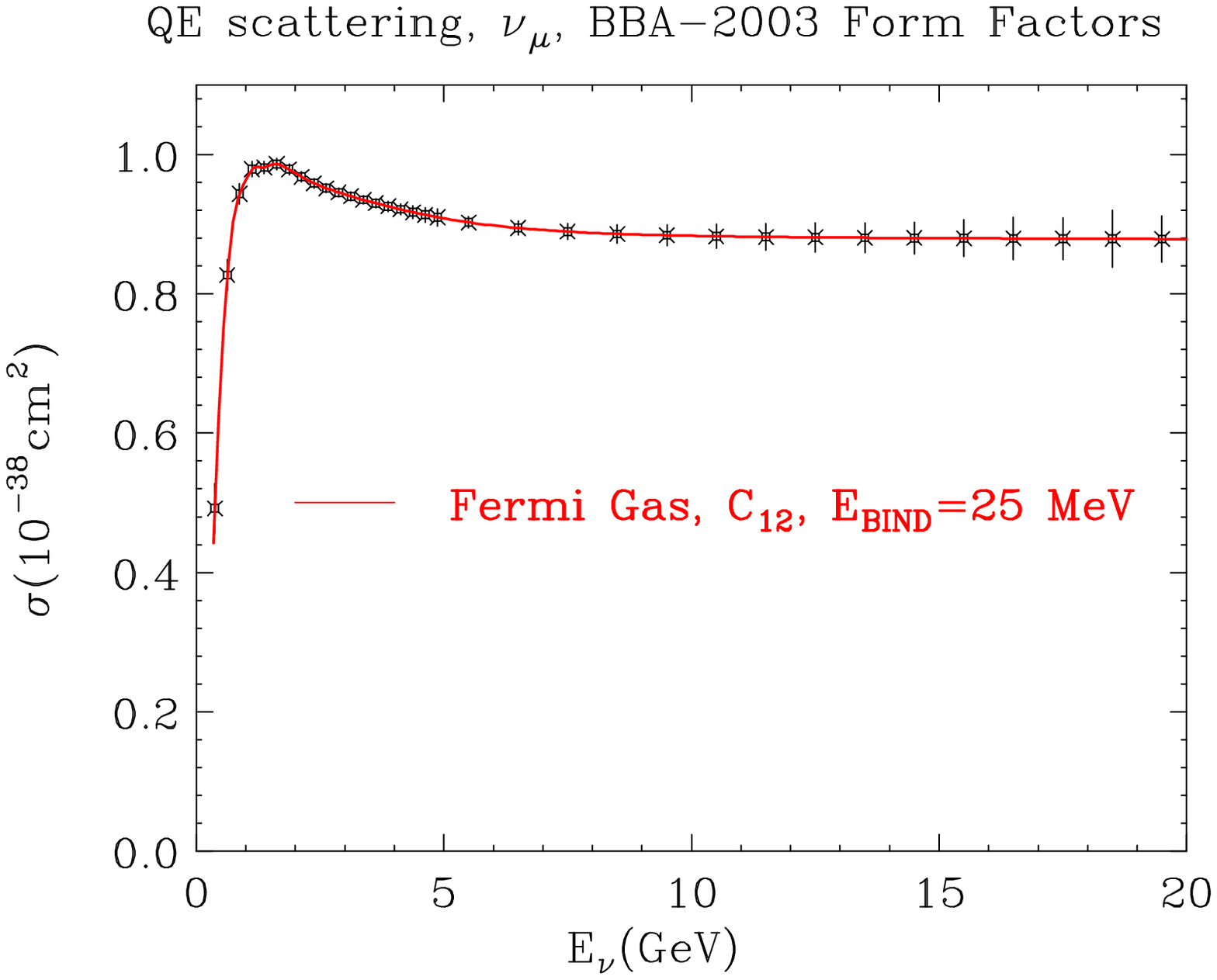,width=5.in}}
\caption
{Cross section for \minerva\ assuming a 3 ton fiducial volume,
4 year run, perfect resolution, 100\% detection efficiency,
BBA-2003 form factors with  $M_A$ = 1.00 GeV, and the  Fermi gas model.}
\label{sigma_JhaKJhaJ_minerva}
\end{figure*}

\minerva\ will measure these quasi-elastic
cross-sections with samples exceeding earlier (mostly) bubble-chamber experiments
by two orders of magnitude. \minerva\ will also perform the first precision measurement of
nucleon form-factors for $Q^2 > 1~\hbox{(GeV/c)}^2$ using neutrinos.

Consistent and up-to-date treatment of the vector and axial-vector form-factors
which characterize the nucleon weak current is essential to a realistic
cross-section calculation.  \minerva\ collaborators have been active in this area for some time\cite{budd}.
Recent parameterizations and fits published by Budd, Bodek and Arrington are hereafter
referred to as ``BBA-2003" results.  The curves in Figures~\ref{elas_JhaKJhaJ_nu}
and~\ref{elas_JhaKJhaJ_nub} are based on BBA-2003 form-factors, with the
axial form-factor mass parameter set to $M_A=1.00~\hbox{GeV}/c^2$. The
solid curves are calculated without nuclear corrections, while the dashed curves
include a Fermi gas model.  The dotted curves are calculations for Carbon
nuclei and include Fermi motion, Pauli blocking, and the effect of nuclear binding
on the nucleon form-factors as modeled by Tsushima \mbox{{\em et al.}\cite{Tsushima_03}.}

Figure~\ref{sigma_JhaKJhaJ_minerva} shows  predictions  for the cross section
assuming the BBA form factors, with the axial mass = 1.00. The number of events
assumes a  3 ton fiducial volume, out of the 6 tons of completely
active target. We assume the efficiency for detections is 100\% and perfect
resolution. We would take all events for which there is a recoil proton
traversing at least one plane or for which the recoil proton is absorbed in the
nucleus. Therefore, efficiency and acceptance would be 100\% since we have a
hermetic detector with a side muon/hadron absrober. The error in the
quasi-elastic cross section is then the statistical error and the error from the
subtraction of resonance events which, through nuclear effects and close-in
scatters, simulate the quasi-elastic signature.. This background uncertainty is
not shown. Not shown also is the 4\% normalization flux error.

Nuclear effects reduce the calculated cross-sections by $\ge$10\%;
this sensitivity to the details of nuclear physics
shows that an understanding of final-state nuclear effects is essential to
interpretation of quasi-elastic neutrino data.  
As as fine-grained tracking calorimeter, \minerva\ is designed to
facilitate systematic comparison of quasi-elastic scattering (and other exclusive channels)
on a variety of nuclear targets, providing a vastly improved empirical foundation for theoretical
models of these important effects.

\subsection{Form-factors in Quasi-elastic Scattering}

\minerva's large quasi-elastic samples will probe the $Q^2$ response
of the weak nucleon current with unprecedented
accuracy.  The underlying V-A structure of this current
include vector and axial-vector form-factors whose $Q^{2}$ response is approximately
described by dipole forms.  The essential formalism is given by\cite{Lle_72}

\begin{eqnarray*}
   \lefteqn{<p(p_2)|J_{\lambda}^+|n(p_1)>  =   }
   \nonumber \\ &
\overline{u}(p_2)\left[
    \gamma_{\lambda}F_V^1(q^2)
    +\frac{\D i\sigma_{\lambda\nu}q^{\nu}{\xi}F_V^2(q^2)}{\D 2M}
+\gamma_{\lambda}\gamma_5F_A(q^2)
+\frac{\D q_{\lambda}\gamma_5F_P(q^2)}{\D M} \right]u(p_1),
\end{eqnarray*}

\noindent where $q=k_{\nu}-k_{\mu}$, $\xi=(\mu_p-1)-\mu_n$, and
$M=(m_p+m_n)/2$.  Here, $\mu_p$ and $\mu_n$ are the
proton and neutron magnetic moments.
It is assumed that second-class currents are absent, hence
the scalar ($F_V^3)$ and tensor ($F_A^3$) form-factors
do not appear.

\noindent The form-factors $F^1_V(q^2)$ and  ${\xi}F^2_V(q^2)$
are given by:
$$ F^1_V(q^2)=
\frac{G_E^V(q^2)-\frac{\D q^2}{\D 4M^2}G_M^V(q^2)}{1-\frac{\D q^2}{\D 4M^2}},
~~~{\xi}F^2_V(q^2) =\frac{G_M^V(q^2)-G_E^V(q^2)}{1-\frac{\D q^2}{\D 4M^2}}.
$$

\noindent According to the conserved vector current (CVC) hypothesis, $G_E^V(q^2)$ and $ G_M^V(q^2)$ are directly
related to form-factors determined from electron scattering
$G_E^p(q^2)$, $G_E^n(q^2)$, $G_M^p(q^2)$, and $G_M^n(q^2)$:
$$
G_E^V(q^2)=G_E^p(q^2)-G_E^n(q^2),
~~~G_M^V(q^2)=G_M^p(q^2)-G_M^n(q^2).
$$

\noindent The axial ($F_A$) and pseudoscalar ($F_P$) form-factors
are
$$ F_A(q^2)=\frac{g_A}{\left(1-\frac{\D q^2}{\D M_A^2}\right)^2 },
~~F_P(q^2)=\frac{2M^2F_A(q^2)}{M_{\pi}^2-q^2}. $$

\noindent In the differential cross-section,
$F_P(q^2)$ is multiplied by $(m_l/M)^2$, consequently its contribution
to muon neutrino interactions is very small, except below 0.2~GeV.
In general, the axial form-factor $F_A(q^2)$ can only be extracted from
quasi-elastic neutrino scattering; at low $q^2$, however, its behavior
can also be inferred from pion electroproduction data.

Until recently, it has been universally assumed that the
form-factors' $q^2$ dependence is described by the dipole approximation.   For example,
the vector form factors are normally expressed:

$$ G_D(q^2)=\frac{1}{\left(1-\frac{\D q^2}{\D M_V^2}\right)^2
},~~M_V^2=0.71~GeV^2$$
$$
G_E^p=G_D(q^2),~~~G_E^n=0,
~~~G_M^p={\mu_p}G_D(q^2),~~~ G_M^n={\mu_n}G_D(q^2).
$$
\noindent As discussed below, the dipole parameterization is far from perfect, and
\minerva\ will be able to measure deviations from this form.


\subsubsection{Vector form-factor discrepancy at high $Q^{2}$}

Electron scattering experiments at SLAC and Jefferson Lab (JLab) have
measured the proton and neutron electromagnetic (vector)
form-factors with high precision.
The vector form-factors can be determined from electron
scattering cross-sections using
the standard Rosenbluth separation technique\cite{JRA_03},
which is sensitive to radiative corrections, or from polarization
measurements using the newer polarization transfer technique\cite{halla}.
Polarization measurements do not directly measure form-factors,
but rather the ratio $G_E$/$G_M$.
These form-factors can be related to their counterparts in quasi-elastic
neutrino scattering by the CVC hypothesis.
Naturally, more accurate form-factors translate directly to improved calculations
of neutrino quasi-elastic cross-sections.

\begin{figure}[htbp!]
\centerline{\psfig{figure=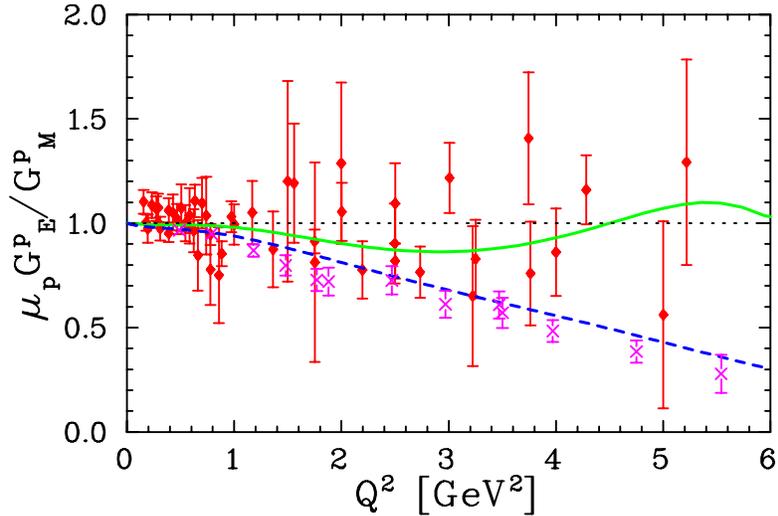,width=4.0in}}
\caption[Ratio of $G_E^p$ to $G_M^p$ showing discrepancy between two techniques]{Ratio of $G_E^p$ to $G_M^p$ as
extracted by Rosenbluth
separation measurements (diamonds) and polarization
measurements(crosses).
The data are in clear disagreement at high $Q^2$.
}
\label{show_gepgmp}
\end{figure}

Recently, discrepancies in electron scattering measurements of
some vector form-factors have appeared; study of quasi-elastic reactions
in \minerva\ may help reveal the origin these discrepacies.
Figure~\ref{show_gepgmp} shows the BBA-2003
fits to $\mu_p G_E^p/G_M^p$.
There appears to be a difference between the two different methods of measuring this ratio.
The fit including only cross-section data ({\em i.e.} Rosenbluth separation)
is roughly flat in $Q^2$ and is consistent with form-factor scaling. This
is expected if the electric charge and magnetization
distributions in the proton are the same. However,
the newer polarization transfer technique
yields a much lower ratio at high $Q^2$, and indicates
a difference between the electric charge and
magnetization distributions.
The polarization transfer technique is believed to be
more reliable and less sensitive to radiative effects
from two-photon corrections.

If the electric charge and magnetization distributions
of the proton are indeed different, a test of the axial form-factor's
high-$Q^2$ shape can provide important new input to help resolve differences between
electron scattering measurements.  As discussed below, \minerva\ will be able to
accurately measure the high-$Q^2$ behavior of $F_A$.


\subsubsection{Form-factor deviations from dipole behavior}

\begin{figure}[htbp!]
\centerline{\psfig{figure=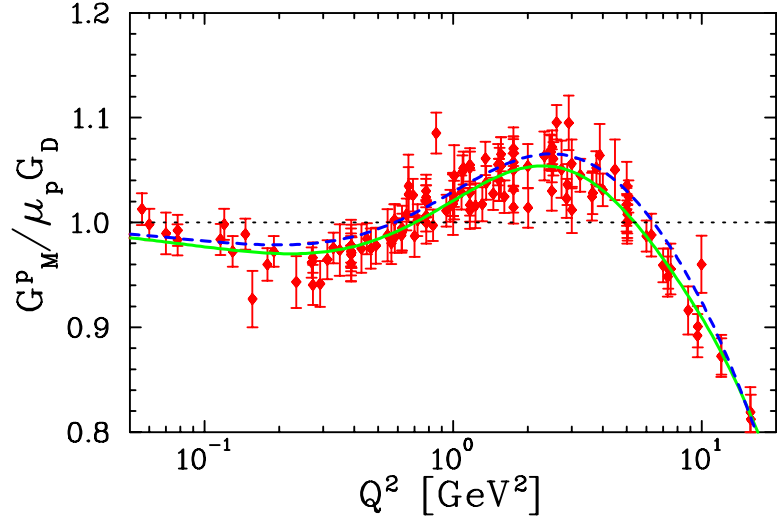,width=4.0in}}
\caption[Deviation of $G_M^p$ from dipole form]{BBA-2003 fits to
$G_M^p/{\mu}_{p}G_D$.
The departure from 1.0 indicates deviation from a pure
dipole form; the
deviation is quite pronounced for $Q^{2} > 1~\hbox{(GeV/c)}^2$. }
\label{show_gmp}
\end{figure}

Electron scattering shows that dipole amplitudes provide only a
first-order description of form-factor behavior at high $Q^{2}$.
Figure~\ref{show_gmp} shows the deviation of $G_M^p$ from dipolar
$Q^{2}$ dependence.  In general, these deviations are different
for each of the form factors.

\begin{figure}[htbp!]
\centerline{\psfig{figure=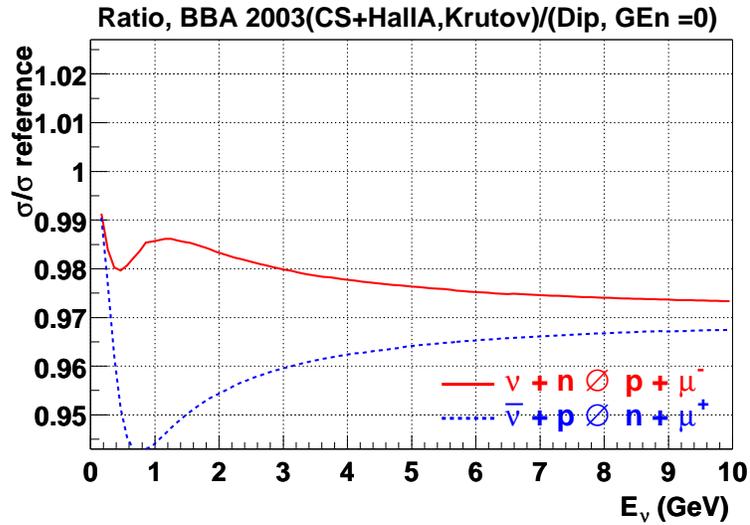,width=4.0in}}
\caption[Comparison between BBA-2003 cross-sections and dipole with $G_E^n=0$]
{Ratio of the neutrino and anti-neutrino
quasi-elastic cross-sections calculated with BBA-2003 form-factors to the
simple dipole approximation with $G_E^n=0$.}
\label{ratio_JhaKJhaJ_D0DD}
\end{figure}

%
%
%
%
%

Figure~\ref{ratio_JhaKJhaJ_D0DD} shows the ratio of the
BPA-2003 neutrino and anti-neutrino quasi-elastic cross-sections to the prediction using
dipole vector form-factors (with $G_E^n=0$ and $M_A$ kept fixed).
This plot shows that the importance accurately parameterizing the form-factors.
In {\minerva}, it will be possible to study the $Q^2$ dependence of the form-factors beyond the
simple dipole dipole approximation which has been assumed
by all previous neutrino experiments.

\subsection{Axial Form-factor of the Nucleon}

    Electron scattering experiments continue to provide increasingly
precise measurements of the nucleon vector form-factors.
Neutrino scattering, however, remains the only practical
route to comparable precision for the axial form-factors, in particular
$F_A(Q^2)$.  The fall-off of the form-factor strength with
increasing $Q^{2}$ is traditionally parameterized using an effective
axial-vector mass $M_A$.  Its value is known to be $\approx 1.00~\hbox{GeV}/c^2$
to an accuracy of perhaps 5\%.
This value agrees with the theoretically-corrected value
from pion electroproduction\cite{Bernard_01}, \mbox{$1.014 \pm 0.016~\hbox{GeV}/c^2$}.
Uncertainty in the value of $M_A$ contributes directly to uncertainty
in the total quasi-elastic cross-section.   

%
%
%

\begin{figure}[htb!]
\centerline{\psfig{figure=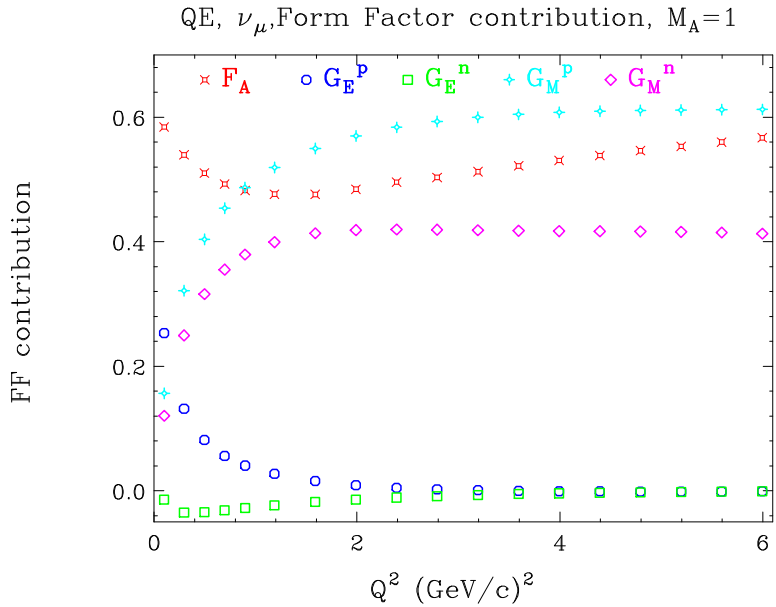,width=4.0in}}
   \caption[Contributions to the $Q^2$ distribution of quasi-elastic
scattering in \minerva.]
{Fractional contributions of $G_M^p$,$G_M^n$,$G_E^p$, $G_E^n$ and
$F_A$ to the $Q^2$ distributions for quasi-elastic
neutrino (top) and anti-neutrino (bottom) samples
with the \numi\ beam.  Because of interference terms, the
sum of the fractions does not necessarily add up to 100\%.}
\label{FFcontributions}
\end{figure}

The fractional contributions of $F_A$,
$G_M^p$,$G_M^n$,$G_E^p$, and $G_E^n$
to the $Q^2$ distribution for
quasi-elastic neutrino and anti-neutrino scattering with
the \numi\ beam are shown in
Figure~\ref{FFcontributions}.
The contributions are determined by comparing
the BBA-2003 cross-sections with and without each of the form-factors included.
\minerva\ will be the first
systematic study of $F_A$, which accounts for roughly half of the quasi-elastic cross-section, over the entire range of $Q^2$ shown in the figure.

\subsubsection{Vector form-factors and $M_A$}

Earlier neutrino measurements, mostly bubble-chamber experiments
on Deuterium, extracted
$M_A$ using the best inputs and models available at the time. Changing
these assumptions changes the extracted value of $M_A$. Hence,
precision measurement of $M_A$ requires starting with the best possible vector
form-factors, coupling constants, and other parameters.

\begin{figure}[htb]
\centerline{\psfig{figure=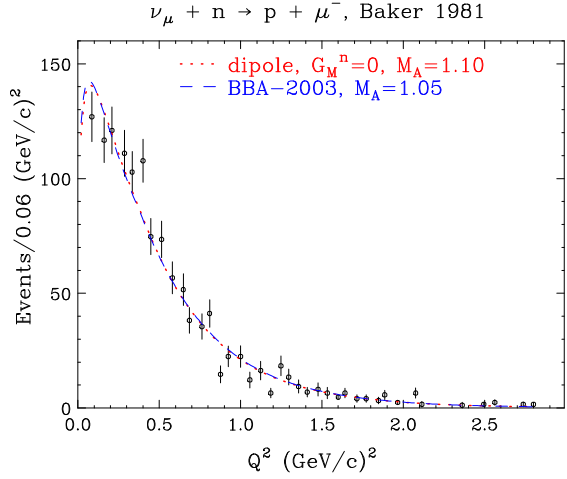,width=2.91in}}
\caption[Comparison of $Q^2$ distributions for neutrinos with two sets
of form-factors]{Comparison of $Q^2$ distributions using two different
sets of form-factors. The data are from Baker {\em et al.}\cite{Baker:1981su}.
The dotted curve uses dipole form-factors with $G_E^n=0$ and
$M_A=1.10~\hbox{GeV}/c^2$. The dashed curve uses more recent BBA-2003 form-factors
and $M_A=1.05~\hbox{GeV}/c^2$.  It is essential to use the best possible
information on vector form-factors from electron scattering experiments
when extracting the axial form-factor from neutrino data.}
\label{Baker_d0dd_110_JhaKJhaJ_105}
\end{figure}

\begin{figure}
\centerline{\psfig{figure=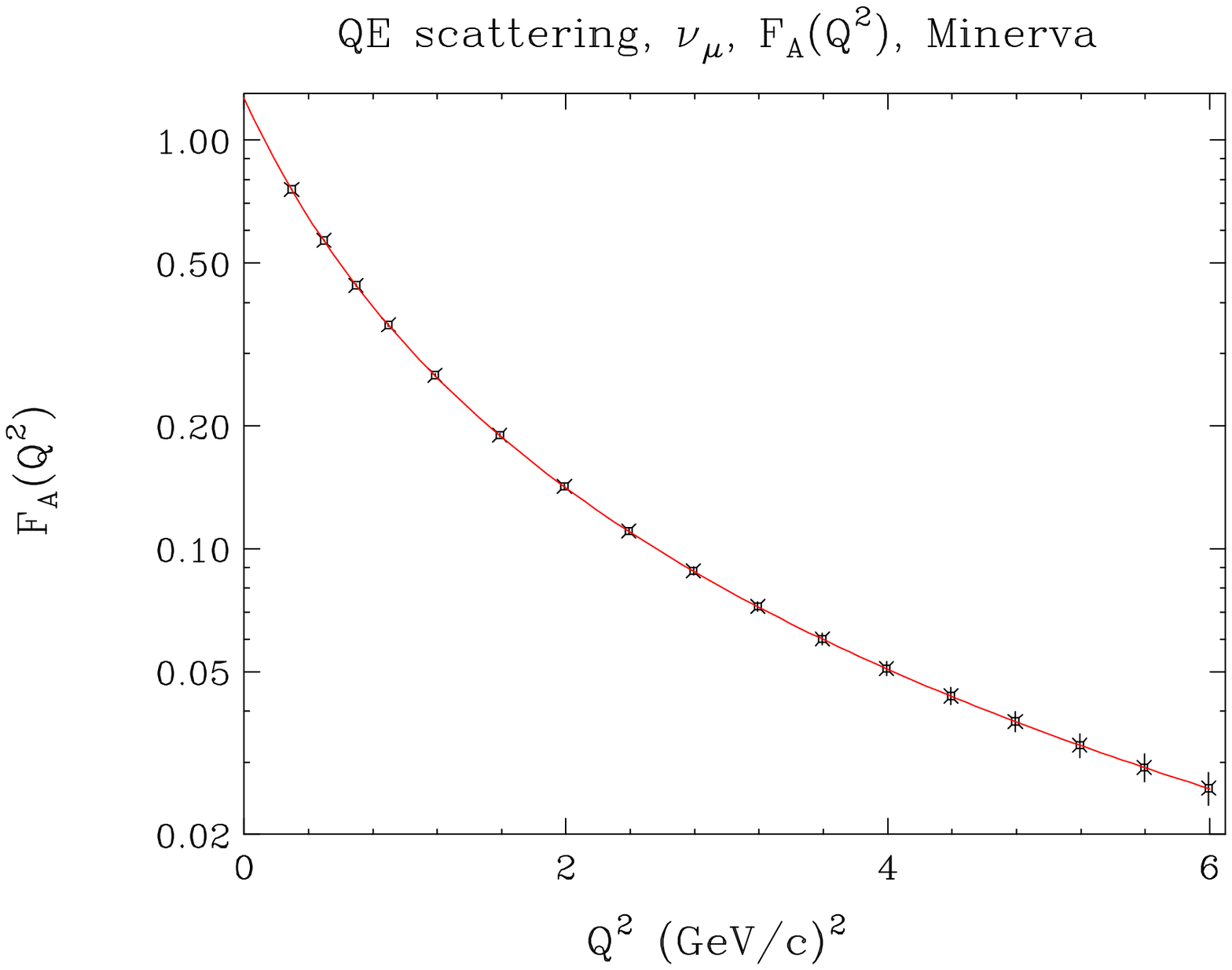,width=4.in}}
\caption[Extraction of $F_A$ in \minerva]{Estimation of $F_{A}$ from a
sample of Monte Carlo neutrino quasi-elastic events recorded in the
{\minerva} active Carbon target.  Here, a pure dipole form for $F_{A}$ is
assumed, with $M_{A} = 1~\hbox{GeV}/c^2$.
The simulated sample and error bars correspond to four years of \numi\ running.}
\label{log:ps}
\end{figure}

Figure~\ref{Baker_d0dd_110_JhaKJhaJ_105} shows the $Q^2$ distribution
from the Baker {\em et al.}\cite{Baker:1981su} neutrino experiment
compared to the dipole form-factor approximation with $G_E^n =0$ and $M_A =1.100~\hbox{GeV}/c^2$.
Also shown are BBA-2003 predictions with $M_A =1.050~\hbox{GeV}/c^2$.
Use of more accurate electromagnetic form-factors
requires a different $M_A$ value to describe the same $Q^2$
distribution.
Thus, with the same value of $g_A$, adopting the dipole approximation (and $G_E^n =0$)
instead of the BBA-2003 form-factors may lead to an error in $M_A$ of $0.050~\hbox{GeV}/c^2$.

\subsubsection{Measurement of the axial form-factor in {\minerva}}

Current and future high-statistics
neutrino experiments at low energies ({\em e.g.} K2K, MiniBooNE, J-PARCnu and
{\minerva}) use an active nuclear target such as scintillator (mostly Carbon) or water (mostly Oxygen).
The maximum $Q^2$ values that can be achieved with incident neutrino energies of
0.5, 1.0, 1.5 and 2~GeV are 0.5, 1.2, 2.1 and 2.9~$\hbox{(GeV/c)}^2$, respectively.
Since K2K, MiniBooNE and J-PARCnu energies are in the 0.7--1.0~GeV range, these
experiments probe the
low $Q^2< 1 ~\hbox{(GeV/c)}^2$ region where nuclear effects are large
(see Figures~\ref{pauli} and~\ref{binding})
and where the free-nucleon axial form-factor is known rather well from neutrino data
on Deuterium (see  Figure~\ref{Baker_d0dd_110_JhaKJhaJ_105}).
The low $Q^2$ ($Q^2< 1 ~\hbox{(GeV/c)}^2$)  MiniBooNE and K2K experiments have
begun to investigate the various nuclear effects in Carbon and Oxygen.

At higher $Q^2$, as shown by the BBA-2003 fits,
the dipole approximation for vector form-factors can be in error by a
factor of two when \mbox{$Q^2> 2~\hbox{(GeV/c)}^2$.}
There is clearly no reason to assume the dipole form will be any better 
for the axial form-factor.
   As shown in Figure~\ref{Baker_d0dd_110_JhaKJhaJ_105}
   there is very little data for the axial
   form-factor in the high-$Q^2$ region (where nuclear effects are
   smaller).  Both the low-$Q^2$ ($Q^2< 1 ~\hbox{(GeV/c)}^2$) and
high-$Q^2$ ($Q^2> 2 ~\hbox{(GeV/c)}^2$) regions
are accessible in higher-energy experiments like \minerva,
which can span the 2--8~GeV neutrino energy range.
\minerva's measurement of the axial form-factor at high $Q^2$
will be essential to a complete understanding of the vector and
axial structure of the neutron and proton.

\begin{figure}
\centerline{\psfig{figure=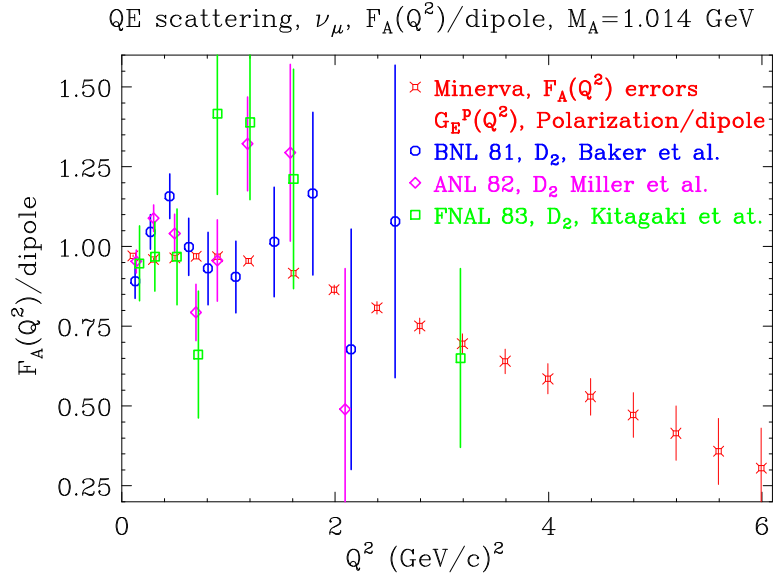,width=4.in}}
\caption[\minerva\ sensitivity to $F_A/F_A(\hbox{Dipole})$]{
Extracted ratio $F_{A}$/$F_{A}$(Dipole) from a
sample of Monte Carlo quasi-elastic interactions 
recorded in the {\minerva} active Carbon target,
from a four-year exposure in the \numi\ beam. 
The {\minerva} points assume this ratio is described by
the ratio of $G_E^p$(Polarization)/$G_E^p$(dipole). $F_A$ is extracted 
from deuterium bubble chamber experiments 
using the $d\sigma/dq^2$ from the papers of FNAL 1983~\cite{Kitagaki_83}
BNL 1981~\cite{Baker:1981su}, and ANL 1982~\cite{Miller_82} }
\label{f_a_polar:ps}
\end{figure}

\begin{figure}
\centerline{\psfig{figure=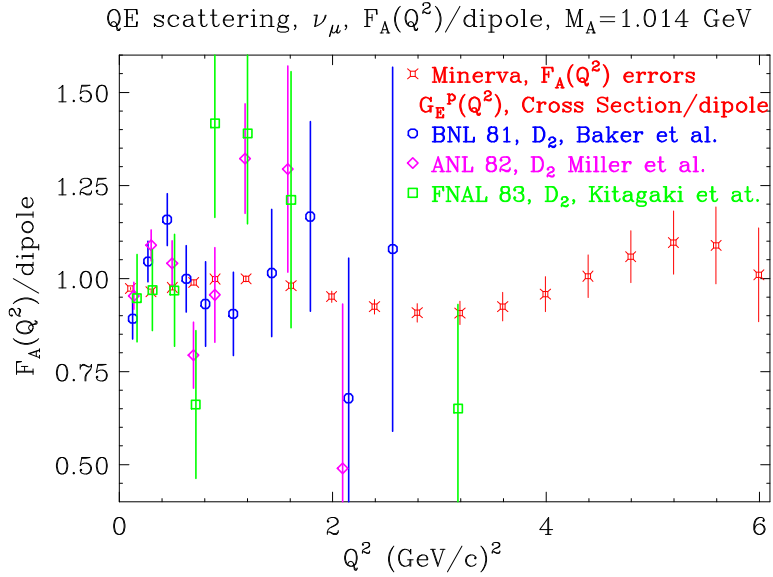,width=4.in}}
\caption[\minerva\ sensitivity to $F_A/F_A(\hbox{Dipole})$]{ Extracted ratio
$F_{A}$/$F_{A}$(Dipole) from a sample of Monte Carlo quasi-elastic interactions
recorded in the {\minerva} active Carbon target, from a four-year exposure in
the \numi\ beam.  The {\minerva} points assume this ratio is described by the
ratio of $G_E^p$(Cross-Section)/$G_E^p$(dipole), which was the accepted
result for $G_E^p$ before new polarization transfer measurements. The extracted
values of $F_A$ for the deuterium bubble chamber experiements are the same as
the previous figure} 
\label{f_a_cross_sect:ps} 
\end{figure}

Figure~\ref{log:ps} shows the extracted values and errors
of $F_{A}$ in bins of $Q^{2}$ from a sample of simulated quasi-elastic
interactions in the {\minerva} active Carbon target,
for a four-year exposure in the \numi\ beam.
Clearly the high-$Q^{2}$ regime, which is inaccessible to
K2K, MinibooNE and J-PARCnu, will be well-resolved in {\minerva}.
  Figure~\ref{f_a_polar:ps} and ~\ref{f_a_cross_sect:ps} 
show these results as a ratio of 
$F_{A}$/$F_{A}$(Dipole), demonstrating \minerva's ability to 
distinguish between different models of $F_A$.
  \minerva\ will be able to measure the axial nucleon form-factor 
with precision comparable to vector form-factor measurements at JLab.

   The plot for $F_A(q^2)$ is done by writing out
the cross section as a quadratic function
of $F_A(q^2)$. The coefficents of $F_A(q^2)$ are functions
of $E_{\nu}$ and $q^2$. The constant
coefficient is a function of the measured
(or predicted for MINER${\nu}$A) cross section
in a $q^2$ bin, as well as being a function
of $E_{\nu}$ and $q^2$. The coefficients
are integrated in energy over the flux and the $q^2$
region of the bin. The $q^2$ of the bin is determined by
bin centering wrt $d\sigma/dq^2$.
For the extraction of $F_A(q^2)$ for \minerva\, we assume
a dipole for $F_A(q^2)$ with a value of $M_A$ given
by pion electro-production $M_A$= 1.014, BBA form factors, 
100\% detection efficency and perfect resolution. 

   For the extraction of data from Miller,
Baker, and Kitagaki, we use the BBA form
factors to determine the coefficients.
Their plots of $d{\sigma}/dq^2$ is
used to extract $F_A(q^2)$.
The overall nomalization of their
data is not given. (In addition
they probably do not know their normalization to
better than 10\%). Hence, we set the overall
normalization of
$F_A(q^2)$ by getting the overall normalization from
the cross section using a dipole $F_A(q^2)$
with $M_A$=1.014. Since \minerva will measure
a normalized cross section, \minerva will not need
to determine the overall normalization of
$F_A(q^2)$ by assuming a form for $F_A(q^2)$.

We recognize that at low $Q^2$, there are large nuclear
corrections and the recoil proton cannot be well measured.
This is why for the measurement of $F_(Q^2)$ we plan on
integrating the axial form factor from $Q^2$=0 to
$Q^2$=0.2 to 0.4 GeV. Since the axial form factor is
known at $Q^2$=0 from neutron decay to be 1.26 with high
precision, this is not an issue. We show two different
models for $F_A$ as a function of $Q^2$ (which differ by
a factor of 5 at high $Q^2$, as indicated by $G^p_E$/$G^p_M$
data.  At high $Q^2$, the proton track is long and
ID'd in the scintillator, and very high $Q^2$ protons
also stop in the side or downstream absorber, since
we have hermetic detector, the acceptance is 100\%.
  Here again, the background from MISId' delta's
needs to be considered. Or alternatively, tighter de/dx
cuts on the proton track (thus trading efficiencies
for pion rejection).

\begin{figure}[h!]
\centerline{\epsfig{figure=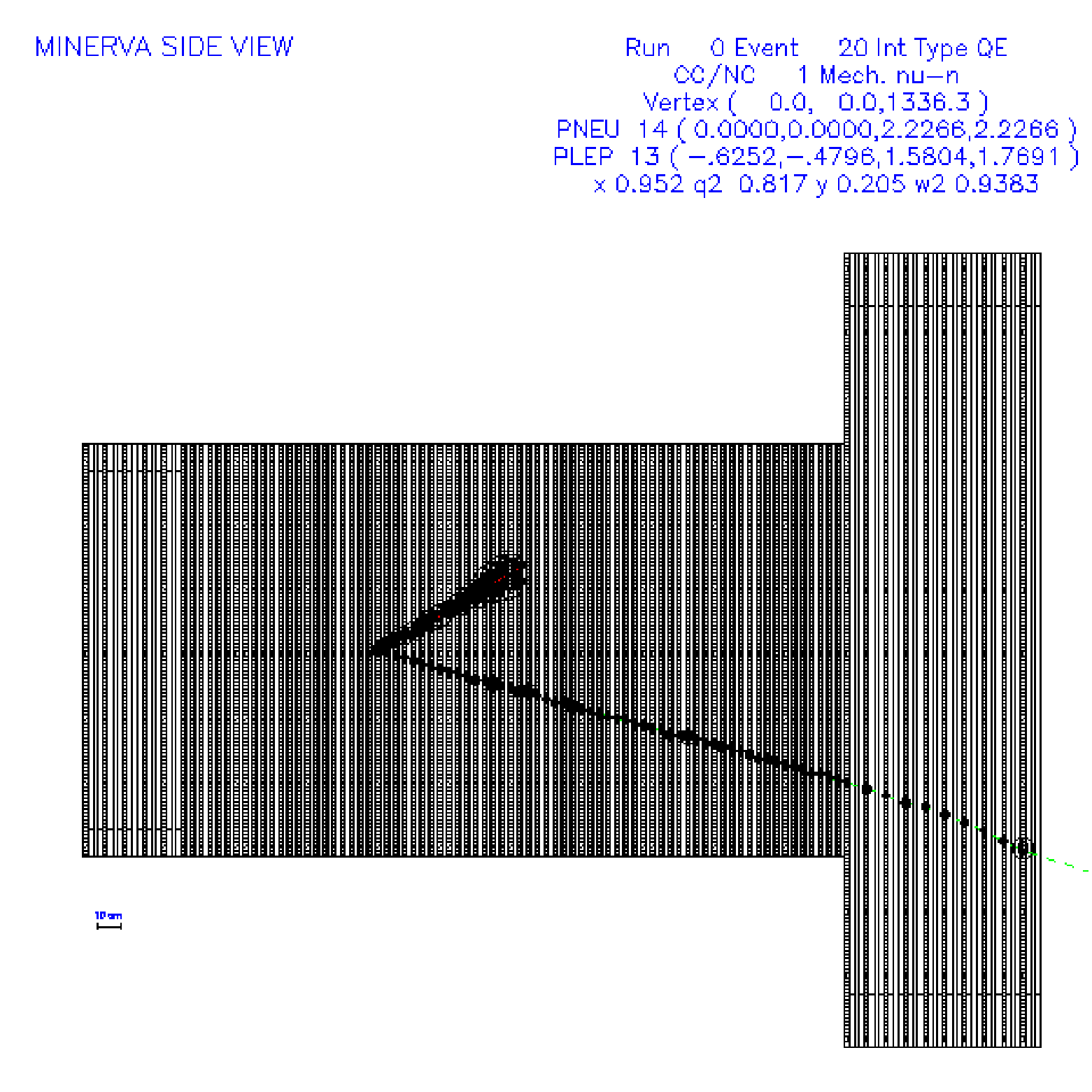,width=\textwidth}}
\caption[Simulated charged-current quasi-elastic interaction]{A simulated charged-current 
quasi-elastic interaction in \minerva. The proton (upper) and muon (lower) tracks are
well resolved. In this display, hit size is proportional to energy loss within a strip.
The increased energy loss of the proton as it slows and stops is clear. Note that for
clarity the outer detector has not been drawn.}
\label{fig:qedisplay}
\end{figure}

Figure~\ref{fig:qedisplay} shows a typical quasi-elastic event, as simulated in \minerva.
 

\subsection{Nuclear Effects in Quasi-elastic Scattering}

\subsubsection{Fermi gas model}

\begin{figure}[htb]
\centerline{\psfig{figure=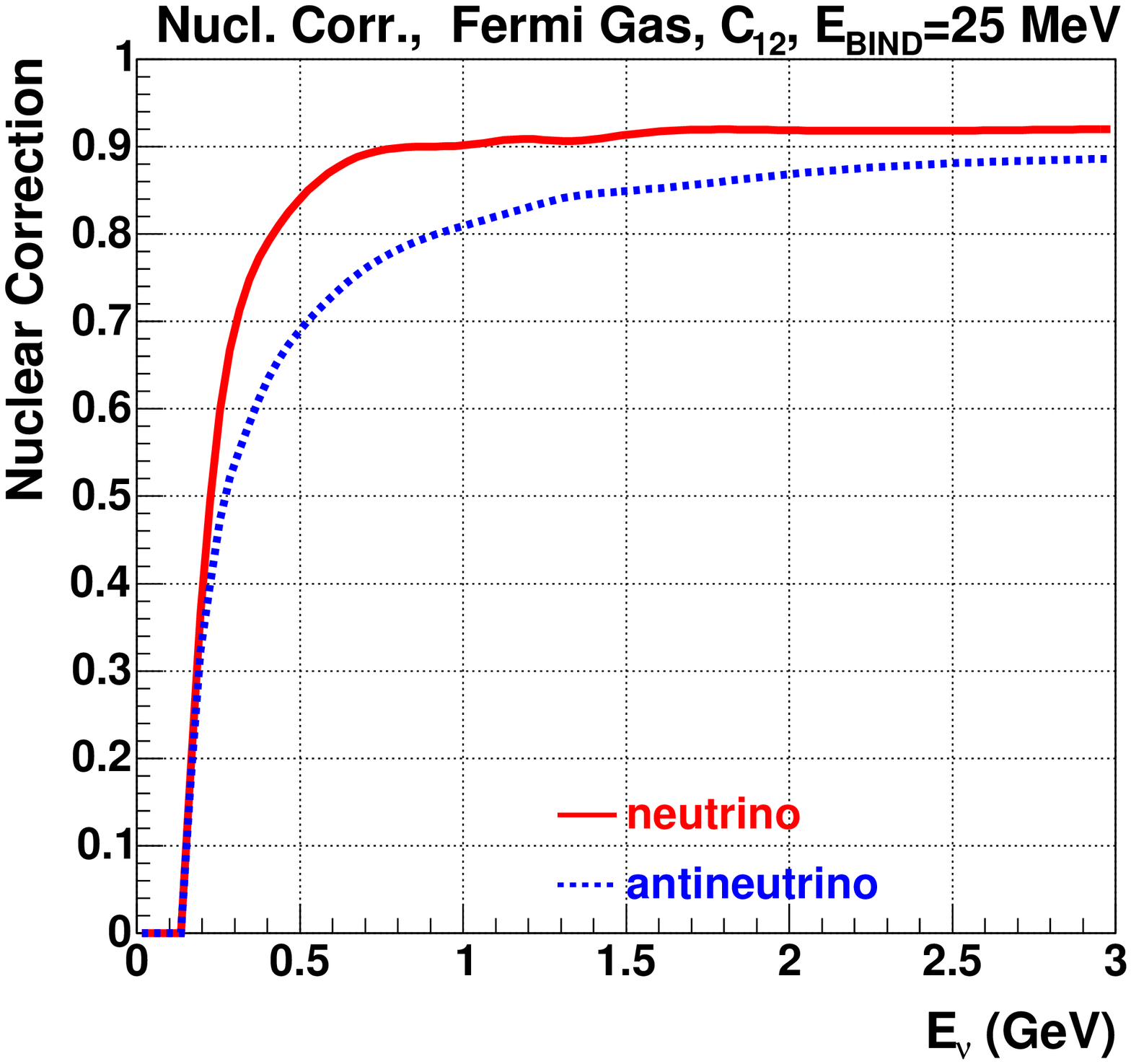,width=2.91in}}

\caption[Suppression of bound cross-sections in Fermi gas model]{Pauli
suppression in
a Fermi gas model for Carbon with binding energy $\epsilon=25~\hbox{MeV}$ and Fermi momentum
$k_f = 220~\hbox{MeV/c}$.  A similar suppression is
expected for quasi-elastic reactions in \minerva. }
\label{pauli}
\end{figure}

\begin{figure}[htb!]
\centerline{\psfig{figure=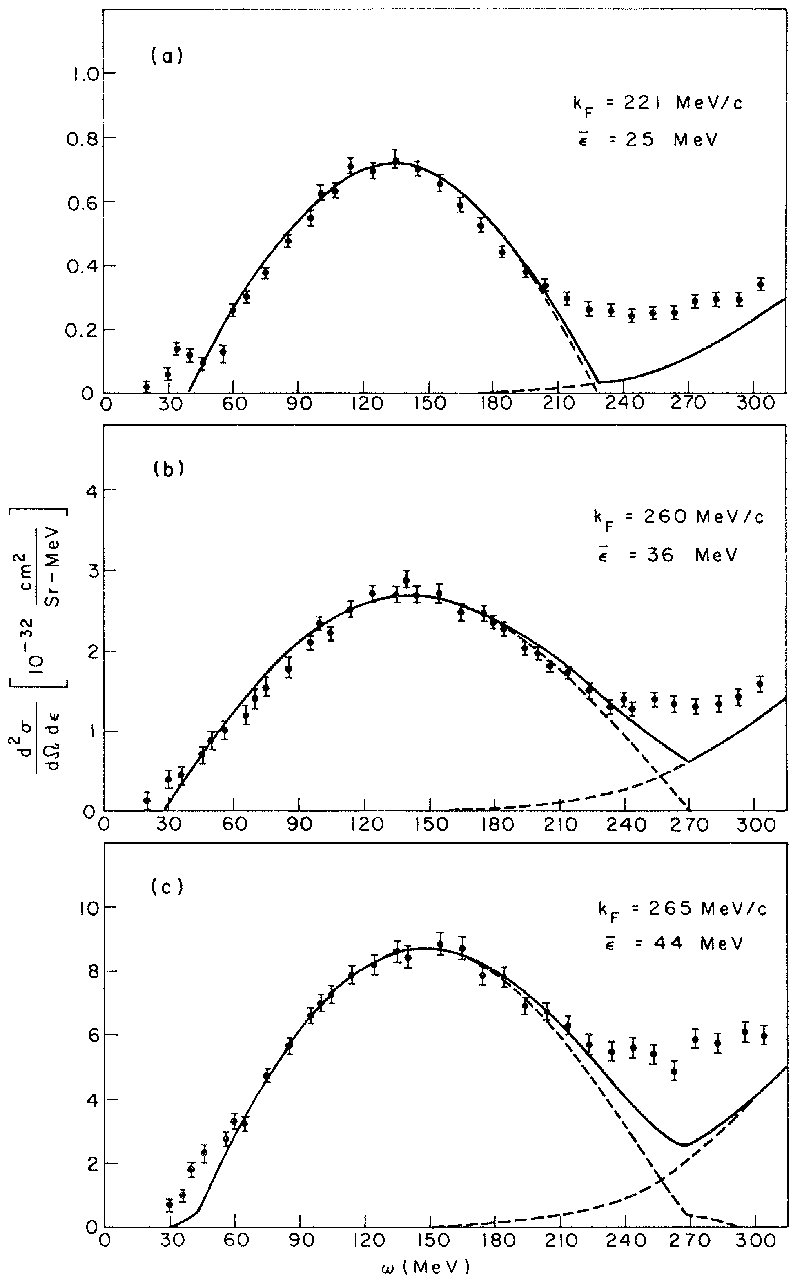,width=2.91in}}
      \caption[Fermi gas model parameters extracted from
electron-scattering data]
      {Extraction of Fermi gas model parameters,
      the effective Fermi momentum $k_f$ and nuclear
      binding energy $\epsilon$,
      from 500~MeV electron scattering data\cite{Smith_72}.
      Distributions shown correspond to scattering from (a) Carbon, (b) Nickel, and (c) Lead.}
\label{monizPRL}
\end{figure}

There are three important nuclear effects in
quasi-elastic scattering from nuclear targets:
Fermi motion, Pauli blocking, and corrections to the nucleon form factors due
to distortion of the nucleon's size and its pion cloud in the nucleus.
Figure~\ref{pauli} shows the nuclear suppression versus $E_\nu$ from
a NUANCE\cite{Casper_02} calculation\cite{Zeller_03} using the
Smith and Moniz\cite{Smith_72} Fermi gas model
for Carbon. This nuclear model includes Pauli blocking and Fermi
motion but not final state interactions.
The Fermi gas model uses a nuclear
binding energy $\epsilon = 25~\hbox{MeV}$ and
Fermi momentum $k_f=220~\hbox{MeV/c}$.
Figure~\ref{monizPRL} from Moniz et. al.\cite{Smith_72}
shows how the effective $k_f$ and nuclear
potential binding energy $\epsilon$
(within a Fermi-gas
   model) for various nuclei is determined from
   electron scattering data. The effective $k_f$
   is extracted from the width of the scattered electron
   energy distribution, and the binding energy $\epsilon$ 
from the shifted location of the quasi-elastic peak.

\subsubsection{Bound nucleon form-factors}

\begin{figure}[htb!]
\centerline{\psfig{figure=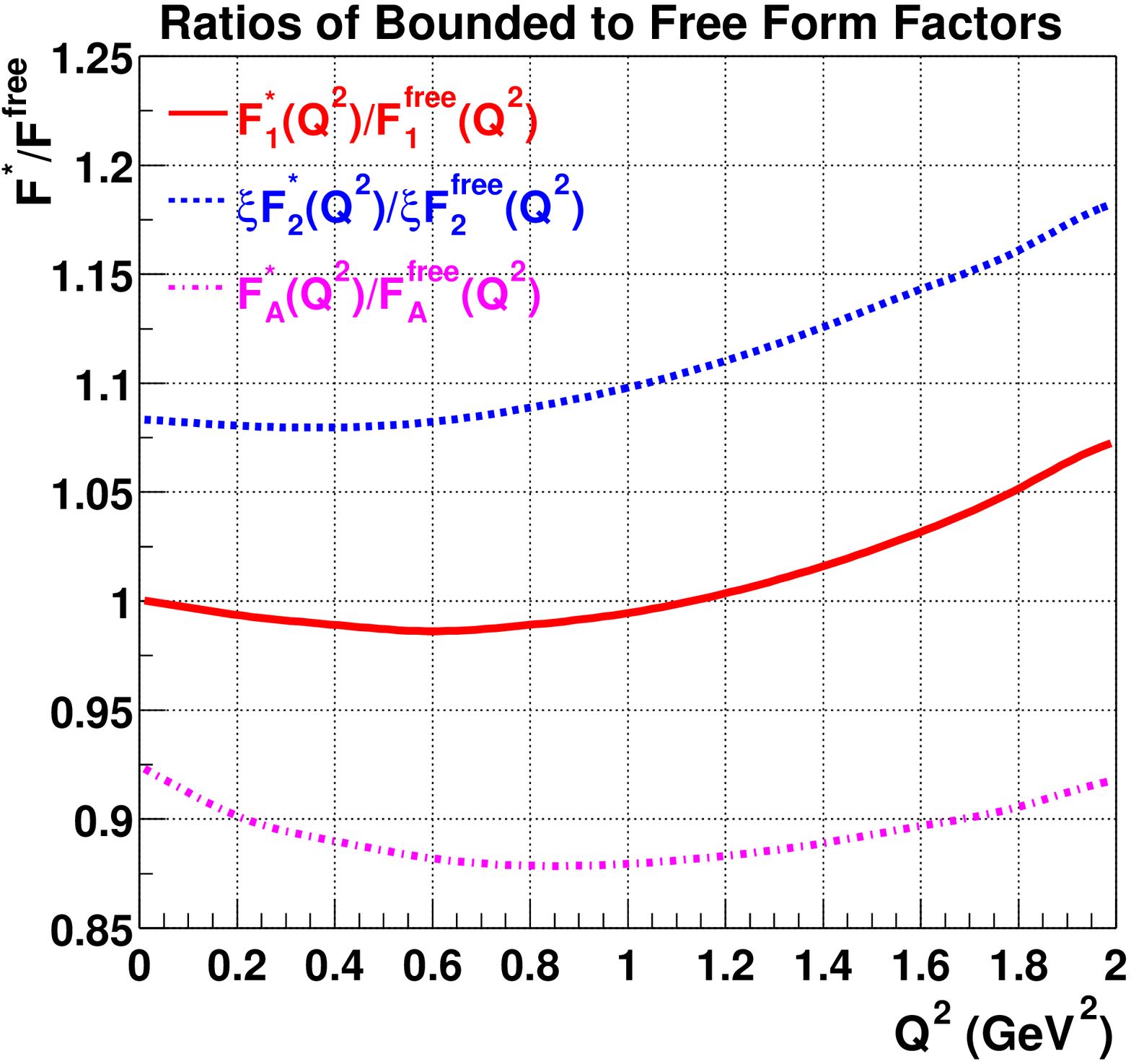,width=2.91in}}
\caption[Ratio of bound to free nucleon form-factors $F_1$, $F_2$, and $F_A$]{
The ratio of bound (in Carbon) to free nucleon
form-factors for $F_1$, $F_2$, and $F_A$ from ref ~\cite{Tsushima_03}.
Binding effects on the form factors
are expected to be small at higher $Q^2$ (therefore, this
model is not valid for $Q^2 > 1~\hbox{(GeV/c)}^{2}$).}
\label{binding}
\end{figure}

The predicted distortions of nucleon form-factors due to nuclear binding are shown in
Figure~\ref{binding} as the ratios of $F_1$, $F_2$, and $F_A$ for bound and free nucleons.  With
a variety of nuclear targets, \minerva\ will be able to compare measured form-factors for a range of
light to heavy nuclei.

\subsubsection{Intra-nuclear rescattering}

In neutrino experiments, detection of the recoil nucleon helps
distinguish quasi-elastic scattering from inelastic reactions.
Knowledge of the probability for outgoing protons to reinteract
with the target remnant is therefore highly desirable.
Similarly, quasi-elastic
scattering with nucleons in the high-momentum tail
of the nuclear spectral function needs to be understood.
More sophisticated treatments than the simple
Fermi gas model are required.  Conversely,
inelastic reactions may be misidentified
as quasi-elastic if a final-state pion is absorbed in
the nucleus.  Because of its constrained kinematics, low-energy
neutrino-oscillation experiments use the quasi-elastic channel
to measure the (oscillated) neutrino energy spectrum at the
far detector; the uncertainty in estimation of this background due
to proton intra-nuclear rescattering is currently an important source of systematic error
in the K2K experiment.

The best way to study these effects is to analyze
electron scattering on nuclear targets (including
the hadronic final states) and test the effects of the
experimental cuts on the final-state nucleons.
\minerva\ can address proton intra-nuclear rescattering
by comparing nuclear binding effects in
neutrino scattering on Carbon to electron data in similar kinematic regions.
Indeed, \minerva\ members
will be working with the CLAS collaboration to
study hadronic final states in electron scattering
on nuclear targets using existing JLab Hall~B data.
This analysis will
allow theoretical models used in both electron and neutrino
experiments to be tested.
Other work in progress, with the Ghent\cite{belgium} nuclear physics group,
will develop the theoretical tools needed to extract
the axial form-factor of the nucleon using \minerva\ quasi-elastic data on Carbon.
The ultimate aim is to perform nearly identical analyses on both neutrino and
electron scattering data in the same range of $Q^{2}$.

%


%% file: resonant.tex
\section{Resonance-Mediated Processes}
\label{sect:resonant}

\begin{figure}[hbtp!]
\centerline{
{\includegraphics[width=\textwidth, bb=25 170 535 670]{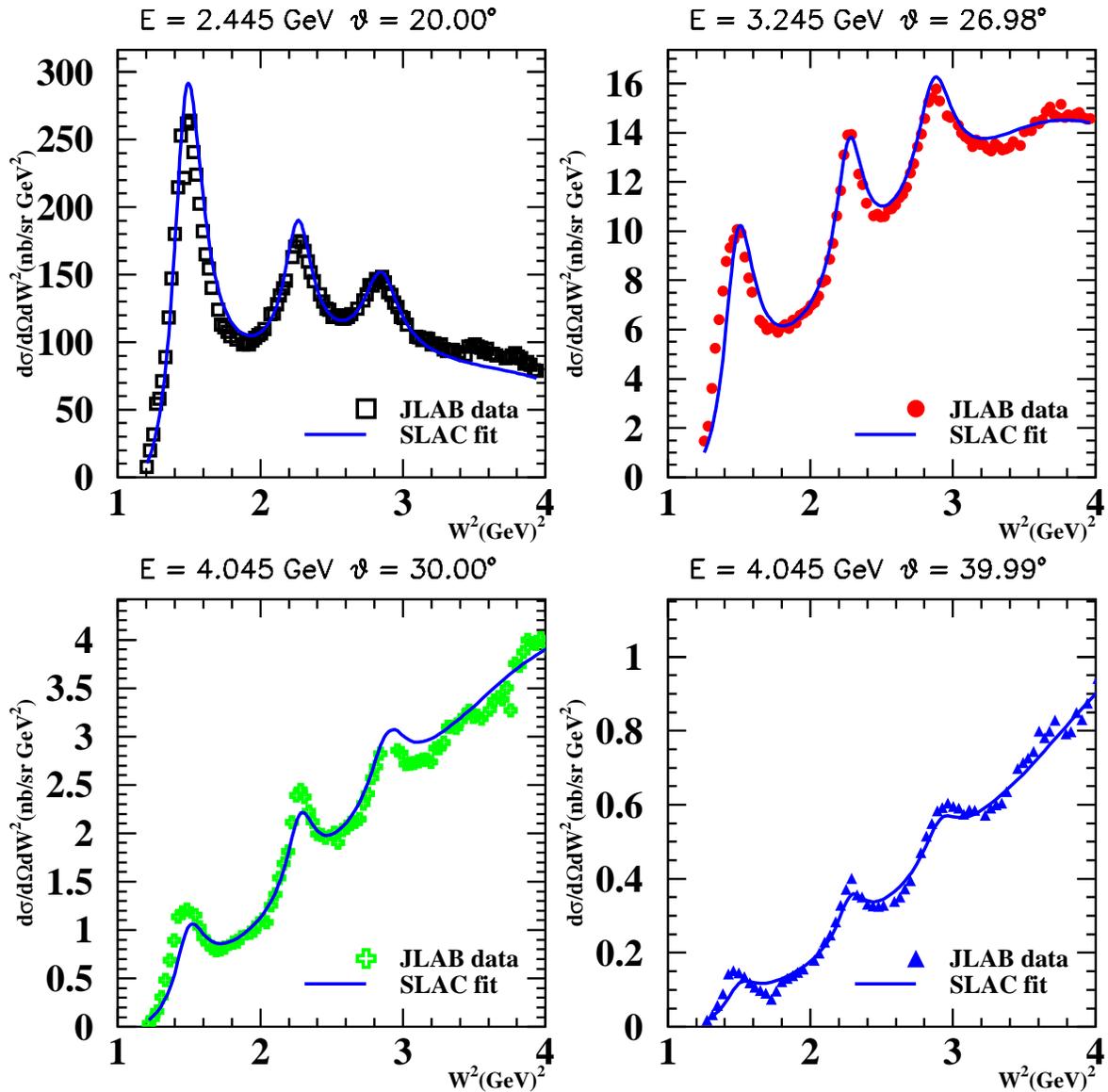}}}
\caption[Inclusive electron scattering showing the $\Delta$ and other resonances]{Inclusive electron scattering showing the $\Delta$ and higher resonances.
$Q^2$ at the $\Delta$ peak is approximately 0.5, 1.5, 2.5 and $3.5~({\rm GeV}/c)^2$ for the four spectra, respectively}
\label{resonant:eeres}
\end{figure}

Inclusive electron scattering cross-sections with hadronic mass $W<2~\hbox{GeV}$ exhibit
peaks corresponding to the $\Delta(1232)$ and
higher resonances at low $Q^2$ (see Figure~\ref{resonant:eeres}).  This resonant structure is also present
in neutrino scattering, although there is little data in this
region.  In addition to the natural interest in probing the nucleon weak current and
axial structure via neutrino-induced
resonance production, a better understanding of this process is essential for interpreting
modern neutrino-oscillation and nucleon-decay experiments.  This is particularly true for neutrinos
in the region around 1~GeV, where single-pion production comprises about 30\%
of the total charged-current cross-section.

In this kinematic region, neutrino Monte Carlo programs have relied on
early theoretical predictions by Rein and Sehgal\cite{reinsehgal}.
Recently Sato, Uno and Lee\cite{satounolee} have extended a model of
$\Delta$-mediated pion electroproduction to neutrino reactions.  Also,
Paschos and collaborators, using the formalism of Schreiner and von
Hippel\cite{shreinervonhippel} have included the effects of pion
rescattering and absorption for resonance production in nuclei.

\subsection{Overview of Resonant Electroproduction}

In electron scattering, the behavior of the $\Delta(1232)$ transition
form-factor is considered to be a primary indicator of the onset of 
perturbative QCD (pQCD). 
The $Q^2$ behavior expected for a resonant spin-flip transition is dramatically different
from the helicity conservation characteristic of perturbative descriptions.
Comparison of the measured elastic and resonant form-factors
reveals\cite{rcar,rsto,rstlr,rgast,rgakr,HallC} that while the nucleon and higher-mass resonant
form-factors appear to approach the predicted $Q^{-4}$ leading-order pQCD behavior
around $Q^2 = 2~\hbox{(GeV/c)}^2$, the $\Delta(1232)$ transition form-factor decreases more rapidly with $Q^2$.
One possible explanation\cite{rcar} is that helicity-nonconserving processes are dominating.
The $\Delta$ excitation is primarily a spin-flip transition at low momentum transfer,
in which the helicity-nonconserving $A_{3\over2}$ amplitude is dominant\cite{rrmdav}.
If the leading order $A_{1\over2}$ helicity-conserving amplitude were also supressed at large
momentum transfers, the quantity $Q^4F$ would decrease as a function of $Q^2$. 

Electromagnetic helicity matrix-elements correspond to transitions in which
the initial state has helicity $\lambda$ and the final states have helicity $\lambda'$.
Transitions between a nucleon state $|N>$ and a resonant
state $|R>$ can be expressed in terms of dimensionless helicity
matrix-elements\cite{rcar}:

\begin{equation} 
G_{\lambda} =
{{1\over{2M}}<R,\lambda'|\epsilon^{\mu}J_{\mu}|N,\lambda>}
\end{equation}

\noindent In this equation, the polarization vectors $\epsilon^{+,-,0}$
correspond
to right- and left-circularly polarized photons, and longitudinally
polarized photons, respectively.  Following the formalism used by
Stoler\cite{rsto} and others, the differential cross-section
may be written in terms of longitudinal and transverse
form-factors $G_E$ and $G_T$, as follows:

\begin{equation}
{d^2 \sigma \over d\Omega dE' } = \sigma_mf_{rec}
\left[ {{|G_E|^2 + \tau^*|G_T|^2}\over{1+\tau*}}
+ 2\tau^*|G_T|^2 \tan^2\left({\theta \over 2}\right)\right] R(W)
\end {equation}

\noindent $G_E$ and $G_T$ are analogous to the Sachs form-factors for elastic scattering. In
terms of the dimensionless helicity elements above,

\begin{equation}
G_E = G_0
\end{equation}
and
\begin{equation}
\tau^*|G_T|^2 = {1\over2}(|G_+|^2 + |G_-|^2)
\end{equation}
where
\begin{equation}
\tau^* = {{\nu^2}\over{Q^2}}
\end{equation}
\noindent The recoil factor $f_{rec}$ is given by
\begin{equation}
{f_{rec}} = {{E'}\over {E_{\circ}}}
\end{equation}
\noindent $R(W)$ is the familiar Breit-Wigner
expression\cite{rbwf} for the line-shape as a function of energy:
\begin{equation}
R(W) = {{2\pi^{-1}W_RM\Gamma_R}
\over{(W^2-W_R^2)^2+W_R^2\Gamma_R^2}}
\end{equation}
\noindent The mass and width of the resonance are $W_R$ and $\Gamma_R$. 
 
Helicity is conserved in vector interactions of free, relativistic fermions.
In the limit that a spin-$1\over 2$ parton is massless and free, its
helicity must be conserved in interactions with a vector gluon or photon.
At sufficient momentum transfer, the constituent quarks within a
hadron can indeed be treated as massless and free, and the hadron helicity can be
replaced by the sum of its constituent quark helicities\cite{rral,rfarr}.
Therefore, at high $Q^2$, hadron helicity should also be conserved.

For resonant electroproduction, the scattering can be analyzed in the 
Breit frame of the $\lambda = 3/2$ $\Delta$ resonance.
The incoming virtual photon can have positive, zero, or negative helicity.
The outgoing resonance helicity can be calculated from angular momentum
conservation\cite{rcarl}:
\begin{equation}
\lambda_{\Delta} = \lambda_{\gamma}- \lambda_N
\end{equation}
Hadron helicity is conserved when the incoming photon helicity is positive,
and the $\Delta$ excitation
emerges with the same helicity (1/2) as the initial
nucleon state. This is described by the helicity amplitude $A_{1\over2}$
given by:
\begin{equation}
A_{1\over2} = \sqrt{{2\pi\alpha}\over{\kappa}}G_+
\end{equation}
$\kappa$ is the energy of an
equivalent on-mass-shell (real) photon producing a
final mass state W:
\begin{equation}
\kappa = {(W^2-M^2)/{2M}}
\end{equation}
Helicity is not conserved when $A_{3\over2}$, given by
\begin{equation}
A_{3\over2} = \sqrt{{2\pi\alpha}\over{\kappa}}G_-
\end{equation}
is the dominant amplitude.
 
In terms of helicity amplitudes a dimensionless form-factor F may be defined 
where:
\begin{equation}
F^2 = |G_T(Q^2)|^2 = {1\over{4\pi\alpha}}{{2M}\over{Q^2}}
(W_R^2 - M_N^2)|A_H(Q^2)|^2
\end{equation}
Here,
\begin{equation}
|A_H(Q^2)|^2 = |A_{1\over2}(Q^2)|^2 + |A_{3\over2}(Q^2)|^2
\end{equation}
At high $Q^2$, the helicity
conserving amplitude should
dominate the helicity-nonconserving amplitude.
$A_{3\over2}$ should be small compared to $A_{1\over2}$
according to pQCD.
  
In leading-order pQCD, two gluons are exchanged among the three pointlike
quarks. These gluon exchanges ensure that the final quarks, like the initial
ones, have low relative momenta, so that no powers of $Q^2$ come from the
wave functions.
Form-factors calculated in the light-cone frame take the form \cite{rsto}:
\begin{equation}
F(Q^2) = \int^{1}_{0}\int^{1}_{0}dxdy\Phi(x)^*T_H\Phi(y) 
\end{equation}
where $x$ and $y$ are the initial and final longitudinal momentum fractions.
$\Phi(x)$ and $\Phi(y)$ are the corresponding quark distribution amplitudes
and $T_H$ is the transition operator which is evaluated over
all possible leading-order diagrams.
This leads to the
dimensional scaling rule\cite{rble}
\begin{equation}
G_+\propto A_{1\over2}\propto Q^{-3}, 
\end{equation}
or
\begin{equation}
F\propto Q^{-4}
\end{equation}
 
This $Q^2$ dependence of the helicity
amplitudes may be established up to factors involving $\ln (Q^2)$\cite{rvzak}.
At high $Q^2$, where the quark helicities are conserved,
\begin{equation}
G_+ \propto Q^{-3}
\end{equation}
\begin{equation}
G_0 \propto ({m \over Q})G_+
\end{equation}
and
\begin{equation}
G_- \propto ({{m^2}\over{Q^2}})G_+
\end{equation}
The prediction that $F(Q^2)\propto{1/{Q^4}}$ if $G_+$ is
dominant can be understood by combining the above with the definitions
of $A_{3\over2}$ and $A_{1\over2}$ in the dimensionless form-factor expression.

In addition to this $Q^2$ dependence of the transition form-factors,
pQCD makes definite predictions about the relative contributions of 
the magnetic dipole $M_{1+}$, electric quadrupole $E_{1+}$, and Coulomb
quadrupole $S_{1+}$ amplitudes. In quark models at low $Q^2$, 
the ${\rm N}-\Delta$ transition is primarily due to a single quark spin-flip,
requiring the $M_{1+}$ to be the dominant contribution\cite{becchi}.
At very low $Q^2$, near zero, experiments have confirmed this prediction, 
evaluating $E_{1+}$ and $M_{1+}$ at the resonance position. However, as 
noted, only helicity-conserving amplitudes should contribute 
at high $Q^2$, which leads to the prediction that the ratio 
$E_{1+} / M_{1+} = 1$. Results from Jefferson Lab\cite{HallC} indicate that
hadron helicity is not yet conserved at $Q^2 = 4$ GeV$^2$, finding 
the transition form-factor $F$ to be decreasing faster than $Q^{-4}$ and
continued $M_{1+}$ dominance. However, while
pQCD apparently does not yet describe resonant excitation at these momentum 
transfers, it is not clear how constituent 
quark models can be appropriate at such high $Q^2$ values, and regardless,
no single model describes all of the data well. The Delta resonance, then,
remains an object of intense study at facilities like Jefferson Lab and Mainz,
with future experiments planned.

\subsection{Weak Resonance Excitation}

Sato and Lee\cite{satolee} have developed a dynamical model for pion photo- and
electroproduction near the $\Delta$ resonance which is used to
extract ${\rm N}-\Delta$ transition form-factors. Through this work, the
afore-mentioned discrepenacy between the $\Delta$ transiton form-factor as
calculated from a constituent quark model and the measured transition
form-factor (a difference of about 35\%) has been understood
by including a dynamical pion cloud effect. Recently this work has been
extended by Sato, Uno and Lee to weak pion production\cite{satounolee}.
They show that the renormalized axial ${\rm N}-\Delta$ form-factor
contains large dynamical pion cloud effects which are crucial in obtaining
agreement with the available data (in this case, on Hydrogen and Deuterium). 
Contrary to previous observations,
they conclude that the ${\rm N}-\Delta$ transitions predicted by the
constituent quark model are consistent with existing neutrino-induced pion 
production data in the $\Delta$ region. It is interesting to note that the 
pion cloud effect on the axial  ${\rm N}-\Delta$ form-factor is mainly to 
increase the magnitude. On the other hand, both the magnitude and the slope
of the $M_{1+}$ are significanlty changed by including pion cloud effects. 
The authors cite the need for more extensive and precise data on 
neutrino-induced pion-production reactions to test their model and to pin down the 
$Q^2$-dependence of the axial-vector ${\rm N}-\Delta$ transition form-factor -
data which {\minerva} can certainly provide. 

{\minerva} will measure scattering on nuclei, at least in the first years without 
a hydrogen target, and comparison to
improved data on a free proton target will not be possible. Still,
as discussed in Section~\ref{sect:duality}, the average 
$Q^2$ dependence of the cross-sections (and, hence, structure functions and
form-factors)
will be magnified by the Fermi smearing of the resonant enhancements.
It should be possible to map out the 
$Q^2$-dependence of the axial-vector ${\rm N}-\Delta$ form-factor.
The work of Sato, Uno and Lee can be used as Monte Carlo
input for {\minerva}, and should be essential to predictions
of $\Delta$ excitation in nuclei which can be 
compared directly with {\minerva} data.

%



\subsection{Nuclear Effects}

Neutrino experiments rely heavily on detailed Monte Carlos to simulate
the response of the rather complicated target / detector systems involved. 
The {\minerva} simulation will be greatly enhanced by accurate descriptions 
of the nuclear effects involved. The majority of hadrons produced in 
inelastic scattering are pions, and so the nuclear attenuation of these must
be taken into account. In considering hadron attentuation results from 
HERMES, Gaskell\cite{dave} suggests that a good first step is the one time 
scale parameterization, which goes as $(1-z)\nu$. The A-dependence could then
be taken into account via a simple $A^{2/3}$ scaling in $(1 - R_A)$, where
$R_A$ is the ratio of cross-section on nucleus A to deuterium.

Another relevant nuclear effect, currently being applied in neutrino event 
generators  for protons but not pions, is termed color 
transparency (CT). Color transparency, first conjectured by 
Mueller and Brodsky~\cite{ct1} refers to the suppression of final 
(and initial) state interactions of hadrons with the nuclear medium in 
exclusive processes at high momentum transfers. 
CT is an effect of QCD, related to the presence of
non-abelian color degrees of freedom underlying strongly interacting matter. 
The basic idea is that, under the right conditions, three quarks (in the case
of the proton), each of which would normally interact strongly with the nuclear
medium, can form an object that passes undisturbed through the nuclear
medium. This small object would be color neutral outside of a small radius in 
order not to radiate gluons. Unambiguous observation of CT would provide a 
new means to study the strong interaction in nuclei.

Several measurements of the transparency of the nuclear medium to high energy 
protons have been carried out in the last decade. 
At Jefferson Lab, CT searches have 
concentrated on the quasi-elastic $A(e,e^{\prime}p)$ reaction which has several 
advantages in the search for CT. To date, A(e,e'p) experiments at 
SLAC~\cite{ne18} and JLab~\cite{ct4} have found no evidence for the onset of 
CT at momentum transfers up to $8.1~{\rm (GeV/c)}^2$.  However, there is some 
potential evidence for CT in $A(p,2p)$ data from Brookhaven \cite{BNL1,BNL2}.

It has been suggested 
that the onset of CT will be sooner in a $q\bar{q}$ system than in a 
three-quark system. Thus, the next best reaction in the expectation
of CT is the $A(e,e^{\prime}\pi)$ reaction.  Current theoretical 
calculations suggest that most of this CT effect should be seen around 
$Q^2 = 10~({\rm GeV}/c)^2$, well within the {\minerva} kinematic range. This
effect has not yet been considered in neutrino Monte Carlos, nor has it
been well studied in other processes. However, it will be well-measured in the
Jefferson Lab kinematic regime prior to {\minerva}\cite{dutta}, and
should then be incorporable into the Monte Carlo. 

\subsection{Exclusive Channels}

While there is a large body of inclusive $(e,e^{\prime})$ scattering data in the
resonance region on
hydrogen, deuterium and nuclei, more exclusive measurements have been
rare until recently.  With JLab $p(e,e^{\prime}p)\pi^0$ spectrometer
measurements\cite{rsto,HallC}, the CLAS $N^*$ program\cite{nstarclas} and
CLAS ${}^{12}{\rm C}(e,e^{\prime}X)$ data, more exclusive
reactions are becoming available.  This data will help to ``calibrate''
the vector current part of weak resonance/meson production models and to
extend Delta resonance models such as that of Sato, Uno and Lee to higher
resonances.  These exlusive measurements are also naturally of interest
because even to make inclusive measurements with neutrinos, the full final
state must be observed and reconstructed.  With the expected statistics
and resolution of {\minerva}, it should thus be possible to extract much more
information about resonances than what is available in the inclusive channel.

\begin{figure}[h]
\centerline{
{\includegraphics[height=0.9\textwidth,angle=-90]{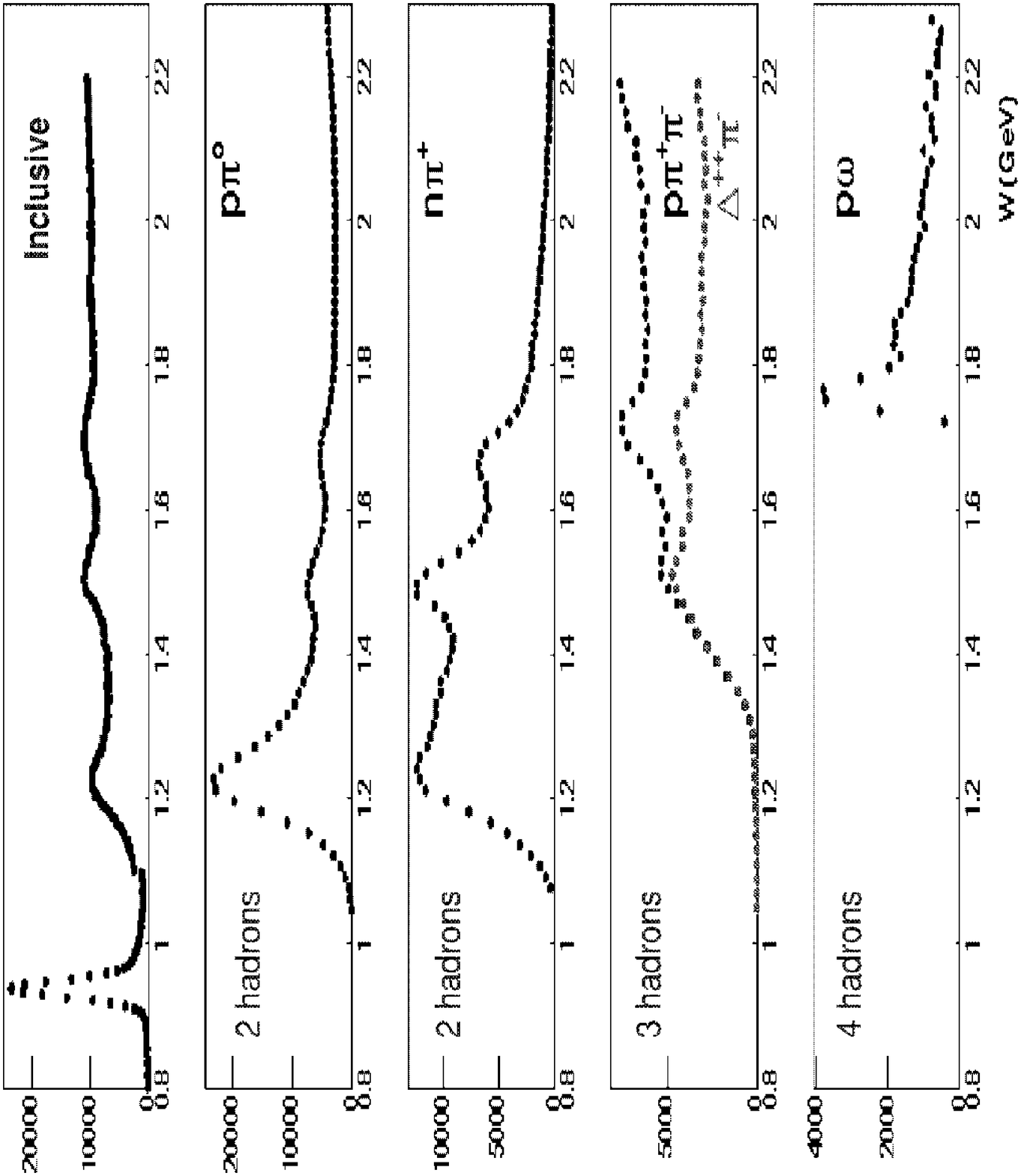}}}
\caption[Invariant-mass spectra from $p(e,e^{\prime}X)$]{Invariant mass spectra from $p(e,e^{\prime}X)$ demonstrating the
multi-hadron reconstruction capability in the JLab CLAS spectrometer.\cite{brooks}}
\label{resonant:brooksmultihad}
\end{figure}

Figure~\ref{resonant:brooksmultihad} from the CLAS\cite{brooks} is an
illustration of the type of just part of the information available when
one or more reaction fragments are detected in resonance region electron
scattering.  One item of interest in this data is a peak observed near
$W=1.72~{\rm GeV}$ in the spectrum for the $p \pi^+ \pi^-$ final state.  While
an analysis of the angular distribution of this peak gives quantum numbers
that agree with the PDG $N^*_{3/2^+} (1720)$ state, the observed hadronic
properties (coupling amplitudes) of this resonance are quite different
from what is predicted from the PDG state.  This
illustrates that electro-weak excitiation of baryon resonances is an
active field and that {\minerva} measurements are timely.

\begin{figure}[h]
\centerline{
{\includegraphics[width=0.45\textwidth]{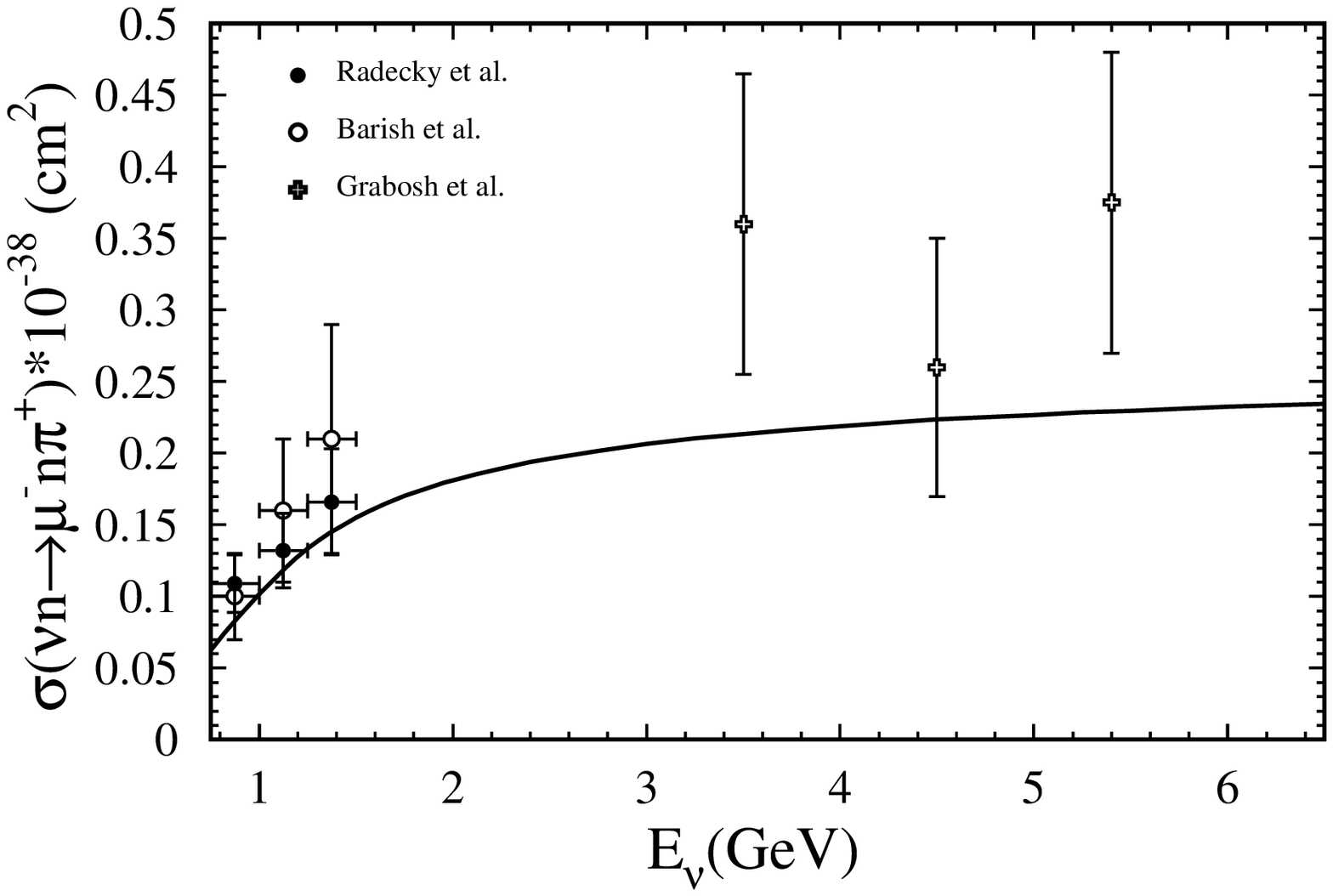}}
{\includegraphics[width=0.45\textwidth]{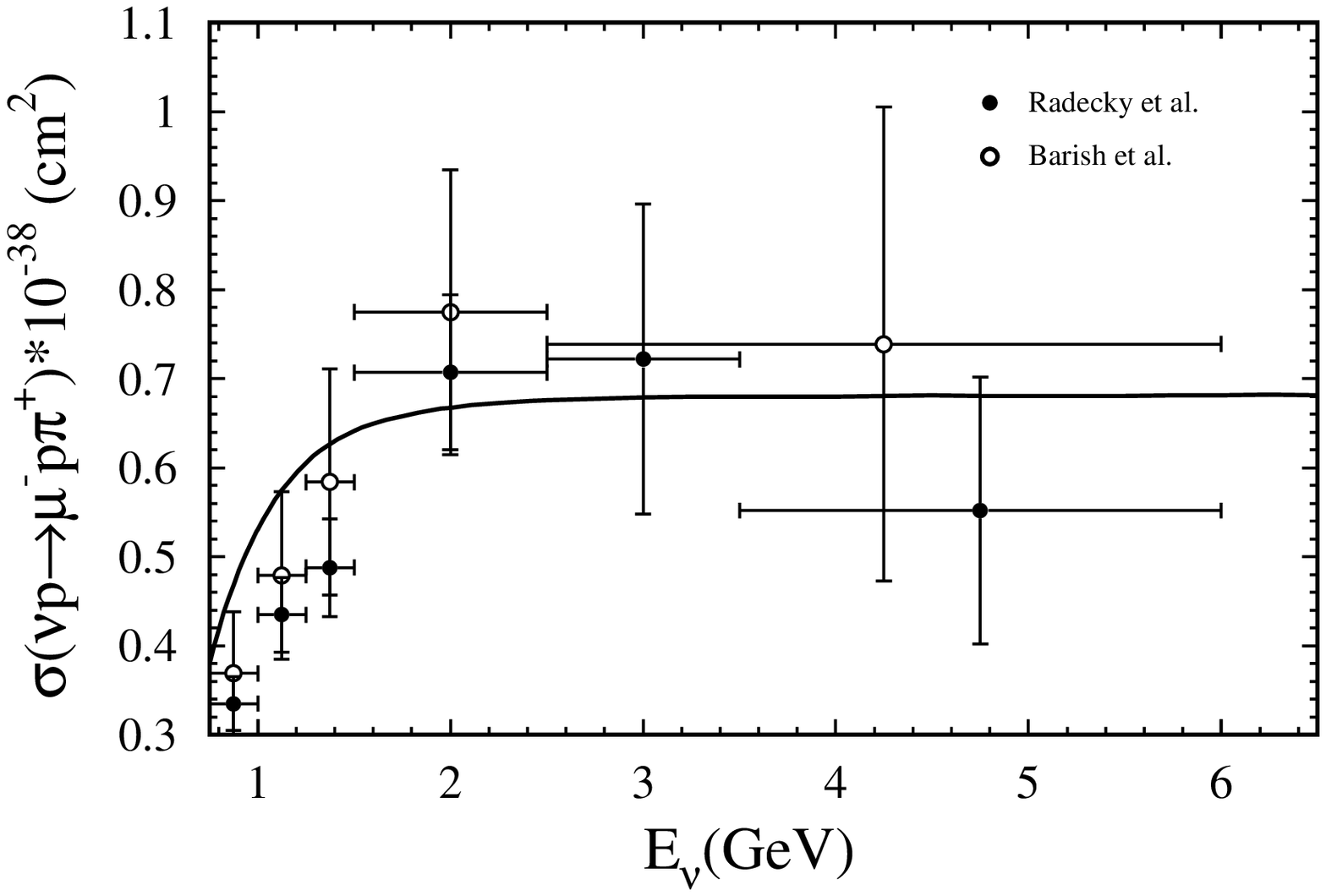}}}
\caption{Total pion production cross-sections.}
\label{resonant:totalpi}
\end{figure}

\begin{figure}[h]
\centerline{
{\includegraphics[width=0.8\textwidth]{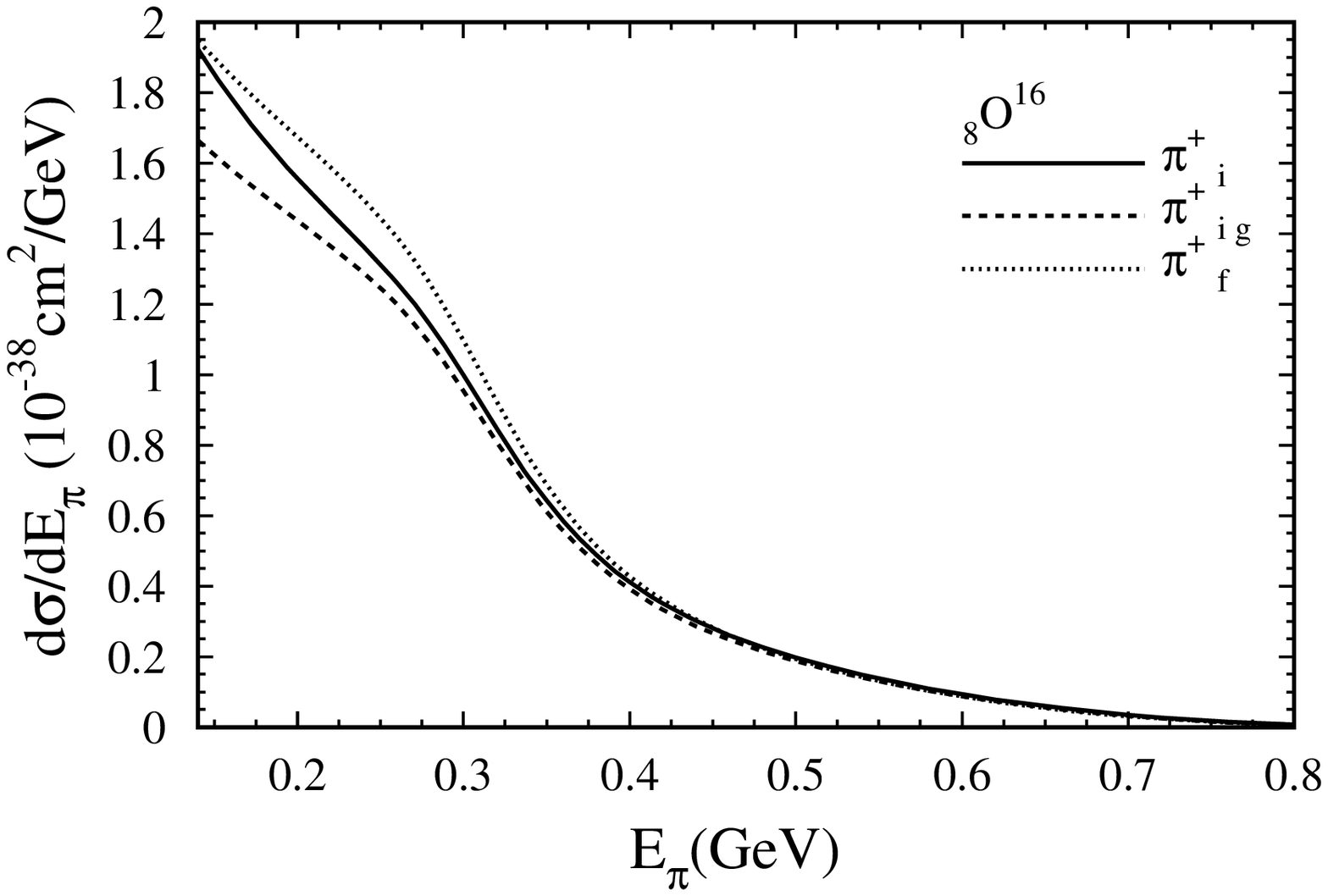}}}
\caption[$\pi^+$ energy distribution for $\nu_\mu$--O scattering]{Predicted $\pi^+$ energy distribution for $\numu$ scattering on ${}^{16}{\rm O}$ of Paschos et. al.\cite{Paschos}.}
\label{resonant:paschos}
\end{figure}

\subsection{Expected Results}

As shown in Tables~\ref{tab:channelRates} and~\ref{tab:xq2res}, over
250,000 resonance production events, with useful statistics to at least
$Q^2 = 4~({\rm GeV}/c)^2$, are estimated for 4 years of running.
Approximately 40\% of these events have $W < 1.4~\hbox{GeV}$, so a
good mix of events mediated by the $\Delta(1232)$ and higher resonances
will be obtained.  
The resonance production measurements in {\minerva} can be grouped
into several categories:

\begin{enumerate}
\item Measurement of inclusive $(\numu, \mu^{-})$ and $(\nu, \nu)$
  spectra in the resonance
  region:
  As is done in the deep-inelastic region,
  this implies extraction of structure functions which can
  be compared to structure functions and form-factors from electron
  scattering. The  experimental method is the
  same as for DIS events.  For each event, we sum up the energies and
  momenta of the muon and all final state hadrons (either pions and
  nucleons, or all nucleons if the pions from resonance decay get absorbed
  on their way out of the nucleus) to get the total
  neutrino energy, and calculate $Q^2$ and $W$, $y$ etc. This
  kind of analysis will be done as a function of $W$
  for the entire resonance region.

  Although these measurements are done on a nucleus, we will be able to
  compare results to resonance production predictions (such as
  Rein Seghal\cite{reinsehgal}) on nucleons, with some guidance from
  electron scattering.  Because inclusive measurements are a sum over all
  final states, nuclear effects should be primarily limited to
  Fermi motion and some Pauli Blocking.   Despite the Fermi motion and
  resolution of {\minerva}, the Delta resonance will still be clear so
  it's form factor
  as a function of $Q^2$ can be measured.  The
  higher resonances will be smeared out, but still can be compared to
  "smoothed" behaviour of resonance models and to predictions from
  duality.  One practical result of measurements above the Delta resonanance
  may be to modify the amount of
  non-resonant background in the resonance region models used in neutrino
  event generators.

\item Examination of specific final state reaction products (single pion
production,
  inclusive pion spectra):  Specific final states, through the reactions:
  $(\numu, \mu^- p \pi^+)$, 
  $(\numu, \mu^- n \pi^+)$, 
  $(\numu, \mu^- p \pi^0)$, 
  $(\numu, \numu^- n \pi^0)$, 
  $(\numu, \numu^- p \pi^0)$, and
  $(\numu, \numu^- n \pi^+)$, 
  are useful in
  selection of a specific final state isospin.

  These final state measurements will rely on
  an improved understanding of final state interactions and will
  benefit from electron scattering hadron transparency studies and CLAS
  ${}^{12}{\rm C}(e,e^{\prime}X)$ data (which includes
  $(e, e^{\prime} p \pi^0)$, and $(e, e^{\prime} n \pi^+)$, and
  which are
  equivalent to two of the above neutrino reactions).  With these inputs, we
  will be able to map out the $Q^2$ dependence of the Axial vector ${\rm N}-\Delta$
  form factor.
  But even without this better understanding, angular distributions should
  be less affected by final state interactions than overall cross
  sections.
  Thus, we will
  be able to extract ratios of weak transition amplitudes to compare to
  similar electron scattering amplitudes

  As measurement of detailed angular distributions of these final states
  is possible, the data on nuclear targets can also be used to study the
  feasibility of doing a phase shift analysis of the data if a hydrogen
  target is used in later phases of this experiment.  This phase shift
  analysis, recommended by Sato, Uno and Lee\cite{satounolee}, like the
  JLab CLAS $N^*$ program, would be aimed at extracting the $N-\Delta$ form
  factor model independently and providing a better understanding of
  low-lying nucleon resonances.
\item Using resonance production as a tool to study final state
  interactions:
  Having a selection of nuclear targets helps here as the $A$
  dependence of the various reactions channels listed above can be
  studied.  Another analysis that can be done along these lines is to
  measure inclusive pion spectra.  Paschos et.al.\cite{Paschos}
  combine resonance
  production and final state interactions to make predictions of pion
  spectra from neutrino scattering on nuclei.
  These spectra (Fig.~\ref{resonant:paschos} can be easily convoluted with
  neutrino beam energy distributions to
  produce pion energy distributions that can be directly compared with our
  data.

\end{enumerate}

\subsubsection{Complementary studies at JLab}

The analysis of the above types of measurements
will be closely coordinated with complementary experiments
at Jefferson Laboratory (which are led by members of the
{\minerva} collaboration). The following are the Jefferson
Laboratory electron scattering
experiments in Hall C
that are connected with measurments
of  inclusive scattering in the resonance region at {\minerva}.

\begin{enumerate}
\item JLab hydrogen experiment E94-110 (investigates inclusive $F_2$ and
$R$
    in the resonance region). C.E. Keppel spokesperson (data already
    taken).
\item JLab deuterium  experiment E02-109,
investigates inclusive $F_2$ and $R$
    in the resonance region. C.E. Keppel, M. E. Christy,
spokespersons (approved to run in 2004).
\item JLab experiment E99-118 investigates nuclear the dependence
of $F_2$ and  $R$ at low $Q^2$
for high values of $W$. A. Brull, C.E. Keppel spokespersons (data already
taken).
\item Jlab experiment E02-103 hydrogen and
deuterium resonance $F_2$ data at  high $Q^2$ approved by Jlab PAC24 to
run in 2004
(J. Arrington, spokesperon)
 \item Jlab experiment E04-001 to investigate $F_2$ and $R$
    in the resonance region with nuclear targets.  A. Bodek and C. E.
    Keppel, spokespersons  (proposed to run in  Hall C together
    with E02-109 in 2004) to provide vector resonance form
    factors and $R$ on the same nuclear targets that are used in
    neutrino experiments (e.g. Carbon, Iron, Lead).
\end{enumerate}

 The following are collaborative programs between
 the electron scattering community that are connected with
measurements of final states in the quasielastic region
and in the region of the first resonance  at {\minerva}.

\begin{enumerate}
\item Steve Manly (Rochester) and Will Brooks (Jlab) program to use
    existing Hall B CLAS data at Jefferson Laboratory to
    study hadronic final states in electron scattering
    on nuclear targets
     (e.g. Carbon).
\item Work with the Argonne group of Lee to model
     first resonance production in the region of the first
      resonance and also Ghent
    nuclear physics group in Belgium~\cite{belgium},
    to model
    both electron and neutrino induced final states. In addition,
    there are other theoretical efforts (e.g. Sakuda and Paschos\cite{Paschos}) on
    nuclear effects for the hadronic final states in the region
    of the first resonance.
\item Comparison of electron scattering data (primarily proton and pion
  transparency measurements) to final state interaction models used in
  neutrino event generators such as NUANCE and NEUGEN\cite{wood02}.
\end{enumerate}

\subsubsection{Resonant form-factors and structure functions}

The analysis of inclusive data in the resonance region with {\minerva}
will be done using the standard structure function analysis
techniques. The sum of neutrino and antinuetrino differential
cross-sections is used to do a Rosenbluth separation and
extract $F_2$ and $R$ for a Carbon target.  The difference between
neutrino and anti-neutrino differential cross-sections is used
to extract the structure function $xF_3$.  The nuclear effects
in the resonance region at low values of $Q^2$ are not well understood.
Electron scattering
data show that duality works at $Q^2$ greater than $1~({\rm GeV}/c)^2$ for
hydrogen and deutrium targets. In addition, there are indications
that the nuclear effects also scale with the Nachtman scaling variable.
However, these observations have not been tested in neutrino scattering, nor
have they been tested in neutrino or electron scattering at lower
values of $Q^2$.  The information from Jefferson Laboratory
proposal E03-110 will provide this information for nuclear targets
for the vector structure functions. {\minerva} in turn will be
able to extend these duality studies to the axial vector structure functions.

At present, the axial form-factor for the first resonance
is not very well known. {\minerva} will have a very
high statistics sample in this region, which is equivalent
to the sample  for quasilelastic scattering
described earlier. However, since {\minerva} data is on a Carbon target,
nuclear effects must be understood. The theoretical tools used
to model the nuclear effects in Carbon for the
final state particles in the region of the first resonance
in neutrino scattering, will be tested with CLAS Hall B
electron scattering Jefferson Lab data on Carbon and other electron
scattering data.

\subsubsection{Single-pion final states}

Using the angular distribution in the exclusive final states
 $\numu p \rightarrow \mu^- \pi^+ p$, we plan
to fit for the resonant and non-resonant amplitudes. The extracted
non-resonant amplitude should be consistent with the measured
value of $R$ in this region (extracted from the inclusive
scattering sample).

By using both neutrino and anti-neutrino data
{\minerva} can investigate transitions into isospin 3/2 states $\Delta^{++}$ and$\Delta^-$.
An analysis of the ratios of various final states.  $p \pi^+$, $n \pi^+$ 
and $p \pi_0$ will provide additional information.
As mentioned ealrier,  we plan to
do a comparison of resonance production with electron
scattering on free
nucleons to Hall B CLAS data with bound nucleons in Carbon.
Within {\minerva} itself, we can compare the reactions
$\nu p \rightarrow \nu n \pi^+$ and
$\nu n \rightarrow \nu p \pi^-$
on bound nucleons directly, and investigate additional channels
in order to better understand the effects of pion and nucleon
rescattering.

\begin{figure}[h]
\centerline{
{\includegraphics[width=\textwidth,bb=45 155 525 650]{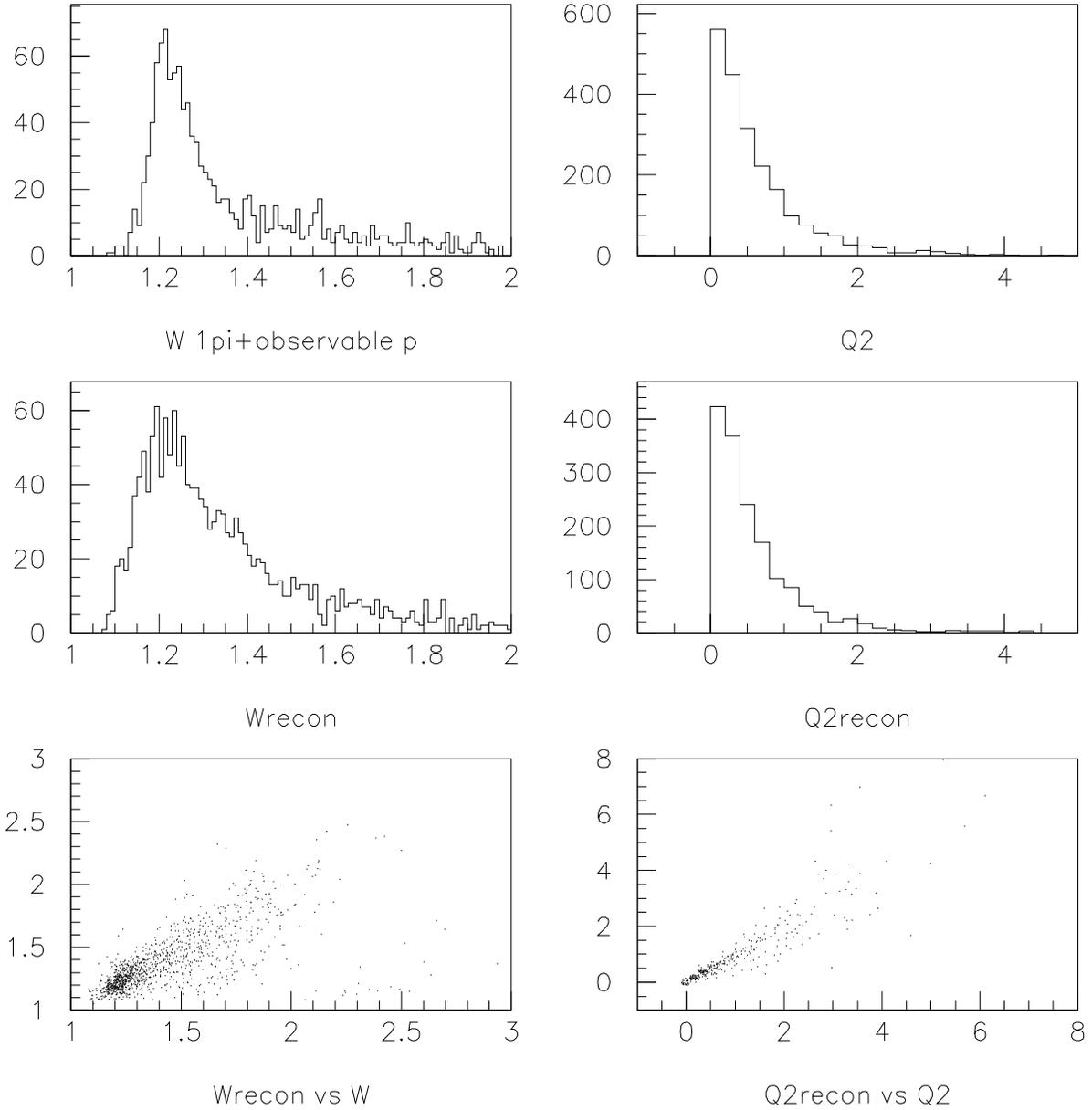}}}
\caption[$W$ and $Q^2$ reconstruction for single-$\pi$ events]{$W$ and $Q^2$ reconstruction for events with a single $\pi^+$.
  Top row are ``true'' $W$ and $Q^2$ distributions from the {\minerva}
  Monte Carlo.  The second row are the reconstructed distributions
  assuming hadron energy resolutions from Figure~\ref{resonant:pires}.
  The invariant mass of the pion and highest energy proton give $W$ which
  along with the muon energy and direction gives sufficient information to
  reconstruct $Q^2$.
  Bottom row shows the correlation between the ``true'' and reconstructed
  quantities.}
\label{resonant:deltares}
\end{figure}

\begin{figure}[tb]
\center
\epsfxsize=110mm\leavevmode
\epsffile{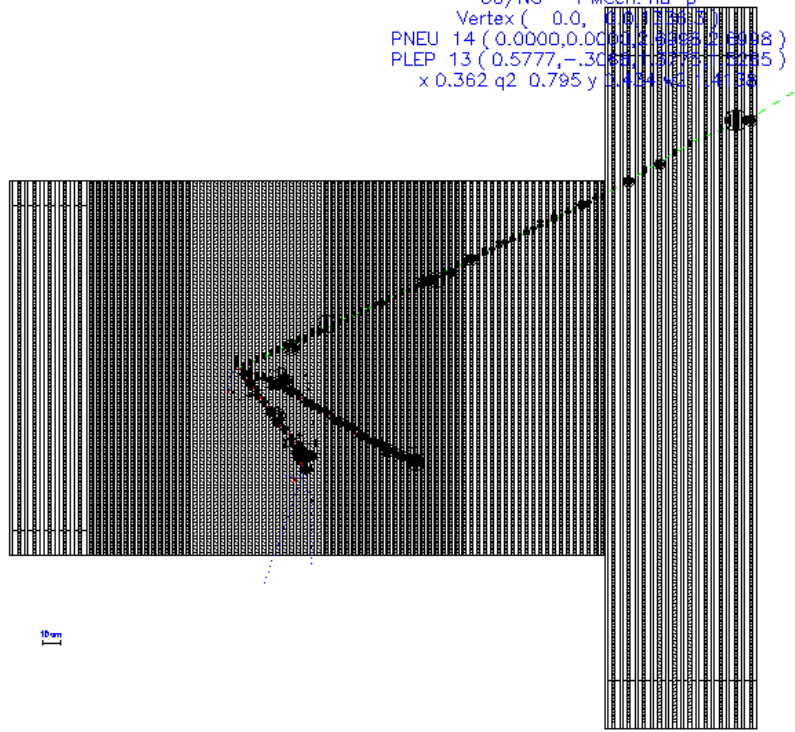}
\caption[$\Delta^{++}$ production]{$\Delta^{++}$ production and decay in a charged-current
neutrino interaction in the \minerva\ detector. Shown are (top track) the muon and (middle and bottom track) 
the pion and proton produced in the decay. Energy deposition is shown by hit size. For clarity the outer
barrel is not shown.}
\label{fig:resdisplay}
\end{figure}

\begin{figure}[h]
\centerline{
{\includegraphics[width=\textwidth,bb=40 160 555 670]{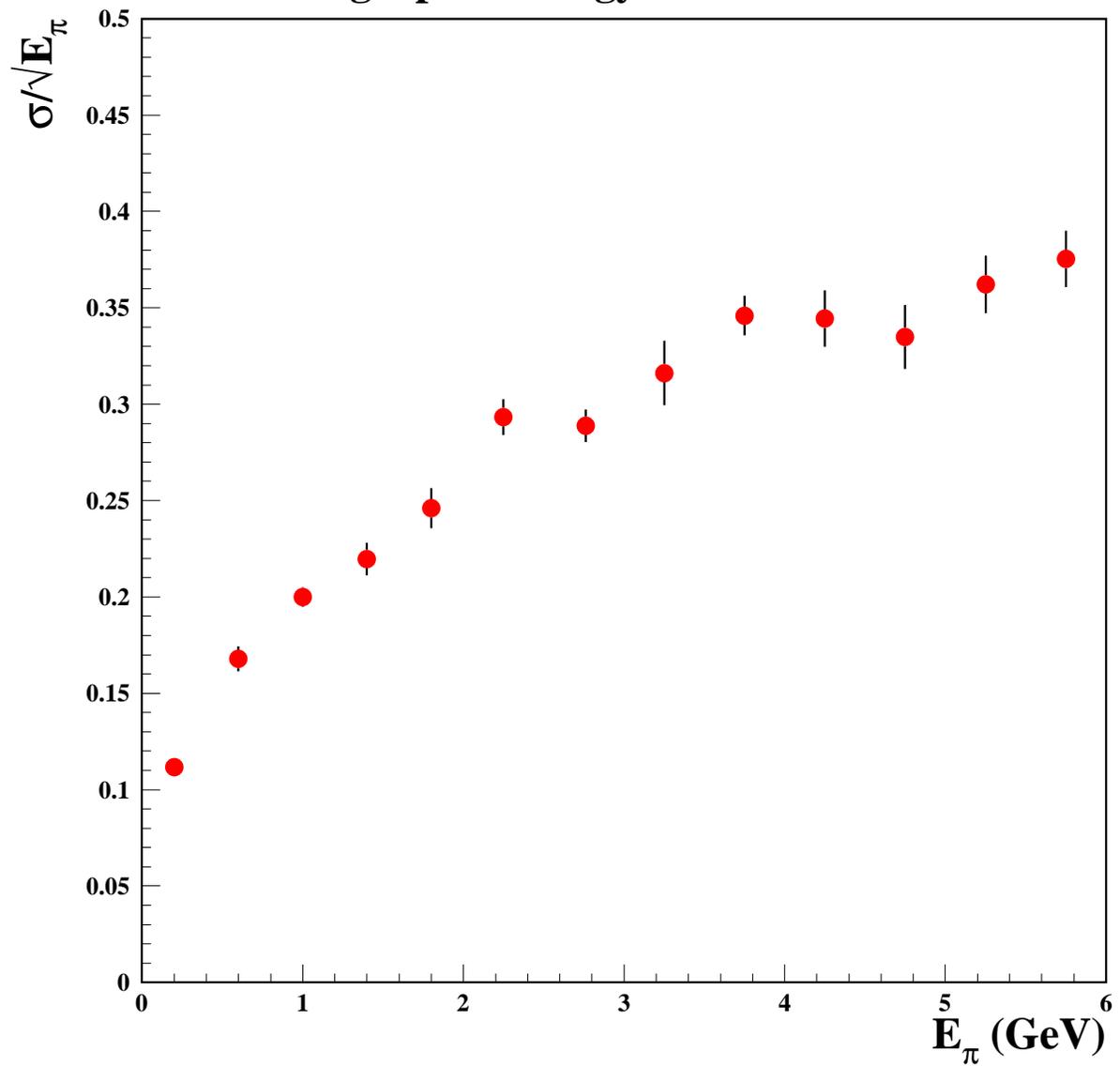}}}
\caption{Single charged pion resolution derived from {\minerva} Monte Carlo.}
\label{resonant:pires}
\end{figure}

\begin{figure}[h]
\centerline{
{\includegraphics[width=0.9\textwidth]{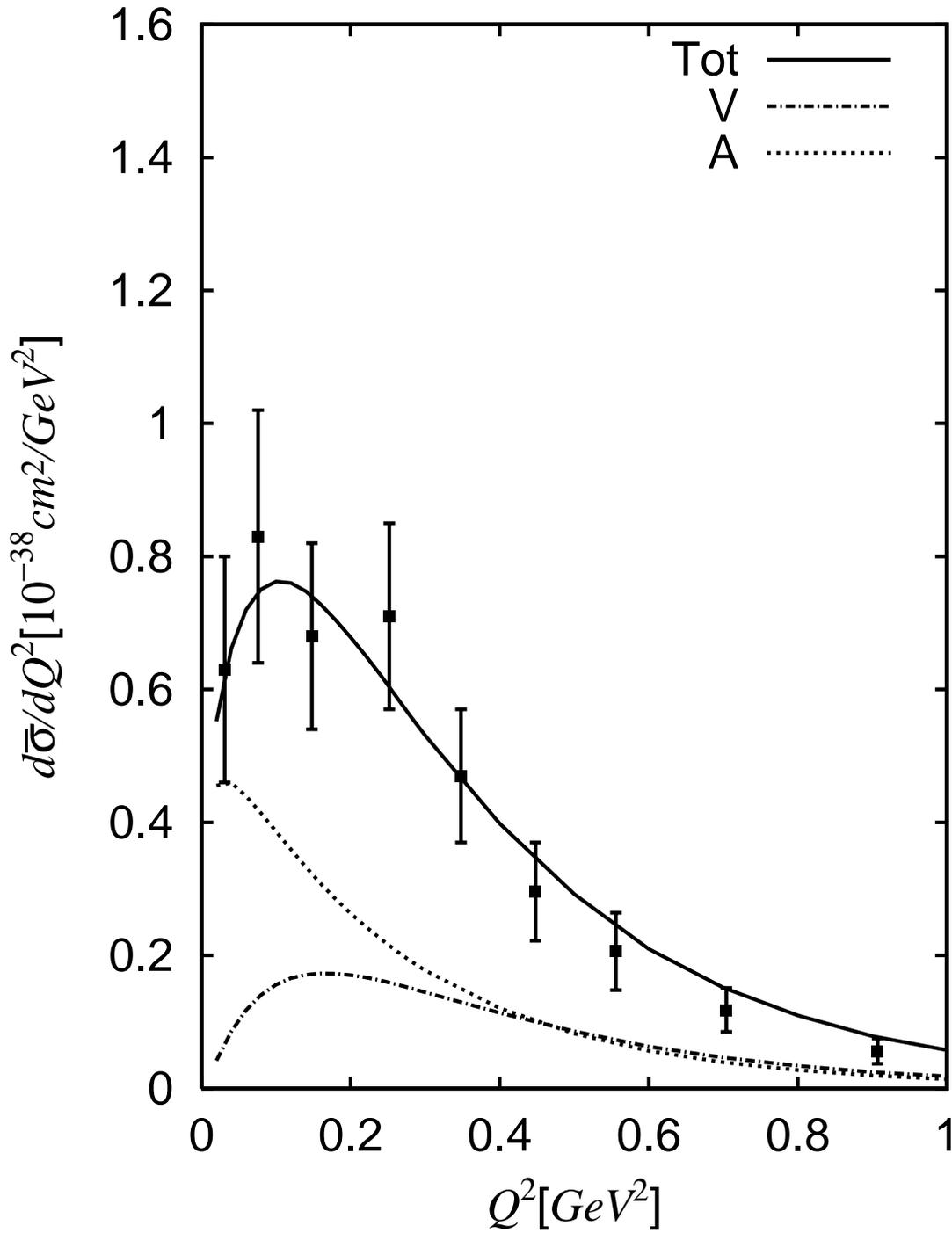}}}
\caption[Differential cross-section $d\sigma/dQ^2$ for single-pion production]{Differential cross-section $d\sigma/dQ^2$ ($10^{-38} {\rm
  cm}^2/{\rm GeV}^2$) of $p(\numu, \mu^- \pi^+)p$
  averaged over neutrino energies.  Calculations from
  Ref.~\cite{satounolee}, data from Ref.~\cite{Barish:1979pj}.}
\label{resonant:satounoleefig6}
\end{figure}

{\minerva} is expected to have good resolution for single pion events in
the resonance region ($W < 2$~{\rm GeV}).
Figure~\ref{resonant:deltares} shows $Q^2$ and $W$ distributions of single
pion events from ${\rm CH}_2$ in the {\minerva} Monte Carlo along with reconstructed
distributions that take into account {\minerva}'s energy resolution for
hadrons (Figure~\ref{resonant:pires}).  While the Fermi motion in nuclei
washes out higher resonances, it is clear that Delta events can be readily
identified and separated from higher resonances.
This  expected resolution
implies that a differential cross-section $d\sigma/dQ^2$ for Delta
production on Carbon 
equivalent to that for Hydrogen (Figure~\ref{resonant:satounoleefig6}) can
be obtained with high statistics. Figure \ref{fig:resdisplay}
shows an example of a charged-current neutrino interaction producing a $\Delta^{++}$ which
decays to a pion and proton. Distinct muon, proton and pion tracks are all visible showing
that the resonance can be well reconstructed.

%% file: coherent.tex
\section{Coherent Neutrino-Nucleus Scattering}
\label{sect:coherent}

The \minerva\ experiment has the potential to dramatically improve our
knowledge of the dynamics of coherent neutrino-nucleus scattering.  This
process, in which the neutrino scatters coherently  from the entire nucleus
with small energy transfer, leaves a  relatively clean experimental signature
and has been studied in  both charged-current ($\nu_\mu + A \rightarrow \mu^- +
\pi^+$) and  neutral-current ($\nu_\mu + A \rightarrow \nu + \pi^o$)
interactions of neutrinos and anti-neutrinos.  Although the interaction rates
are typically an order of magnitude or more lower than other single-pion
production mechanisms, the  distinct kinematic characteristics of these events
allow them  to be identified.  Because the outgoing pion generally  follows the
incoming neutrino direction closely, this  reaction is an important background
to searches for  $\nu_\mu \rightarrow \nu_e$ oscillation, as these events can
easily mimic the oscillation signature of a single energetic electron shower. 
Neutral-current coherent production will be discussed in more detail  in
Section~\ref{section:theta13}; here we limit our attention to the
charged-current channel where the kinematics can be fully measured and the
underlying dynamics explored.  

\begin{figure}[bhp] \center
\epsfxsize=90mm
\leavevmode 
\epsffile{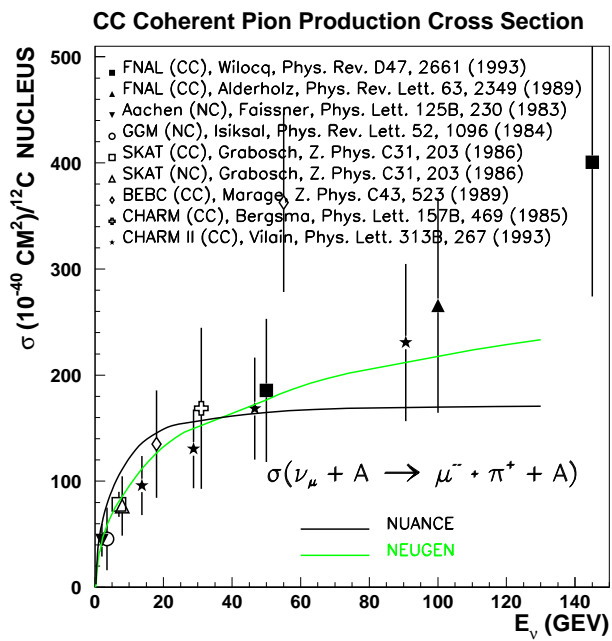} 
\caption[Charged-current
coherent scattering cross-sections]{Charged-current neutrino--carbon coherent
cross-sections.   Results have all been scaled to carbon assuming an A$^{1/3}$
dependence,  and $\sigma(CC)=2\sigma(NC)$ \cite{Zeller:Nuint02}.}
\label{fig:coh_cc} 
\end{figure}

\subsection{Theory}

It is well known from electron scattering that at low  $Q^2$ and high $\nu$,
vector mesons are abundantly produced through diffractive mechanisms.   These
interactions are  interpreted as fluctuation of the virtual photon intermediary
into  a virtual meson with the same quantum numbers, which by the uncertainty
principle can travel a length 

\begin{equation}
l \sim \frac{\nu}{Q^2 + m^2} 
\end{equation}

where $m$ is the mass of the meson in question.  For the weak  current, similar
fluctuations can occur, into both vector- and axial-vector mesons.   From the
Adler relation and ``partially-conserved axial current" (PCAC) hypothesis,
it is known that the hadronic current at low  $Q^2$ is proportional to the
pion field.  The hadronic properties of the weak current in these kinematic
regions have been investigated through the  study of nuclear shadowing at low
$x$ and the coherent  production of $\pi$, $\rho$, and $a_1$ mesons.  Coherent
scattering therefore  allows investigation of the PCAC hypothesis and  hadron
dominance models of the weak  current in detail \cite{Kopeliovich:1993ym}.  

A number of calculations of coherent scattering, involving  substantially
different procedures and assumptions,  have been made  over the past thirty
years\cite{Rein:1983pf,Belkov:1987hn,Piketty:1970sq,Paschos:2003hs}. These
calculations factorize the problem in terms of the hadron-like component of the
weak current and the scattering of this hadron with the nucleus.   The
calculations assume PCAC as a starting point but  quickly diverge when it comes
to the number of hadronic states  required to describe the weak current and how
the  hadron--nucleus scattering should be treated.   The Rein-Sehgal  model,
used by both NUANCE and NEUGEN, describes the weak current only in terms of the
pion field; the $Q^2$ dependence of the  cross-section is assumed to have a
dipole form.  Other calculations rely on meson-dominance
models\cite{Piketty:1970sq} which include  the dominant contributions from the
$\rho$ and $a_1$ mesons.       Figure~\ref{fig:coh_cc} shows the coherent
charged-current cross-section
 as a function of energy, compared to the
model by Rein and Sehgal as implemented in NEUGEN and the calculation in
\cite{Paschos:2003hs}. 

\subsection{Experimental Signatures}
The kinematics of coherent scattering are quite distinct compared
to the more common deep-inelastic and resonant interactions.  Because the coherence
condition requires that the nucleus remain intact, low-energy transfers to the nuclear
system, $|t|$, are needed.  Events are generally defined as 
coherent by making cuts on the number of prongs emerging from the 
event vertex followed by an examination of the t distribution, where t 
is approximated by:
\begin{equation}
-|t| = -(q - p_\pi)^2 =
(\Sigma_i (E_i - p_i^{||}))^2 - (\Sigma_i(p_i^\perp))^2
\end{equation}
With its excellent tracking capabilities, the \minerva\ inner
detector can measure this kinematic variable well.
\begin{figure}[tb]
\center
\epsfxsize=90mm\leavevmode
\epsffile{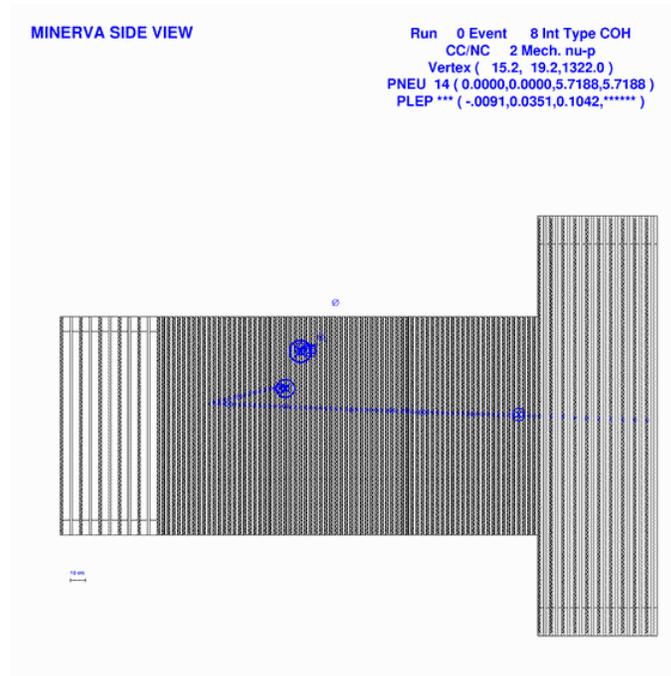}

\caption[Charged-current coherent scattering in \minerva]{A charged-current
coherent event in the inner tracking  detector of \minerva.  For clarity the
outer barrel detector is not shown. }

\label{fig:coh_pic1}
\end{figure}

Figure~\ref{fig:coh_pic1} shows an event display of a coherent charged-current
interaction in the \minerva\ inner tracking detector.  Distinct muon
and pion tracks are clearly visible and the vertex location is well defined.
\begin{table}
\begin{center}
\begin{tabular}{|l|c|c|l|r|c|}
\hline
Experiment & Reaction & Energy (GeV) & A & Signal & Ref \\
\hline
Aachen-Padova  & NC  & 2     &  27  &  360 & \cite{Faissner:1983ng} \\
Gargamelle     & NC  & 2     &  30  &  101 & \cite{Isiksal:1984vh} \\
CHARM          & NC  & 20-30 &  20  &  715 & \cite{Bergsma:1985qy} \\
CHARM II       & CC  & 20-30 &  20  & 1379 & \cite{Vilain:1993sf} \\
BEBC (WA59)    & CC  & 5-100 &  20  &  158 & \cite{Allport:1989cq,Marage:1986cy} \\
SKAT           & CC (NC)  & 3-20  &  30  &   71 (14)& \cite{Grabosch:1986mt} \\
FNAL 15'       & NC  & 2-100 &  20  &   28 & \cite{Baltay:1986cv} \\
FNAL 15' E180  & CC  & 10-100&  20  &   61 & \cite{Ammosov:1987kr} \\
FNAL 15' E632  & CC  & 10-100&  20  &   52 & \cite{Willocq:1993fv} \\ 
\hline
\end{tabular}
\label{tab:cohexp}
\end{center}
\caption[Existing data on coherent pion production]{Existing measurements on coherent pion production\cite{Kopeliovich:1993ym}.}
\end{table}

\subsection{Expected Results}

To determine the ability of the \minerva\ experiment to measure the charged current
coherent cross section, a Monte Carlo study was carried out using the GEANT 
detector simulation described elsewhere in this proposal.  Analysis cuts were 
tuned on a sample of coherent interactions
corresponding to that expected in a 
3 ton fiducial volume for the integrated 4 year run (24630 events).  
Events were generated 
according to the appropriate mix of low, medium, and high energy running.  
This study used the Rein-Seghal \cite{Rein:1983pf} 
model of coherent production, as implemented in NEUGEN3. A 20k low-energy 
beam event sample was used for background determination.  This sample 
included the appropriate mix of NC and CC events.  Based on published 
bubble chamber analyses, it is expected that charged current reactions are
the largest contributor to background processes, in particular quasi-elastic 
and delta production reactions where the baryon is not observed or is 
misidentified as a pion.
To isolate a sample of coherent interactions, a series of cuts are placed 
on event topology and kinematics.  

Topological Cuts:  an initial set of cuts are applied to isolate a sample 
of events which contain only a muon and charged pion.  These cuts are based
on the hit-level and truth  information as provided by the GEANT simulation.   
\begin{enumerate}  
\item {\bf 2 Charged Tracks:}
The event is required to have 2 visible 
charged tracks emerging from the event vertex. 
A track is assumed to be visible if it produces at least 8 hits which are 
due to this track alone.  
\item {\bf Track Identification:}
The two tracks must be identified as a muon and pion.  The muon track is taken
to be the most energetic track in the event which does not undergo hadronic
interactions.  The pion track is identified by the presence of a hadronic 
interaction.  The pion track is required not to have ionization characteristic
of a stopping proton (which is assumed can be identified 95\% of the time).
\item {\bf $\pi^o$/neutron Energy:}
Because the \minerva\ detector is nearly hermetic we have also assumed that
neutral particles will produce visible activity which can be associated with 
the event and cause it to be identified as not coherent.  Events with more than
500 MeV of neutral energy ($\pi^o$ or neutron) produced in the initial 
neutrino interaction are rejected. 
\item {\bf Track Separation:} 
In order to make good measurements of the two tracks, it is required that the 
interaction point of the pion be greater than 30 cm from the vertex and that
at this interaction point at least 4 strips separate the two tracks in at least
one view.  
\end{enumerate}

Kinematic Cuts:  because of the very different kinematics between coherent 
and background reactions, cuts on kinematic variables are very effective at 
isolating the final sample.  In this analysis, the true pion and muon 
4-momenta were used as the reconstruction values.  For the final event 
rates we reduce our overall signal sample by 0.65 to roughly account 
for this assumption.  
\begin{enumerate}  
\item { x$<0.2$:} 
A cut is made requiring that Bjorken-x (as reconstructed from the observed pion 
and muon 4-momenta) be less than 0.2.  This cut eliminates
a large amount of the background coming from quasi-elastic reactions which 
have x$\sim$1.  
\item { t$<0.2$GeV$^2$:}
The most powerful variable for the identification of coherent events is the
square of the 4-momentum transfer to the nucleus.  
The previous expression relating t to the observed particles in the event is used 
as the estimator of this quantity.   
\item { $p_\pi$> 600 MeV:}
Requiring $p_\pi>600$ MeV effectively eliminates backgrounds from delta production 
which tend to produce lower energy pions.  
\end{enumerate}

\begin{table}
\begin{center}
\begin{tabular}{|l|r|r|}
\hline
Cut                    &   Signal Sample & Background Sample \\
\hline
                       &        5000     &             10000 \\
2 Charged Tracks       &        3856     &              3693 \\
Track Identification   &        3124     &              3360 \\
$\pi^o$/neutron Energy &        3124     &              1744 \\
Track Separation       &        2420     &               500 \\
x$<$0.2                  &        2223     &               100 \\
t$<$0.2                  &        2223     &                19 \\
$p_\pi<600$ MeV        &        1721     &                12 \\
\hline
\end{tabular}
\label{tab:cohcuts}
\end{center}
\caption[Coherent cuts]{Analysis cuts to isolate a sample of coherent interactions.  
The cuts are described in the text.}
\end{table}

\begin{figure}[tb]
\center
\epsfxsize=140mm\leavevmode
\epsffile{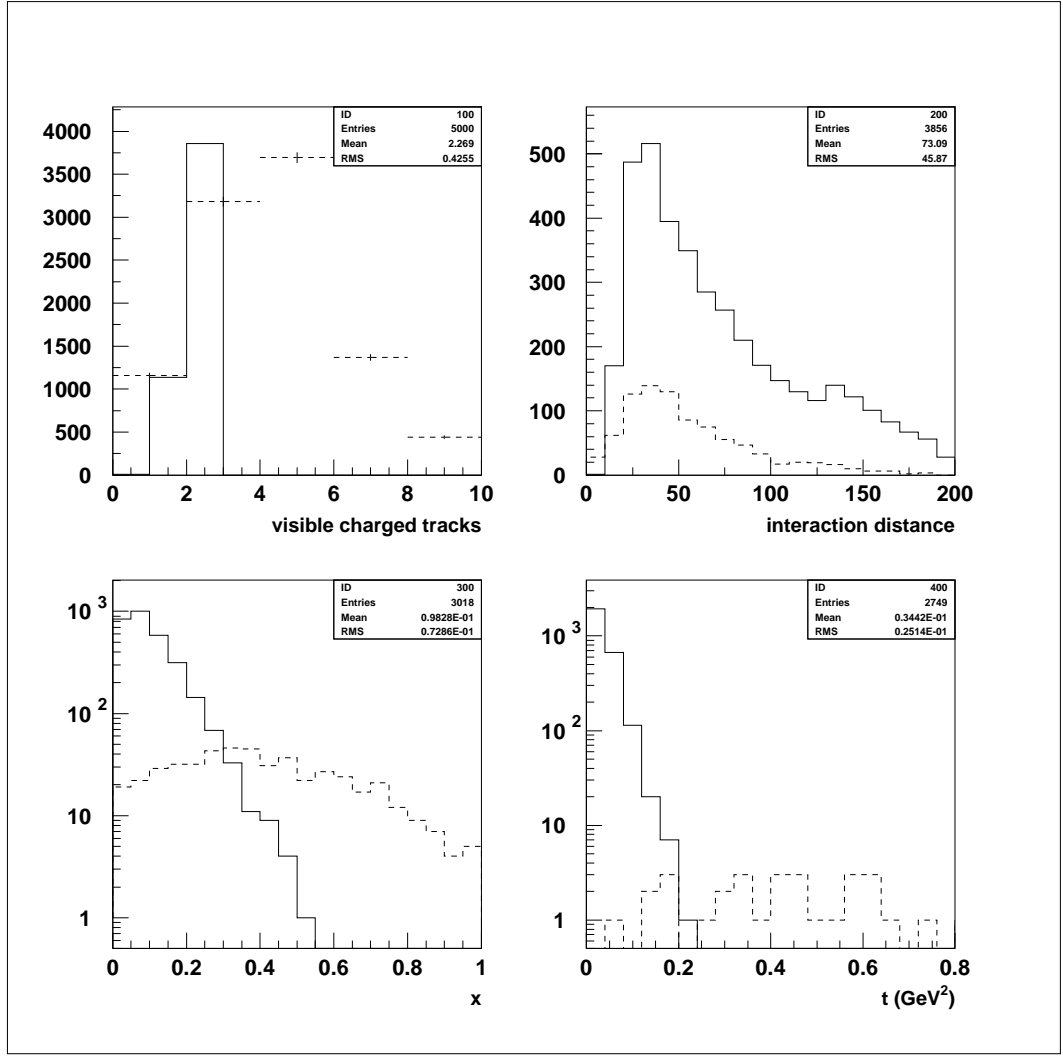}
\caption{Topological and kinematic quantities used to define the coherent sample.  In all plots
the solid histogram is the coherent sample and the dashed histogram are background processes.  
The relative normalizations of the two distributions in the initial plot is arbitrary, subsequent
plots show the effect of the applied cuts.   Top Left:  Visible charged tracks.  Top Right: 
Distance between the event vertex and the location of the pion interaction (in cm).  Bottom Left:
Bjorken-x as computed from the true pion and muon 4-momenta.  Bottom Right:  
Square of the 4-momentum transfer to the nucleus (in GeV$^2$) as calculated from the pion and 
muon 4-momenta.}
\label{fig:cohcuts}
\end{figure}

\begin{figure}[tb]
\center
\epsfxsize=90mm\leavevmode
\epsffile{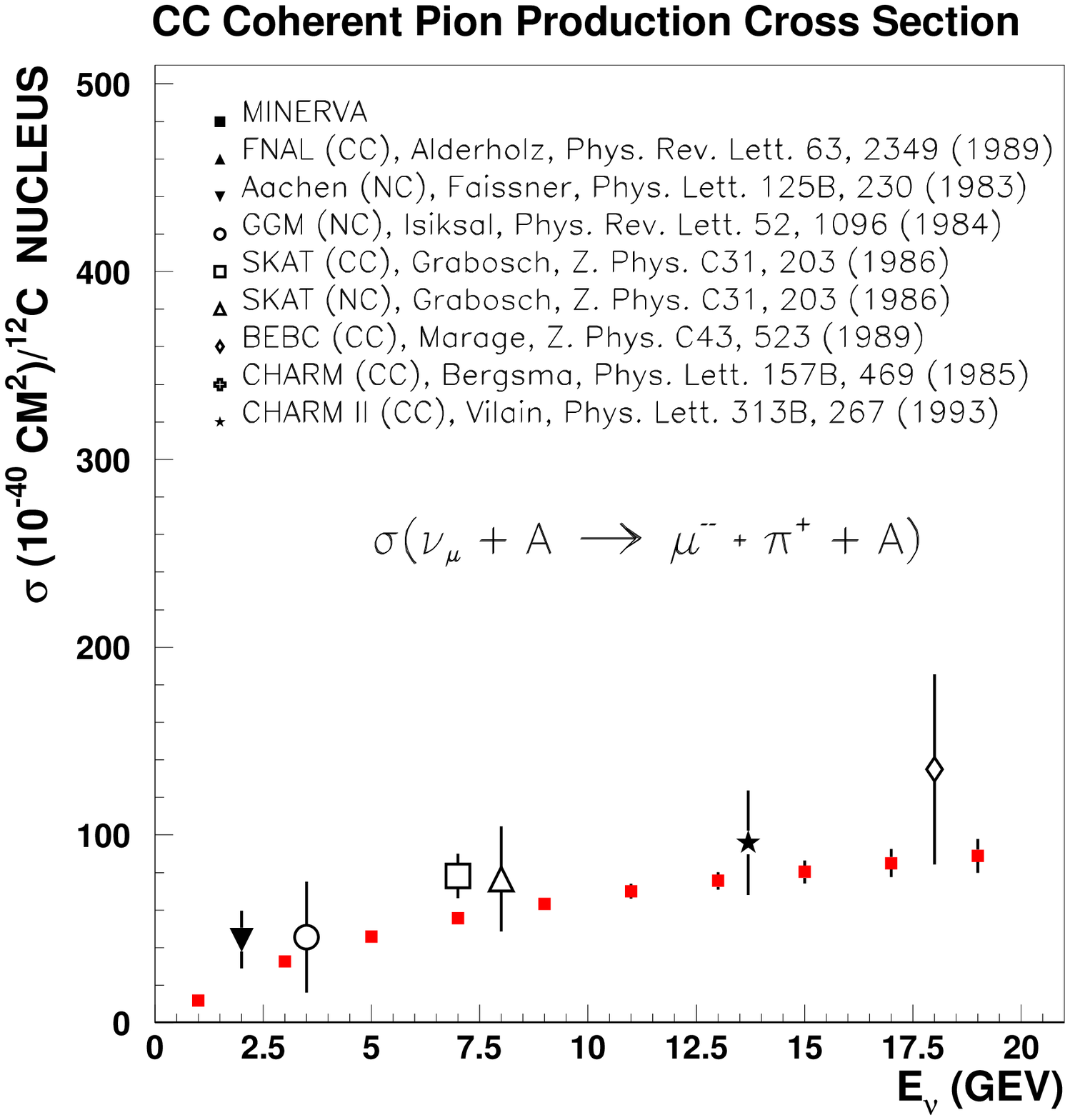}
\caption{Coherent cross-sections as measured by \minerva\ compared with existing 
published results.  \minerva\ errors here are statistical only. }
\label{fig:cohminerva}
\end{figure}

Applying this set of cuts to our signal sample we find that 7698 signal 
events pass all cuts, which gives an overall efficiency of 31\%.  
Applying the factor 0.65 to account for the fact that we have 
not used fully reconstructed quantities for our kinematic cuts gives us a final event 
sample of 5004 events.  Applying these cuts 
to the background sample we find that 12 events out of 20k 
pass all cuts.  Normalized to the total event rate this gives an expected background of 4400 
events.    
We note that in this analysis other important variables for background rejection, 
related to associated activity around the vertex, were not used.   
Figure \ref{fig:cohminerva} shows the expected precision of the \minerva\ 
measurement as a function of neutrino energy.  Here we have only included 
the statistical error on the signal and assumed that the measured value is that 
predicted by Rein-Seghal.  

\begin{figure}[tb]
\center
\epsfxsize=90mm\leavevmode
\epsffile{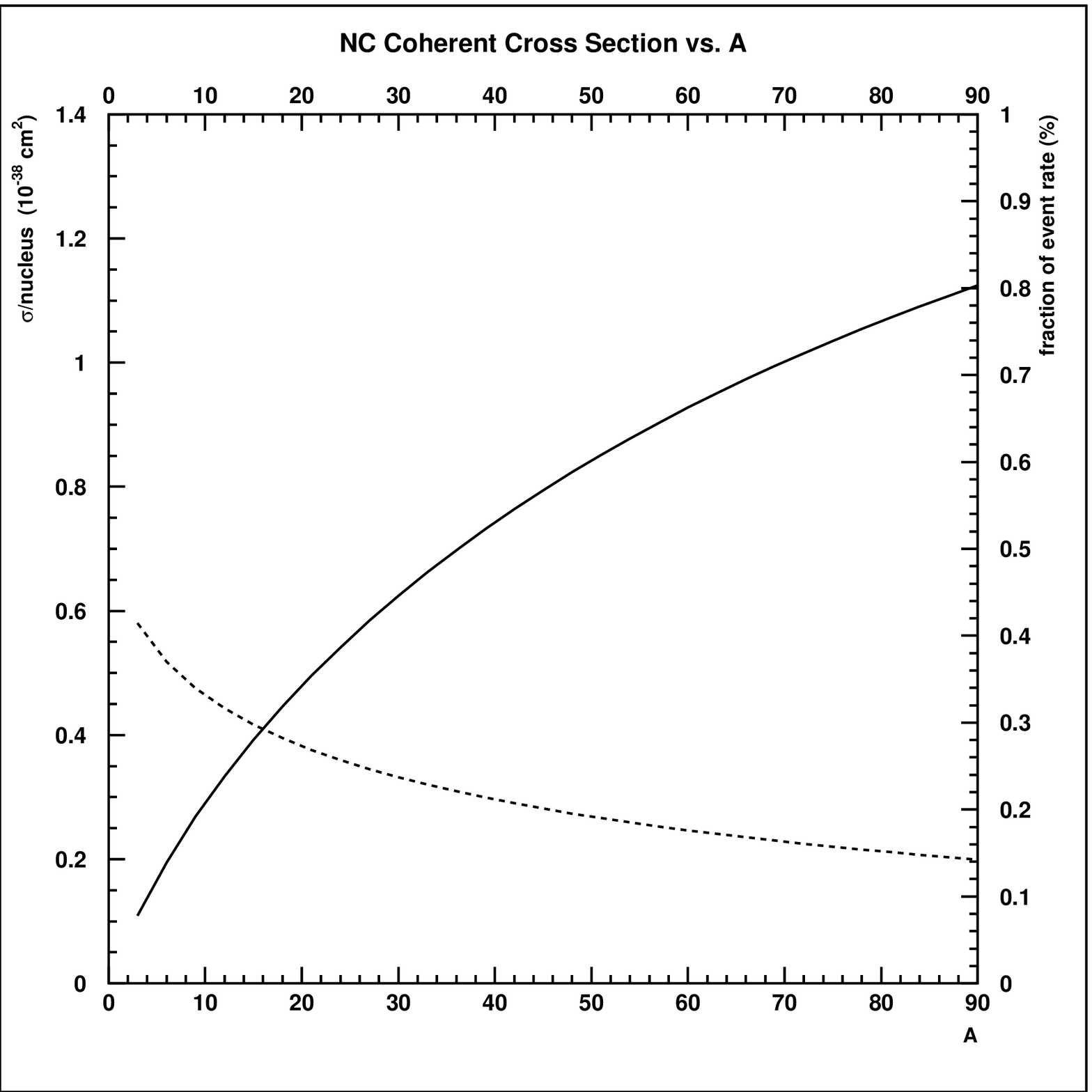}
\caption{Coherent cross-sections as a function of atomic number.}
\label{fig:coh_a}
\end{figure}

Another task for \minerva\ will be comparison  of reaction rates for lead and
carbon.  The expected yield  from lead will be $\approx$ 1800 charged-current
events,  assuming the same efficiency.   The A dependence of the cross-section
depends mainly on the model assumed for the hadron--nucleus interaction, and 
serves as a crucial test for that component of the predictions.  No experiment
to date has been able to perform this comparison.  For reference, the predicted
ratio of carbon to lead NC  cross-sections at  10 GeV  in the Rein-Sehgal and
Paschos models are 0.223 and 0.259, respectively  \cite{Paschos-K}.  
Figure~\ref{fig:coh_a} shows the predicted A-dependence according to the model
of Rein and Sehgal.  

%
%

%% file: strangeCharm.tex
\section{Strangeness and Charm Production}
\label{sect:strangeCharm}
\subsection{Overview}
 
    The \minerva\ experiment in the \numi\ near hall will allow 
high-statistics studies of the rich complexion
of exclusive-channel strange-particle production reactions accessible 
in the $1 \le E_{\nu} \le$ 8~GeV energy regime.  We propose precision 
measurement of cross sections $\sigma(E_{\nu})$ of exclusive associated-production
reactions ($\Delta S = 0$) and Cabbibo-suppressed $\Delta S = 1$ reactions.
The $\Delta S$ weak hadronic current will be mapped out in detail, 
including its $q^{2}$ dependence, resonant structure, and polarizations of produced
hyperons, to elucidate its coupling strengths and form-factors.
A panoramic experimental delineation of all near-threshold \numu--N strangeness 
production processes is envisaged which will
motivate renewed efforts to formulate detailed 
models of these reactions.  The resulting picture will have ramifications
in other areas of particle physics, for example in estimation of atmospheric neutrino
$\Delta S$ backgrounds for nucleon-decay searches at megaton-year sensitivities.
A \minerva\ exposure will also enable searches for new processes, e.g.
unusual baryon resonances such as the recently reported candidate pentaquark state in 
${\rm K}^{+}$n and ${\rm K}^{0}_{s}$p systems, and neutral-current 
strangeness-changing reactions.  Extended running of the \numi\ beam will 
allow \anumu~ exposures that will provide valuable complementary 
data for many neutrino topics.  Anti-neutrino exposure will facilitate study 
of $\Delta S = 1$ single-hyperon production $(\Lambda, \Sigma,{\rm Y}^{*})$.   
Study of hyperon reactions will greatly extend 
the ${\rm q}^{2}$ range over which the weak
interaction form-factors which govern hyperon beta-decay can be examined.  Thus a much
better determination of the form-factors - especially of the three axial form-factors -
will be possible.   Hyperon polarization will provide additional analyzing power
here, and the analysis will be free of the 'missing neutrino' problem which has hindered
examination of the underlying V-A structure using semi-leptonic hyperon decays.
As a natural extension of strange-particle production studies, we will search for
strangeness production which accompanies dilepton processes.   Such reactions have, in
previous neutrino experiments, served as gateways to the study of charmed baryon production.

The NOMAD experiment\cite{nomad} has studied inclusive strange-particle
production extensively. \minerva\ will not improve significantly on those results, and the
physics motivation for attempting to do so is unclear. \minerva will focus instead on 
exclusive channels; this is relatively unexplored territory, with
the potential for high impact on future physics.

\subsection{Neutrino Strangeness Production Near Threshold }
\label{sect:strangeProd}

   In the threshold regime $1 \le E_{\nu} \le$ 8~GeV, 
neutrino interactions involving strangeness production yield 
final states containing either 
one or two strange particles.  
Exclusive \neut--N channels comprise 
three categories, distinguished by reaction type 
(charged-current (CC) or neutral-current (NC)) and the
net change in strangeness 
$\Delta S = S_{f} - S_{i}$ of the hadronic system (either  $\Delta S = 0$ or 
$\Delta S = 1$). 

The first category comprises charged-current   
$\Delta S = 0$ reactions initiated by~\numu.  These are associated-production
reactions where a strangeness +1 meson (K$^+$ or K$^0$) 
is produced together with a
strangeness -1 hyperon ($\Lambda$ or $\Sigma^{\pm,0}$) 
or meson (K$^-$ or $\overline{\rm K}^0$).
Reactions of this category include:
\begin{eqnarray}
\nu_\mu {\rm n} & \rightarrow & \mu^- {\rm K}^+ \Lambda^0 \label{sc:1}\\
\nu_\mu {\rm n} & \rightarrow & \mu^- \pi^0 {\rm K}^+ \Lambda^0 \label{sc:2}\\
\nu_\mu {\rm n} & \rightarrow & \mu^- \pi^+ {\rm K}^0 \Lambda^0 \label{sc:3} \\
\nu_\mu {\rm n} & \rightarrow & \mu^- {\rm K}^- {\rm K}^+ {\rm p} \label{sc:4}\\
\nu_\mu {\rm p} & \rightarrow & \mu^- {\rm K}^+ \label{sc:5}
 \overline{\rm K}^0 \pi^0 {\rm p}
\end{eqnarray}

\noindent
Among charged-current $\Delta S = 0$ reactions, reaction (\ref{sc:1}) 
has the largest cross-section.
This reaction, and reactions (\ref{sc:2}) and (\ref{sc:3}) 
as well, may proceed predominantly via N$^*$ production followed by strong decay into
K$\Lambda$.

Charged-current $\Delta S$ = 1 reactions make up a second category.  For 
\numu  reactions, the resulting
final states contain single strange
K-mesons.  The reaction cross-sections  
are Cabbibo-suppressed relative to $\Delta S$ = 0 reactions involving
similar hadronic masses.  Additionally the $\Delta S = \Delta Q$ selection rule applies,
and so the produced mesons are necessarily (K$^+$, K$^0$)
and not (K$^-$, $\overline{\rm K}^0$). 
Reactions of this category include
\begin{eqnarray}
\numu {\rm p} &  \rightarrow &  \mu^-  {\rm K}^+ {\rm p} \label{sc:6}\\
\numu {\rm n} & \rightarrow  & \mu^-  {\rm K}^0  {\rm p} \label{sc:7}\\
\numu {\rm n} & \rightarrow  & \mu^- \pi^+ {\rm K}^0 {\rm n}\label{sc:8}.
\end{eqnarray}

\noindent
Reaction (\ref{sc:6}) has the largest cross-section among $\Delta S = 1$ exclusive
\numu--N reactions.

Note that $\Delta S = \Delta Q$ selection 
restricts $\Delta S = 1$ single-hyperon
production to \anu ~rather than $\nu$ ~reactions, e.g.

\begin{equation}
      \anumu{\rm N} \rightarrow \mu^+ +(\Lambda, \Sigma,{\rm Y}^{*}). \label{sc:9}\end{equation}
\noindent

      Strange-particle $\Delta S = 0$ associated production
can also proceed via neutral-current reactions.  
Observed channels include
\begin{eqnarray}
\numu {\rm p} & \rightarrow & \nu {\rm K}^+ \Lambda^0 \label{sc:10}\\
\numu {\rm n} & \rightarrow & \nu {\rm K}^0 \Lambda^0 \label{sc:11}\\
\numu {\rm n} & \rightarrow & \nu \pi^- {\rm K}^+ \Lambda^0 \label{sc:12}
\end{eqnarray}

\noindent
As with final states of (\ref{sc:1}) - (\ref{sc:3}), 
it is similarly plausible that the
hadronic systems of (\ref{sc:10}) through (\ref{sc:12}) are dominated by intermediate
N$^*$ states.

\subsection{Strangeness Production Measurements at Bubble Chambers}
\label{sect:strangeCharmData}

Cross-sections for many associated-production and $\Delta S = 1$ reactions were obtained
during the 1970's and '80s in experiments using 
large-volume bubble chambers exposed to
accelerator neutrino beams.  Principal experimental programs were the 
\numu\ and \anumu ~exposures of the Gargamelle heavy-liquid (CF$_3$Br) 
bubble chamber at CERN \cite{Deden, Erriques} and the \numu--D$_2$
exposures of the 12-foot diameter 
bubble chamber at Argonne \cite{Barish_74} and of the
7-foot diameter bubble chamber at Brookhaven \cite{Baker_81Strange}.  Typical 
samples involved less than ten observed events per channel,
and cross-sections thereby inferred relate to one or a few bins in $E_\nu$. 
Contemporaneous theoretical/phenomenological treatments 
of reactions (\ref{sc:1}), (\ref{sc:7}), (\ref{sc:8}),
(\ref{sc:10}), and (\ref{sc:11}) can be found in~\cite{Shrock, Amer, Dewan}.

      Cross-section ratios obtained by the 
bubble-chamber experiments provide rough
characterizations of relative rates of occurrence among the strangeness
reaction categories.  For example, the frequency of strange versus
non-strange hadronic final states in charged-current reactions is 
indicated by \cite{Barish_74}:
\begin{equation}
 \frac{\sigma(\nu{\rm N}\rightarrow\mu^-\Lambda {\rm K}^+ + \mu^- {\rm p\ K})}
{\sigma(\nu {\rm N} \rightarrow \mu^- {\rm N} + {\rm pion(s)})}
   = 0.07 \pm 0.04 
\label{sc:13}
\end{equation}

\noindent The relative contribution of neutral-current versus charged-current 
reaction to threshold strangeness production is indicated by \cite{Baker:1981su}:
\begin{equation}
\frac{\sigma(\numu {\rm N} \rightarrow \numu {\rm V}^0 + {\rm anything})}
{\sigma(\numu {\rm N} \rightarrow \mu^- {\rm V}^0 + {\rm anything})}
   = 0.22 \pm 0.14
\label{sc:14}
\end{equation}

and 
\begin{equation}
\frac{\sigma(\nu {\rm K}^+ \Lambda^0 )}
{\sigma(\mu^- {\rm K}^+ \Lambda) + \sigma(\mu^- {\rm K}^+ \Lambda X^0)}
  =  0.18 \pm 0.13
\label{sc:15}
\end{equation}
 
\noindent Perhaps the most significant ``find" arising from 
bubble-chamber survey
experiments was the first observation of CC charmed-baryon production in
a $\Delta S = 1$ final state at BNL\cite{Cazzoli}:
\begin{eqnarray}
\numu {\rm p}   \rightarrow  \mu^- &\Sigma_c^{++}& \label{sc:16}\\
&\Sigma_c^{++} & \rightarrow  \Lambda^+_c \pi^+  \rightarrow  \Lambda^0 \pi^+ \pi^+ \pi^- \pi^+ \label{sc:17}
\end{eqnarray}      

\begin{figure}[htb]
\centerline{\epsfig{file=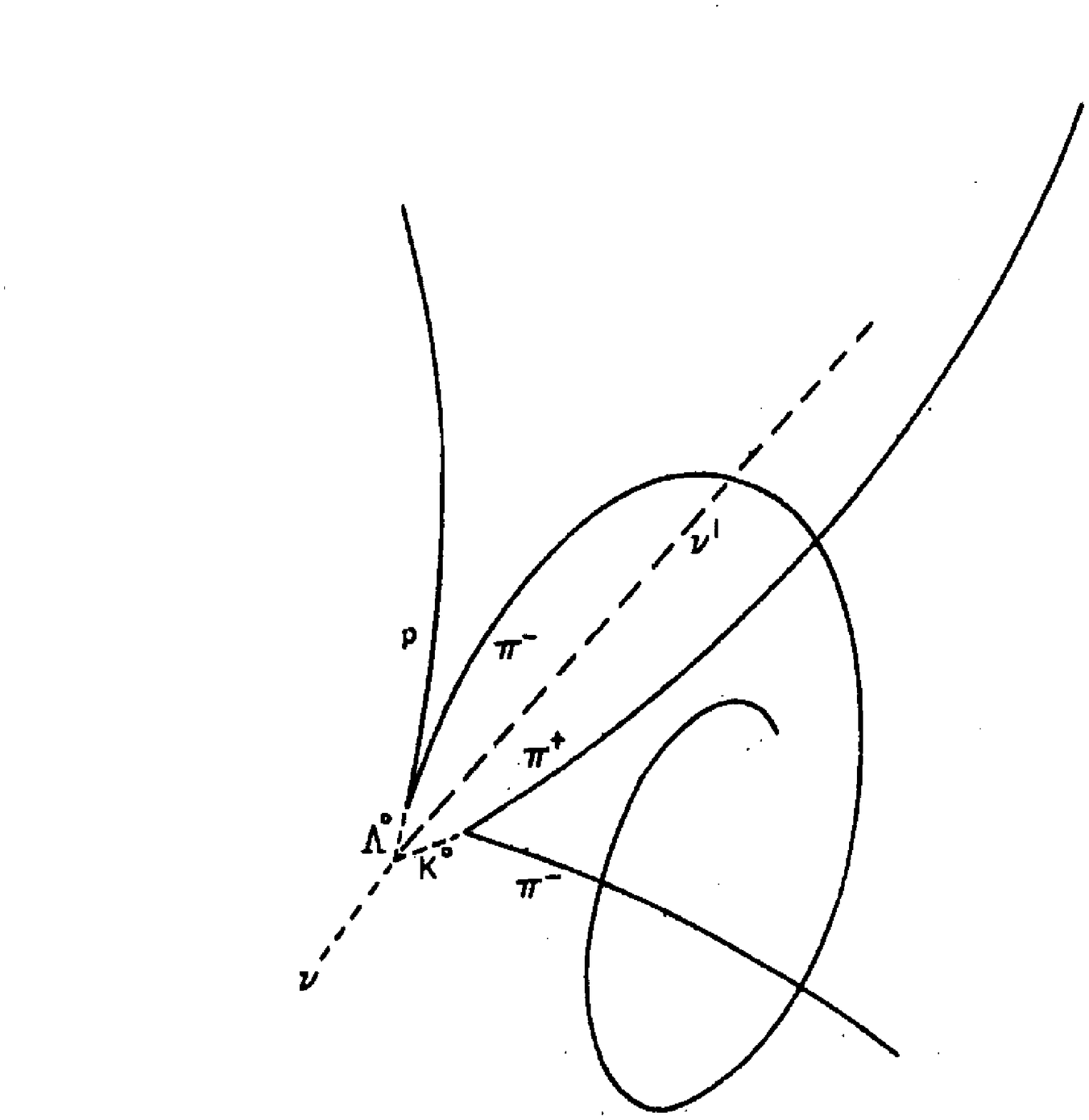, width=9.0cm}}
\caption[Associated production in the ANL bubble chamber]{Trace of photograph from the ANL 12-ft diameter bubble chamber,
of a neutrino neutral-current interaction 
in liquid deuterium yielding NC associated production 
$\nu$K$\Lambda$.  Flight paths to the vee decays of the two strange particles
in the event are 8.0~cm and 5.5~cm in real space.}
\label{fig:fig01}
\end{figure}

That excellent spatial resolution is a prerequisite for study
of neutrino strangeness-production reactions is illustrated by the bubble-chamber
event in Figure~\ref{fig:fig01}.
The figure shows the tracing of a photographic image recorded
by one of four separate camera views of this \numu--n interaction.   
The event shown was the first example of NC
associated strangeness production via reaction (\ref{sc:11}) obtained using the 
deuterium-filled 12-ft diameter bubble chamber at ANL.  
Within the final state, the flight paths  of
the ${\rm K}^{0}_{s}$ and and $\Lambda^{0}$ 
from the primary vertex to their respective ``vee" decay points 
are 8.0~cm and 5.5~cm respectively~\cite{Barish_74}.
Fortunately it should be possible, with the lattice of triangular-cell 
scintillator tracking elements of a fine-grained
detector, to achieve spatial resolutions near bubble-chamber quality (studies currently
predict vertex resolutions of less than 1 centimeter in \minerva\/; see Section \ref{sect:vertex}).
 This capability, together with $dE/dx$ ionization
imaging and momentum determination by ranging and external magnetic
tracking, will allow \minerva\ to explore exclusive 
strangeness-production processes.

\subsection{\minerva\ Samples Amenable to Hypothesis Fitting}

   As described above, the available data on exclusive channel strange particle
production by neutrinos is currently limited 
to samples of few tens of events isolated in bubble
chamber experiments of the 1970's and 80's.  With the proposed \minerva\ program this
data pool can be boosted by two orders-of-magnitude, thereby paving the way for 
comprehensive phenomenological treatments of neutrino strangeness production near
threshold.  In \minerva\, occurrences of \numu N exclusive strangeness production
will comprise only a small fraction of the total event rate in the detector.  However
these events can be readily extracted from the accumulating total sample by exploiting
\minerva\ capabilities:
\begin{itemize}

\item The primary charged particle multiplicities are low for all strangeness production
      channels of interest; this feature is readily discernible event-by-event as result
      of \minerva\'s fine granularity.   

\item A prompt K$^+$ occurs in more than 50\% of exclusive strangeness channels. 
      The subsequent in-detector decay of the K$^+$ some tens 
      of nanoseconds later yields a signature in the light yield versus time profile
      of these events.  The double-peak signature will be unmistakable in low-multiplicity
      events.
      
\item All channels of interest which do not have a K$^+$ meson, have a final state K$^0$ meson.
      In the latter reactions, those K$^0$'s which 
      decay via K$^0_s \rightarrow \pi^+ \pi^-$ can be identified
      by examination of two-particle invariant mass and vertex displacement.

\end{itemize}

 Table \ref{sptable} summarizes the sample populations for exclusive channel reactions
 obtainable by \minerva\ in the initial four-year run with the \numu beam. 
 Note that the listed channel rates 
 are restricted to sub-sets of events for which the imaged final states
 allow kinematic constraints (energy and momentum conservation) to be imposed
 at the primary reaction vertices.  That is, for final states which include
 V$^{0}$ particle(s), the only events tallied for Table \ref{sptable} are those
 for which each V$^{0}$ particle decays into two charged tracks, 
 e.g. $\Lambda^0 \rightarrow $p$\pi^-$ and K$^0_s \rightarrow \pi^+ \pi^-$.
 Additionally, an overall detection efficiency is included
 which is based upon processing experience with strange particle production
 reactions in the ANL 12-ft diameter bubble chamber \numu D exposures~\cite{Mann_86}.
 Thus the sample populations estimated for Table \ref{sptable} represent events which
 can be reconstructed and treated using hypothesis fitting.   Since the incident
 neutrino direction will be known relatively precisely, conservation of four-momentum
 will enable three-constraint fitting to charged current hypotheses and zero-constraint
 fitting to neutral current hypotheses (with Fermi motion of the target nucleon restricted
 to an allowed range in the fit).  Given the various strange particle signatures in the
 reactions of Table \ref{sptable} and given the four-momentum constraints which can
 be imposed at decay as well as at primary vertices, backgrounds from non-strange \numu N
 interactions can be strongly mitigated.  It should be possible to limit
 their contamination of total sample rates to well below 10\% for most channels.

\begin{table}
\begin{center}
\caption{Event populations for kinematically constrainable samples of exclusive-channel
strangeness production reactions, obtainable in a four-year exposure of the three-ton
inner fiducial volume of {\minerva}.}
\label{sptable}
\begin{tabular} { l l r}
\hline
Reaction Type & Exclusive Channel & No. Events ($\ge 0$ constraint) \\
\hline
~~$\Delta$S = 0~ CC & ~$\nu_\mu {\rm n} ~\rightarrow$~ $\mu^-$K$^+ \Lambda^0$  & 10,500~~~~~~~~~ \\
             & ~$\nu_\mu {\rm n} ~\rightarrow$~ $\mu^- \pi^0 $K$^+ \Lambda^0$  &  9,300~~~~~~~~~ \\
             & ~$\nu_\mu {\rm n} ~\rightarrow$~ $\mu^- \pi^+ $K$^0 \Lambda^0$  &  6,300~~~~~~~~~ \\
             & ~$\nu_\mu {\rm n} ~\rightarrow$~ $\mu^- $K$^-$K$^+$ p           &  5,100~~~~~~~~~ \\
             & ~$\nu_\mu {\rm p} ~\rightarrow$~ $\mu^- $K$^0$K$^+ \pi^0$ p     &  1,500~~~~~~~~~ \\
\hline
~~$\Delta$S = 1~ CC & ~$\nu_\mu {\rm p} ~\rightarrow$~ $\mu^-$K$^+$ p          & 15,900~~~~~~~~~ \\
            & ~$\nu_\mu {\rm n} ~\rightarrow$~ $\mu^-$K$^0$p                   &  2,400~~~~~~~~~ \\
            & ~$\nu_\mu {\rm n} ~\rightarrow$~ $\mu^- \pi^+ $K$^0$             &  2,100~~~~~~~~~ \\
\hline
~~$\Delta$S = 0~ NC & ~$\nu_\mu {\rm p} ~\rightarrow$~ $\nu$K$^+ \Lambda^0$    &  3,600~~~~~~~~~ \\
               & ~$\nu_\mu {\rm n} ~\rightarrow$~ $\nu$K$^0 \Lambda^0$         &  1,100~~~~~~~~~ \\
               & ~$\nu_\mu {\rm n} ~\rightarrow$~ $\nu$K$^0 \Lambda^0$         &  2,800~~~~~~~~~ \\
\hline
\end{tabular}
\end{center}
\end{table}

\subsection{Expected Results}

The paragraphs below summarize some specific topics 
involving neutrino strangeness-production processes that can be investigated 
using \minerva.

\subsubsection{Backgrounds to nucleon decay}

Current lifetime lower limits for nucleon decay ($\tau/\beta \ge 10^{33}$ years)
have not diminished hopes for the eventual success 
of supersymmetric grand unification (SUSY GUTs).  Indeed,
there is strong motivation to proceed 
with more ambitious experimental searches.  
For the near future, improved searches will be carried out
by Super--Kamiokande.  Eventually these will be taken up
by a next generation of underground detectors, e.g. by megaton-scale
water Cherenkov experiments such as
Hyper--Kamiokande  and/or UNO\cite{NextGenWC}.

   Continued progress, either by improving limits to
$10^{34}$ year lifetimes or discovery of nucleon decay, hinges
upon improved knowledge of certain neutrino interactions which, when
initiated by atmospheric neutrinos, can imitate nucleon-decay signals.
The most problematic backgrounds to SUSY GUT modes arise via 
neutral-current associated production of strangeness at threshold 
energies.

SUSY GUTs
predict that nucleon-decay modes proceeding via virtual transitions involving 
inter-generational mixing are favored.  Such modes yield final states
containing strangeness $+1$ mesons, e.g.
\begin{eqnarray}
{\rm p} & \rightarrow & \nu {\rm K}^+   \label{sc:18}\\
{\rm n} & \rightarrow & \nu {\rm K}^0  \label{sc:19}
\end{eqnarray}

and possibly
\begin{eqnarray}
{\rm p} & \rightarrow & \mu^+ {\rm K}^0  \label{sc:20}\\
{\rm p} & \rightarrow &  {\rm e}^+ {\rm K}^0 \label{sc:21}
\end{eqnarray}

Decays (\ref{sc:18}) and (\ref{sc:19})  are thought to hold particular 
promise for first observation of baryon instability.

\paragraph{Search for p $\rightarrow \nu {\rm K}^+$}

The leading nucleon-decay search experiment for the next decade 
(and perhaps longer) will be Super--Kamiokande. 
Its successor is also likely to be
an underground water Cherenkov detector with similar resolutions but
a fiducial volume approaching megaton scale.
In Super--Kamiokande, the search for proton
decay mode (\ref{sc:18}) is currently carried out using three different methods, each
motivated by the particulars of
the final-state sequence being sought:

\vspace{0.5cm}
$^{16}${\rm O}(7{\rm p} + {\rm p} + 8{\rm n}) $\rightarrow$
 $^{15}${\rm N} + \g (6.3 {\rm MeV}) \\
\begin{picture}(20,25)(-70,0)
\line(0,1){24}\vector(1,0){20}
\end{picture}
\hspace{2.2cm}  $\nu {\rm K}^+$ \\
\begin{picture}(20,25)(-100,0)
\line(0,1){22}\vector(1,0){20}
\end{picture}
\hspace{3.3cm} $ \left\{ \begin{array}{l}
\mu^+\numu,\hspace{0.2cm} \mu^+ \rightarrow {\rm e}^+\nu\anu \\
\\
\pi^+  + \pi^0 \\
\end{array}
\right. $
\\
\begin{picture}(12,11)(-167,-3)
\line(0,1){10}\vector(1,0){12}
\end{picture}
\noindent \hspace{5.7cm} \g \g \\
\mbox{\hspace{5.4cm}$\mu^+\numu,\hspace{0.2cm} \mu^+ 
\rightarrow {\rm e}^+\nu\anu$} \\
\begin{picture}(12,23)(-139,-32)
\line(0,1){22}\vector(1,0){12}
\end{picture} 

The three Super--Kamiokande approaches to finding proton decay (\ref{sc:18}) are:

\begin{itemize}

\item [\it i)] K$^+ \rightarrow \mu^+\nu$ spectrum search:  
      Looks for an excess of
      single $\mu$-like ring events for which the reconstructed momentum
      $p_\mu$ matches that of two-body K$^+$ decay at rest 
      and the delayed rings accompanied by subsequent 
      $\mu \rightarrow$ e decay showers.  This technique is already background limited.

\item [\it ii)] K$^+ \rightarrow \mu^+ \nu$  gamma search:  
      A candidate event has a signature
      6.3~MeV gamma emitted by the parent nucleus together with a 
      single $\mu$-like ring having $p_\mu$ for a stopped K$^+$ and 
      accompanied by $\mu \rightarrow$ e decay.

\item [\it iii)] K$^+ \rightarrow \pi^+ \pi^0$ search:  Candidates have 
      three rings
      compatible with $\pi^+ \pi^0$ with $\pi^0 \rightarrow \g \g$
      from a stopped
      K$^+$ and with a subsequent $\mu \rightarrow$ e decay signal.

\end{itemize}

\paragraph{Neutrino background for p$ \rightarrow {\rm K}^+ \nu$ }

The combined search sensitivity for 
p$ \rightarrow \nu {\rm K}^+$ is dominated by the prompt gamma method {\it ii)}
for which detection of a 6.3~MeV gamma from 
the nuclear de-excitation chain is crucial.
Assuming this capability will be retained by next-generation 
underground water Cherenkov detectors, there is but one atmospheric
neutrino reaction which may become an irreducible background in the search
for this mode, and that is the neutral-current associated strangeness-production
reaction (\ref{sc:10}).  That is, in an underground water Cherenkov
detector, an atmospheric neutrino of \numu ~or \nue~flavor may
interact with a proton bound in an oxygen nucleus, producing a K$^+$
meson together with a $\Lambda$ hyperon and an (invisible) outgoing 
neutrino.   Subsequently, the  $^{15}$N nucleus which is the remnant of 
the struck $^{16}$O, de-excites producing the 6~MeV signature $\gamma$.
The final state $\Lambda$ is a target fragment and will most always have
low momentum.  When it decays into p$ \pi^-$ as will happen
in two-thirds of reaction (\ref{sc:10}) occurrences, the daughter tracks will 
usually be below Cherenkov threshold and hence invisible.
The final-state K$^+$ will subsequently decay,
usually at rest, to yield a $\mu^+$ or $\pi^+ \pi^0$ signature.
Consequently the detection sequence in a water Cherenkov experiment 
indicated above for proton decay (\ref{sc:18}) is perfectly mimicked:

\vspace{0.5cm}
$\nu$ + $^{16}${\rm O} $\rightarrow \nu + {\rm K}^+ + \Lambda^0$
 + $^{15}${\rm N} + \g (6.3 {\rm MeV}) \\
\begin{picture}(20,25)(-132,0)
\line(0,1){24}\vector(1,0){20}
\end{picture}
\hspace{4.5cm}
\vspace{-0.8cm}
 ${\rm p} \pi^-$ (below \v C threshold) \\
\mbox{\hspace{4.3cm} $ \left\{ \begin{array}{l}
\mu^+\numu,\hspace{0.2cm} \mu^+ \rightarrow {\rm e}^+\nu\anu \\
\\
\pi^+  + \pi^0 \\
\end{array}
\right. $}
\begin{picture}(20,60)(143,-5)
\line(0,1){59}\vector(1,0){20}
\end{picture}
\\
\begin{picture}(12,11)(-172,-3)
\line(0,1){10}\vector(1,0){12}
\end{picture}
\noindent \hspace{5.8cm} \g \g \\
\mbox{\hspace{5.6cm}$\mu^+\numu,\hspace{0.2cm} \mu^+ \rightarrow {\rm e}^+\nu\anu$} \\
\begin{picture}(12,25)(-143,-32)
\line(0,1){24}\vector(1,0){12}
\end{picture}

\noindent

It is crucial for future, and for ongoing proton decay searches as well,
that neutrino background posed by (\ref{sc:10}) and by other neutrino strangeness-production
reactions be quantitatively understood.   Fortunately, the
relevant neutrino strangeness-production cross-sections, including their $E_{\nu}$ dependence,
can be precisely measured in \minerva.
            
\subsubsection{Measurement of $\sigma(\nu \Lambda {\rm K}^+ )$} 

\minerva will measure the exclusive $\Delta S = 0$ neutral-current channel
\begin{equation}
\frac{d\sigma}{dE_{\nu}} (\numu {\rm N} \rightarrow \numu {\rm K}^+ \Lambda),
\label{eq:35}
\label{sc:22}
\end{equation}

\begin{figure}[!tb]
\centerline{\epsfig{file=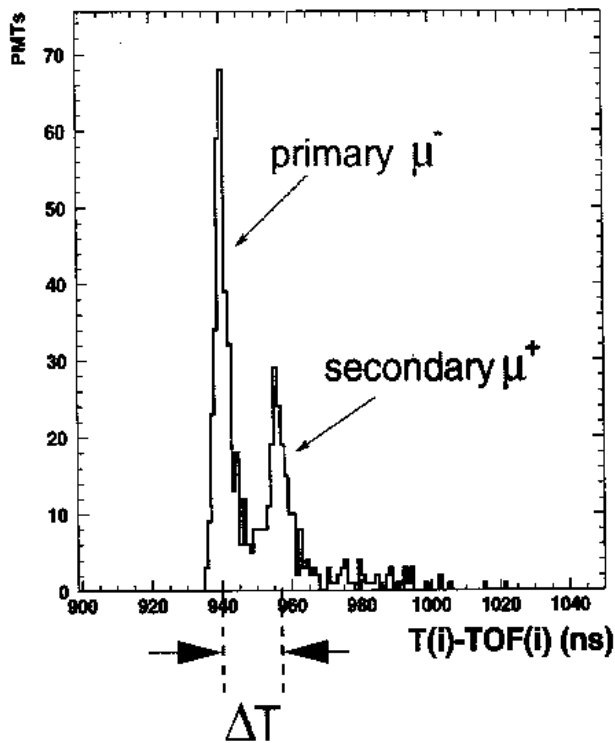, width=8.0cm}}
\caption[Time profile of associated-production event in a water Cherenkov detector]{Time distribution from a neutrino interaction candidate for 
 $\nu n \rightarrow \mu^{-} \Lambda K^{+}$, 
 $K^{+} \rightarrow \mu^{+} \nu$ recorded using
 the 1 kiloton water Cherenkov detector (1KT) at KEK.  Two peaks,
 separated in time by a few tens of nanoseconds, signal the occurrence
 of a $K^{+}$ decay subsequent to the primary charged-current interaction.} 
\label{fig:fig06}
\end{figure}

\noindent from its threshold
at $\approx$ 1~GeV through its rise and leveling off to an energy-independent 
value at $E_{\nu}$ between 10-15~GeV.   For purposes of
comparison and as a valuable check on systematics\cite{Day}, the
$\Delta S = 0$ companion charged-current reaction will also be measured:
\begin{equation}
\frac{d\sigma}{dE_{\nu}} (\numu {\rm n} \rightarrow \mu^- {\rm K}^+ \Lambda).
\label{eq:36}
\label{sc:23}
\end{equation}
      
      The off-line selections required to isolate reactions (\ref{eq:35})
and (\ref{eq:36}) are straightforward.   Assuming the final-state $\Lambda$ decays
into p $\pi^-$ for these reactions, they share the following topological 
attributes:

\begin{itemize}

\item [\it i)]  The reactions have relatively low charged-particle
           multiplicities from the primary vertex region.  Reaction
           (\ref{eq:35}) has three charged prongs, including the two
           daughter tracks from $\Lambda$ decay;
           reaction (\ref{eq:36}) has four charged prongs.

\item [\it ii)] The proton track of $\Lambda$ decay will appear as a short,
 heavily-ionizing track from the vertex region which stops in the scintillator.

\item [\it iii)] The final-state K$^+$ mesons will decay at rest or nearly at
           rest, and consequently a large-angle $\mu^+$ track will result.

\end{itemize}

   The most distinctive signature, however, arises with the time
sequence for light emission in scintillator elements from these
events.   For reaction (\ref{eq:35}) a 
``prompt" signal arises from the two-body decay of the $\Lambda$ into charged
tracks; in reaction (\ref{eq:36}) the prompt burst is enhanced by the presence
of the charged-current $\mu^-$ in the final state.  The prompt signal
is followed by a second signal, delayed by some few tens of nanoseconds,
from two-body decay of the K$^+$.  This timing signature,
taken in conjunction with the three topology attributes above, should yield
clean samples of reactions (\ref{eq:35}) and (\ref{eq:36}).

      The feasibility of exploiting the signature
afforded by the time profile of these reactions is
illustrated in Figure~\ref{fig:fig06}. The figure shows the time 
distribution from Cherenkov light from a candidate event for reaction
(\ref{eq:36}), where the occurrence of two peaks separated by approximately 16~ns
is readily seen~\cite{Mine}. At K2K the effective energy reach 
of the KEK neutrino beam restricts cross-section measurements
to $E_{\nu} \le$ 3~GeV; the atmospheric neutrino flux however
extends to higher energies.  Thus the \numi\ \numu ~beam
operated in the ``low-energy" configuration will enable a
complete picture of $\sigma(E_{\nu})$ for reactions (\ref{eq:35}) and (\ref{eq:36})
to be obtained, providing an observational basis for future proton-decay 
searches to discover or set improved lifetime lower limits on 
decay modes favored by SUSY grand unification models.

\subsubsection{Strangeness-changing neutral currents}

Notably absent from the interaction categories of the previous paragraphs are neutral-current
strangeness-changing reactions.  These have never been observed; their
occurrence at rates accessible in \numi\ would imply new physics 
beyond the Standard Model.  The existing limits on NC 
$\Delta S = 1$ processes are based upon searches for rare K decays. 
Although there are experimental difficulties 
with unambiguous identification of such processes
in neutrino reactions, there is nevertheless an opportunity 
for strangeness-changing NC search in the neutrino
sector.

Hints that an unrecognized type of neutral-current processes may
exist are to be found in discrepancies involving hyperon weak radiative decays. 
These strangeness-changing weak decays have a 
clear disagreements between existing data and a variety
of theoretical models - see \cite{lach},\cite{singer} 
for recent reviews. A long-standing puzzle concerns the large negative asymmetry 
coefficient observed in $\Sigma^+\rightarrow$ p$\gamma$ decay,
the measured value of which contradicts
accepted notions concerning the size of SU(3)-breaking.  To date, all of the 
assorted models invoked to describe these decays - including pole models, 
quark models, skyrmion models, vector meson dominance models and chiral
models - fail to explain either the asymmetries observed or the decay rates of
the various hyperons.  Very recently, measurements of
asymmetries which are large and negative in the $\Xi^0\rightarrow\Sigma^0\gamma$ and
$\Xi^0\rightarrow\Lambda^0\gamma$ decays by the KTeV (Fermilab) and NA48 (CERN)
experiments (\cite{ktev} and \cite{schmidth} respectively), 
run counter to the theoretical predictions for
a sizable positive value \cite{zen}. According to the comprehensive analysis of Gilman and 
Wise \cite{gilman}, the hypothesis that all weak radiative hyperon decays in the 56
multiplet of SU(6) are driven by the single-quark short-distance
transition $s\rightarrow d\gamma$, is untenable.

   A search for strangeness-changing neutral-current
neutrino interactions can usefully clarify the extent to which new physics
parameters may be missing from the analysis of weak radiative hyperon decays.
It is plausible that neutrino reactions, in contrast to hyperon weak decays, may 
provide cleaner signals for a new weak current in as much as the multiloop quark-gluon
diagrams which complicate hyperon decay analysis would be absent.
To hope for such a circumstance is perhaps not unreasonable; after all, the first
clear evidence for existence of the Z$^0$ in the guise of neutrino NC reactions preceded
the direct production of the  Z$^0$ by ten years.
 
   Below we list charged-current neutrino interactions which are examples of hyperon
production; included are two-body final states which represent the inverse of 
hyperon beta-decay.  These CC reactions require exposure of \minerva\
to an $\anumu$ beam.   Also listed are ``companion" NC neutrino reactions which 
yield single final-state hyperons.  The latter include possible strangeness-changing
neutral-current reactions (labeled SCNC), a subset of which could be the focus of
a dedicated search.  Note that the SCNC reactions are in principle accessible with 
either $\numu$ or $\anumu$ beams.

\begin{eqnarray}
\bar{\nu_\mu} + {\rm p} \rightarrow \mu^+ + \Lambda^0 \hspace{3cm}
& {\nu_\mu} + {\rm p} \rightarrow {\nu} + \Sigma^+ \hspace{1.5cm} SCNC \\ 
\rightarrow \mu^+ + \Sigma^0 \hspace{3cm} &  
\hspace{1.2cm} \rightarrow {\nu} + \pi^0 + \Sigma^+ \hspace{0.6cm} SCNC \\ 
\rightarrow \mu^+ + \pi^0 + \Sigma^0 \hspace{2.1cm} &
\hspace{1.3cm} \rightarrow {\nu} + K^0 + {\rm p} \hspace{0.5cm} SCNC \\
\bar{\nu_\mu} + {\rm n} \rightarrow \mu^+ + \Sigma^- \hspace{2.8cm} &
{\nu} + {\rm n} \rightarrow {\nu} + \Lambda^0 \hspace{1.5cm} SCNC \\
\rightarrow \mu^+ + \pi^- + \Lambda^0 \hspace{1.9cm} &
\hspace{1.3cm} \rightarrow {\nu} + \Sigma^0 \hspace{1.5cm} SCNC \\
\vdots \hspace{4.5cm} & \vdots \hspace{2.75cm} \nonumber
\end{eqnarray}

To isolate SCNC interactions, it is of course necessary to
distinguish them amongst the predominate neutrino CC and NC samples.
For certain selected SCNC reactions, this should be feasible.
First and foremost on our list to
identify SCNC process is the absence of a charged lepton since they
are only in the neutral-current reactions in conjunction with only
one strange particle being present. Other methods at our
disposal is the existence of hyperon resonances without an accompanying
meson which requires a highly-segmented detector with excellent 
containment of neutral mesons. Background estimates under the
hyperon resonances can be accurately determined by off-resonance
measurement of p$\pi^-$ states that would then give the accuracy needed for
the resonance region. Events above that expectation would yield limits on the
SCNC processes. Prerequisite detector requirements are
good resonance mass reconstruction, neutral meson containment and a magnetic field, 
knowledge of the sideband backgrounds not going through resonances 
to an accuracy $10 \times$ better than the
resonance search region. All of this is achievable in the current design 
of \minerva.

\subsubsection{Hyperon beta-decay and inverse neutrino processes}

Hyperon beta-decay $A \rightarrow B \hspace{0.1cm} e^{-} \hspace{0.1cm}
\bar{\nu}_{e}  \hspace{0.1cm}$ provides a window onto 
weak hadronic current form-factors and their underlying
structure. 
In the V-A formulation the transition amplitude is:
\begin{equation}M = \frac{G}{\sqrt{2}} <B|J^{\lambda}|A>{\bar{u}_{e}} 
\gamma_{\lambda} (1 + \gamma_{5}) u_{\nu} \end{equation}
The V-A hadronic current can be written as:
\begin{eqnarray*} <B|J^{\lambda}|A> = {\cal C} \hspace{0.1cm} i \hspace{0.2cm}
\bar{u}(B) & \{ & f_{1}(q^{2})\gamma^{\lambda}
+ f_{2}(q^{2}) \frac{\sigma^{\lambda \upsilon}\gamma_{\upsilon}}{M_{A}} +
f_{3}(q^{2}) \frac{q^{\lambda}}{M_{A}} + \\
& [ & g_{1}(q^{2}) \gamma^{\lambda} + g_{2}(q^{2}) 
\frac{\sigma^{\lambda \upsilon} \gamma_{\upsilon}}{M_{A}} + 
g_{3}(q^{2}) \frac{q^{\lambda}}{M_{A}} ]
\hspace{0.1cm} \gamma_{5} \hspace{0.25cm} \} u(A) 
\hspace{1.5cm} \end{eqnarray*}
where ${\cal C}$ is the CKM matrix element and $q$ is the momentum transfer.
There are 3 vector form-factors: $f_1$ (vector), $f_2$ (weak magnetism) and
$f_3$ (an induced scalar); plus 3 axial-vector form-factors: $g_1$ (axial-vector),
$g_2$ (weak electricity) and $g_3$ (an induced pseudo-scalar).

Recent high statistics measurements of
these form-factors using KTeV $\Xi^0$ hyperon beta-decays have been reported\cite{beta_ktev};
the results show that the level of
SU(3) breaking is very small compared to expectations of modern theories\cite{ratclif}. These new 
results have been used to extract the CKM matrix elements V$_{us}$\cite{cabibbo}. 
Similar physics studies can be done with anti-neutrino interactions that produce
hyperons.  The hyperon decays themselves
will have the added feature of a self-analyzing power
of the polarization vector. Thus the fundamental form-factors and
CKM matrix elements will be accessible without the hindrance of 
double solutions due to the missing
neutrino energy.  On the other hand, in \minerva\ there arises the problem
of dealing with the nuclear potentials which
comprise the environment for target nucleons. This consideration might motivate
running with liquid Hydrogen and Deuterium targets in a future program.

Although the simplest beta-decays and their corresponding inverse processes provide
the predominate samples for both hyperon beta-decays and
in $\Delta S =$ 1 neutrino interactions, there are also interesting albeit 
more complicated 4-body beta-decay processes listed
below along with some corresponding strangeness-producing neutrino interactions:

\begin{eqnarray}
\Lambda \rightarrow {\rm p} \pi^0 e^- \bar{\nu} \hspace{3.5cm} 
     & \bar{\nu_\mu} + {\rm p} \rightarrow \mu^+ \Lambda \pi^0 \\
\Sigma^+ \rightarrow {\rm p} \pi^- e^+ {\nu} \hspace{3.5cm} &  \\
\Sigma^+ \rightarrow {\rm p} \pi^- \mu^+ {\nu}  \hspace{3.5cm} &
      \nu_\mu + {\rm p}  \rightarrow  \mu^- \Sigma^+ \pi^+ \label{eq:sig1} \\
& \nu_\mu + {\rm p} \rightarrow \mu^- {\rm p} K^+  \label{eq:sig2} \\
\Sigma^- \rightarrow {\rm p} \pi^- e^- \bar{\nu} \hspace{3.5cm} &
      \bar{\nu_\mu} + {\rm p}  \rightarrow  \mu^+ \Sigma^- \pi^+
\end{eqnarray}

Although none of the 4-body hyperon 
beta-decays have been officially observed, a handful of $\Omega^-$ 
candidate decays may have
been isolated; preliminary results were presented at DPF 2003\cite{beta_e871}. The
theory behind these decays, \cite{pais} and \cite{singh}, with their more complicated
interaction models, were developed in the 60s. With neutrino interactions
these processes should be easier to obtain and hence studied for the strength of
physics interactions. They would allow for
a much more complicated form-factor analysis process that involves 16 variables
in the V-A style, but in the Standard Model would be a way to check the scale of 
the SU(3) breaking. It should also be noted that there are several types of
beta-decays in this list that can be studied in neutrino beams before going to
anti-neutrinos; these are the types shown in Equations \ref{eq:sig1} and \ref{eq:sig2}
and others not listed.

There are also forms of 4-body hyperon beta-decays and likewise neutrino interactions that
are forbidden by $\Delta S = - \Delta Q$. 
These too can be extensively searched for in neutrino-beam running before the necessity for
anti-neutrino beams. While the existence of the forbidden decays would be exciting in
hyperon beams, these interactions in neutrino beams would give information about probing
the forms of the interactions.

Study of $\Delta S = 1$ pentaquark states, like those recently announced\cite{pquark}, 
could be greatly extended here. In regard to these pentaquark states (whether 4-quark and a anti-quark
bound combination or a 
loosely-bound baryon-meson combination similar to mesonic atoms), with the production of
hyperons and mesons together a wealth of combinations can be throughly examined for studying the
full spectrum of the pentaquark family\cite{jaffe} of particles as well as
other exotic quark combinations such as di-baryons.

\subsubsection{Charm production physics}

Historically, most neutrino scattering experiments found their way into
charm-production studies when they investigated opposite or same-sign muon pairs
generated by neutrino interactions. This signal arises because 
many of the charm particles decay with
a muon, giving an extra muon along with the one produced by the CC neutrino reaction.
The decay muons usually differ substantially from the direct CC neutrino
muons in both momentum and angular distribution, but in some
cases it is not possible to discern a difference. In \minerva, with its lower neutrino energy beam,
the production of charm
particles will be suppressed compared to previous high-energy physics experiments.
Hence the reach of \minerva\ will be limited, but its large neutrino flux still allows 
interesting charm physics to be done.

An important contribution \minerva\ will provide in charm production is study
of the cross-section turn-on at or just a few hundred MeV above threshold. This
threshold is very sensitive to the bare charm mass. With the proposed beam
running schedule for \minerva\ we expect $\sim 6500$ charm events for a
three-ton detector over the first five years, with an additional $\sim 3200$
from anti-neutrino beam running for $x_{F} > 0$. Most of these charm events
($\sim$ 65\%) will be produced during the HE beam running configuration. As
noted, these yields depend strongly on the bare charm mass; varying this
parameter by 10\% results in expected yield changes of 30\%.  As discussed
earlier, neutrino experiments measure charm-production parameters by studying 
opposite-sign dimuon production. From preliminary studies, the expected number
of dimuons in \minerva\ over five years is $530 \pm 50$ for a bare charm mass
of 1.3~$\mbox{GeV/c}^2$. For bare charm masses of 1.15~$\mbox{GeV/c}^2$ or
1.45~$\mbox{GeV/c}^2$, the expected yields are $680 \pm 60$ and $420\pm40$
respectively. The yield  assumes charm produced with $x_{F} > 0$ and a lower
momentum cut on the decay muon of 1.5~GeV/c. The errors on the yields include
the error on the average semi-leptonic branching ratio for charm\cite{nutev}
and the error on subtracting the background rate from pion decay. \minerva\ is
at an advantage in being able to determine the sign of the muons via magnetic
tracking. Background rates can be determined by looking for same-sign dimuons.
At \minerva\ beam energies the expected number of background events should be
approximately equal to the signal values. \minerva\ should improve on the
charm-quark mass determination currently set by the NuTeV/CCFR data at
$1.38\pm0.13~\mbox{GeV/c}^2$\cite{nutev}.

%% file: duality.tex
\section{Perturbative/Non-Perturbative Interface}
\label{sect:duality}

\subsection{Parton Distribution Functions}

One obvious reason for the importance of neutrino data in the
extraction of parton distribution functions is the neutrino's ability to
directly resolve the flavor of the nucleon's constituents: $\nu$ interacts
with d, s, $\ubar$ and $\cbar$ while the $\overline{\nu}$ interacts with u,
c, $\dbar$ and $\sbar$. This unique ability of the neutrino to ``taste" only
particular flavors of quarks assists the study of parton distribution
functions. A high-statistics measurement of the partonic structure of the
nucleon is here proposed, using the neutrino's weak probe, to
complement on-going study of this subject with electromagnetic probes at other laboratories.

With the high statistics anticipated in \minerva, as well as the special
attention to minimizing neutrino beam systematics, it should be possible
to determine the individual structure functions
$F_1^{\nu N}(x,Q^2)$, $F_1^{\bar \nu N}(x,Q^2)$, $F_2^{\nu 
N}(x,Q^2)$, $F_2^{\bar \nu N}(x,Q^2)$, x$F_3^{\nu N}(x,Q^2)$ and
x$F_3^{\bar \nu N}(x,Q^2)$ (where N is an isoscalar target) for the first time. 

In leading-order QCD, four
of the structure functions are related to the parton distribution functions (PDFs)
by:

\begin{eqnarray*}
2F_1^{\nu N}(x,Q^2) &=& u(x) + d(x) + s(x) + \bar u(x) + \nonumber \\
                    & & \bar d(x) + \bar c(x)
\nonumber \\
2F_1^{\bar \nu N}(x,Q^2) &=& u(x) + d(x) + c(x) + \bar u(x) + \nonumber \\ 
		    & & \bar d(x) + \bar s(x)
\nonumber \\
xF_3^{\nu N}(x,Q^2) &=& u(x) + d(x) + s(x) - \bar u(x) -  \nonumber \\
 		    & & \bar d(x) -\bar c(x)
\nonumber \\
xF_3^{\bar \nu N}(x,Q^2) &=& u(x) + d(x) + c(x) - \bar u(x)-  \nonumber \\ 
		         & & \bar d(x) - \bar s(x)
\end{eqnarray*}

Taking differences and sums of these structure functions 
allows extraction of individual parton distribution functions in each
$(x, Q^2)$ bin:

\begin{eqnarray*}
2F_1^{\nu N} - 2F_1^{\bar \nu N} &=& [s(x) - \bar s(x)] + [\bar c(x) -
c(x)]
\nonumber \\
2F_1^{\nu N} - xF_3^{\nu N} &=& 2[\bar u(x) + \bar d(x) + \bar c(x)]
\nonumber \\
2F_1^{\bar \nu N} - xF_3^{\bar \nu N} &=& 2[\bar u(x) + \bar d(x) + \bar
s(x)]
\nonumber \\
xF_3^{\nu N} - xF_3^{\bar \nu N} &=& [\bar s(x) + s(x)] - [\bar c(x) +
c(x)]
\end{eqnarray*}

As the order of QCD increases and gluons are taken into consideration, 
global fitting techniques must be applied to extract of the parton
distribution functions.  With the manageable systematic errors
expected for the {\numi} beam, the ability to isolate individual parton
distribution functions will be dramatically improved by measuring the full
set of separate $\nu$ and $\bar \nu$ structure functions with the impressive
statistics possible in this experiment.

There are two primary (related) methods for extracting this full set of
structure functions.  One exploits the varying y behavior of the
coefficients of the structure functions in the expression for the cross
section:

\begin{eqnarray*}
\frac{d^2 \sigma^{\nu (\bar \nu)}}{dx dy} &=&
2\frac{G^2_F M_p E_\nu}{\pi} \Bigl[
x y^2 F_1^{\nu (\bar \nu)}(x,\qsq) + \nonumber \\
& & \left( 1- y - \frac{M_pxy}{2E_\nu} \right)
                  F_2^{\nu (\bar \nu)}(x,\qsq) \pm \nonumber \\
& & y \left( 1-y/2 \right) xF_3^{\nu (\bar \nu)}(x,\qsq)\Bigr],
\end{eqnarray*}

\noindent the other uses the "helicity representation" of the cross section:

\begin{eqnarray*}
\frac{d^2 \sigma^\nu}{dx d\qsq} &=& \frac{G^2_F}{2 \pi x}
\Bigl[\frac{1}{2}
\left( F_2^\nu (x,\qsq) + xF_3^\nu (x,\qsq) \right) + \nonumber \\
& & \frac{(1-y)^2}{2}
\left(F_2^\nu (x,\qsq) - xF_3^\nu (x,\qsq) \right) - \nonumber \\
& &2 y^2 F_L^\nu x,\qsq) \Bigr],
\end{eqnarray*}
and
\begin{eqnarray*}
\frac{d^2 \sigma^(\bar \nu)}{dx d\qsq} &=& \frac{G^2_F}{2 \pi x}
\Bigl[\frac{1}{2} \left( F_2^{\bar \nu} (x,\qsq) - xF_3^{\bar \nu}
(x,\qsq) \right) + \nonumber \\
& &\frac{(1-y)^2}{2} \left(F_2^{\bar \nu} (x,\qsq) +
xF_3^{\bar \nu} (x,\qsq) \right) - \nonumber \\
& & 2 y^2 F_L^{\bar \nu} (x,\qsq) \Bigr],
\end{eqnarray*}

\noindent By plotting the data as a function of $(1-y)^2$ in a given $x-\qsq$
bin, it is possible to extract all six structure functions.

For this sort of parton distribution function study, large
anti-neutrino samples are an imperative.

\subsection{Quark Distributions at Large $x$}

Although a large body of structure function data exists over a wide range
of $x$ and $Q^2$, the region $x > 0.6$ is not well explored.
For $x \geq 0.4$ contributions from the $q\bar q$ sea become
negligible, and the structure functions are dominated by valence quarks.

Knowledge of the valence quark distributions of the nucleon at large $x$
is vital for several reasons.
The simplest SU(6) symmetric quark model predicts that the ratio of $d$
to $u$ quark distributions in the proton is $1/2$, however, the
breaking of this symmetry in nature results in a much smaller ratio.
Various mechanisms have been invoked to explain why the $d(x)$
distribution is softer than $u(x)$.
If the interaction between spectator quarks is dominated by single-gluon exchange, for instance,
the $d$ quark distribution will be suppressed, and the $d/u$ ratio will
tend to zero in the limit $x \to 1$\cite{SPIN0}.
This assumption has been built into most global analyses of parton
distribution functions\cite{GLOBAL}, and has never been tested
independently.
On the other hand, if the dominant reaction mechanism involves
deep-inelastic scattering from a quark with the same spin 
orientation as the nucleon, as predicted by perturbative QCD counting
rules, then $d/u$ tends to $\approx 1/5$ as $x \to 1$\cite{FJ}.

Measurement of structure functions at large $x$ will yield 
insights into the mechanisms responsible for spin-flavor symmetry
breaking. In addition, quark distributions at large $x$ are a crucial input for
estimating backgrounds in searches for new physics beyond the Standard
Model at high energy colliders\cite{KUHL}.

The QCD evolution of parton distribution functions takes high-$x_{Bj}$
pdf's at low $Q^2$ and evolves them down to moderate-and-low $x$ at higher
$Q^2$. This obviously means that one of the larger contributions to background
uncertainties at LHC will be the very poorly-known high-$x$ PDF's
at the lower $Q^2$ values accessible to the {\numi} beam.  The appearance
of an anomaly at high $x$ will be discussed below. First note
that one
problem in studying this problem has been accumulation of sufficient
data at high $x$, off light targets, to extract the PDF's. 
The {\numi} beam will finally yield the necessary statistics to address this
important concern.

Uncertainties in 
current nucleon parton distribution functions at high $x$
are of two types: the ratio
of the light quark PDF's, $d(x)/u(x)$, as $x\to1$, and the role of leading
power corrections (higher twist) in extraction of the high $x$ behavior
of the quarks.

Analyses of present leptoproduction data from hydrogen and deuterium targets
have been unable to pin down the high-$x$ behavior of $d(x)/u(x)$.  Part of the
problem is due to the still unknown nuclear corrections involved in extracting
the "neutron" results from deuterium~\cite{WallyTony}. An analysis by Bodek and
Yang\cite{Yang:1998zb} indicated that the $d(x)/u(x)$ quark ratio approaches
0.2 as $x\to1$. However global QCD analyses of experimental results, such as
the CTEQ fits\cite{Lai:1994bb}, do not indicate the need for this higher value
of $d(x=1)/u(x=1)$.  Besides the statistical and experimental uncertainties in
the existing data, a complication with past experimental results was 
to model nuclear binding effects in the deuterium target  used. These issues
could be avoided with high-statistics exposure of an $H_{2}$ target, which
could directly measure the $d(x)/u(x)$ ratio in protons as $x \to1$ from the
ratio of neutrino-proton to antineutrino-proton cross-sections. Such a
measurement would require only a small correction for the residual sea-quark
contributions at high $x$.

Measurement of quark densities at high-$x_{\rm Bj}$ is closely related to the
question of the leading-power corrections known as ``higher twist effects''.
The $n^{th}$ order higher-twist effects are proportional to $1/Q^{2n}$ and
reflect the transverse momentum of the quarks within the nucleon
and the larger size of the probe as $Q^{2}$ decreases, increasing the
probability of multi-quark participation in an interaction. As for
the $d/u$ ratio, different analyses of higher-twist corrections in
current data leave unresolved issues that new
experimental information would clarify.  Recent work by Yang and Bodek\cite{Bodek-Yang}
seems to indicate that what has been measured as "higher-twist" in charged-lepton
scattering analyses is essentially accounted for by increasing the
order (NNLO) of the perturbative QCD expansion used in the analysis.

The only actual measurements of a higher-twist term in neutrino experiments
have been two low-statistics bubble-chamber experiments: in
Gargamelle\cite{Gargamelletwist} with freon and in BEBC\cite{Varvell:1987qu}
with $NeH_{2}$. Both bubble-chamber analyses are complicated by nuclear
corrections at high-$x$. However, both found a twist-4 contribution
smaller in magnitude than the charged leptoproduction analysis and,
most significantly, negative.

\begin{figure}[htbp]
\epsfysize=2.5in
\centerline{
{\includegraphics[clip=true, bb=127 248 502 602, width=0.8\textwidth]{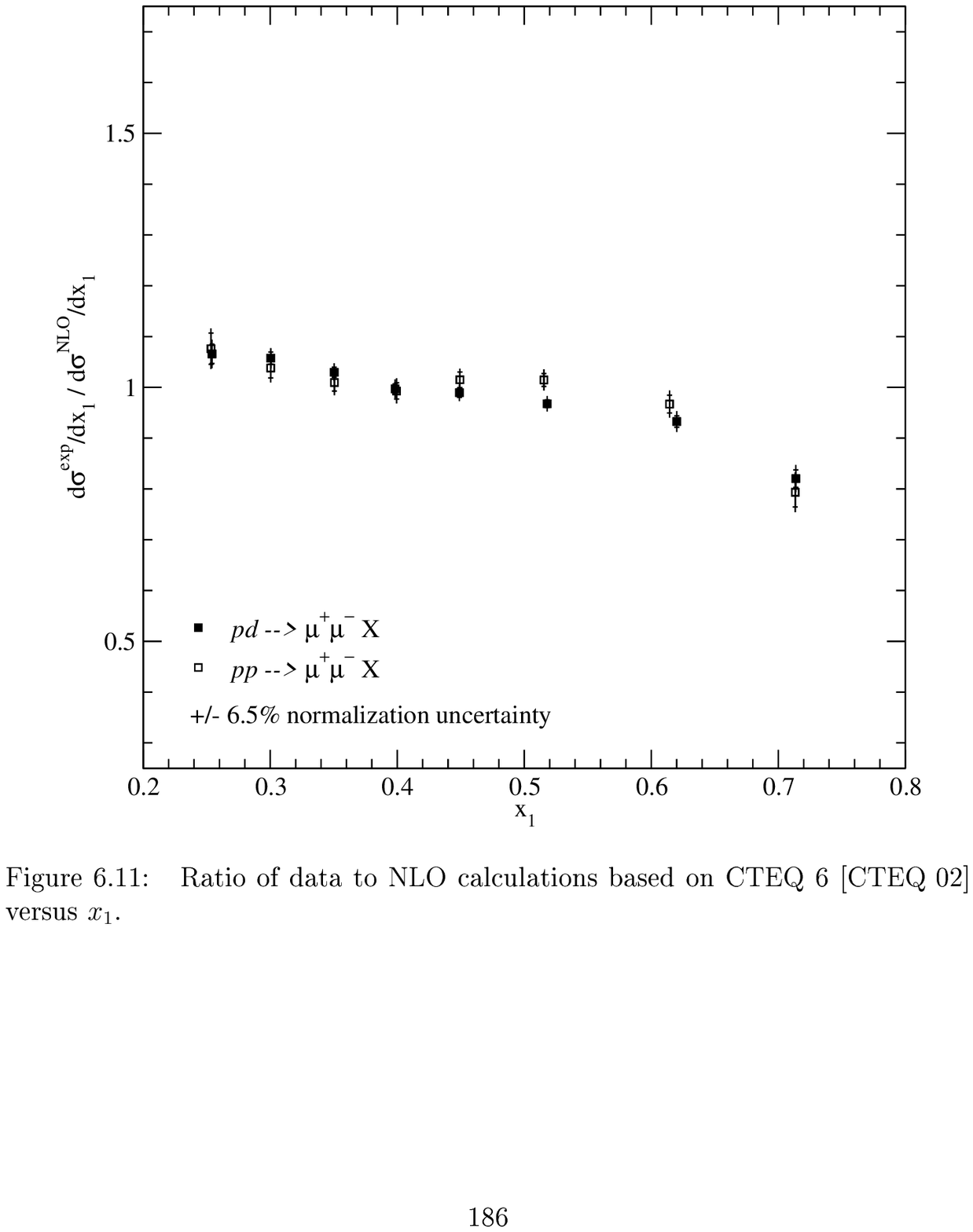}}}
\caption{Ratio of data to predictions based on CTEQ6 NLO pdf's vs $x_1$}
\label{e866}
\end{figure}

\begin{figure}[htbp]
\epsfysize=2.0in
\centerline{\epsffile{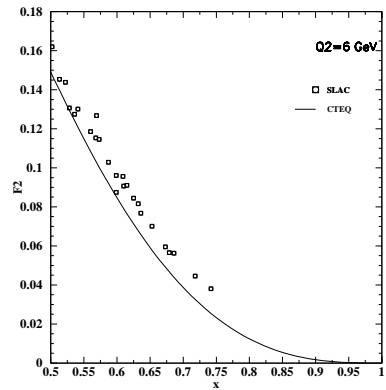}}
\caption[Comparison of SLAC electron scattering results with CTEQ NLO prediction]{Results from SLAC electron scattering experiments compared to the
CTEQ6 NLO prediction at high-$x_{\rm Bj}$}
\label{CK-highx}
\end{figure}

There are several indications that current parameterizations of the PDFs 
are {\bf not} correct at high $x$.  Figure~\ref{e866} shows the ratio 
of measured Drell-Yan pair production\cite{webb} compared to the latest
CTEQ global fits, CTEQ6\cite{cteq6}. The comparison seems to indicate 
that the valence distributions are {\em overestimated} at high-$x_{\rm Bj}$.
This directly contradicts a recent analysis at Jefferson Lab which seems
to indicate that the valence distributions are {\em underestimated} at
high $x$, as shown in Figure~\ref{CK-highx}.

Efforts are underway to understand how the $d(x)/u(x)$ ratio enters into the 
experimental comparison just discussed, and the large sample of high $x$
events in \minerva\ would certainly help clarify these results. 

The principal reason that the $d(x)/u(x)$ ratio is not better known is
the difficulty of accessing the structure of the neutron, due to the 
absence of free neutron targets, and the substantial theoretical uncertainties
associated with extracting information from neutrons bound in nuclei.
To overcome this problem, the BONUS experiment at Jefferson Lab\cite{BONUS}
has been approved to measure the inclusive
electron scattering cross section on an almost-free neutron using the
CEBAF Large Acceptance Spectrometer (CLAS) and a novel recoil detector
with low momentum threshold for protons and high rate capability.
This detector will allow tagging of slow backward-moving spectator
protons with momentum as low as 70 MeV/c in coincidence with the
scattered electron in the reaction $D(e,e^{\prime} p_s)X$.
This will ensure that the electron scattering took place on an almost
free neutron, with its initial four-momentum inferred from the observed
spectator proton spectrum. These measurements will unambiguously provide 
neutron structure measurements, which will thereby also reveal which of the 
available models best describe for instance, on-shell extrapolation for 
neutrons in nuclei. 

It should be stressed that the BONUS experiment at Jefferson Lab
will provide complementary 
information to \minerva\ measurements, overlapping in kinematics, and
on a similar time scale. 
With BONUS and \minerva\ combined, most of the 
questions in large-$x$
nucleon structure, parton distributions, and medium modifications, will be 
solved in the coming decade. BONUS will provide vital input regarding the 
extraction of neutron information from nuclei, while \minerva\ can uniquely
provide flavor decomposition information. 

\subsection{Quark/Hadron Duality}
\label{sect:qhduality}

The description of hadrons in terms of their fundamental quark and gluon 
constituents is one of the major challenges in nuclear physics today. 
While at present the quark and gluon degrees of freedom in QCD cannot describe the
structure and interactions of hadrons directly, in principle it should be just a matter of
convenience whether to describe a process in terms of quark-gluon
or hadronic degrees of freedom.
This idea is referred to as {\em quark/hadron duality}, and means that
one can use either set of complete basis states to describe physical
phenomena.
At high energies, where the interactions between quarks and gluons become
weak and quarks can be considered asymptotically free, an efficient
description of phenomena is afforded in terms of quarks; at low
energies, where the effects of confinement make strongly-coupled QCD
highly non-perturbative and the final state is guaranteed to consist of 
hadrons, it is more efficient to work in terms
of collective degrees of freedom, the physical mesons and baryons.
The duality between quark and hadron descriptions reflects the
relationship between confinement and asymptotic freedom, and is
intimately related to the nature of the transition from non-perturbative
to perturbative QCD. It has been said that (short of the full solution of 
QCD) understanding and 
controlling the accuracy of the quark-hadron duality is one of the most 
important and challenging problems for QCD practitioners today\cite{Shifman}.

\begin{figure}[htbp]
\begin{minipage}[b]{3.0in}
\epsfig{figure=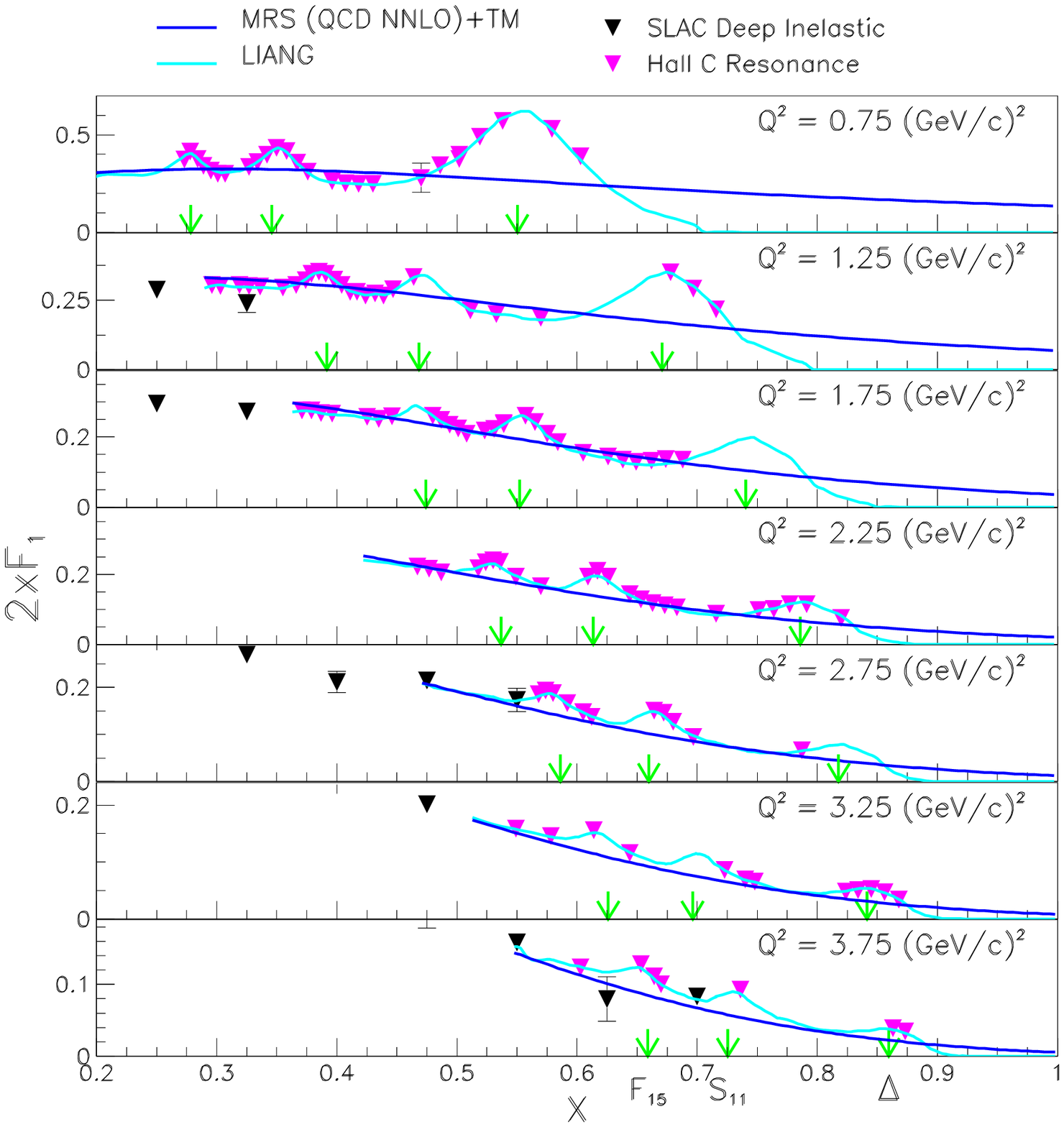, width=7cm}
\end{minipage}
\begin{minipage}[b]{3.0in}
\epsfig{figure=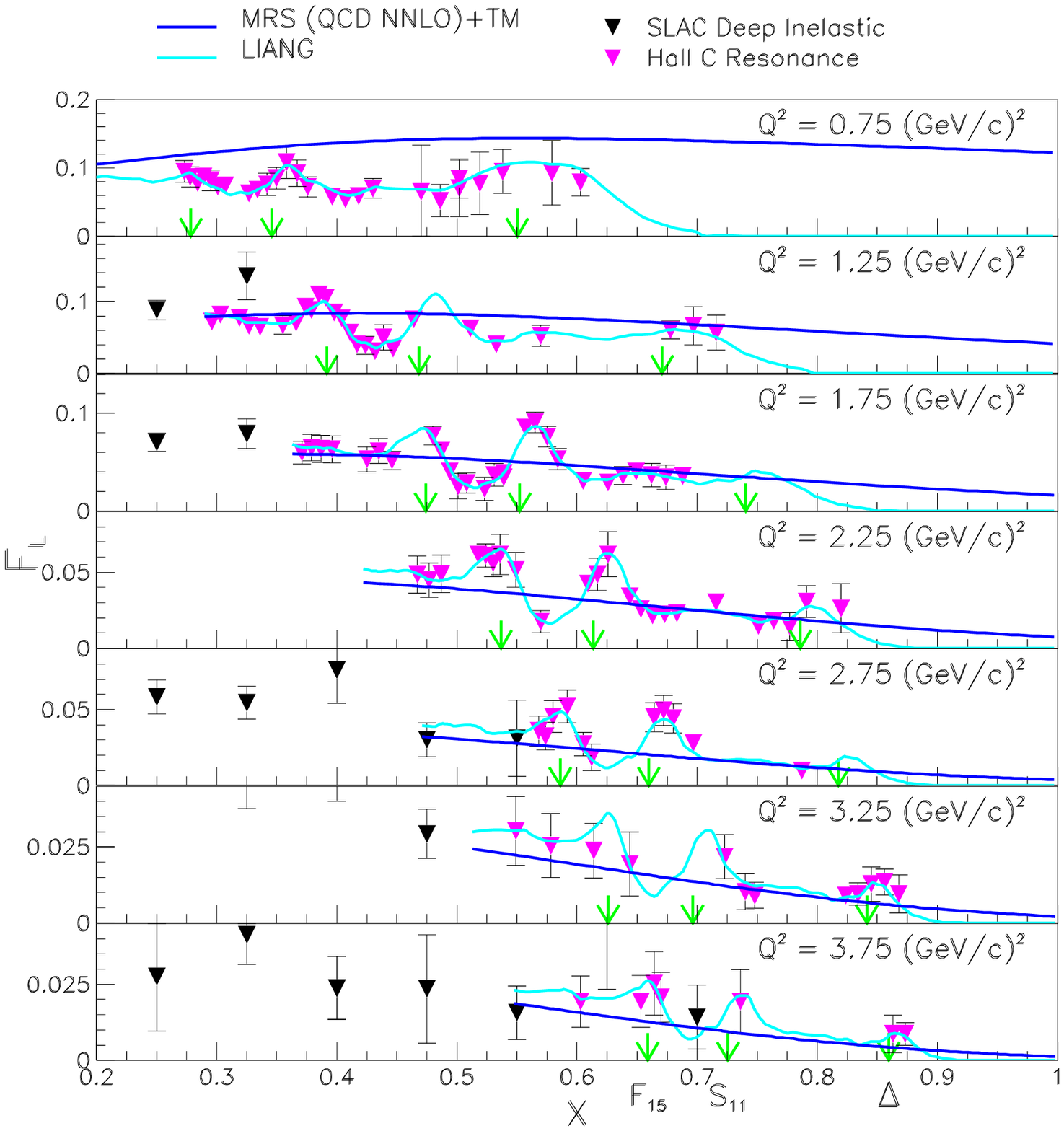, width=7cm}
\end{minipage}
\begin{centering}
\label{BGdual} 
\caption[Nucleon transverse and longitudinal structure functions from e--p data]{
The nucleon transverse ($2xF_1$) and longitudinal ($F_L$) 
structure functions, as measured in electron--proton scattering, 
are plotted as a function of the Bjorken scaling
variable $x$ for the indicated $Q^2$ bins. The curves are a fit to the
resonance data by Liang (light blue), and the parameterization from
MRST\cite{MRS} (dark blue) at next-to-next-to leading order,
corrected for target mass\cite{barbieri}. The data are in the 
resonance (from Hall C at Jefferson Lab\cite{Keppel}, purple) and 
deep-inelastic (from SLAC\cite{Dasu}, black) regimes, as indicated.}
\end{centering}
\end{figure}

Although the duality between quark and hadron descriptions is formally
exact in principle, how duality is manifest, specifically, in different
physical processes and under different kinematical conditions is a
key to understanding the consequences of QCD for hadronic structure.
The phenomenon of duality is quite general in nature and can be
studied in a variety of processes, such as $e^+ e^- \rightarrow$ hadrons,
or semi-leptonic decays of heavy quarks.
Duality in lepton--nucleon scattering, historically called 
Bloom-Gilman duality, links the physics of resonance production to the 
physics of deep-inelastic scaling. This duality is illustrated in  
Figure~\ref{BGdual}, 
where the nucleon transverse ($2xF_1$) and longitudinal
($F_L$) structure functions, measured in electron--proton scattering, 
are plotted as a function of the Bjorken scaling
variable $x$ for the indicated $Q^2$ bins. The curves are a fit to the
resonance data by Liang, and the parton distribution function based
parameterization of the MRST\cite{MRS} group at next-to-next-to leading order,
corrected for target mass\cite{barbieri}. The data are in the 
resonance (from Hall C at Jefferson Lab\cite{Keppel}) and deep-inelastic 
(from SLAC\cite{Dasu}) regimes, as indicated.
Duality appears here in the observation that the hadronic (resonance) 
and quark (scaling) strengths are, on average, equivalent. Moreover, this
is true for all $Q^2$ bins observed, and thus
the perturbative curve (MRST) apparently describes the average $Q^2$ 
dependence of the hadronic, non-perturbative, resonance enhancement region.

The proposed \minerva\ experiment is uniquely poised to provide a wealth of 
data to answer where duality works, in what structure functions,
in what reactions, and for what kinematics. Duality has been 
well-verified for the proton $F_2$ structure function\cite{ioana1},
observed recently in the separated longitudinal and transverse 
unpolarized structure functions\cite{Keppel}, on nucleons and in nuclei\cite{dualnuc},
and in polarized structure functions\cite{g1hermes}. 
While its fundamental cause remains a mystery, duality appears experimentally
to be a non-trivial property of nucleon structure. It is, therefore, crucial
to test it in a variety of reactions -- including neutrino--nucleon and --nucleus
scattering and the structure function $xF_3$. 
Duality studies of electron--deuteron scattering at low $Q^2$ 
found a resemblance to deep-inelastic neutrino--nucleus 
scattering at much higher $Q^2$, indicating potential sensitivity of duality
to the valence quarks\cite{ioana2}. \minerva\ will allow this observation
to be verified and tested for the first
time, as data from similar kinematic regimes but differing in probe and 
interaction (from {\minerva} and Jefferson Lab) may be compared directly. 

It is important to point out that a revolutionary application of duality, 
if one understands the workings of the resonance--deep-inelastic interplay, 
would be to open-up
the region of very high $x$, which has not been possible in any other
experiment. As discussed above, the region of $x \approx 1$ is an important 
testing ground for
understanding of the valence quark structure of the nucleon, and it
will allow us to discriminate between various models for the 
mechanisms of spin-flavor symmetry breaking in the valence quark
distributions of the nucleon. 
A first attempt at such an application is the recent analysis by Bodek and
Yang\cite{Bodek2002}, offering a promising procedure for fitting
$F_2$ in the low $Q^2$, high $x$ region.  
Extrapolating their results through the resonance region yields values of 
$F_2$ consistent with duality arguments and the Jefferson Lab results 
mentioned above. 
In addition, with nuclear targets, duality extensions to large $x$ 
would permit measurements of the nuclear-medium modification 
of the nucleon structure function (nuclear EMC effect) at large $x$, where
deviation of
the ratio of nuclear to nucleon structure functions from unity is largest, and
sensitivity to different nuclear structure models greatest. 

Members of 
the {\minerva} collaboration are currently investigating quark/hadron duality 
in high-statistics electron scattering at Jefferson Lab
with the same or similar
nuclear targets as those proposed for {\minerva}\cite{e03110}. 
This will be followed by a comparison with all 
existing neutrino data, with the aim of continuing these studies with 
the higher statistics {\minerva} neutrino experiment in the future. Note that 
investigation of quark/hadron duality in the axial structure 
functions of nucleons and nuclei with neutrinos also adds a new 
dimension to the previous electron studies. Many issues,
such as nuclear dependencies, should be well understood in advance of the 
{\minerva} data. 

\subsection{QCD Moments}

\begin{figure}[htbp]
\centerline{
{\includegraphics[clip=true, bb=45 400 522 646, width=0.8\textwidth]{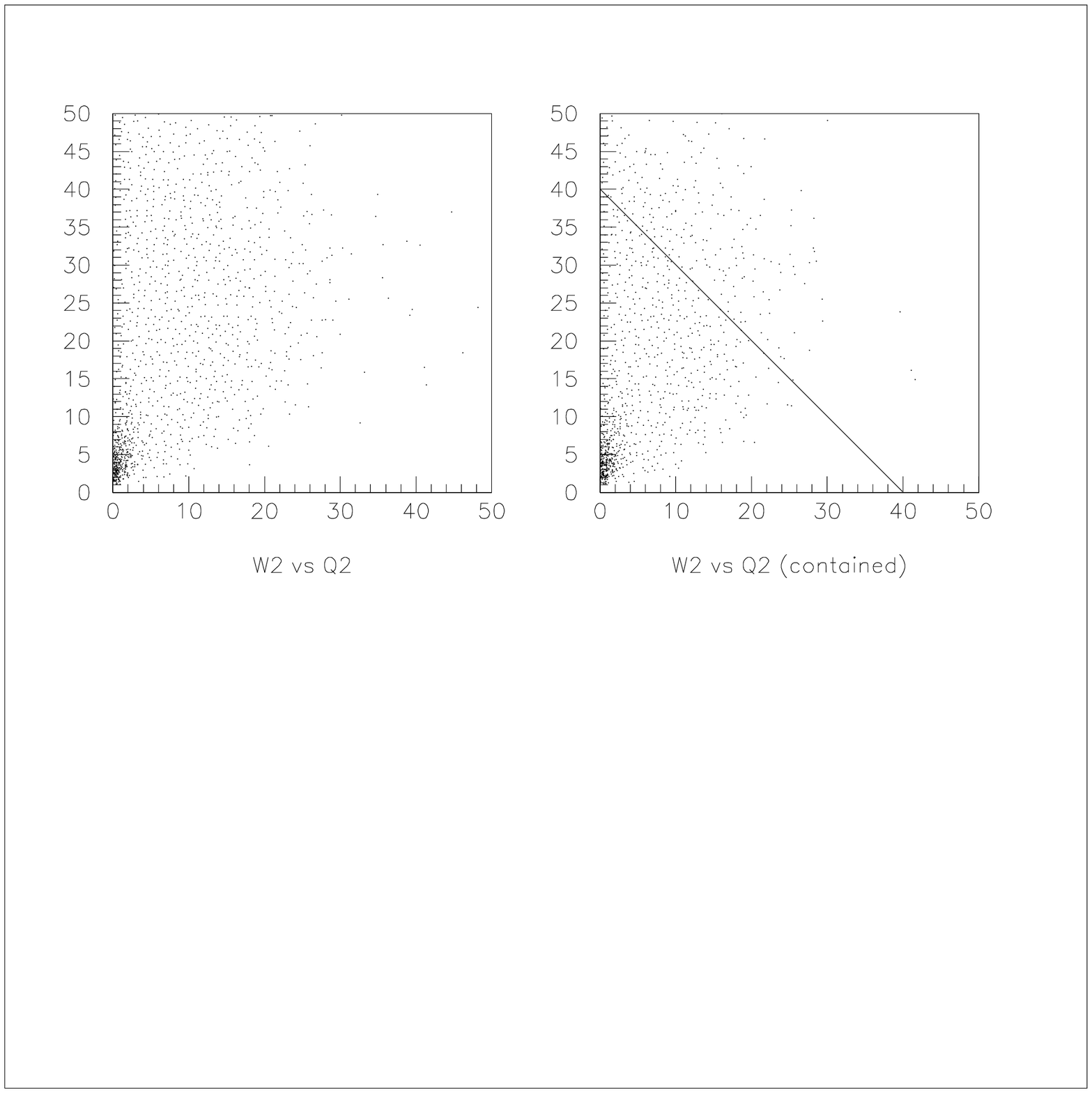}}}
\caption[DIS kinematics and containment in \minerva]{Left: Distribution of DIS events with LE beam in {\minerva} Monte Carlo.  Right: Events where total hadronic energy is contained by {\minerva}.  The line is an estimate of the limit where 50\% of events do not have containment of hadronic energy.}
\label{duality:w2vq2}
\end{figure}

Figure~\ref{duality:w2vq2} depicts the
substantial enhancement in the kinematic domain of precision data
made possible by {\minerva} over a range in $x$ and $Q^2$. This data will serve
a variety of purposes, and address long-standing 
questions regarding structure function behavior at low $Q^2$. Perhaps most 
importantly, the range 
of the data will allow for accurate moments of the structure functions to be
obtained. To obtain a structure function moment, it is necessary to 
integrate over the full range in $x$ at a fixed value of $Q^2$. The 
Cornwall-Norton moment of a structure function $F$, for instance, is expressed
as:

\begin{equation}
\label{eq:cnmom}
M_n^{CN}(Q^2) = \int_0^1 \; dx \; F(x,Q^2) \; x^{n-2}.
\end{equation}

The moments
are fundamental quantities, calculable in QCD and recently  
calculated in lattice QCD at $Q^2 = 4~\hbox{(GeV/c)}^2$ for valence distributions\cite{lattice}.
If duality is shown to hold, the proposed 
data may provide one of the few available quantities that can be directly 
compared to lattice QCD calculations.

Bloom-Gilman duality can be formulated in the language of an operator
product expansion (OPE) of QCD moments of structure functions, in which
contributions are organized according to powers of $1/Q^2$.
The leading terms are associated with free quark scattering, and are
responsible for the scaling of the structure function.
The $1/Q^2$ ``higher twist'' terms involve interactions between 
quarks and gluons and hence reflect elements of confinement dynamics. 
Duality measurements have
been explained in terms of a weak $Q^2$ dependence of the low moments of 
the structure functions\cite{DGP}. This is interpreted
within the OPE as indicating that non-leading, $1/Q^2$-suppressed,
higher-twist interaction terms do not play a major role even at low $Q^2$
($\approx 1$~GeV$^2$). It is this interpretation that facilitates 
comparison to lattice calculations, as the latter have no higher twist 
effects included.

Large-$x$ (resonance region) data become increasingly important for higher-order
moments due to the $n-2$ weighting of the moment. 
At n=6, for example, the resonance and large $x$ region above 
$x = 0.7$ make up 70\% of the Cornwall-Norton moment of $F_2$ at $Q^2 = 10~\hbox{(GeV/c)}^2$.
The contribution is larger at $Q^2 = 4~\hbox{(GeV/c)}^2$, where
lattice calculations are available. As noted above, there currently
exist little to no neutrino resonance cross-section
data in the resonance region or at larger $x$, while such
data will be easily obtainable with {\minerva}.

It is important to reiterate that, regardless of duality or OPE arguments, 
the experimental values for the moments can
be unambiguiously obtained with {\minerva}. For example, it is straightforward
to note, from Figure~\ref{duality:w2vq2}, that even the low-energy beam 
provides data covering a large range in $x$ (or $W^2$) for each $Q^2$ value 
up to $10~\hbox{(GeV/c)}^2$. The higher-energy beams will complement this sensitivity, extending 
the $Q^2$ range over which moments can be obtained, and adding statistics to
the much of the region covered by the low-energy beam. While comparable 
coverage can be obtained by combining electron and muon scattering data from
a multitude of laboratories, {\minerva} will uniquely provide, for instance, 
the $xF_3$ structure function, valence sensitivity (necessary to
current lattice comparisions), and flavor decomposition.

\subsection{Expected Results}

The proposed studies of structure function moments and quark/hadron duality
are straightforward with the proposed {\minerva} experiment. These topics
do not have the demanding experimental constraints that many of the other
proposed topics do. While it is crucial to understand the projected
$W^2$ or $x$ resolution, for 
instance, in studying resonance production behavior, duality studies and 
moment extractions average over these kinematic variables and are therefore
virtually insensitive to resolution issues. The expected {\minerva} resolutions 
are more than
adequate both to form integrals such as that in Equation~\ref{eq:cnmom} and
to study duality using data comparison with perturbative predictions, as in 
Figure~\ref{BGdual}.

Additionally, it is not necessary
to isolate specific production processes in these studies. It is only total cross-section
averages that are of interest, making {\minerva} 
essential to this effort in neutrino scattering.  

\begin{figure}[htbp]
\centerline{
{\includegraphics[width=0.7\textwidth]
{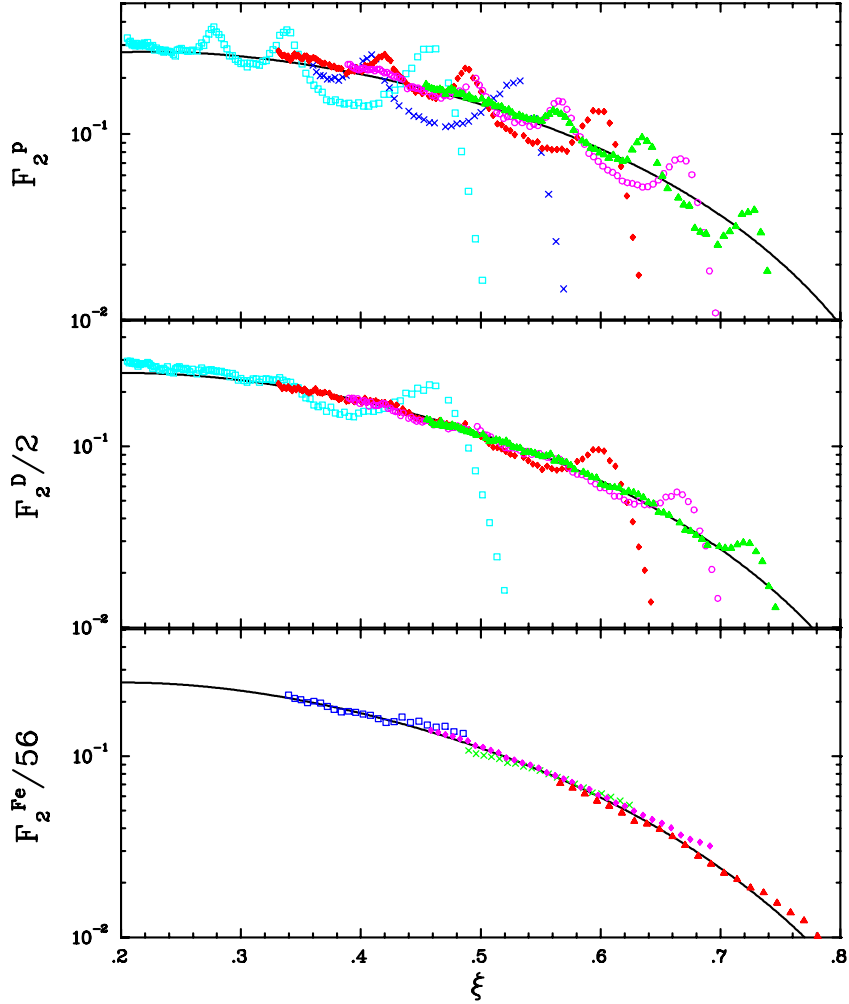}}}

\caption[Structure function $F_2$ data in the resonance region on various
nuclei]{Structure function $F_2$ data in the resonance region on  Hydrogen
(top), Deuterium (center),  and Iron (bottom) covering a range in $0.8 < Q^2 <
3.3~\hbox{(GeV/c)}^2$, and  plotted as a function of the  Nachtmann scaling
variable $\xi$. The elastic (quasi-elastic) peaks have been  removed. The
curves are the MRST and NMC parameterizations of the structure function, with a
model of the EMC effect applied for Iron.}

\label{xiscale}
\end{figure}

It has been observed from studies of quark/hadron duality
using nuclear targets that data in the resonance regime scale even more 
obviously when smeared by the nuclear Fermi momentum\cite{dualnuc}, 
as shown in Figure~\ref{xiscale}. In Hydrogen, the resonance peaks 
are prominent, while they are much less so in Deuterium, and completely
smeared away in Iron. In all cases, however, the resonance region averages to 
the scaling curves. In Iron, the smearing is such that the resonance data 
and scaling curves overlap completely; the nucleus performs the duality
averaging. 

With concerns about nuclear effects removed, then, there remain 
no impediments to studying duality for the first time
in neutrino scattering with \minerva. Similarly, extractions of higher-twist
contributions and studies of evolution for parton distribution function
extraction through the $Q^2$ dependence of the structure functions will not
be rendered ambiguous through the utilization of nuclear targets. 
 
\begin{figure}[hbtp]
\epsfysize=2.5in
\centerline{
{\includegraphics[width=0.7\textwidth]{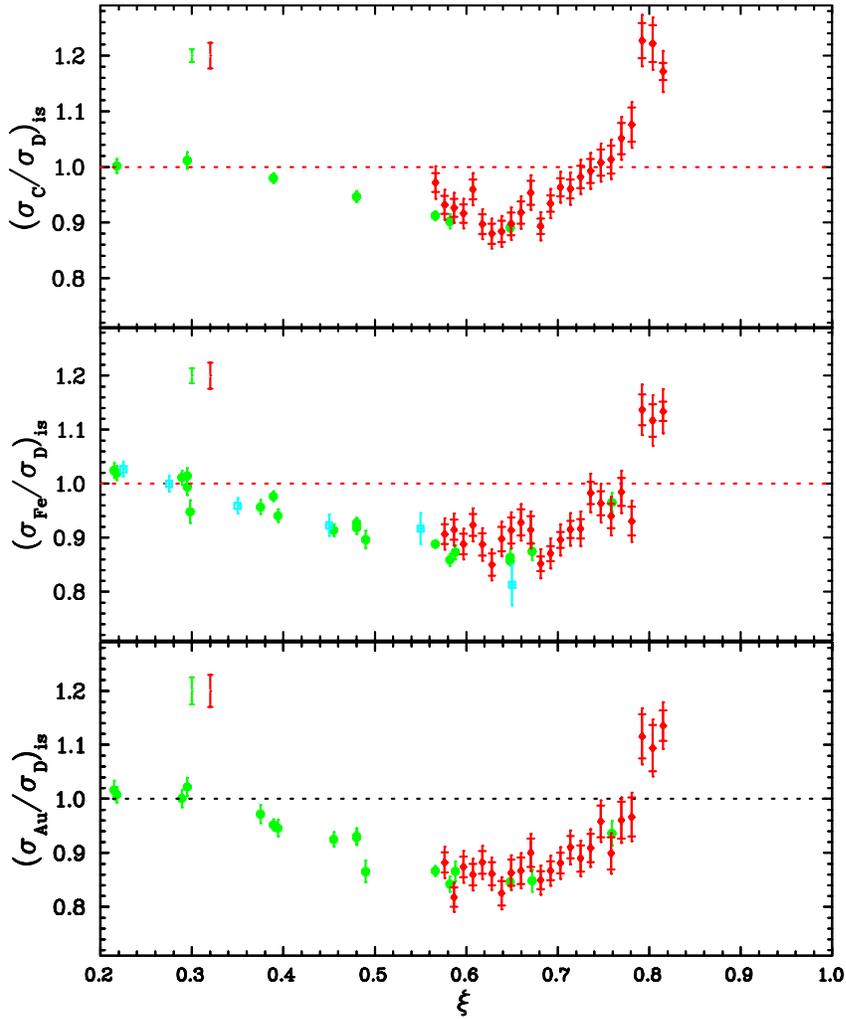}}}
\caption[Ratio of electron-nucleus scattering data to electron-deuterium]{Ratio of electron-nucleus scattering data 
(from top to bottom, Carbon, Iron, Gold) to
that obtained from Deuterium scattering, for data in both the resonance (red)
and deep inelastic (green) regimes. The data are plotted as a function of the
Nachtmann scaling variable $\xi$, allowing direct comparison of high $W^2$,
$Q^2$ DIS data to lower $W^2, Q^2$ resonance data.}
\label{nucdualfig}
\end{figure}

Most strikingly, it appears that the nuclear effects at large $x$ are the same
in the resonance and scaling regimes, as evidenced by  
Figure~\ref{nucdualfig} from\cite{dualnuc}, where the cross-section ratios
of carbon, iron, and gold to deuterium obtained in the resonance region
(red) are the same as those obtained in the deep-inelastic regime (green). 
Whatever the underlying cause for medium modifications to the structure 
functions as measured in nuclei, it is the same apparently for both hadronic 
(resonance) and scaling observables. Since the large $x$ region of the EMC
effect is ubiquitously attributed to Fermi motion in the nucleus, \minerva\ neutrino data should
yield similar A-dependent
results as the electron data in the figure. That is, it is expected (and 
will be tested) that the proposed data in 
the larger $x$ and resonance regions will have the same EMC effect as data 
at higher $W^2$.
Therefore, \minerva\ data at large $x$ can be used 
for parton distribution studies,
higher twist analyses, and nucleon structure studies with minimal nuclear
extraction uncertainties. 

The expected numbers of events for the resonance and deep inelastic regimes 
are tabulated in Section~\ref{sect:numiSample}.  These will make possible all of the studies here 
discussed in the perturbative and non-preturbative transition region of 
larger $x$ and lower $Q^2$ values. This is an exciting regime, with many 
unanswered
problems both interesting on their own and of import to other high energy 
applications.

%% file: gpd.tex
\section{Generalized Parton Distributions}
\label{sect:gpd}

One of the main goals of subatomic physics is to 
understand the structure of hadrons, and in particular 
the structure of the nucleon.   The primary approach to this problem
has been through measurement of the nucleon form-factors, 
with (quasi-)elastic scattering (for $Q^2$ up to a few (GeV/c)$^2$), parton 
densities, through inclusive deep-inelastic scattering (DIS), and 
distribution amplitudes, through exclusive processes.  However, the usual 
parton densities extracted from DIS are only sensitive 
to the longitudinal component of the parton distributions and 
do not give information on the transverse component, or 
other contributions to the nucleon angular momentum. 

\subsection{The Nucleon Spin Puzzle and GPDs}
 
In the late 1980's, results from polarized DIS showed that a 
relatively small fraction, about 20\%, of the nucleon spin is carried 
by the valence quarks.  The obvious candidates for the missing 
spin were the quark and gluon orbital momentum and gluon helicity. 
However, information on those quantities cannot be 
 extracted from DIS. 
 
In 1997, Ji\cite{Ji97,Ji97b} showed that a new class of nucleon observables, 
which he called ``off-forward parton distributions", could be used to 
determine the spin structure of the nucleon.  This work, along with 
developments by others,  especially Radyuskin\cite{Rady96,Rady96b} and Collins\cite{Collins}
showed that these distributions, now called generalized parton 
distributions (GPDs), had the potential to give a full three-dimensional 
picture of the nucleon structure. This exciting development has led to 
an immense amount of theoretical work in the last 
few years.  Short reviews can be found in~\cite{Rady02,Vander} and 
a comprehensive review can be found in~\cite{Diehl}. 
 
\begin{figure}[htbp]  
\centerline{\epsfig{file=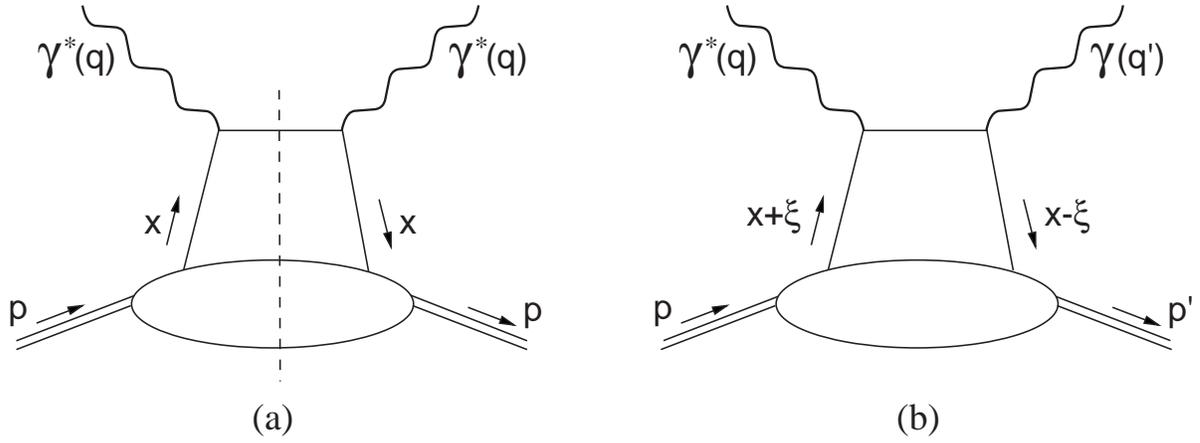,width=\linewidth}}
\caption[Forward virtual Compton and ``Handbag" diagrams]{(a) Forward virtual Compton amplitude which describes the DIS  
 {\em cross-section}   
via the optical theorem ($x_B=x$);  
(b) Handbag diagram occurring in the DVCS {\em amplitude}.  
}  
\label{handbag}  
\end{figure}

Ji showed that in leading twist there are four GPDs, which he called 
$H$, $\tilde{H}$, $E$, and $\tilde{E}$, for each quark flavor.  $H$ and 
$\tilde{H}$ 
are nucleon helicity-conserving amplitudes and $E$ and $\tilde{E}$ are 
helicity-flipping amplitudes. The GPDs are 
functions of $x$, $\xi$ (a factor determining the ``off-forwardness" 
of the reaction), and the total momentum-transfer squared, $t$.   The GPDs 
can be accessed experimentally through reactions proceeding via 
the ``handbag" diagram shown in Figure~\ref{handbag}. 
 
\subsection{Deeply-virtual Compton Scattering}

\begin{figure}[htbp]
\centerline{\epsfig{file=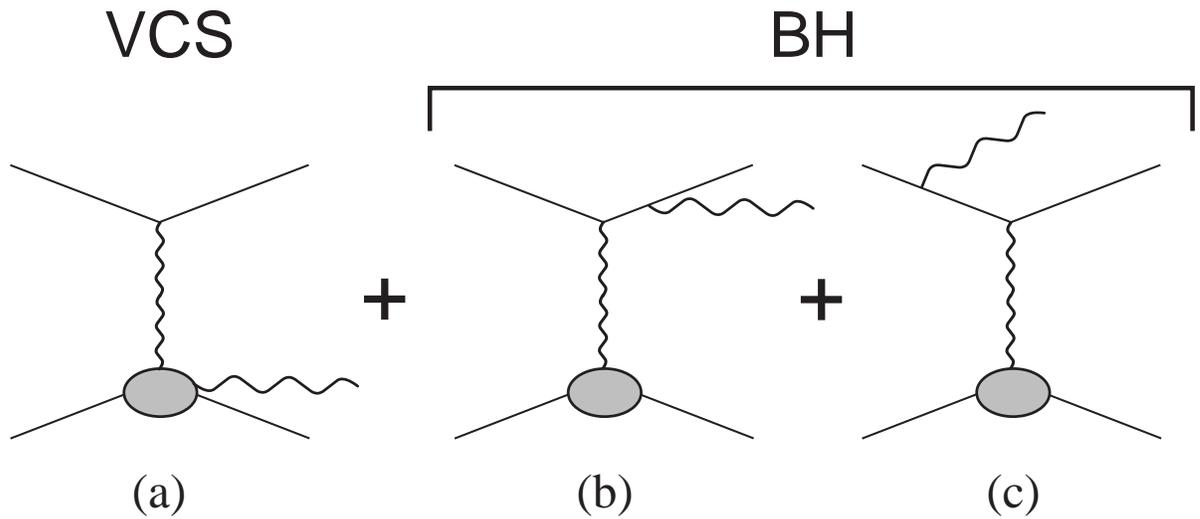,width=\linewidth}}
\caption[The DVCS process and interfering diagrams]{The DVCS process (a) along with the interfering Bethe-Heitler   
diagrams (b) and (c).}  
\label{dvcsbh}  
\end{figure}

The most promising reaction to measure GPDs 
identified so far is deeply-virtual Compton scattering (DVCS).  The DVCS 
reaction 
is shown in Figure~\ref{dvcsbh}a.  An interesting feature of DVCS is that it 
can interfere with the Bethe-Heitler process, Figure~\ref{dvcsbh}b, which is 
completely calculable in terms of the nucleon elastic form-factors.  This 
interference causes an asymmetry in the 
azimuthal distribution of the scattered proton allowing some quantities 
to be determined that would otherwise require a polarized target. 
However, DVCS involves a combination of the 
four GPD amplitudes, which cannot be separated using DVCS alone.  Some 
complementary information can also be obtained from nucleon form-factor 
measurements and deep exclusive meson electroproduction. 

 
\begin{figure}[htbp]  
\begin{minipage}[b]{6.0in}  
\begin{center}  
\epsfig{file=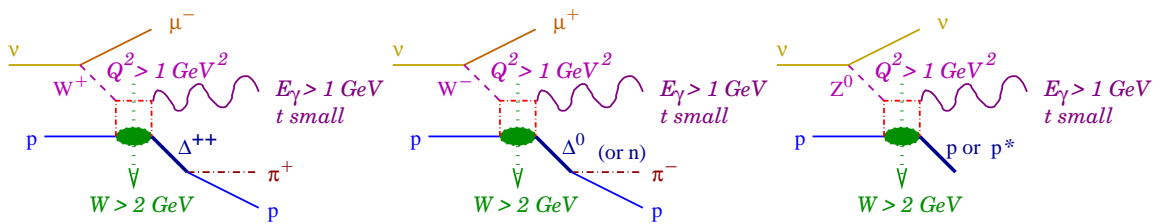,width=\linewidth}  
 \caption{Reactions sensitive to GPDs in neutrino
scattering.}  
\label{gpds}  
\end{center}  
\end{minipage}\hfill  
\end{figure}

Neutrino scattering provides a very similar reaction to DVCS.  In this 
case, the virtual exchange is of a $W^{\pm}$ with the production of 
an energetic photon, a $\mu^{\pm}$, with either a recoiling nucleon or 
nucleon resonance, as shown in Figure~\ref{gpds}.  
This ``weak DVCS" reaction is very promising 
theoretically 
because it provides access to different GPDs than DVCS.  It will 
help resolve the individual flavors, e.g. $d$ in neutrino scattering 
and $u$ in anti-neutrino scattering, and the interference of the $V$ and $A$ 
currents will give access to C-odd combinations of GPDs. 

\subsection{Measurement of GPDs in \minerva}
 
Studies of the weak DVCS reaction are currently underway by A.~Radyuskin, 
A.~Psaker, and W.~Melnitchouk.  One very encouraging result so far is 
that the $q\bar{q}$ equivalent of the polarized structure function $g_1$ 
can be measured {\em without} using polarized targets.   This would allow 
separation of the valence and sea parts of the spin-dependent GPDs, 
and help determine the role of the axial anomaly in the proton spin puzzle. 
In addition, although the 
Bethe-Heitler process is suppressed, it is still present and the 
interference with it would allow measurement of individual GPDs.  
 
The estimated cross-section for weak DVCS is about $10^{-39}~\hbox{cm}^2$.  For neutrino energies in the
5--10~GeV range, 
this would yield a few hundred events/year for a 1~GeV-wide bin in 
neutrino energy.  Although most of the events will be from nucleons in 
carbon, any nuclear modifications are expected to be small except at 
very small or very large $x$.

At least one other reaction, the hard exclusive production 
of the $D_s$ has also been proposed\cite{Lehm01} as a probe of GPDs. 
This reaction is sensitive to the gluon structure of the nucleus. 
Unfortunately, the cross-section for this reaction 
(estimated at $10^{-5}~\hbox{pb}$ for $Q^2 > 12~\hbox{(GeV/c)}^2$), is too small 
to be 
measured with precision in \minerva.  Nevertheless, over the entire 
run perhaps a few hundred events would be observed over all values of 
$Q^2$, which would give some information on the gluon GPDs. 

%% file: nuclear.tex
\section{Studying Nuclear Effects with Neutrinos}
\label{sect:nuclear}

In most neutrino scattering experiments, massive
nuclear target/detectors are necessary to obtain useful reaction rates. Neutrino-oscillation
experiments, despite the extremely intense beams designed for them,
must also use very massive Iron, water or other nuclear target/detectors, since
they are located hundreds of kilometers from the production point. 
Analysis of neutrino reactions within nuclear media requires an understanding
of certain processes which are absent in neutrino scattering on free nucleons;
these processes involve the so-called ``nuclear effects". Two general
categories of such effects can be distinguished.  Effects  comprising a first
category include:

\begin{itemize}

\item The target nucleon is moving within the nucleus and, when incoming
neutrino energies are $\leq$ 2~GeV, the initial target
energy and momentum must be accounted for using simulations which include
either a target Fermi gas model or, preferably, nucleon spectral functions.

\item Certain final states are excluded as a result of Pauli blocking among
identical nucleons.

\item  The resulting final state may undergo final state interactions
(FSI), including re-scattering and absorption; these interactions may
significantly alter the observed final-state configuration and measured energy.
\end{itemize}

The first two effects are either already included in Monte Carlos or are 
currently being examined in collaboration with nuclear theorists and will soon
be included.  The third effect is perhaps the most troublesome for current
and future neutrino experiments. There is a dearth of data for which nuclear
effects on specific hadronic final states (the fragmentation functions) have
been isolated, whether for neutrino or charged-lepton beams.  These effects are likely to be sizable
for neutrino energies producing a large
fraction of elastic and resonant final states\cite{Paschos}.

\begin{figure}[h]
\centerline{
\epsfxsize=\textwidth
\epsffile{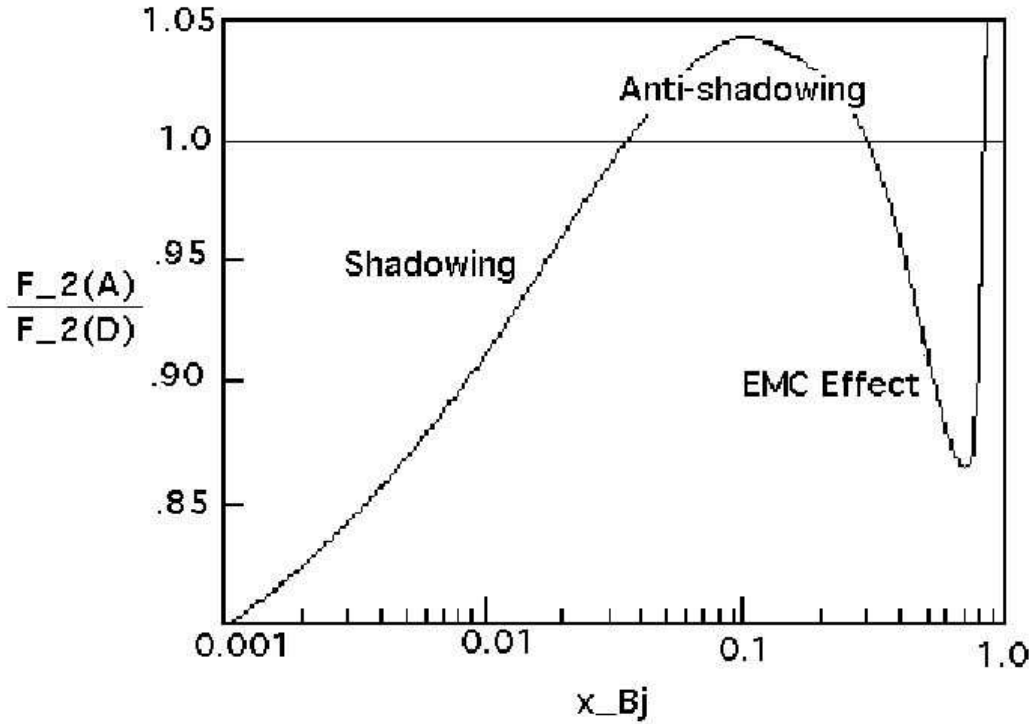}}
\caption[$F_2(\hbox{Nuclear})/F_2(\hbox{D}_2)$ vs. $x_{Bj}$]{The trend of the ratio of F$_2$ measured with 
a heavy nuclear target to F$_2$ measured using deuterium, for charged-lepton scattering,
as a function of x$_{Bj}$.}
\label{nuc_allx}
\end{figure}

A second category of nuclear effects are those by which the neutrino
interaction probability on nuclei is modified relative to that for free nucleons.
These effects occur across a wide range of
neutrino energies and are normally categorized by the x$_{Bj}$ of the quark
involved in the scattering, and the Q$^2$ of the intermediate vector boson
exchanged.  Nuclear effects of this type have been
extensively studied in deep-inelastic scattering (DIS) measurements of structure functions using  muon and
electron beams.  For example, Figure~\ref{nuc_allx} shows the ratio
of the structure function  $F_2$ measured on a heavy nuclear target to
$F_2$ measured for Deuterium. 

With neutrino beams, these nuclear effects have only been studied with
low-statistics in bubble-chamber experiments. 

\subsection{Final-state Interactions}

Distortions which result from FSI depend on the particle type and
energy.  Of primary concern are effects involving final-state nucleons and
pions.  For nucleons, rescattering is the major effect, resulting in {\it i)}
change of direction and energy loss,  {\it ii)} production of secondary
nucleons, or {\it iii)} neutron or proton pickup leading to deuteron emission. 
For pions, FSI can also lead to scattering with possible nucleon emission.  The
pions can charge exchange or be totally absorbed leading to emission
of nucleons only.  In all of these cases, particles that escape from the
nucleus have lower energy than the initial, primary particle, and the
redistributed energy information is lost due to detector thresholds.

The most reliable information on FSI for nucleons comes from transparency
measurements in $(e,e'p)$ reactions on nuclei. Transparency, defined as the
probability of escaping the nucleus {\bf without} interaction, is measured by
detecting the scattered electron and integrating the protons detected within
the quasi-elastic peak. The most recent results quoted for protons in the
energy range~0.5 to 4~GeV are about 60\% for C, to 40\% for Pb\cite{garrow02},
with very little energy-dependence.  The composition and energy distribution of
the final-state particles is not well measured.  Two-proton and proton-neutron
final states should dominate, with each nucleon having about half the total
energy of the initial nucleon.

Pion interactions, especially for pions below a few hundred MeV, are dominated
by the $\Delta$ resonance.  The data on FSI can be inferred from reactions of
free pions on nuclei.  There is little specific data for pions
resulting from $\Delta$ or other resonant particle production in the nucleus.

The significant feature of pion reactions is the strong  absorption component -
both inelastic scatters and ``true" absorption, when the pion disappears from
the final state. The absorption component comprises about two-thirds of  the
total cross-section. True absoprtion ranges from about 25\% of the total cross-section
for C to nearly 40\% for Pb, for both positive and negative pions in the 100--300~MeV range\cite{ashery}.  Inelastic cross-sections are
comparable, although generally smaller for heavier nuclei.  Because of the
strong absorption component, pions in this energy range should escape the nucleus only about
50\% of the time.

Several experiments have found pion absorption to be a fairly
complicated process in complex nuclei\cite{rans,lads}.   Although the first
step is believed to be absorption on an isospin-0 np pair (quasi-deuteron), even
in a nucleus as light as carbon the final state is dominated by three-nucleon
emission.  For  heavier nuclei, the final state has a large component of four-nucleon
emission.  Even $\pi^{+}$ absorption usually includes emission of a
neutron, and of course $\pi^{-}$ absorption is dominated by neutron emission. 

There is very little information on pion reaction cross-sections for energies
above about 500~MeV.  The total pion--nucleon cross-section drops significantly
from the 200~mb resonant peak at 200~MeV to around 30~mb for energies
above 500~MeV.   Since this is not significantly different than the
nucleon-nucleon cross-section, pion transparency should be 
comparable to proton transparency at higher energies, i.e. approximately half
the pions will react through either scattering or absorption.

Interactions of 1--10~GeV neutrinos will produce pions with a wide
range of energies.  It should be noted that backward decay of the $\Delta$ resonance
can produce rather low-energy pions, because the
decay pions have a velocity in the $\Delta$ rest frame comparable to that of the
$\Delta$ in the lab.

The large absorption cross section (100--200~mb for C, 400--600~mb in Fe) for
100-300~MeV pions means that even pions that escape the nucleus may interact
again, with absorption rates of a percent/cm in scintillator.

There are other effects which influence the observed transparency of produced
secondaries.  As described in the Nuclear Effects section of Chapter
\ref{sect:resonant}, the quantum effect of hadron formation length and the QCD
effect of color transparency can increase the probability that a secondary
escapes the nucleus without undergoing FSI.  These effects are proportional to
the energy and Q$^2$ transfer and will not influence the transparency of low
momentun secondaries.

\subsection{Nuclear Effects and Interaction Probabilities}

\minerva\ will provide the setting for a
systematic, precision study of cross-sections and, with sufficient $\nubar$, 
structure functions, on a variety of nuclear targets.  Briefly reviewing the
nuclear effects on measured structure functions (directly proportional to the
cross-sections) as a function of $x_{Bj}$ reveals:

\subsubsection{Low-x: Nuclear shadowing}

  In the shadowing region, $x<0.1$, there are several areas where neutrino
scattering can provide new insights compared to charged-lepton probes.
``Shadowing" is a phenomenon which occurs in nuclear targets and is
characterized by reduction of the cross-section per nucleon for larger-A nuclei,
such as Fe, compared to smaller-A nuclei such as D$_2$. A recent summary
of theoretical interpretation of this effect is presented in~\cite{Frankfurt:2002kd}.

Vector-meson dominance (VMD) is the accepted explanation for shadowing in the
scattering of charged leptons off nuclei (i.e. $\mu/e - A$) for $Q^{2} \leq \sim 5~\hbox{GeV}^{2}$.
In essence, the incoming boson dissociates into a $q\qbar$
pair which interacts strongly with the nucleus as a meson.  Due to the V-A nature of the
weak interaction, neutrino scattering should involve not
only a VMD effect but additional contributions from axial-vector mesons
such as the $a_{1}$. Other sources of nuclear shadowing (mainly in larger
nuclei) involve gluon recombination from neighbors of the struck
nucleon, shifting the parton distributions toward higher values of $x$.  At
higher $\qsq$, shadowing is  dominated by Pomeron exchange in diffractive
scattering.  

\begin{figure}[h]
\centerline{
\epsffile{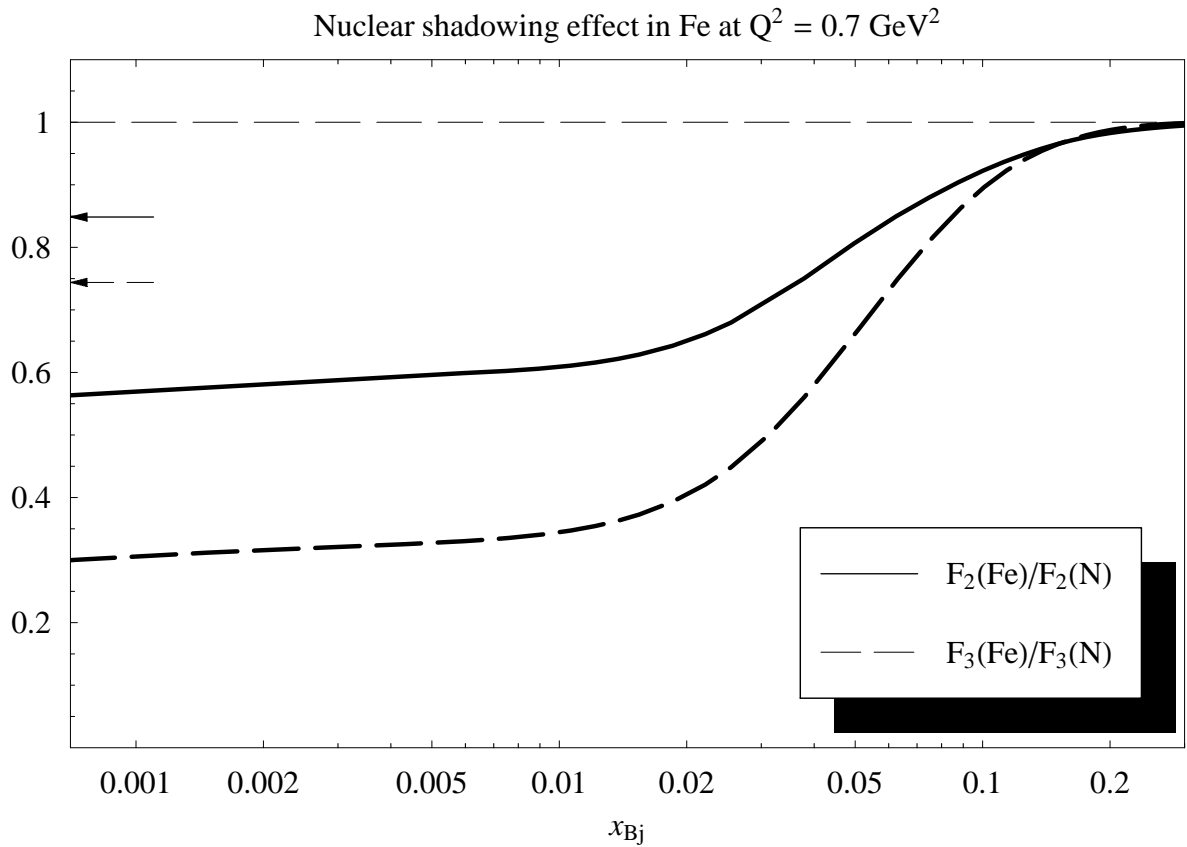}}
\caption[Shadowing effects in Iron at $Q^2 = 0.7 \hbox{(GeV/c)}^2$]{Expected shadowing effects off an Fe target at $Q^{2} = 0.7~\hbox{(GeV/c)}^2$ with Kulagin's non-perturbative parton model emphasizing
the difference in shadowing for F$_2$ and xF$_3$.  The arrows in the
vicinity of R = 0.8 indicate the expected shadowing strength at
$Q^{2} = 15~\hbox{(GeV/c)}^2$.}
\label{Kulaginfig}
\end{figure}

A quantitative analysis of neutrino shadowing effects by Kulagin\cite{Kulagin}
uses a non-perturbative parton model to predict shadowing effects in $\nu$--A
scattering.  As illustrated in Figure~\ref{Kulaginfig}, which predicts the
ratio of scattering off Fe to scattering off D$_2$, shadowing effects with
neutrinos should be dramatic at low $\qsq$ (the kinematic
region of the \numi\ neutrino beam) and still significant even at large $\qsq$. 
Kulagin also attempts to determine the quark-flavor dependence of shadowing
effects by separately predicting the shadowing observed in $F_{2}(x,Q^{2})$
(sum of all quarks) and $xF_{3}(x,Q^{2})$ (valance quarks only). These
predictions should be testable in \minerva.

\subsubsection{Mid-x: Anti-shadowing and the EMC effect}

  Drell-Yan experiments have also measured nuclear effects and their results
are quite similar to DIS experiments in the shadowing region.  However, in the
anti-shadowing region where $R_{A}$, the ratio of scattering off a nucleus A to
scattering off Deuterium,  makes a statistically-significant excursion above
1.0 in DIS, Drell-Yan experiments see no effect.  This could indicate a
difference in nuclear effects between valence and sea quarks as also
predicted by Kulagin.

  Eskola et al.\cite{Eskola} have quantified this difference using a model
which predicts that the differences between nuclear effects in
$xF_{3}(x,Q^{2})$ and $F_{2}(x,Q^{2})$, identified by Kulagin in the shadowing
region, should persist through the anti-shadowing region as well. More recent
work by Kumano\cite{Kumano} supports these findings using  different fitting
techniques.

  Based on the various theoretical explanations for the anti-shadowing and EMC
effects existing today, the measured effects could be considerably different
for neutrinos.  Neutrino scattering results would help clarify the theoretical
understanding of this phenomenon.

\subsubsection{High-x: Multi-quark cluster effects}

  Analyses from DIS experiments of $F_{2}(x,Q^{2})$ in the ``Fermi-motion"
region $x \geq 0.7$ have used few-nucleon correlation and multi-quark
cluster models to fit the data. These models boost the momentum of
some quarks, which translates into a high-$x$ tail of $F_{2}(x,Q^{2})$ that should
behave as $e^{-ax}$. However, fits to $\mu - C$\cite{Benvenuti:1994bb} and
$\nu - Fe$\cite{Vakili:1999qt} scattering have obtained
two different values for the fitted constant $a$: $a = 16.5 \pm 0.5$ and $a =
8.3 \pm 0.7 \pm 0.7$ (systematic), respectively. This is surprising
because any few-nucleon-correlation or multi-quark
effects should have already saturated by Carbon.   A high-statistics data
sample, off several nuclear targets, could go a long way towards resolving the
dependence of the value of $a$ on the nucleus and lepton probe.

\subsection{Measuring Nuclear Effects in \minerva} 

To study nuclear effects in \minerva, Fe and Pb nuclear targets will be
installed upstream of the pure scintillator active detector which, essentially,
acts as a carbon target.    Two configurations are currently being examined. 
One would have (upstream to downstream) three 2.5~cm Fe plates, each plate
followed by a module of active scintillator detector.   Following this would be
six 0.8~cm Pb plates (equal radiation thickness to the Fe) again separated by
scintillator modules.  This would give just over 1 ton of each target.  The
second possible configuration involves a total of six planes only, with each
plane divided transversely into Fe and Pb segments. As one proceeds upstream to
downstream, the Fe and Pb exchange sides on each of the six planes.  As always,
a scintillator module separates each of the six planes.  This configuration
would also translate to just over 1~ton of each target.  For the standard
four-year run described in Section~\ref{sect:runPlan}, \minerva\ would collect
940~K events on Fe and Pb and 2.8~M events on the C within the fiducial volume
of the scintillator.

\minerva's goals in measuring nuclear effects can be summarized as
follows:

\begin{itemize}

\item Measure final-state multiplicities, and hence absorption probabilities, as a function of A with incoming
$\nu$;

\item measure the visible hadron energy distribution as a function of target
to determine relative energy loss due to FSI;

\item measure $\sigma(x_{Bj})$ for each nuclear target to compare
$x_{Bj}$-dependent nuclear effects with $\nu$ and charged lepton.

\item With sufficient $\nubar$, measure the nuclear effects on F$_2$(x,Q$^2$)
and xF$_3$(x,Q$^2$) to determine whether sea and valence quarks
are affected differently.

\end{itemize}

\subsubsection {Multiplicities and visible hadron energy}

\begin{figure}[h]
\epsfysize=4.0in
\centerline{
\epsffile{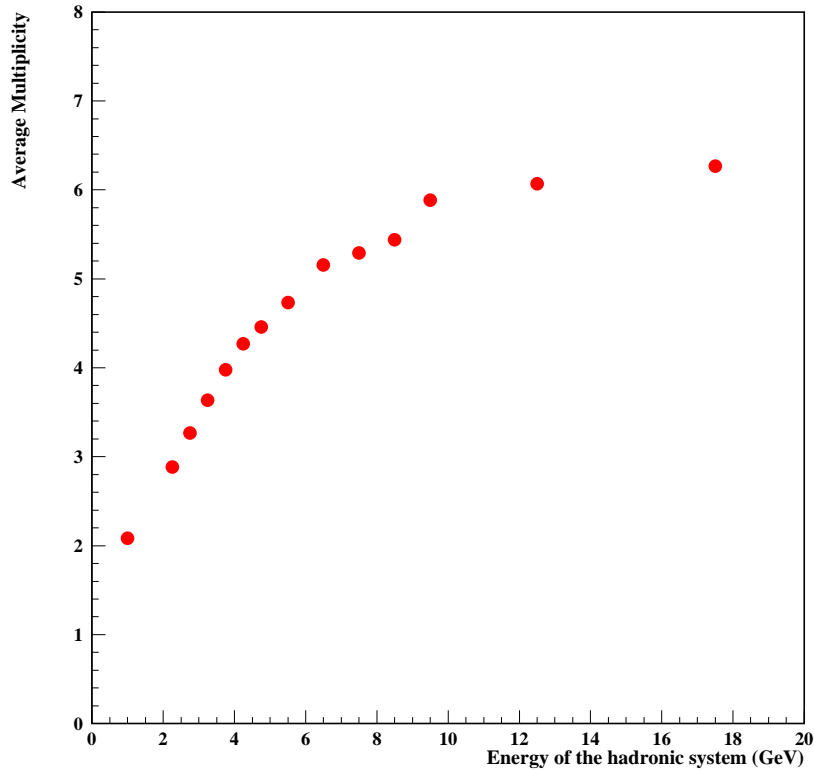}}
\caption[Average multiplicity vs. hadronic energy]{The average multiplicty, excluding neutrons, as a function of the
hadron energy of the event.  The distribution is predicted by the NEUGEN
Monte Carlo without turning on FSI.}
\label{multfig}
\end{figure}

\begin{figure}[htb]
\epsfysize=4.5in
\centerline{
\epsffile{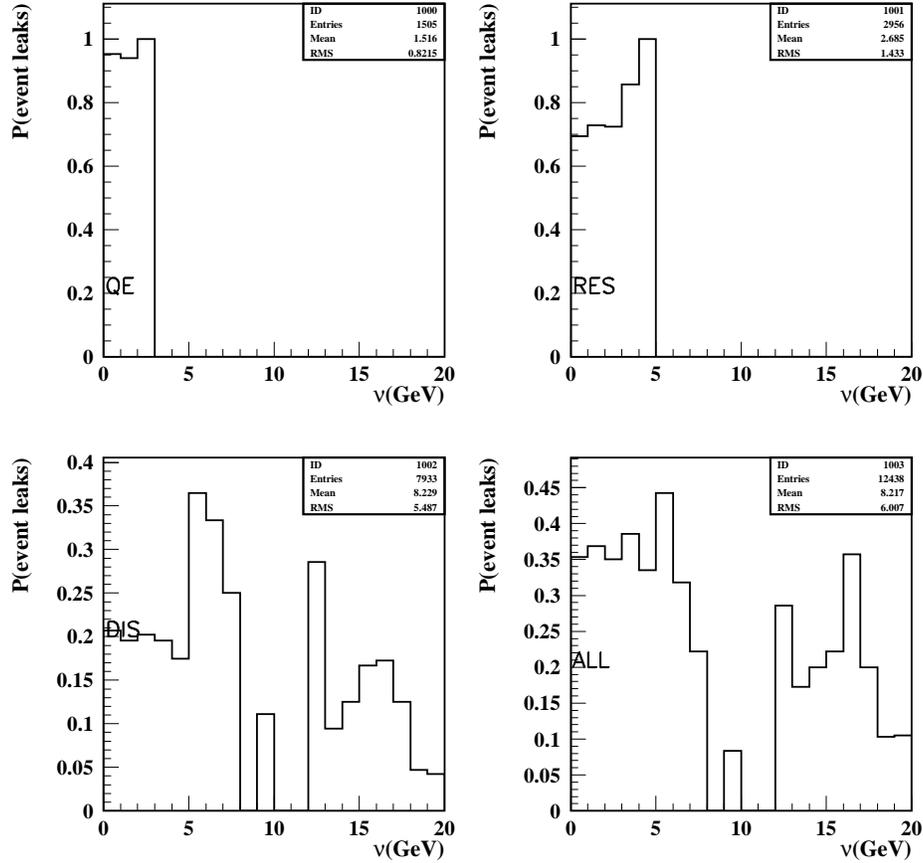}}
\caption[Fraction of fully-contained nuclear target events vs. $W$]{The fraction of events which are fully contained within the active
scintillator detector downstream of the nuclear targets as a function of the
total energy of the hadronic system.  The distributions are for quasi-elastic,
resonant, DIS and all reactions, as noted.}
\label{contained}
\end{figure}

The expected average multiplicity of neutrino events as a function of $E_H$, 
with no nuclear effects, is shown in Figure~\ref{multfig}. As mentioned
earlier, FSI will perturb this distribution via pion absorption and hadron
re-scattering in the nuclear medium.   FSI will also distort the initial hadron
energy, transfered by the intermediate vector boson, yielding less visible
energy in the detector.   Restricting the study to events where all particles
stop within the 2~m of active scintillator downstream of the nuclear targets
will permit measurement of the hadron energy by range to within  a few
percent.  The sample of events meeting these criteria is a function of the
hadron energy $E_H$, and is shown in Figure~\ref{contained}.  As can be seen,
even at higher values of the hadron energy $\nu$, around 20\% of the events
have all secondary tracks contained within the active scintillator volume. With
nearly one million events on each nuclear target in the four-year run, there
will be sufficient statistics to determine the nuclear dependence of both
multiplicities and visible hadronic energy. 

\subsubsection {x$_{Bj}$-dependent nuclear effects}

Just over 16\% of the total event sample has $x_{Bj} \leq 0.10$.  The
(approximate) statistical accuracy for measurements of the nuclear effects
in the ratios of Fe to C events at small x (shadowing region) are
summarized in the following table. The columns designated DIS indicate that a
cut has been made to retain only events with $W \geq 2.0~\hbox{GeV}/c^2$ and $Q^2 \geq 1.0~\hbox{(GeV/c)}^2$.
For the \minerva\ DIS analysis, the first three bins could be combined
into two bins to reduce statistical errors.

\begin{table}[hbt]
\begin{center}
\begin{tabular}{@{}lllll}
\hline
\multicolumn{5}{c}{Ratio Fe/C: $\sim$ Statistical Errors} \\ \hline \\
x$_{Bj}$ & \minerva  & \minerva   \\
         & 4-year       &  DIS          \\ \cline{1-5}
0.0 - .01 & 1.4 \% & 20 \%   \\
.01 - .02 & 1.1   & 8  \\
.02 - .03 & 1.0   & 5     \\
.03 - .04 & 1.0   & 3      \\
.04 - .05 & 0.9   & 2.5       \\
.05 - .06 & 0.9   & 2.1     \\
.06 - .07 & 0.8   & 1.8     \\ \cline{1-5}
\end{tabular}
\end{center}
\caption[Statistical errors on Fe/C ratio vs. $x_{Bj}$ bin]{Statistical errors on the ratio of
fully-contained Iron to Carbon events, assuming the level of shadowing predicted by the model of Kulagin, as
a function of the $x_{Bj}$ bin.}
\label{tab:nucStatErr}
\end{table}

\begin{figure}[hbtp] 
\epsfysize=4.5in
\centerline{ \epsffile{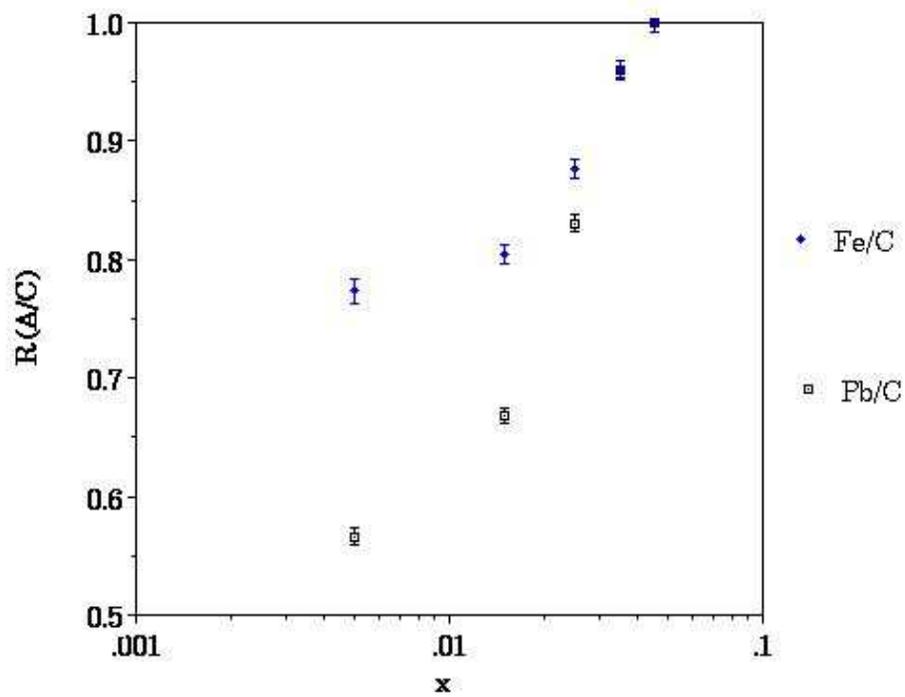}} 
\caption[Ratio of shadowing effects off Pb, Fe and C targets]{Kulagin's
predicted ratio of shadowing effects off Pb, Fe and C targets with the expected
errors from all events from the 4-year run.} 
\label{Kulagin-MINOS} 
\end{figure}

Assuming the level of shadowing predicted by Kulagin, the measured ratio
of Fe/C and Pb/C, with statistical errors corresponding to the data accumulated
during the 4-year run, is shown in Figure~\ref{Kulagin-MINOS}.  The ratios
plotted are for all events.  The statistical errors would increase, as indicated
in Table~\ref{tab:nucStatErr}, after making a DIS cut. 

The baseline 4-year run would be adequate to achieve
the physics goals of the nuclear effects study, although some would be limited
by the kinematic reach of the neutrino beam energies used for MINOS
running and the minimal $\nubar$ exposure planned for MINOS.



\subsection{Nuclear Effects and Determination of $\stw$}

There have been many attempts to explain the recent NuTeV  
\cite{Zeller:2001hh} measurement of $\stw$, which is 3$\sigma$ away from the 
Standard Model prediction.  Among the most persuasive are the unknown nuclear
corrections involving neutrinos\cite{Kovalenko:2002xe}.   \minerva\
will be able to directly measure the ratio NC/CC on  various
nuclear targets to explore these nuclear effects experimentally.

%% file: oscillation.tex
\section{\minerva\ and Oscillation Measurements} 
\label{sect:oscillation}

Over the past decade neutrinos have moved to center stage in the field
of particle physics with the discovery of neutrino oscillation.  
Following on the initial discovery of solar and atmospheric neutrino oscillation
are a new generation of high-precision long-baseline experiments dedicated to mapping out the 
neutrino mixing matrix and mass hierarchy in detail.  In this section we address
some of the ways in which the measurements made by \minerva\
can help these ambitious and expensive experiments 
achieve maximum senstivity.    

\subsection{Neutrino Oscillation Landscape}
One accelerator-based experiment to explore the atmospheric oscillation sector has already begun,
and several more are in the construction phase.  
The K2K experiment in Japan has seen evidence for 
$\nu_\mu$ disappearance, and expects to double its sample of about 
50 events over the next year\cite{k2kosc}.  The MINOS experiment, with a much larger
expected event sample, will make the first precision 
measurement of the atmospheric mass splitting, again through 
$\nu_\mu$ disappearance\cite{Adamson:2002mj}.  Finally, 
the OPERA and ICARUS experiments in Europe will 
attempt to further confirm the $\nu_\mu \to \nu_\tau$ oscillation hypothesis by reconstructing
actual $\nu_\tau$ charged-current interactions in a beam produced as 
$\nu_\mu$.  The solar sector is being addressed by novel detection techniques of 
solar neutrinos themselves, and 
the KamLAND experiment, which uses a number of reactors as its 
anti-neutrino source\cite{kamland}.   
If confirmed by MiniBooNE\cite{McGregor:ds}, the LSND anomaly would dramatically affect the 
lines of inquiry for future experiments, demanding 
precise oscillation measurements with both long and 
short baselines, and hence both 1~GeV and several-GeV neutrino beams.  

One reason for the flurry of recent activity in neutrino physics is 
that non-zero neutrino masses and mixing have profound
implications not only for the origin of flavor in the universe, but 
possibly also the origin of the matter-antimatter asymmetry.  
Because the lepton mixing matrix
seems to have large off-diagonal elements, 
leptonic CP violation could be much larger than observed in the 
quark sector, and may be large enough to explain the
matter/anti-matter asymmetry that we see today.

A three-generation neutrino mixing matrix can be described by three 
independent mixing angles ($\theta_{12},\theta_{23},\theta_{13}$) 
and a CP-violating phase ($\delta_{CP}$).  The standard 
notation for this matrix, which transforms between the flavor 
and mass eigenstates is as follows:
\begin{eqnarray} 
\left( {\begin{array}{*{20}c}
   {\nu _e }  \\
   {\nu _\mu  }  \\
   {\nu _\tau  }  \\
\end{array}} \right) = U\left( {\begin{array}{*{20}c}
   {\nu _1 }  \\
   {\nu _2 }  \\
   {\nu _3 }  \\
\end{array}} \right)
\end{eqnarray} 
where if $s_{ij}=\sin\theta_{ij}, c_{ij}=\cos\theta_{ij}$, then $U$ can 
be expressed as three rotation matrices:  
\begin{eqnarray*}
 U = \left( {\begin{array}{*{20}c}
   1 & 0 & 0  \\
   0 & {c_{23} } & {s_{23} }  \\
   0 & { - s_{23} } & {c_{23} }  \\
\end{array}} \right)\left( {\begin{array}{*{20}c}
   {c_{13} } & 0 & {s_{13} e^{i\delta } }  \\
   0 & 1 & 0  \\
   { - s_{13} e^{ - i\delta } } & 0 & {c_{13} }  \\
\end{array}} \right)\left( {\begin{array}{*{20}c}
   {c_{12} } & {s_{12} } & 0  \\
   { - s_{12} } & {c_{12} } & 0  \\
   0 & 0 & 1  \\
\end{array}} \right)
\end{eqnarray*}

In this parameterization, the $\nu_\mu \to \nu_\tau$ oscillation probability, 
which describes atmospheric neutrino disappearance, can be expressed: 
\begin{equation} 
P (\nu_\mu \to \nu_\mu) = 1 - \cos^4\theta_{13}\sin^2 2\theta_{23}
\sin^2 \left( \frac{\Delta m_{23}^2 L}{4E} \right) 
\end{equation} 

The solar (electron) neutrino disappearance, which has been further confirmed
by the KamLand reactor (electron anti-)neutrino 
experiment (with average baseline 100~km), can be expressed as: 
\begin{equation} 
P(\nu_e \to \nu_e) = 1 - \sin^2 2\theta_{12}
\sin^2 \left( \frac{\Delta m_{12}^2 L}{4E} \right) 
\end{equation} 

The measurements in the solar and atmospheric sectors have shown that the 
mixing angles $\theta_{12}$ and $\theta_{23}$ are large, but there
remains one  
mixing angle which has not been determined, $\theta_{13}$.  
This angle would be manifest by electron neutrino disappearance 
a few kilometers from a reactor, or electron neutrino {\emph appearance} in a
few-GeV $\nu_\mu$ beam a several hundred kilometers from an accelerator.
In the latter case the oscillation probability in vacuum is 
\begin{equation}
P(\nu_\mu \to \nu_e) = \sin^2\theta_{23}\sin^2 2\theta_{13}\sin^2 
\left( \frac{\Delta m^2_{32} L}{4E} \right) +... 
\label{eqn:mue} 
\end{equation}
where the missing terms are suppressed by at least one factor of 
$\Delta m^2_{12}/\Delta m^2_{23}$.  

Although reactors can play an important 
role in discovering non-zero $\theta_{13}$, this field will rely
on accelerator experiments to eventually search for CP violation 
and determine the mass hierarchy.  For example, the asymmetry in neutrino
and anti-neutrino oscillation probabilities, in the absence of matter effects, 
is (to first order): 
\begin{eqnarray} 
\label{eqn:cpasym} 
\frac{P(\nu_\mu \to \nu_e)-P(\bar\nu_\mu \to \bar\nu_e)}
{P(\nu_\mu \to \nu_e)+P(\bar\nu_\mu \to \bar\nu_e)} 
 &\approx&  
\frac{\Delta m^2_{12}L}{E}\frac{\sin\delta}{\sin\theta_{13}} < 1
\end{eqnarray} 

When electron neutrinos pass through the earth they can scatter off electrons,
which creates an additional potential not present for muon or tau neutrinos\cite{MSW}. 
This additional potential means the effective mixing angle and oscillation
length is changed from equation~\ref{eqn:mue}, and is changed differently 
for neutrinos and antineutrinos.  Moreover, the sign of the asymmetry
is determined by whether $\Delta m^2_{23}$ is positive or negative.  
The asymmetry in $\nu_\mu \to \nu_e$ oscillation probabilities 
due to matter 
effects, in the limit of $\Delta m^2_{12}$ being zero, is (to first order, when $E<E_R$) 
\begin{eqnarray} 
\label{eqn:matter} 
\frac{P(\nu_\mu \to \nu_e)-P(\bar\nu_\mu \to \bar\nu_e)}
{P(\nu_\mu \to \nu_e)+P(\bar\nu_\mu \to \bar\nu_e)} 
 &=& 
\frac{2E}{E_R}\left( 1 - \left[ \frac{\pi}{2} \right]^2 \frac{E-E_{om}}{E} 
\right) \\ 
\nonumber
E_R &=& \frac{\Delta m^2_{23}}{2\sqrt{2}G_F\rho_e} \approx 11GeV \\
\nonumber
E_{om} &=& \frac{\Delta m^2 L}{2\pi}
\end{eqnarray} 

To measure this asymmetry, oscillation 
experiments will need to search for electron neutrino appearance 
in muon neutrino (and anti-neutrino) beams, and to measure the 
atmospheric $\Delta m^2$ precisely, future (and current) experiments will need to 
measure the muon neutrino survival probability with corresponding precision.  Both kinds of 
experiments will require extremely long baselines, as well as near and 
far detectors to make the actual probability measurements.  
Even with an identical near detector, oscillation measurements will 
require reliable neutrino interaction models.  For $\nu_\mu$
disappearance measurements, these models will be used to determine the 
mixing parameters from measured distributions in near and far detectors. 
For $\nu_e$ appearance measurements, these
models will be used to predict the far detector backgrounds based on data from  
a near detector.  In both cases, the 
measurements are complicated by the fact that 
the far detector's $\nu_\mu$ charged-current event spectrum is  
dramatically different from the near detector's, 
due to the large, energy-dependent $\nu_\mu$ disappearance probability. 

\subsection{Benefits of \minerva\ to Oscillation Experiments} 

With its fine grained, fully-active 
inner detector, electromagnetic and hadronic calorimetry, and excellent 
muon measurement capabilities, \minerva\ will have  
superb pattern recognition, energy resolution, and efficiency. 
These abilities, coupled with the high flux of the \numi\ beam make 
possible a host of improvements that will directly assist oscillation experiments.
Indeed, most of the physics areas discussed in this proposal, from the 
precision measurement of quasi-elastic form factors, to exclusive channel
studies, to coherent production, to nuclear physics, have the potential 
to improve oscillation measurements in one way or another.   In particular,
improving our knowledge in these areas may help avoid problems 
in the analysis of oscillation data that are difficult to foresee at this 
point. 

In practice, this benefit will be realized in the development of improved
neutrino event generators that encapsulate the information learned from \minerva\ 
and provide a powerful and portable resource for all future neutrino experiments.  
The primary authors of two of the most widely-used, publicly 
available event generators are actively involved in \minerva\ and 
see the development of such a `next generation' generator as one of the 
principal tangible benefits of this experiment.   

The remainder of this chapter will focus on two specific areas where 
\minerva\ can aid oscillation experiments.  One is 
determination of the ``neutrino energy calibration'', important
for $\Delta m^2_{23}$ measurements, and the second is measurement
of backgrounds to $\nu_e$ appearance, in a search for $\theta_{13}$.

While this section focuses on MINOS and the proposed \numi\ off-axis
experiment, \minerva\ will undoubtedly benefit other future oscillation experiments as well, including the proposed J-PARCnu project in Japan.  The J-PARCnu beam energy is matched to its shorter baseline (and is therefore lower), but neutrino energy calibration and neutral-current $\pi^0$ backgrounds are as essential to J-PARCnu as they are at \numi. Neutrino energy reconstruction in J-PARCnu (as in K2K) is limited by knowledge of the non quasi-elastic background induced by inelastic reactions which feed down from higher neutrino energies.  Similarly, most neutral-current background to $\nu_e$ appearance in J-PARCnu originates from the high-energy tail of the beam.  Although \minerva\ obviously cannot directly measure the J-PARCnu beam (as it can for \numi), its somewhat higher energy reach arguably makes it better-suited for minimizing J-PARCnu systematic uncertainties from neutrino-interaction physics than the J-PARCnu beam itself. 

\subsection{$\Delta m^2_{23}$ Measurements} 

As an example of the importance of neutrino interaction physics 
to oscillation experiments, consider 
a measurement of the ``atmospheric'' mass splitting, which is the 
primary (but not only) goal of the MINOS experiment.  

In a long-baseline experiment there are
two limiting cases: a far detector without a near detector (and maximum systematic uncertainties) and a far detector with a perfect near detector (and negligible systematic uncertainties).
For a far detector without a near detector, the predicted ``no oscillation" distributions
are determined by integrating the flux and cross section over a smearing function with takes into 
account detector acceptance, reconstruction inefficiencies and measurement resolution.  
In this case the determination of oscillation parameters can only be as good as 
the understanding of 
the beam, neutrino interaction cross-sections, detector performance, and reconstruction.  
In the other idealized extreme, identical near and far detectors see an identical spectrum.  In this case, allsources of systematic uncertainty cancel in the near/far comparison. 

For MINOS, the near detector will help reduce many of the important systematic
errors, but the situation is not quite as perfect as ideal case.  The
beams (oscillated vs. unoscillated and point source vs. line source), 
detector shapes (octagon vs. ``squashed octagon"), photodetectors
(Hammamatsu M16 vs. M64), electronics (IDE vs. QIE), reconstruction (single neutrino event per
readout vs. multiple neutrino events per readout), and beam-related backgrounds
all differ between far and near.  Great effort in the areas of calibration, reconstruction, and
analysis is invested to understand and correct for 
these small near-far differences.  In addition, dedicated measurements of hadron
production on the \numi\ target by the MIPP experiment 
should reduce the uncertainties on the near to far flux ratio to the few percent level.  

\subsection{Neutrino Energy Calibration}
Analysis of near detector data aims to predict 
expected rates and spectra in the far detector in the absence of oscillation, and for
different values of $\theta_{23}$ and $\Delta m^2_{23}$.  
Differences between these predictions and real data will be used to
fit the oscillation parameters.  A crucial link in this prodcedure is the translation between measured
energy and the neutrino energy, as this quantity is directly related to the oscillation
probability.     
Even the most elaborate suite of near- and far-detector calibrations
can at best characterize the response of the detector to known incoming 
{\em charged} particles.  
There is no equivalent test beam for ``neutrino calibration"; 
this final step requires appeal to a model. 
A reliable understanding of the spectrum and multiplicity of particles produced
in neutrino interactions is indispensible for reconstructing the true neutrino energy
from the visible energy measured in a calorimeter.  
There are 
several areas where these models have large uncertainties, 
which \minerva\ could help to reduce:

\subsubsection{Charged and neutral pion production}
As MINOS responds differently to electromagnetic and hadronic showers, it will be essential to estimate the relative abundances
of charged and neutral particles.  
These abundances are determined by isospin amplitudes
at each point in phase space.  As explained in Section~\ref{sect:generators}, 
these amplitudes necessarily include a resonant component from the low invariant
mass region where specific exclusive channels are produced, and the deep-inelastic regime
where scattering is described by quark transitions in the framework of the parton model.
The predictions therefore depend on the resonance model adopted, 
the model for fragmentation of inclusive quark final-states (particularly at low invariant mass where standard
models like the JETSET string model are not applicable), and the treatment of 
the deep-inelastic/resonant overlap region.   

\subsubsection{Charged particle multiplicities}
Neutrino energy reconstruction 
also depends on the charged particle multiplicity, as the rest energy
of pions disappears via nuclear absorption and the neutrinos produced in $\pi$ and $\mu$ decays.  
Correction for these losses is therefore also related to the model(s)
of charged pion production.    

\subsubsection{Intra-nuclear scattering}
\begin{figure}[hptb!]
\centerline{\epsfig{file=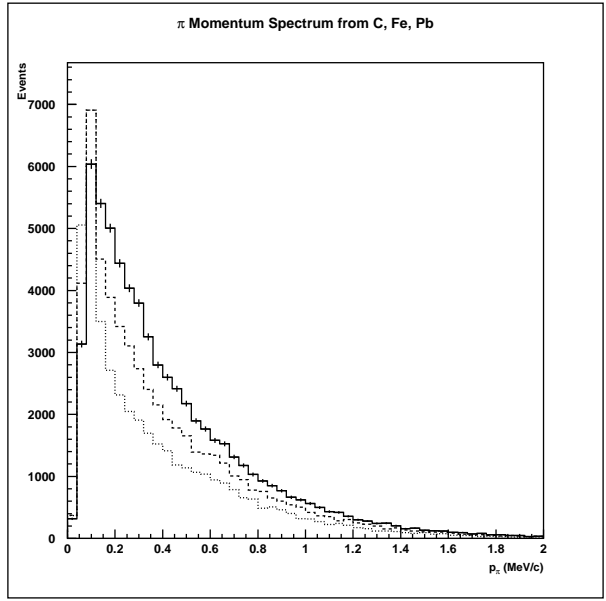, width=0.5\textwidth}}
\caption[$\pi^+$ momenta for 3~GeV charged-current interactions on Carbon, Iron and Lead]{Simulated $\pi^+$ momenta for 100,000 3~GeV \numu\
charged-current interactions on Carbon, Iron, and Lead.  } 
\label{fig:cfepb}
\end{figure}

At \numi\ energies, intra-nuclear scattering can result in 
large distortions of the hadronic multiplicities, 
angular distributions and total 
energies.  A feeling for the variation of intranuclear rescattering
can be gleaned from Figure~\ref{fig:cfepb}, which shows the $\pi^+$ spectra
from 3~GeV neutrino interactions on three different target nuclei:  Carbon, Iron, and Lead.
Little data is available to constrain or validate models of these effects.  
Measurement of these processes requires a $4\pi$ detector, hence
electron scattering data has limited utility.

\begin{figure}[htb]
\centerline{\epsfig{file=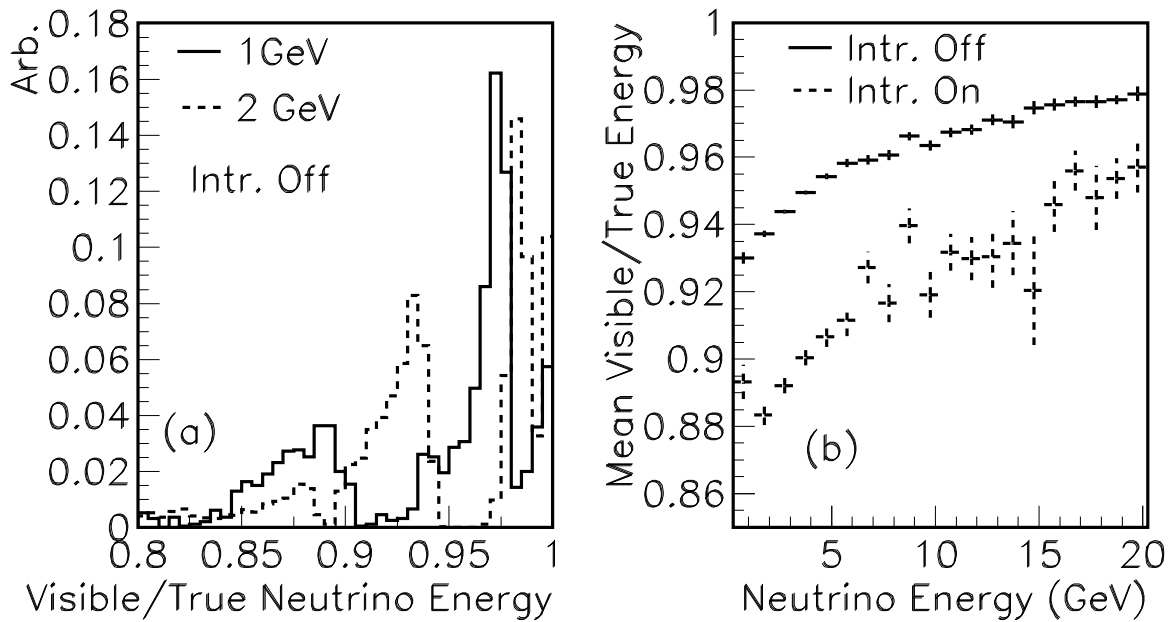, width=\textwidth}}
\caption[Nuclear rescattering effects on reconstructed neutrino energy]{(a) Distribution of the ratio of visible to true 
neutrino energy for 1~GeV and 2~GeV charged-current neutrino interactions
(b) the average of that ratio versus true neutrino energy, for two 
different models:  one withoug intra-nuclear rescattering, 
and one where rescattering in Carbon is simulated.} 
\label{fig:oscfig1}
\end{figure} 

\subsubsection{Expected results}
Figure~\ref{fig:oscfig1}(a) shows the ``neutrino energy 
resolution'' for 1~GeV and 2~GeV charged-current interactions in 
a perfect calorimeter that measures the 
{\em kinetic} energy of charged pions and baryons but the {\em total}
energy of photons and electrons.  The structure in the distribution 
results from production of one or more charged pions, and with corresponding amounts of
lost rest energy.  
Figure \ref{fig:oscfig1}(b) shows the average ratio of visible to true neutrino 
energy versus neutrino energy for the same detector.  For the solid lines
there is no intranuclear rescattering, while for the dashed lines rescattering in 
a carbon target is assumed.  Both models show that a correction as large as 10\% 
is required, which translates into a $\Delta m_{23}^2$ uncertainty equivalent to the
statistical error of MINOS\cite{minos5yr}.  

\begin{figure}[hptb!]
\centerline{\epsfig{file=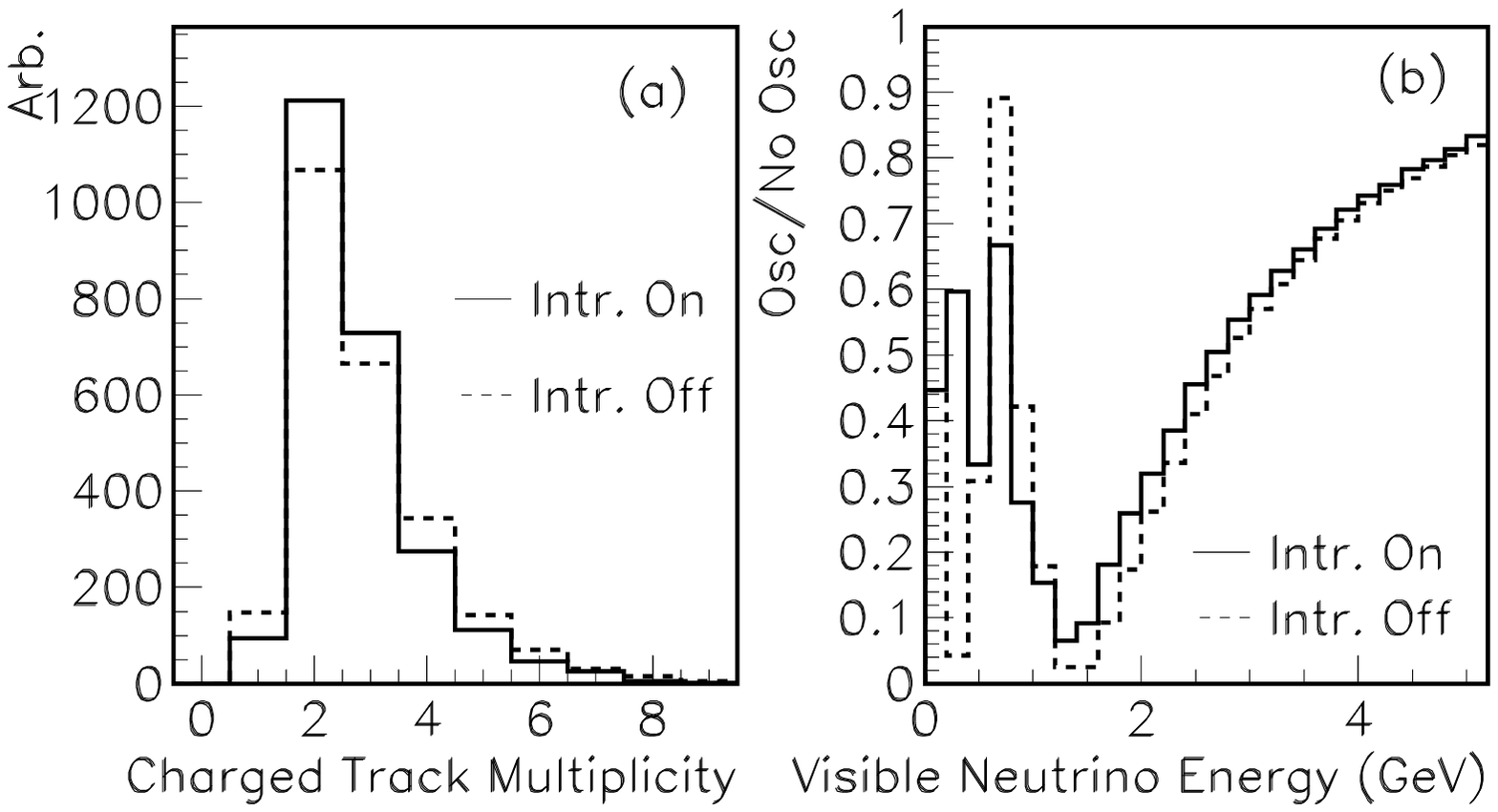, width=\textwidth}}
\caption[Nuclear rescattering effects on charged-particle multiplicity and event spectra]{(a) Charged particle multiplicity for the two 
models of nuclear rescattering discussed in the text, 
(b) Ratio of oscillated and unoscillated event spectra
for those two models, assuming $\Delta m^2{23} = 2.5\times 10^{-3}~\hbox{eV}^2$
} 
\label{fig:oscfig2}
\end{figure}

\minerva\ can play an important role in the MINOS $\Delta m^2$ 
analysis by measuring the charged-pion multiplicity as a function of
visible neutrino energy.
Data from a variety of nuclear targets (including Iron) will strongly constrain and
redundantly validate rescattering models.
Figure \ref{fig:oscfig2}(a) shows the charged paticle 
multiplicity for $\nu_\mu$ charged-current interactions 
in the \numi\ low-energy beam, for the two models described in
Figure~\ref{fig:oscfig1}(b).  Finally, Figure~\ref{fig:oscfig2}(b) shows the 
difference between the oscillated and unoscillated $\nu_\mu$ interaction spectra
for the two models of nuclear rescattering.  Note that the 
differences below 1~GeV are enormous, but even at 3 or 4~GeV they are 
sizeable.  

%
%

\begin{figure}[hptb!]
\centerline{\epsfig{file=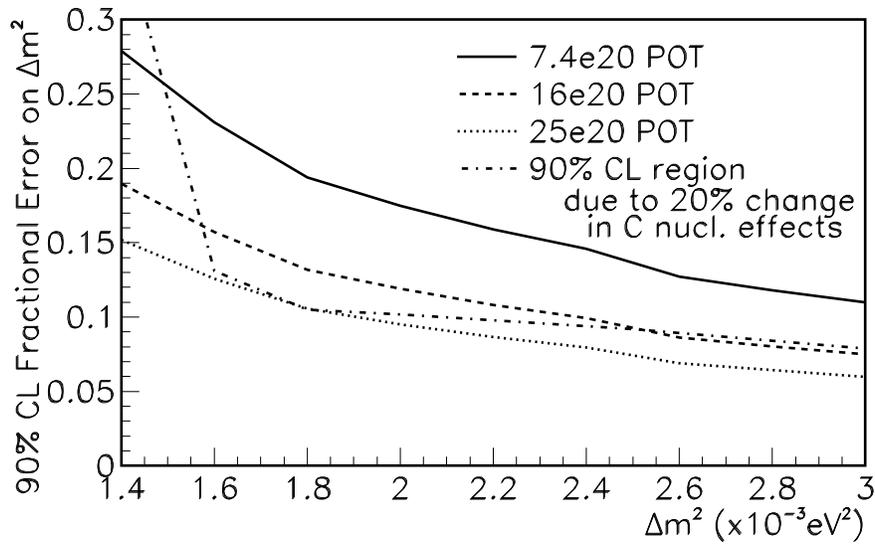, width=.75\textwidth}}
\caption[Nuclear rescattering effects on a MINOS-like 
$\Delta m^2$ measurement]{Fractional size of the $90\%$ confidence
level contour at $\sin^2 2\theta=1$, due to MINOS statistical 
errors for different exposures, and due to a 20\% uncertainty 
on ``neutrino energy resolution'' which would come about 
due to nuclear rescattering effects. }
\label{fig:oscfig2b}
\end{figure}

To quantify the effect this would have on a $\Delta m^2$ measurement,
a toy monte carlo was used to approximate the MINOS $\Delta m^2$
analysis, including neutral current contamination and cuts to reduce
that contamination.  Figure \ref{fig:oscfig2b} shows the fractional
size of the 90\% confidence level contour region due to a 20\%
uncertainty in the total ``neutrino energy loss''.  Also shown in the
figure is the size of the 90\% CL region at $\sin^2 2\theta=1$, versus
$\Delta m^2$ from a more complete MINOS simulation (\cite{minos5yr}).
Although this systematic error would be lower than the statistical
error for the lowest exposure of protons on target, it is far from
negligible, and dominates for values of $\Delta m^2$ below 1.5$eV^2$
for all exposures.  Without \minerva\ the error due to nuclear effects
may be roughly this size, but with even a small amount of data on
several targets \minerva\ would be able to model this effect to much
better than $10\%$ of itself, making this effect negligible.  Future
$\Delta m^2$ measurements (such at those proposed at J-PARC or NuMI
Off Axis), which expect to achieve $1 \sigma$ statistical errors
closer to the 1\% level rather than the 3\% level, will have to
understand this effect (among others) even more precisely.

\subsection{$\theta_{13}$ Measurements} 
\label{section:theta13}

A longer-term goal in oscillation physics is to probe leptonic CP violation and the neutrino mass hierarchy by comparing measurements of $\nu_\mu \to \nu_e$ and $\bar\nu_\mu \to \bar\nu_e$ 
oscillation.  These measurements are
particularly challenging because of backgrounds. Conventional 
neutrino beams always include $\nu_e$ contamination from muon and kaon decays in the beamline. In addition,
neutral-current interactions produce $\pi^0$ which can be mistaken for electron appearance in the far detector.

\begin{figure}[tbp]
\centerline{\epsfig{file=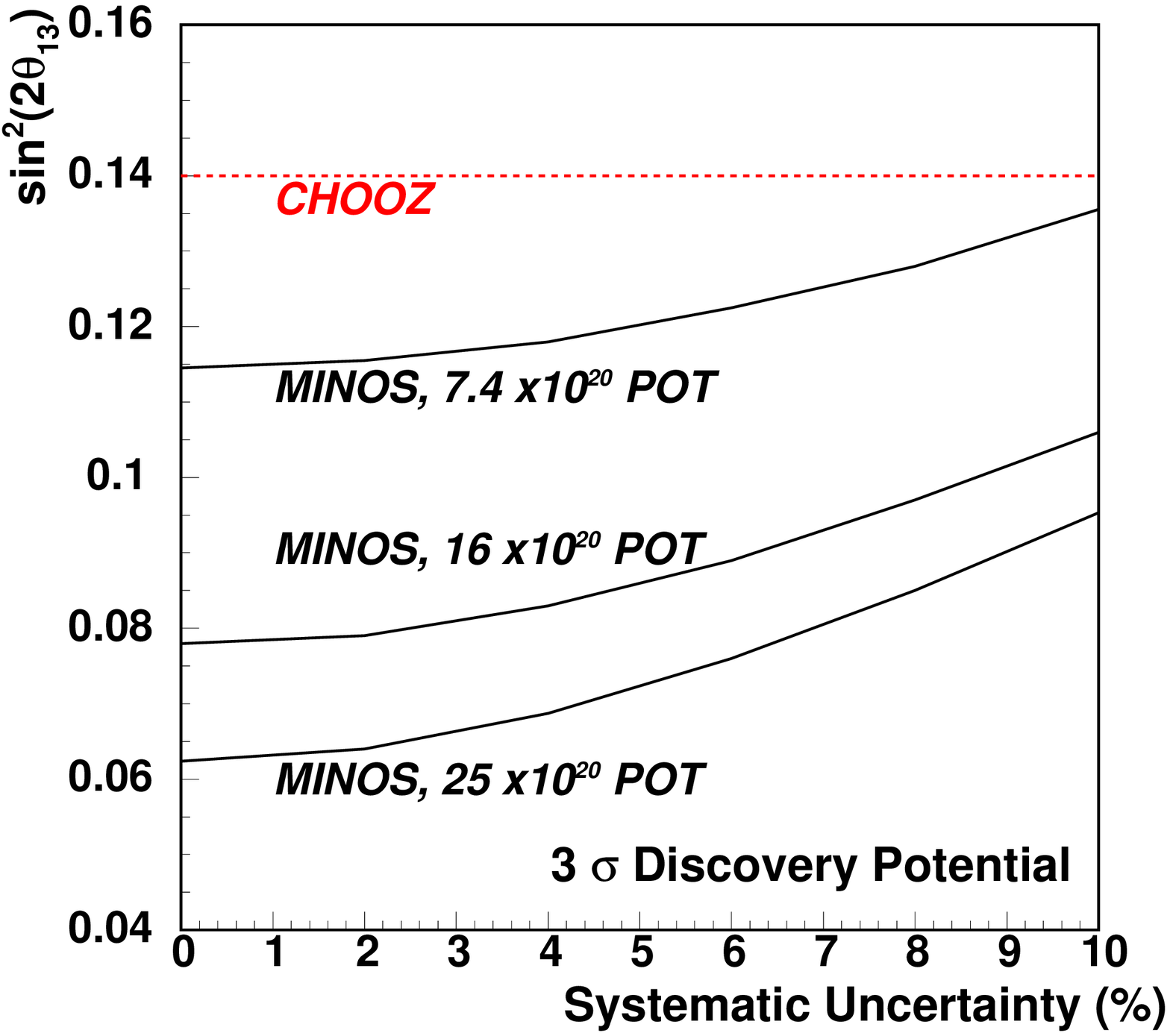, width=.5\textwidth}}
\caption[MINOS 3$\sigma$ sensitivity to $\theta_{13}$ vs. systematic background uncertainty]{MINOS 3$\sigma$ sensitivity to non-zero $\theta_{13}$ 
(assuming $\Delta m^2_{23} = 2.5\times 10^{-3}~\hbox{eV}^2$), as a 
function of the systematic uncertainty on the background, for 
several possible integrated proton luminosities.} 
\label{fig:oscfig3}
\end{figure} 

In MINOS, neutral-current backgrounds should be several times  
the intrinsic $\nu_e$ contamination, and $\theta_{13}$ sensitivity
depends strongly on the assumed systematic uncertainties, as shown in Figure~\ref{fig:oscfig3}.
MINOS is not optimized to separate neutral-current interactions
from $\nu_e$ charged-current interactions, and the background for 
this search is roughly 75\% deep-inelastic $\pi^0$ production, 15\%
intrinsic $\nu_e$, and 10\% $\nu_\mu$ charged-current interactions\cite{minosnue}.  

Because MINOS can only hope to achieve, at best, a factor two improvement
in $\sin^2 \theta_{13}$ sensitivity over existing limits, other experiments better
suited to this search have been proposed.  One proposed experiment
would use a much finer-grained detector off-axis from the \numi\ beamline,
where the neutrino spectrum is much narrower and thus the signal to background ratio is higher. 
In this off-axis experiment, the dominant background would be 
intrinsic $\nu_e$, although neutral-current 
background is comparable, and the $\nu_\mu$ charged-currents are considerably smaller\cite{numioa}.

Since both MINOS and the off-axis experiment need to measure the 
intrinsic $\nu_e$ and neutral-current background rates, 
they each require near detectors similar to their respective far detectors.
Measuring neutral-current backgrounds directly is difficult at a near detector location, as $\nu_\mu$ charged-current interactions are far more abundant in the unoscillated beam.

\subsection{$\nu_e$ Appearance Backgrounds}

\minerva\ can measure all three types of $\nu_e$ appearance backgrounds. Thanks to superior segmentation,
\minerva\ can isolate a very clean $\nu_e$ charged-current sample and directly measure the $\nu_e$ flux.  Similarly, with its excellent $\pi^0$ identification, energy ($\sigma_E = 6\%/\sqrt{E}$), and 
angular resolution, \minerva\ can map out all the processes that produce neutral pions.
Finally, with a fully active detector and good timing resolution, $\nu_\mu$ charged-current
backgrounds can also be identified in \minerva\ by exploiting the delayed muon-decay signature that is
unavailable to oscillation detectors.  The remainder of this section sketches two possible analyses illustrating both \minerva's potentially decisive ability to isolate appearance backgrounds, and its impressive resolution.

\subsubsection{Beam $\nu_e$}
\label{sect:oscBeamNue}
\begin{figure}[htbp]
\centerline{\epsfig{file=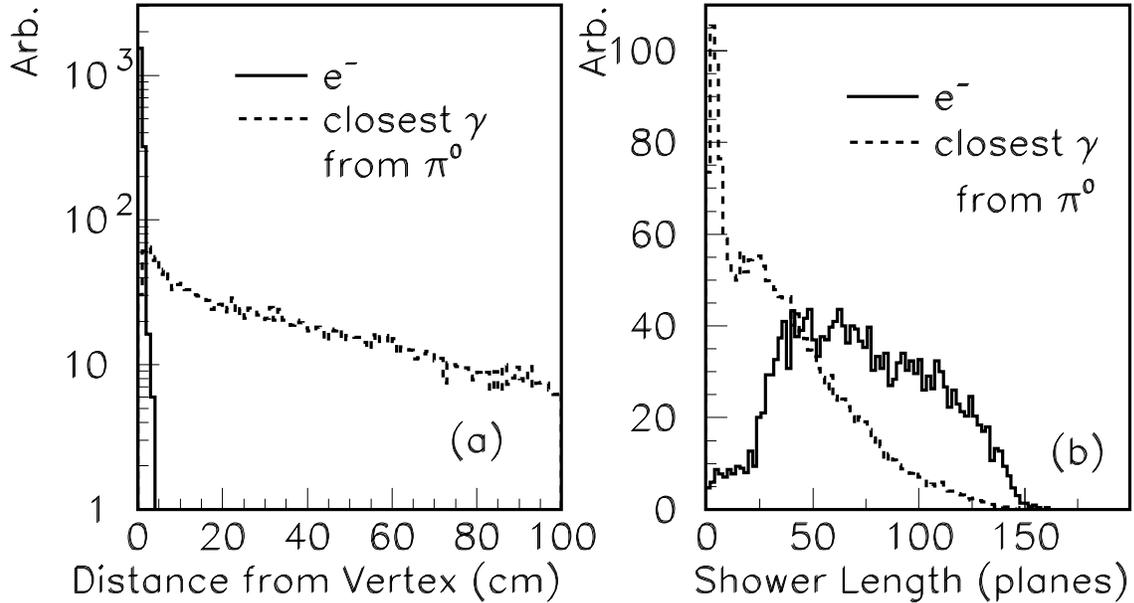, width=\textwidth}}
\caption[Topological electron/$\pi^0$ separation in \minerva]{(a) The distance in centimeters between the neutrino 
vertex, which can be determined from a proton track, and 
the start of the most upstream electromagnetic shower, for 
both electrons and photons from neutral pions. (b) The 
shower length in units of scintillator planes, for 
electrons and neutral pions.} 
\label{fig:elgamdif}
\end{figure}

The cleanest signature for $\nu_e$ charged-current interactions in \minerva\ will be the 
presence of an electromagnetic shower originating near 
a proton track.  Figure~\ref{fig:elgamdif} shows the 
distance between the electromagnetic shower origin and the true primary
vertex for charged-current $\nu_e$ interactions and $\pi^0$ production. The figure
also shows the length of the showers, measured in \minerva\ scintillator planes, or
1.75~cm of polystyrene.  For neutral pions the length is from the beginning of the first showering photon 
to the end of the second one.

The MIPP experiment\cite{MIPP} will reduce the uncertainty on the $\nu_e/\nu_\mu$
flux ratio to roughly 5\%, but to determine the 
true intrinsic $\nu_e$ background, the uncertainty on the quasi-elastic (Section~\ref{sect:quasielastic})
and resonant (Section~\ref{sect:resonant}) cross-sections must also be taken into account.
 
\begin{figure}[htbp]
\centerline{\epsfig{file=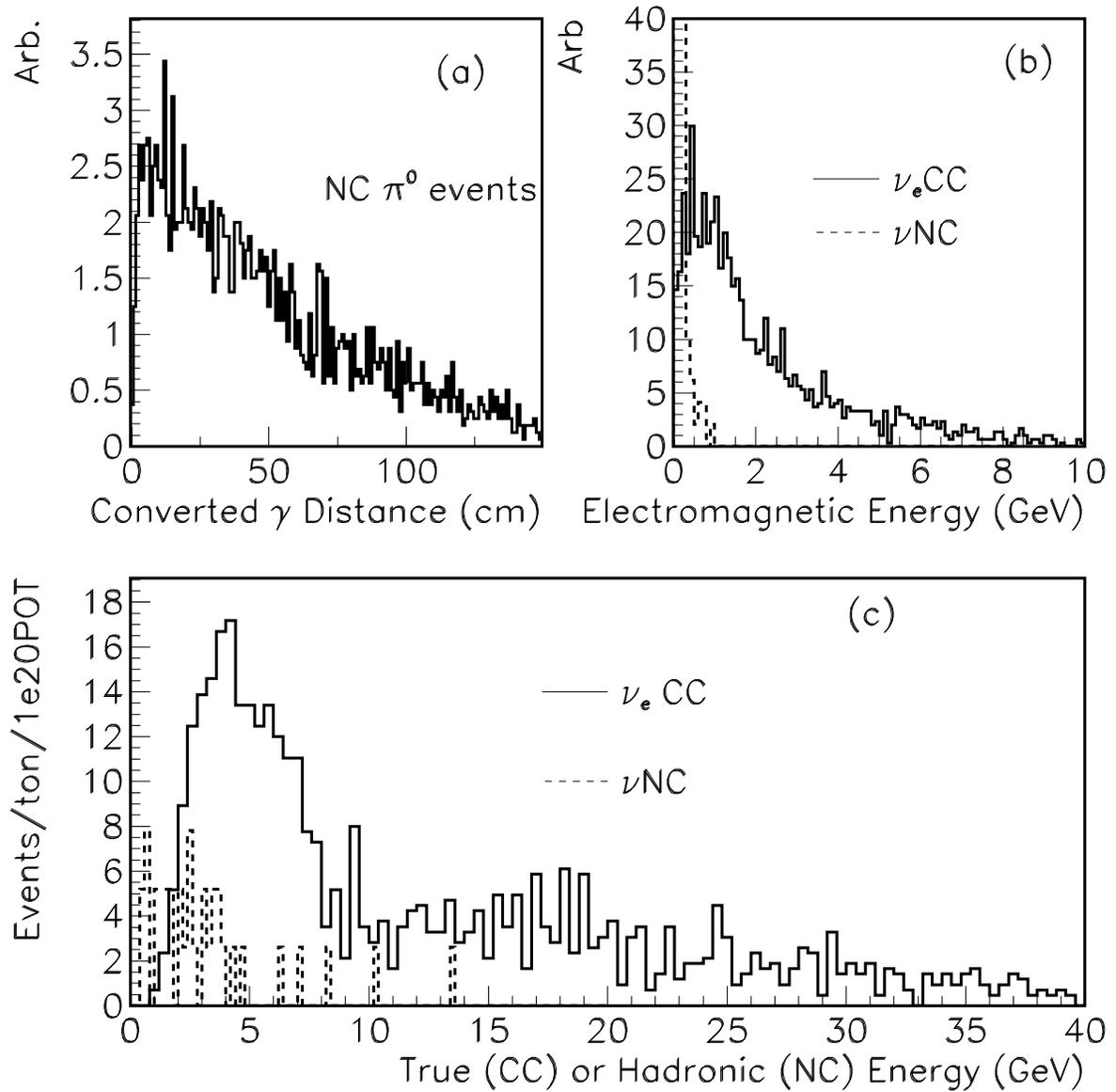, width=\textwidth}} 
\caption[$\nu_e$ flux measurement in \minerva]{Variables to
 identify $\nu_e$ charged current events in \minerva\ :
 (a)Distance between vertex (determined by proton track) and
 most upstream converted photon from $\pi^0$ decay, and
 (b)Energy in electromagnetic cluster, for $\nu_e$ charged-current
 and all neutral-current events.  (c) True (hadronic) energy for $\nu_e$
 charged-current (all neutral-current) events, after applying the simple cuts described in the text.}
\label{fig:nuemeas}
\end{figure}

With a simple analysis that requires a proton track and 
an electromagnetic shower depositing over $0.5~\hbox{GeV}$ in the detector 
and starting within 2 planes of the proton track, 
\minerva\ would collect roughly 1500 charged-current $\nu_e$ events per year
in a 3-ton fiducial volume, with a neutral-current background about a third the size of the $\nu_e$ signal. 
Figure~\ref{fig:nuemeas} shows the resulting energy spectra for 
the $\nu_e$ signal and neutral-current background.  
Further cuts to remove events with an identifiable second photon 
cluster from $\pi^0$ decay could reduce this background even further.  

\subsubsection{Neutral-current $\pi^0$ production}

\begin{figure}[h!]
\centerline{\epsfig{figure=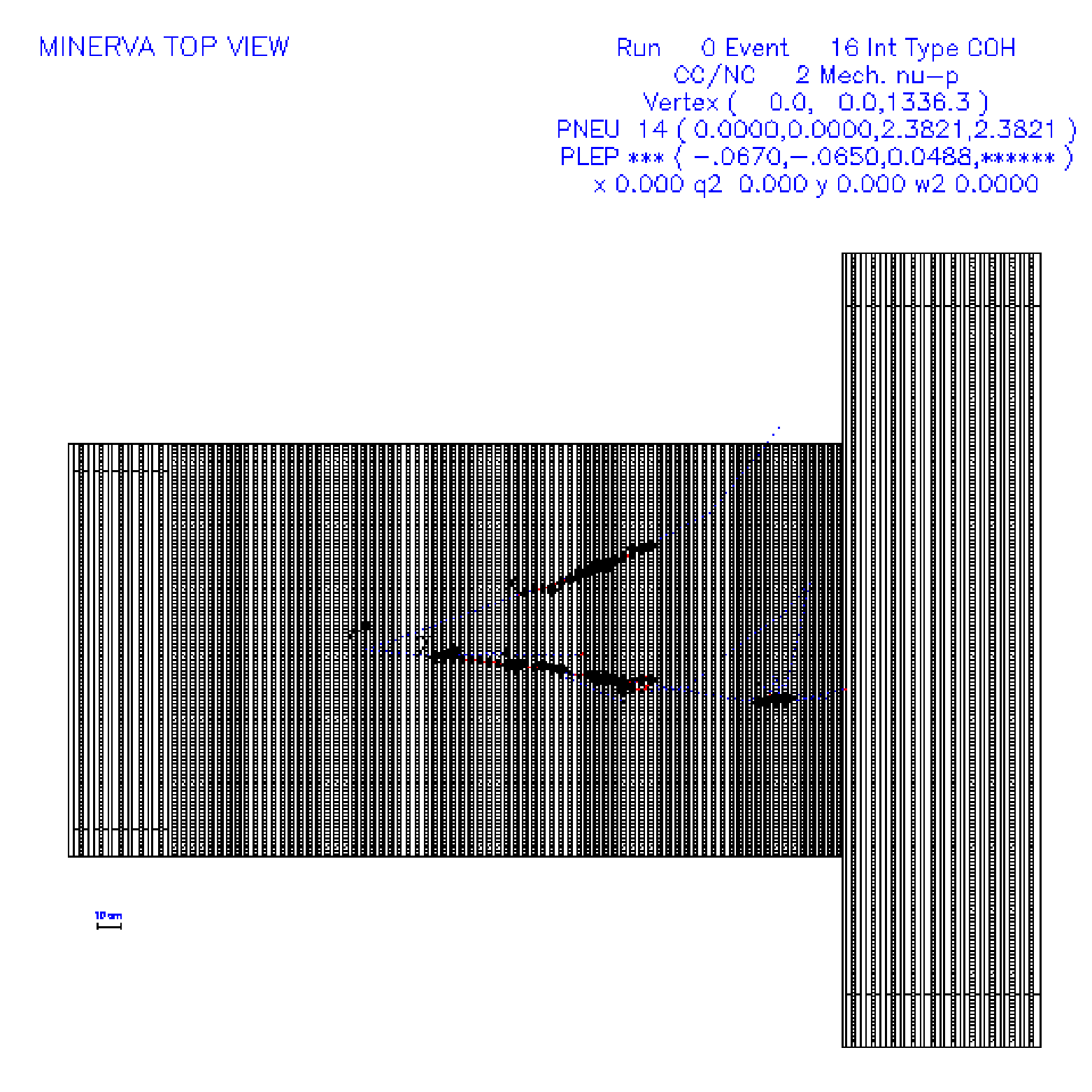,width=\textwidth}}
\caption[Simulated neutral-current $\pi^0$-production event]{A simulated neutral-current coherent $\pi^0$ production event in \minerva.  The position of the $\pi0$ decay vertex can be determined accurately by extrapolating the two photons backward.  Notice that both photons pass through a number of planes before beginning to shower, distinguishing them from electrons.}
\label{fig:ncpi0Evt}
\end{figure}

Neutral-current $\pi^0$ production can occur through a number of mechanisms - resonant production, coherent production, and deep-inelastic scattering.  Figure~\ref{fig:ncpi0Evt} shows a striking example of \minerva's response to coherent $\pi^0$ production.

Coherent $\pi^0$ production is a dangerous $\nu_e$ appearance background, because the neutral pion is produced along the direction of the incoming neutrino, and carries away most of the neutrino's energy.
See Section~\ref{sect:coherent} for a complete discussion of this process. 
Its cross-section uncertainty is $\pm 50\%$ or worse, and it has not been measured accurately
at 2~GeV,  the relevant energy for the \numi\ off-axis experiment.  

Production of $\Delta$ (and other nucleon resonances) is another mechanism for faking a
$\nu_e$ signal, since their decay products often include $\pi^0$.  Neutral pions
from resonance decay are not as energetic or collinear as those produced coherently, but 
their angular distribution mimics that of the signal.  Resonant $\pi^0$ are particularly
susceptible to final-state nuclear interaction and rescattering, which will be studied in detail
by \minerva\ using charged-current reactions (see Sections~\ref{sect:resonant} and~\ref{sect:nuclear}).

\begin{figure}[p!]
\centerline{\epsfig{figure=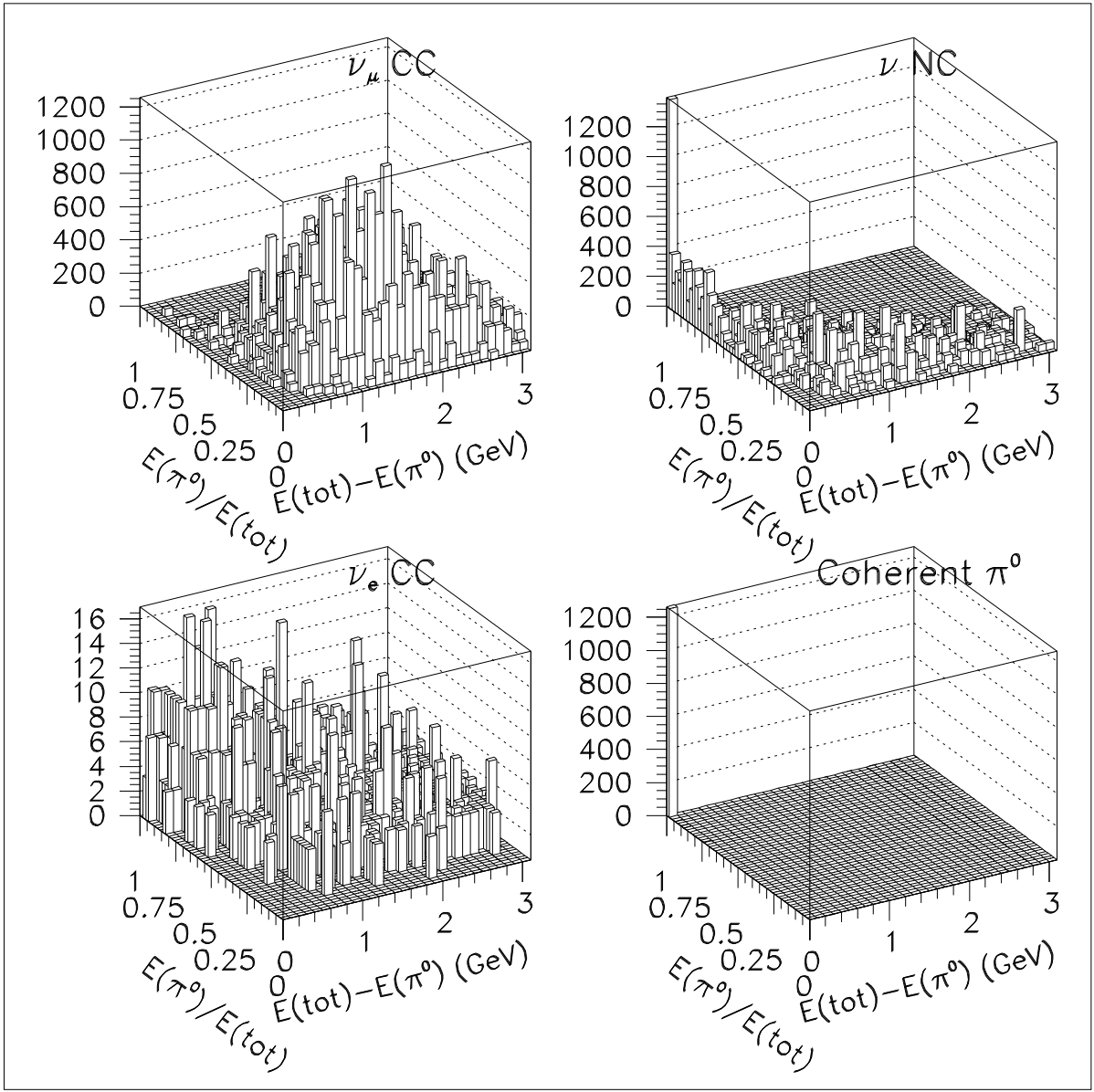,width=\textwidth}}
\caption[Selection of neutral-current single-$\pi^0$ production]{Selection of neutral-current single-$\pi^0$ production.  The variables plotted are the fraction of visible energy carried by the $\pi^0$ candidate ($E_\pi/E_{tot}$) and the residual energy $E_{tot}-E_\pi$.  The left-hand plots show the backgrounds from $\nu_\mu$(top) and $\nu_e$(bottom). The plot at top right shows the same distribution for true neutral-current $\pi^0$ production, and the lower right shows the subset from coherent scattering. In the neutral-current plots, notice the dramatic concentration of the coherent $\pi^0$ signal in a single bin, in the left-most corner of the graph.  All samples shown are normalized to a 3 ton-yr exposure of \minerva.}
\label{fig:pi0lego}
\end{figure}

As a proof-of-concept, a sample of neutral-current single-$\pi^0$ events has been selected using simple cuts.  For events with two well-separated electromagnetic clusters ($E_\pi \equiv E_1 + E_2$), each passing through at least six planes of the fully-active region, requiring $E_\pi/E_{tot} > 90\%$ and $E_{tot}-E_\pi < 100~\hbox{MeV}$ efficiently isolates a neutral-current $\pi^0$ sample, as shown in Figure~\ref{fig:pi0lego}.  After these cuts, the contamination of $\nu_e$ and $\nu_\mu$ charged-current interactions (combined) is less than 1\%.  The resulting sample contains about 2400 neutral-current $\pi^0$ events per 3~ton-yr, of which half are resonant and half coherent.

\begin{figure}[hptb]
\centerline{\epsfig{figure=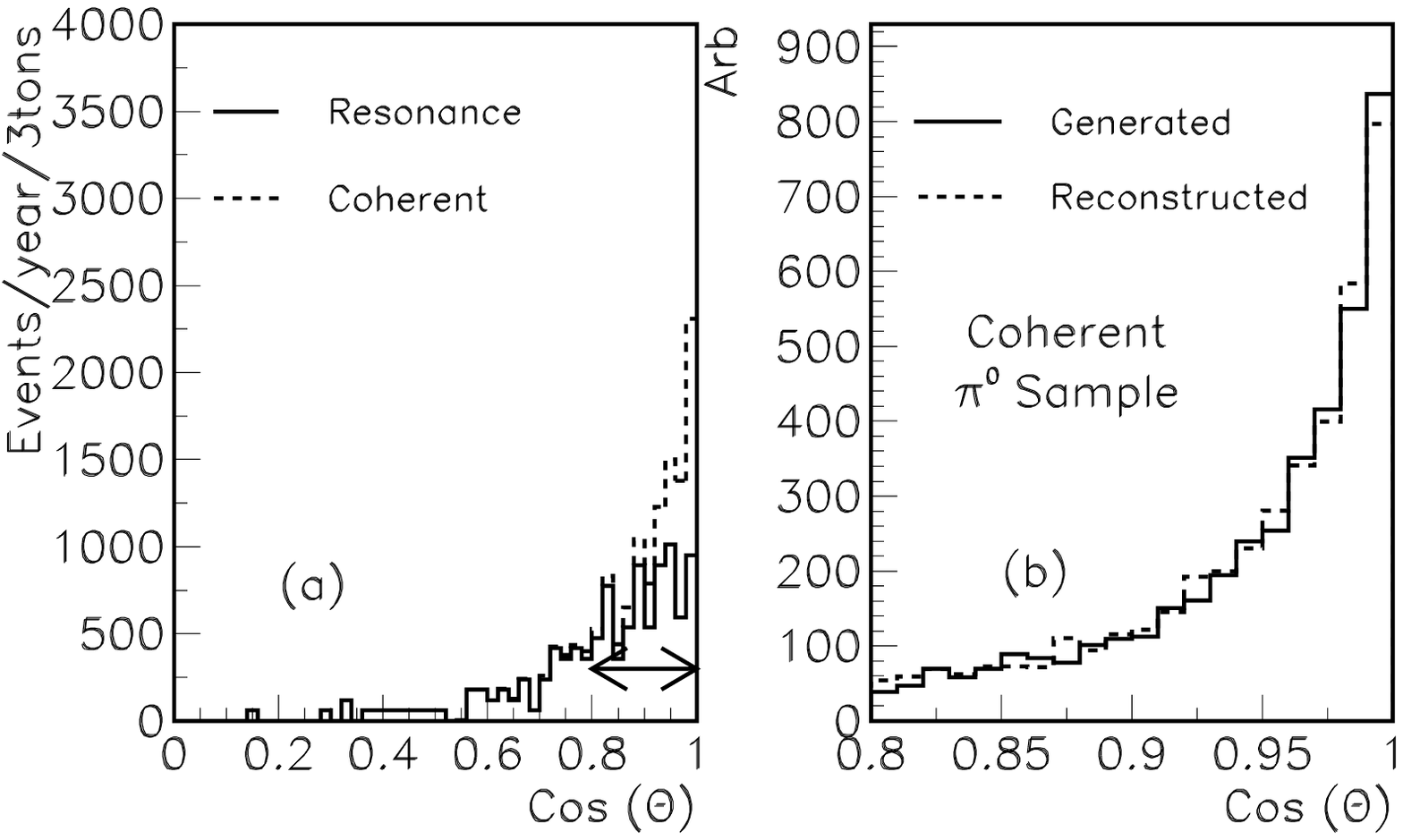,width=\textwidth}}
\caption[Angular distribution of neutral-current single-$\pi^0$ production sample]{Angular distribution of neutral-current single-$\pi^0$ sample.  The plot at left shows all events passing the cuts on $E_\pi/E_{tot}$ and
$E_{tot}-E_\pi$ described in the text, broken down into coherent and resonant reactions. The coherent sample is strongly forward-peaked. The plot at right is a close-up of the forward region comparing the true and reconstructed $\pi^0$ angular distributions from the beam direction. The distributions are nearly identical, highlighting the \minerva's excellent angular resolution.}
\label{fig:cohang}
\end{figure}

Coherent and resonant interactions can be cleanly separated by cutting on the $\pi^0$ angle to the beam direction, as shown in Figure~\ref{fig:cohang}, which also highlights \minerva's excellent $\pi^0$ angular resolution.  The overall efficiency for selecting coherent neutral-current $\pi^0$ is about 40\%.

Finally, some $\nu_\mu \to \nu_e$ backgrounds in oscillation 
experiments will come from deep-inelastic scattering, 
although that sample is easily isolated from the other two processes in \minerva\
because of the high multiplicity.  
Since the mean hadron multiplicity in deep-inelastic scattering is large, and the $\pi^0$ angular distribution rather flat, this channel is less likely to contribute background to a $\nu_e$ search than the other two.  On the other hand the cross-section for deep-inelastic scattering is larger, even at $E_\nu = 2~\hbox{GeV}$ than for either resonant or coherent production, and most deep-inelastic interactions in MINOS or the \numi\ off-axis experiment will fall into the poorly-understood $W \sim 2~\hbox{GeV/c}^2$ transition region at the border of resonant production (see Sections~\ref{sect:duality}), and fragmentation of low-$W$ hadronic systems is not well-modeled by existing simulations like PYTHIA.

%% file: numi.tex
\section{The \numi\ Near Experimental Hall}
\label{sect:numihall}
\input{numiHall}

%% file: numiHall.tex
The MINOS Near Detector Hall\cite{MINOS_TRD_NDHALL} is a fully-outfitted experimental facility
that can accomodate the \minerva\ experiment with a limited
number of additions to the infrastructure.

The detector hall is 45~m long, 9.5~m wide, 9.6~m high, with
its upstream end just over 1~km from the NuMI target, at a depth of 
106~m below grade.  The MINOS detector
will be installed towards the downstream end of the hall, leaving a
free space upstream amounting to, roughly, a cylinder 26~m in length and
3~m in radius.  The neutrino beam centerline will descend at a slope of
$3.3^\circ$, and enter the MINOS detector at a height of 3~m from the hall
floor. The hall has been excavated, and is currently
being outfitted for the MINOS near detector, with beneficial occupancy
expected in early March, 2004.

Ground water is pumped from the NuMI/MINOS complex at a rate of approximately 
320--400~gallons (1300--1600~l) per minute.  The hall floors and walls may 
be damp in places, and a drip ceiling will need to be extended upstream of 
the MINOS detector to protect \minerva.  The
air will be held at a temperature of between $60^\circ$~F and $70^\circ$~F
($15^\circ$~C and $21^\circ$~C), and 60\% relative humidity.

\subsection{Utilities}
\label{sect:utilities}

The MINOS Service Building on the surface houses the access shaft to
the Near Detector Hall, and is the entry point for electrical, cooling,
and data services to the hall.  A 15-ton capacity crane, with a hook
height of 18.5~feet (5.66~m), will be used to lower the 3.47~T MINOS detector
planes to the hall.  MINOS Detector planes will be moved within the hall 
using an overhead 15-ton crane, with 22~foot (6.7~m) hook height and a coverage
along the beam axis of approximately 40~m.  

Quiet power to the hall is provided by a 750~KVA transformer at the
surface, which branches to a 45~KVA transformer 
for the muon monitoring alcoves, and two 75~KVA transformers for the Near 
Detector hall.  The power needs of the MINOS detector account for the 
capacity of the 4 panelboards served by the two 75~KVA transformers, so 
additional panelboards for \minerva\ will likely be needed.  
The current estimate for \minerva\ electronics and high voltage power is less than
5000~W.  It appears that overall capacity for the additional load exists within the 
MINOS hall, but this needs to be verified in detail.

\minerva's main non-quiet power need is for the magnet coils, with
an estimated ohmic power loss of 30~kW.  The MINOS magnet coil power supply
will be served by a 480~V line with 400~A capacity, but will require less
than 80~kW of power.  This should leave ample capacity for the addition of
a power supply for the \minerva\ coil on the same line.

The heat sink for the MINOS LCW cooling circuit is the flux of ground
water collected in the MINOS sump.  The cooling is adequate for MINOS,
with an output water temperature of 70$^\circ$F.  This should be sufficient
to absorb the heat load of the \minerva\ magnet, but would likely be too
warm to effectively cool the front end electronics.  The relatively low
heat load of the \minerva\ electronics would likely be absorbed without
problems by the MINOS hall air conditioning.

\subsection{Detector Placement}

\begin{figure}[bthp]
\center
\epsfxsize=\linewidth\leavevmode
\epsffile{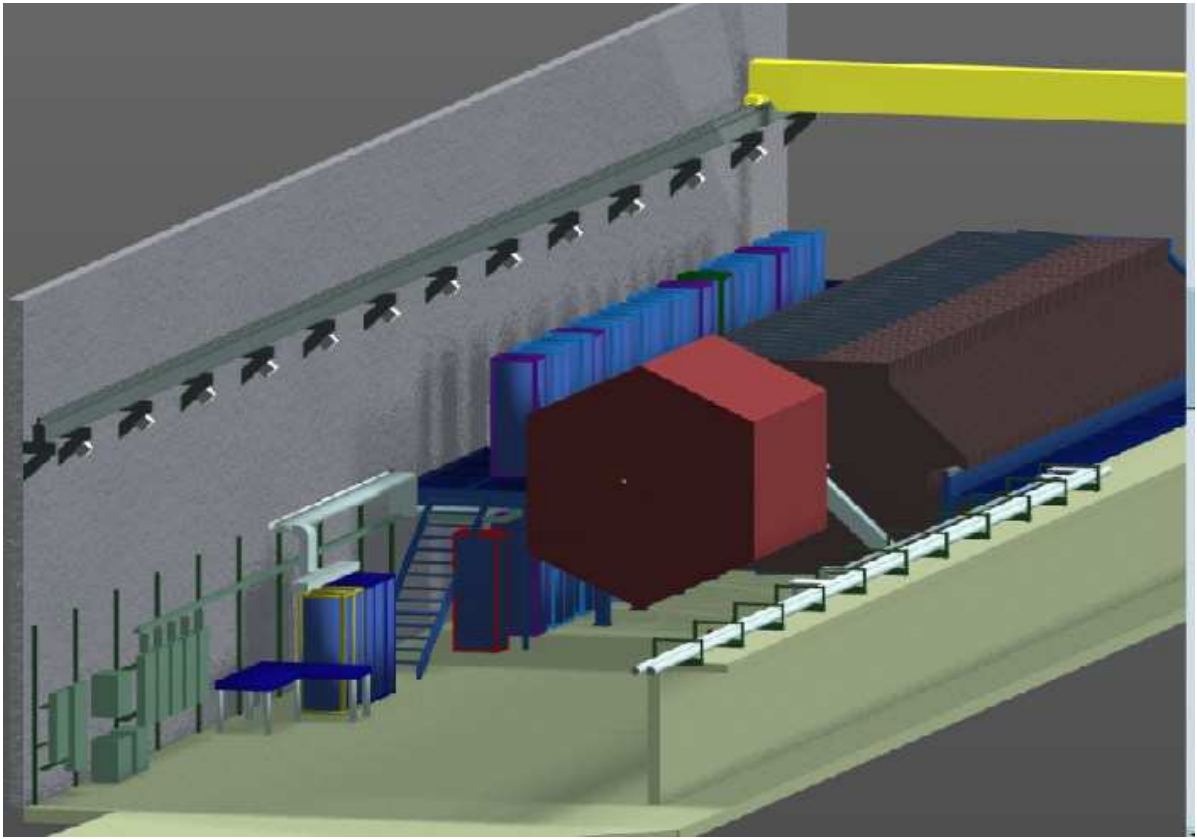}
\caption[Isometric view of \minerva]{
View of the proposed \minerva\ detector, and the MINOS detector, looking 
dowstream. 
}
\label{fig:det_iso}
\end{figure}

\begin{figure}[ph]
\center
\epsfysize=0.9\textheight\leavevmode
\epsffile{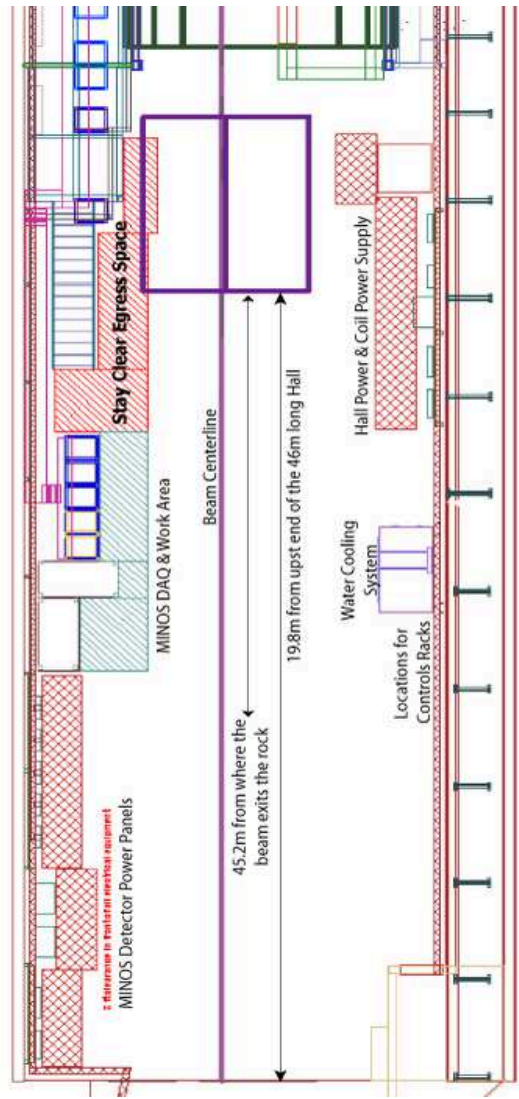}
\caption[Plan view of \minerva]{
Plan view of the \minerva\ detector (purple outline near top of figure).}
\label{fig:det_plan}
\end{figure}

\begin{figure}[ph]
\center
\epsfxsize=\linewidth\leavevmode
\epsffile{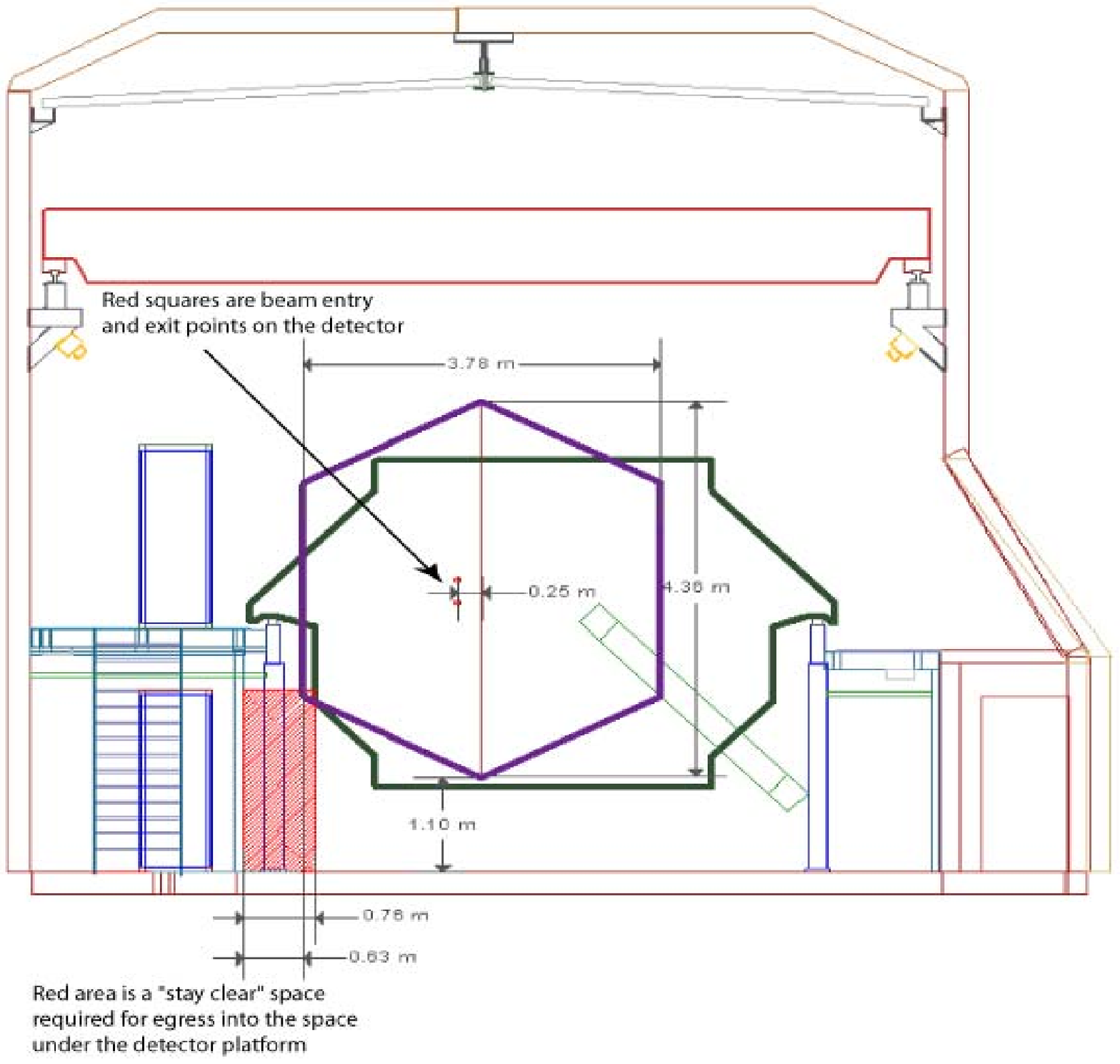}
\caption[Front view of \minerva]{
Front view of the \minerva\ detector.}
\label{fig:det_xsec}
\end{figure}

\minerva\ will be placed with its downstream end 1.75~m upstream
of MINOS.   This will leave sufficient work space between the two detectors,
and avoid interfering with the MINOS coil, which extends approximately 1.5~m
upstream of MINOS, to the lower right in the view of Figure~\ref{fig:det_iso}.
To have the beam axis intersect the detector axis close to the
center of the active plastic target, the lower vertex of the \minerva\ detector
would be placed 1.10~m off the hall floor.  The beam
centerline would enter the detector at an elevation of 3.4~m from the 
floor (Figures~\ref{fig:det_plan} and~\ref{fig:det_xsec}).

\minerva\ will impinge slightly on a ``stay clear'' egress space
for the lower MINOS detector electronics racks.  This could be resolved by
either raising \minerva\ by less than 10~cm, or by rearranging
the layout of the upstream part of the MINOS electronics platform and
stairs.

\subsection{Impact on MINOS}

The impact on MINOS of the heat load and power consumption of the \minerva\ 
detector can be made negligible through relatively minor additions to the hall
infrastructure. The presence of the detector in the neutrino beam will cause an
increase in the rate of activity in the MINOS detector, particularly in the
first 20 planes forming the MINOS veto region.  With the current design of the
\minerva\ detector, the expected event rate in the detector is $\approx$ 1.4 CC
events / $10^{13}$ POT.  For a spill of 2.5 x $10^{13}$ POT this is 3.4 CC
events plus an additional 1.0 NC event.  Since, in addition, the vectors of all
particles leaving \minerva\, with a trajectory heading towards MINOS, will be
made available to MINOS when \minerva\ is taking data, this should be a
managable situation. 

\subsection{MARS Simulation of Radiation Flux}

The intense neutrino beam will create a fluence of other particles due to
neutrino interactions in the rock surrounding the experimental hall. Several
physics topics are sensitive to background interactions caused, particularly,
by neutrons. A MARS14-based model\cite{MARS} has been created to estimate 
non-neutrino background in the detector. The model includes the rock
surrounding the experimental hall, and the \minerva\ detector located upstream
of MINOS. Both the detectors are positioned on the \numi\ axis.  The \minerva\
detector is simulated as described in Section~\ref{sect:detector}, but with the
magnetic field ignored.

\begin{figure}[htb!]
\begin{minipage}[t]{0.49\linewidth}
\begin{center}
\epsfig{figure=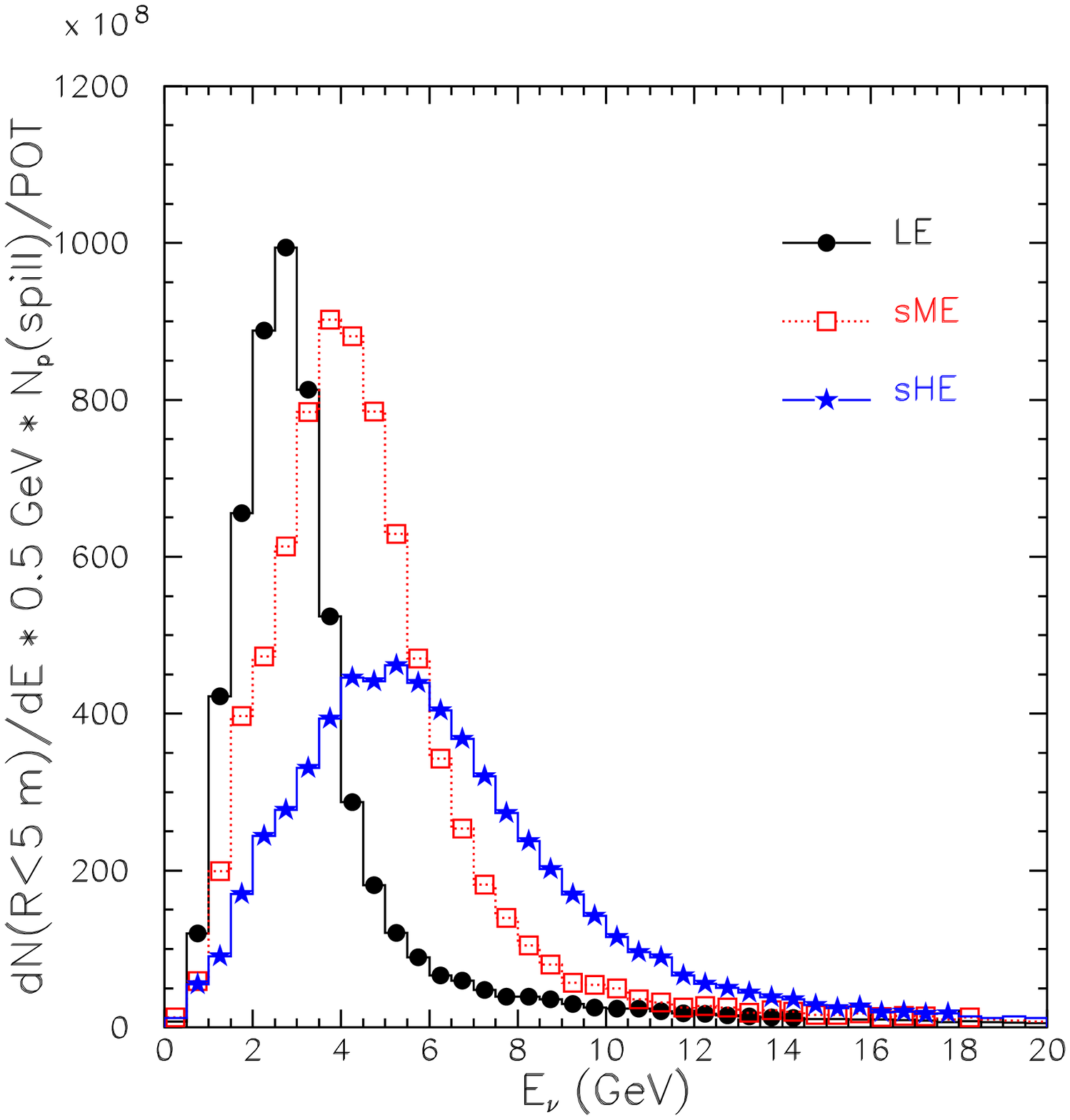, width=\linewidth}
\caption[$\nu_\mu$ energy spectrum within 5~m of \numi\ axis]{Muon neutrino energy distributions at the near detector hall
within a distance of 5~m around the axis of \numi.}
\label{fig:energy5m}
\end{center}
\end{minipage}
\hfill
\begin{minipage}[t]{0.49\linewidth}
\begin{center}
\epsfig{figure=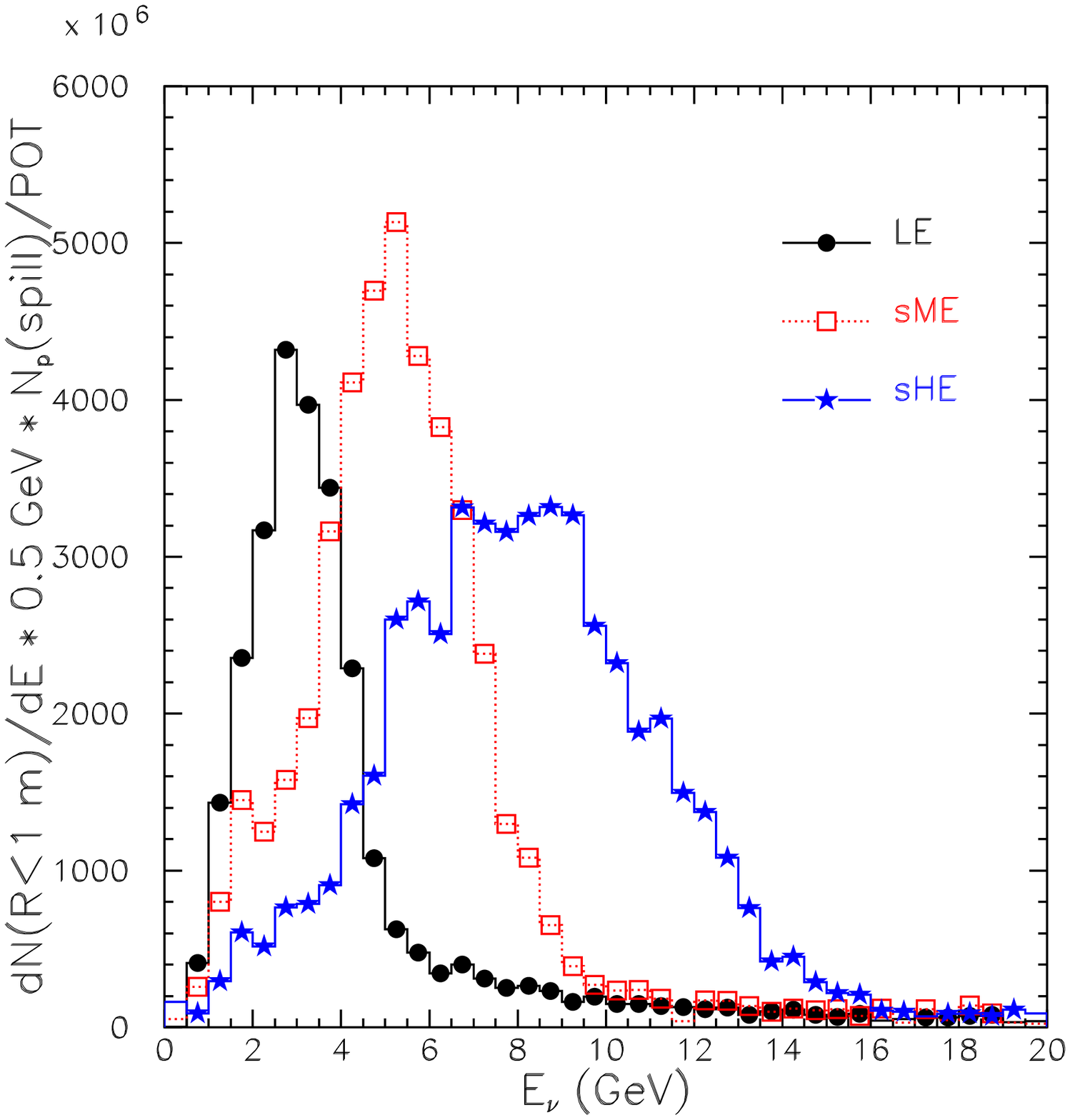, width=\linewidth}
\caption[$\nu_\mu$ energy spectrum within 1~m of \numi\ axis]{Muon neutrino energy distributions at the near detector hall
within a distance of 1~m around the axis of \numi.}
\label{fig:energy1m}
\end{center}
\end{minipage}
\end{figure}

\begin{figure}[htb!]
\begin{minipage}[t]{0.49\linewidth}
\begin{center}
\epsfig{figure=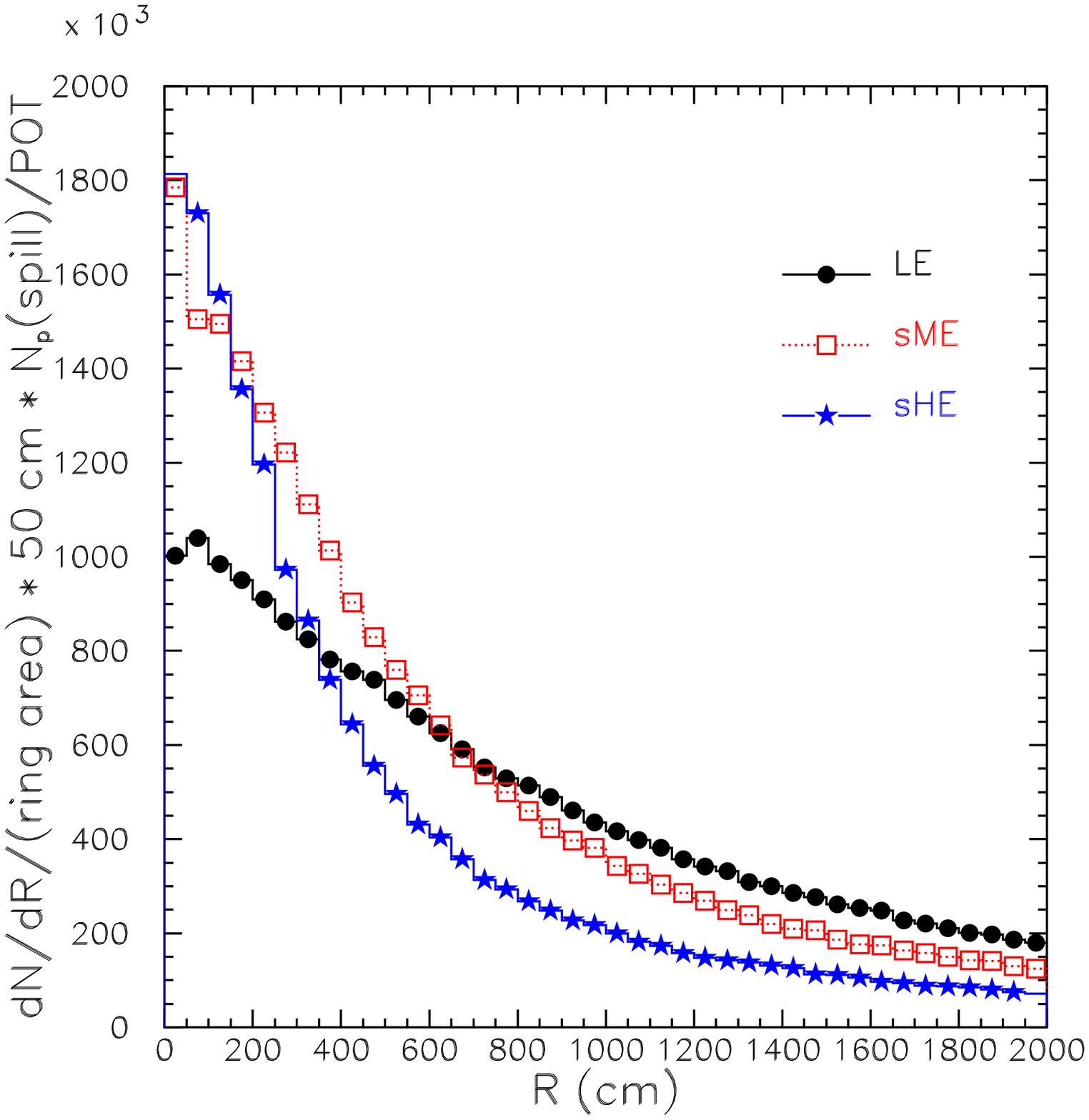, width=\linewidth}
\caption[$\nu_\mu$ radial distribution at near detector hall]{Muon neutrino radial distributions at the near
detector hall. The origin corresponds to the \numi\ axis.}
\label{fig:rdens}
\end{center}
\end{minipage}
\hfill
\begin{minipage}[t]{0.49\linewidth}
\begin{center}
\epsfig{figure=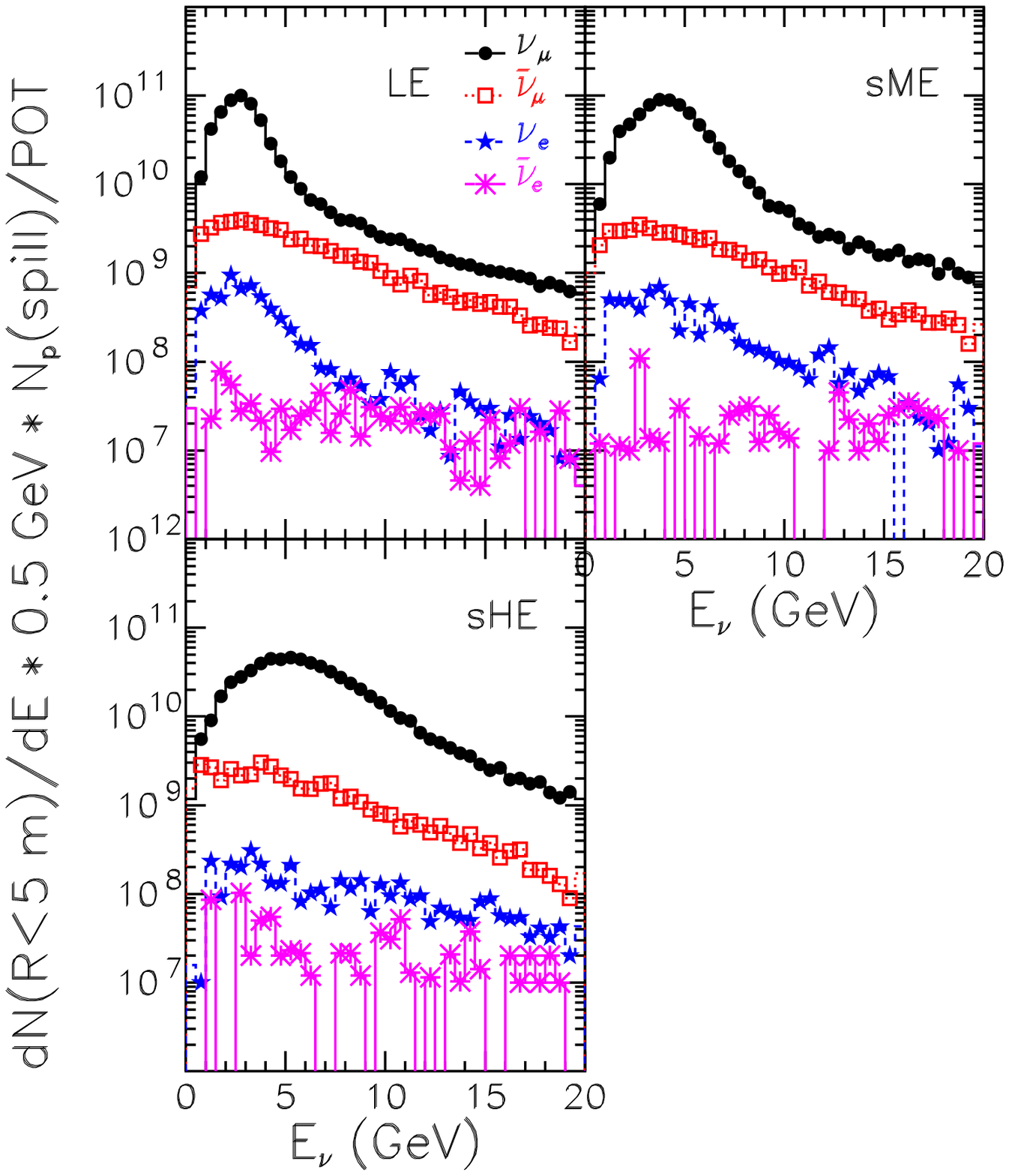, width=\linewidth}
\caption{Neutrino beam components.}
\label{fig:beamcomp}
\end{center}
\end{minipage}
\end{figure}

The muon neutrino energy distributions used in this simulation for the Low,
Medium, and High Energy beam configurations are shown in
Figures~\ref{fig:energy5m}--\ref{fig:rdens}. There is an admixture of the other
types of neutrinos from $\pi^-$, kaon and muon decays shown in 
(Figure~\ref{fig:beamcomp}).

The MARS14 neutrino interaction model tracks the energy and angle of 
final state neutrinos, hadrons, $e^{\pm}$, and $\mu^{\pm}$ from neutrino
interactions. These particles, along with the showers they initiate, are
transported through the user-defined geometry and energy deposition and 
dose are estimated. MARS distinguishes four types of neutrinos: $\nu_{\mu}$,
$\overline{\nu}_{\mu}$, $\nu_e$ and $\overline{\nu}_e$, all of which
are present in the NuMI beam (Figure~\ref{fig:beamcomp}).  The interactions
included in the model are listed in Table~\ref{tab:radsim_interact}.

\begin{table}[bth]
\vskip 1cm
\begin{center}
\begin{tabular}[h]{|c|c|c|} 
\hline
1) & $\nu_{\ell} N \to \ell^+ X$         & $\overline{\nu}_{\ell} N \to \ell^- X$ \\ \hline
2) & $\nu_{\ell} N \to \nu_{\ell} X$     & $\overline{\nu}_{\ell} N \to \overline{\nu}_{\ell} X$ \\ \hline
3) & $\nu_{\ell} p \to \ell^+ n$         & $\overline{\nu}_{\ell} n \to \ell^- p$ \\ \hline
4) & $\nu_{\ell} p \to \nu_{\ell} p$     & $\overline{\nu}_{\ell} p \to \overline{\nu}_{\ell} p$ \\ \hline
5) & $\nu_{\ell} n \to \nu_{\ell} n$     & $\overline{\nu}_{\ell} n \to \overline{\nu}_{\ell} n$ \\ \hline
6) & $\nu_{\ell} e^- \to \nu_{\ell} e^-$ & $\overline{\nu}_{\ell} e^- \to \overline{\nu}_{\ell} e^-$ \\ \hline
7) & $\nu_{\ell} e^- \to \nu_{e} \ell^-$ &  \\ \hline
8) & $\nu_{\ell} A \to \nu_{\ell} A$     & $\overline{\nu}_{\ell} A \to \overline{\nu}_{\ell} A$ \\
\hline
\end{tabular}
\end{center}
\caption[MARS model interactions for  $\nu_{\ell}$ and $\overline{\nu}_{\ell}$]{MARS model interactions for  $\nu_{\ell}$ and $\overline{\nu}_{\ell}$,
where $\ell=\{e,\mu\}$.}
\label{tab:radsim_interact}
\end{table}

The model is described in detail in~\cite{ref:Mars_nuInteractions}.
Some notable features of the model include:\\
$\bullet$ Recoil of the target is simulated only for elastic
interactions. \\
$\bullet$ The model does not include 
inelastic neutrino interactions which produce pions via resonances. 
The cross sections for such processes are relatively small however 
compared to ones for the deep-inelastic (DIS) and coherent
elastic scattering (process 8 in Table~\ref{tab:radsim_interact}).\\
$\bullet$ For charged-current DIS 
(process 1 in Table~\ref{tab:radsim_interact}), the process of hadronization 
is simplified. Once the momentum of lepton is decided, the total
momentum is balanced by a single pion, which is
forced to undergo a deep-inelastic interaction in the same nucleus.
This coarse ``hadronization'' is justifiable since we are interested
in certain gross averages over the showers.\\

\newcommand{\Lower}[1]{\smash{\lower 1.5ex \hbox{#1}}}

\begin{table}[t]
\begin{center}
\begin{tabular}{c|c|c|c|c|c|c|c}
\hline
\Lower{Particle} & \Lower{$E_{th}$ (MeV)} & \multicolumn{6}{c}{flux ($10^{-5} cm^{-2}$)} \\ \cline{3-8}

           &      & LE tot & LE sig & sME tot & sME sig & sHE tot  & sHE sig \\ \hline

 n         & 0.1  &  3.10  & 0.0465 & 6.50    & 0.0986  & 8.77     & 0.1351  \\
           &  20  &  1.26  & 0.0271 & 2.45    & 0.0586  & 3.48     & 0.0822  \\
           & 100  &  0.49  & 0.0140 & 0.95    & 0.0258  & 1.35     & 0.0355  \\ \hline

 Charged   & 0.1  &  0.66  & 0.0523 & 1.39    & 0.1059  & 2.04     & 0.1534  \\
 hadrons   &  20  &  0.73  & 0.0522 & 1.34    & 0.1012  & 1.93     & 0.1550  \\
           & 100  &  0.55  & 0.0505 & 1.10    & 0.0877  & 1.57     & 0.1220  \\ \hline

 $\gamma$  & 0.1  &  15.94 & 0.3150 & 33.64   & 0.6033  & 46.26    & 0.9879  \\
           &  20  &  1.08  & 0.0583 & 1.63    & 0.1254  & 2.20     & 0.1608  \\
           & 100  &  0.26  & 0.0238 & 0.41    & 0.0493  & 0.51     & 0.0710  \\ \hline

 $e^{\pm}$ & 0.1  &  1.28  & 0.0614 & 2.11    & 0.1272  & 2.93     & 0.1554  \\
           &  20  &  0.44  & 0.0436 & 0.83    & 0.0717  & 1.16     & 0.1038  \\
           & 100  &  0.16  & 0.0163 & 0.26    & 0.0318  & 0.32     & 0.0480  \\ \hline

 $\mu^{\pm}$& 0.1 &  1.43  & 0.0206 & 2.61    & 0.0416  & 3.08     & 0.0493  \\
           &  20  &  1.41  & 0.0206 & 2.58    & 0.0417  & 3.09     & 0.0491  \\
           & 100  &  1.40  & 0.0190 & 2.59    & 0.0397  & 3.07     & 0.0472  \\ \hline
\end{tabular}

\caption{Particle fluxes averaged over the active target.$E_{th}$ is a
threshold kinetic energy used in the simulations.}

\label{tab:fluxes}
\end{center}
\end{table}

\subsection{Fluxes in \minerva} 

Particle fluxes in the scintillator part of detector were calculated for three
beam configurations and various threshold energies (Table~\ref{tab:fluxes}).
All the fluxes are given for one spill with the beam intensity of $2 \times
10^{13}$ protons on target/spill. The units are $10^{-5} \hbox{cm}^{-2}$. The
total integrated path-length of a given type of particle is obtained by
multiplying the flux by the fiducial volume of the detector.  The columns "XX
tot" and "XX sig" refer to particles coming from "all" sources or from only
the central volume of scintillator respectively (signal).



%% file: monteCarlo.tex
\section{Monte Carlo Studies and Performance}
\label{sect:MC}
This section outlines the event simulation and reconstruction software used to optimize the detector's design and quantify its physics capabilities. Much of this software has been borrowed from other experiments, where it has been thoroughly validated.  The detector simulation and reconstruction software has been developed specifically for \minerva, but is based on widely-used libraries and algorithms.

\input{monteCarloGenerators}

\input{monteCarloBeam}

\input{monteCarloDetsim}

\input{monteCarloLightsim}

\input{monteCarloReconstruction}

%% file: monteCarloGenerators.tex
\subsection{Event Generators}
\label{sect:generators}

The \minerva\ simulation software interfaces with two event generators that model neutrino interactions with matter: NEUGEN\cite{NEUGEN} and NUANCE\cite{Casper_02}.  NEUGEN was originally designed for the Soudan~2 experiment and is now the primary neutrino generator
for the MINOS experiment. NUANCE was developed for the IMB experiment and is currently used by the Super-Kamiokande, K2K, MiniBooNE and SNO collaborations. Both have evolved from ``proprietary" programs designed for atmospheric neutrino studies into freely-available, general-purpose utilities that aim to model neutrino scattering over a wide range of energies and for different nuclear targets.  Total charged-current cross-sections calculated by NUANCE (Figure~\ref{fig:total}) and NEUGEN (Figure~\ref{fig:neugen}) appear elsewhere in this proposal.  As the results of the two generators agree with each other (to within the depressingly large range of uncertainties in available data)\cite{Zeller:Nuint02}, they have been used interchangeably for the present studies.

As in the past, future studies of neutrino oscillation and searches for nucleon decay will rely heavily on the best possible description of neutrino interactions with matter.  Neutrino event generators are tools which encapsulate our understanding of this physics in an easily usable and portable form.  Practically, they serve two related functions: to allow the rates of different reactions with the experimental target to be calculated, by providing total exclusive and inclusive cross-sections, and to simulate the dynamics of individual scattering events, by sampling the differential cross-sections.  Many comparable packages are available to the collider physics community, and have been incrementally improved for decades, forming a common basis for discussion of different models and phenomena.  One important goal of \minerva\ is to improve the quality of neutrino Monte Carlo event generators, and thereby enhance the physics reach of many future experiments.

\minerva\ will attack this problem from both experimental and theoretical directions.  Experimentally, \minerva\ will make definitive measurements of dozens of exclusive and inclusive cross-sections, across the range of energies most important for future oscillation and nucleon-decay experiments, with a well-controlled flux, and on a variety of nuclear targets.  The era of 25\% uncertainties and marginally-consistent cross-section data for even the simplest neutrino reactions will end with \minerva; for the first time it will be possible to validate the details, and not merely the gross features, of competing models.

At the same time, \minerva\ will be a natural focus of attention for theorists and phenomenologists developing these models.  NEUGEN and NUANCE are two of the most sophisticated neutrino-physics simulations in the world, but NUANCE models quasi-elastic scattering with the 1972 calculation of Smith and Moniz\cite{Smith_72}, and both programs use the Rein--Sehgal\cite{Rein:1980wg} resonant production model which dates from 1981.  That no other widely-accepted models for these, the most fundamental neutrino--nucleon reactions, have emerged in the last quarter century is sobering evidence that an experiment like \minerva\ is long overdue.  New, high-quality data is the surest way to catalyze theoretical ingenuity, and \minerva\ will provide the former in abundance.  Through our contacts with these theorists, and ability to translate well-tested, state-of-the-art models into universally-available and widely-adopted software, \minerva\ will serve as a conduit for expertise from a diverse collection of disciplines into the high-energy neutrino physics community.\footnote{This trend is already beginning, thanks to collaborative work sparked by the NUINT series of workshops.  The BBA--2003 quasi-elastic form-factor fits (see Chapter~\ref{sect:quasielastic}) and Bodek--Yang duality-inspired model of deep-inelastic scattering (Section~\ref{sect:qhduality}) have recently been implemented in NUANCE, and NEUGEN is exploring Benhar's spectral-function approach\cite{Benhar:2003ka} to nuclear binding effects.}

%
%
%
%
%
%

%% file: monteCarloBeam.tex
\subsection{Beam Simulation}

The neutrino fluxes used in the simulation are derived from the GNuMI\cite{GNuMI} program developed
for the MINOS experiment. GNuMI is a full GEANT\cite{GEANT}-based simulation of the
NuMI beamline.  As with all current neutrino beams, neutrinos arise from decay of
$\pi$, $K$ and $\mu$ mesons that originate in collisions of a proton beam on a production target.
GNuMI simulates all aspects of the neutrino beamline. Protons are fired into the target
and the interaction products are transported through the focusing and filtering elements
of the beam.  Appropriate care is taken to ensure that the description of the 
beamline's geometry is as complete as possible, and that meson decays proceed with the
correct kinematics and branching ratios.  For this proposal, all fluxes in \minerva\ are taken from
the official tables used by the MINOS collaboration.

%% file: monteCarloDetsim.tex
\subsection{Detector Simulation}

The simulation of neutrino interactions in the MINERvA detector is carried out by
a GEANT-based Monte Carlo program. This program combines a flexible description of
the detector geometry, the NuMI neutrino beam flux from the beam simulation, neutrino
interaction physics from either of the two generators and simulation of the scintillator
response with the standard tracking and particle interaction routines available in GEANT.

\subsubsection{Interface to the GNuMI flux}

 The output of the GNuMI simulation of the beamline is a set of files recording the neutrino
flux in 0.5 GeV bins for a nominal number of protons on target. The flux files are in a standard
format and hence can be interchanged with no additional modifications to the code. In this way different
beam  configurations can be easily studied. An option exists to generate interactions with
a flat energy spectrum. In this case, beam weights are stored in an output ntuple. This is particularly
useful if one wishes to study the effect of different beam configurations without furthur Monte Carlo
running.

\subsubsection{Interface to the event generators}

 The Monte Carlo simulation program can be configured to accept neutrino interactions from either
NEUGEN3 or NUANCE. The results of a neutrino interaction can be passed to the simulation in a number
of ways. By default, the event generation routines in NEUGEN3 are usually called from within the simulation
itself. In this mode, the code chooses a neutrino energy from the flux files, samples the
density of material along the neutrino path; chooses a vertex and nucleus type, calls the
kinematics generator and inserts the list of particles thus obtained into the GEANT data structures.
This is not the only mode of generation. As a stand-alone generator, NUANCE provides events in 
either a text or ntuple format and so provision is made to read in events from a 
standard external format. NEUGEN3 has been modified to write out events
in the same format, so that the results of both generators may be compared in a consistent manner.

\subsubsection{Geometry}

 Flexibility drives the design of the detector geometry code. The size, segmentation, 
material and shape of 
all components of the detector can be set and altered almost entirely from the input datacards. 
The detector is logically divided into longitudinal
sections. Each section can have different dimensions, strip sizes and absorber widths. In addition
the absorbers in each section can be be constructed from segments of differing material and
widths. The geometry description is sufficiently abstract that minor changes in detector design
may be accommodated merely by changing the datacard, allowing for fast detector reconfiguration
and easy bookkeeping.

\subsubsection{Hits and digitizations}

Particles are tracked through the GEANT geometry in the standard manner. When a particle 
traverses a sensitive detector volume the particle type, volume identifier, entrance and exit
points and energy deposition (including Landau and other fluctuations) are recorded as a hit. 
When GEANT has finished tracking the event, the hits are considered and converted to digitizations. 
There are as many digitizations as there are strips hit. Multiple hits on a single strip are 
condensed into one digitization, although information on which tracks contributed to the 
digitization is stored. These digitizations are then passed to the event reconstruction program.

%% file: monteCarloLightsim.tex
\subsection{Photon Transport Simulations}
\label{sec:lightsim}
The \minerva\ detector simulation assumes ``ideal" light collection, and records the raw energy deposited in each channel.  During event reconstruction, the energy deposit is converted to a number of detected photo-electrons.  The scale factor between energy deposited and expected photo-electrons detected is determined by a standalone optical simulation validated for the MINOS experiment; the expected number of photo-electrons is smeared by Poisson statistics, and a 10\% channel-to-channel Gaussian smearing reflecting a conservative estimate of remaining systematics after calibration and attenuation corrections.

\input{light_mc}

%% file: light_mc.tex
\begin{figure}[tb]
\center
\epsfxsize=90mm\leavevmode
\epsffile{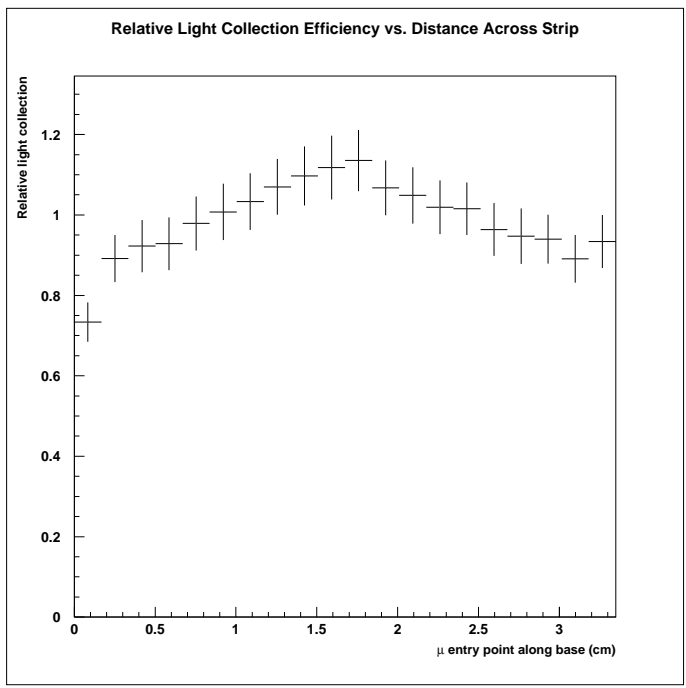}
\caption[Light collection efficiency vs. position across strip]{Relative light collection efficiency 
across the 3.35 cm 
triangular width of the scintillator extrusion.}
\label{fig:rel_light}
\end{figure}
\begin{figure}[tb]
\center
\epsfxsize=90mm\leavevmode
\epsffile{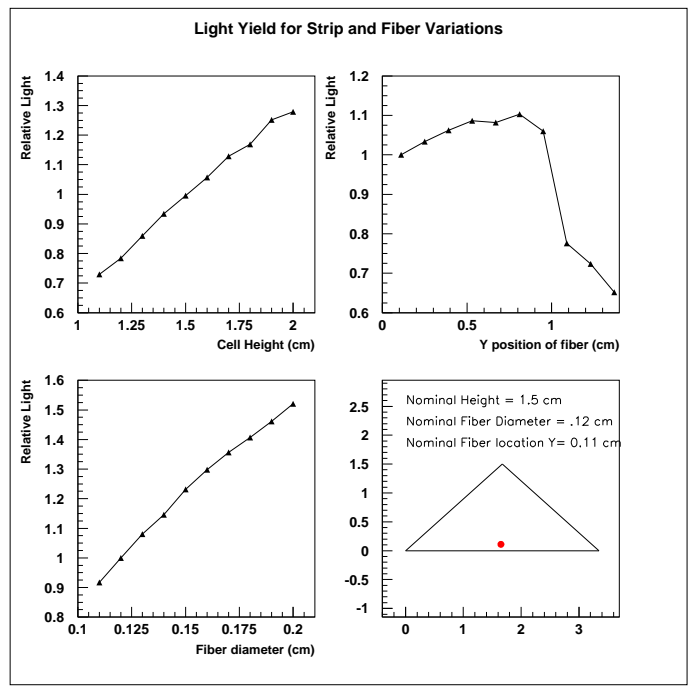}
\caption[Light yield for different strip and fiber dimensions]{ Relative light yield for different strip widths and 
fiber diameters. }
\label{fig:lighttable}
\end{figure}

In addition to the GEANT Monte Carlo, a photon transport Monte Carlo written by 
Keith Ruddick\cite{Ruddick} for the MINOS experiment 
was used to optimize the strip and fiber dimensions.
The average light yield from a MINOS scintillator module is 
4.25 photo-electrons/MIP at a distance of 4 meters, 
and attenuation in the fiber is well described in terms of a double exponential
\cite{Mualem}: 
\begin{equation}
N(x) = A ( e^{-x/90~\hbox{cm}} + e^{-x/700~\hbox{cm}} ) 
\end{equation}  
The photon transport Monte Carlo (LITEYLDX) is used to calculate the number of photons trapped in the
fiber for a MIP entering at a particular position and for a given 
configuration of strip geometry, fiber diameter, and fiber placement.
This information is then used to determine a relative light collection efficiency for a particular
configuration compared to MINOS strips.  With the overall normalization and attenuation curve from 
MINOS one can then calculate the amount of light for any particular configuration.  
Figure \ref{fig:lighttable}, for instance, shows the relative light output for triangular 
extrusions when the strip thickness, fiber diameter and fiber placement are varied.  
As expected, light output is nearly proportional to the strip thickness, and is greatest when 
the fiber is placed at the center of gravity of the strip.  Having a fiber in a groove at the 
edge only results in a 9\% drop in the light level.   
Figure \ref{fig:rel_light} shows the relative light collection efficiency for a triangular 
extrusion where the entry point of the minimum ionizing particle is varied across the 
strip width, and indicates that the collection efficiency varies by $\pm 10$\% over the strip width.    

%% file: monteCarloReconstruction.tex
\subsection{Event Reconstruction}

The output of the detector simulation comprises a list of digitizations for each strip. We have developed a basic reconstruction program to take this list and reconstruct the tracks and vertices in an event.

\subsubsection{Pattern recognition}

Development of a fully-realistic pattern-recognition algorithm to associate hits to track candidates was not undertaken, in view of the manpower and time available.  We are confident that the three-dimensional XUXV modular design of the detector, and its relatively modest occupancy, will allow highly-efficient pattern recognition and track identification.  Visual inspection of events through the graphical interface of the detector simulation program reinforces this conclusion. For our design studies, we have adopted ``omniscient" pattern recognition based on Monte Carlo truth information. All hits generated by a given track (ignoring channels with overlap) are used to reconstruct it.

\subsubsection{Coordinate reconstruction}

Tracks generating hits in at least six scintillator planes of the inner detector, including three planes of the X view, are reconstructed.  Coordinates are estimated from the raw, smeared digitizations, using only planes which have one or two strips hit.  Tracks at high angles to the detector axis may pass through more than two strips in a single plane, and it should be possible to recover these higher-multiplicity hits with a more sophisticated algorithm.  For single hits, the coordinate is taken as the center of the strip.  For dual hits, the position is interpolated using the charge-sharing between between strips, with a small geometrical correction based on the estimated crossing angle.

The coordinate resolution for a large test sample of single and double hits can be measured directly using the residuals obtained when each coordinate is excluded, in turn, from the track's fit.  This coordinate resolution is parameterized as a function of the track's crossing angle, and used to assign errors to coordinates in the fitter.

\subsubsection{Track reconstruction}

Reconstructed coordinates are used to fit each track using a Kalman filter algorithm\cite{Fruhwirth}.  For this proposal, tracking performance has only been studied in the non-magnetic region of the detector; the track model is perforce a strictly linear one.  Neglect of the magnetic field is justified because mission-critical resolutions are determined by performance of the fully-active (non-magnetized) volume, and since coordinate resolution for the strips should not depend on the presence of a magnetic field.  The momentum resolution for  charged tracks in a magnetic field can be reliably estimated from the coordinate resolution, momentum and field strength\cite{pdg}.  As long tracks may pass through many radiation lengths of scintillator and absorbing material, the Kalman filter's ability to correctly account for multiple Coulomb scattering (``process noise") is essential.  The algorithm can optionally be used to exclude outliers from the fit.

\begin{figure}[pthb] 
\centerline{\psfig{file=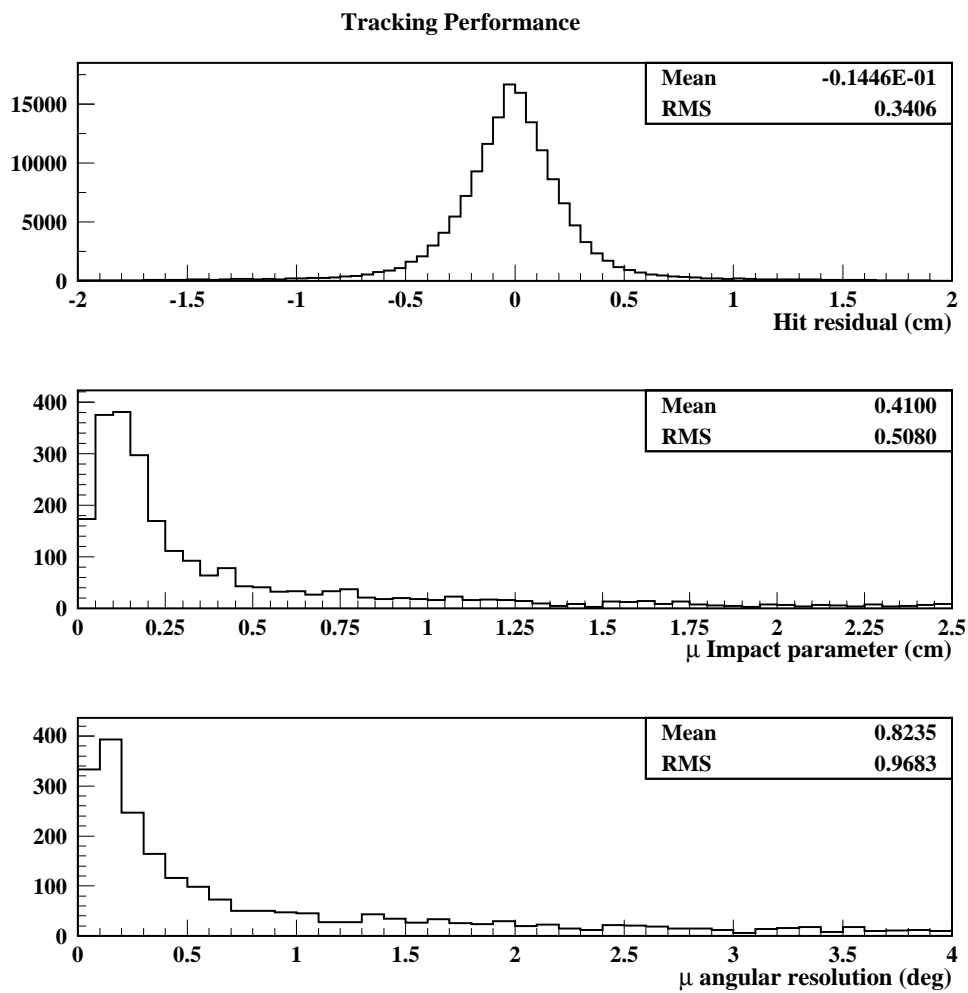,width=5in}}
\caption[Tracking performance for muons from quasi-elastic interactions]{Performance of the tracking algorithm on muons from
from a sample of simulated charged-current quasi-elastic interactions. Shown
are (top) the hit residuals, (middle) the impact parameter of the muon
with the vertex and (bottom) the muon angular resolution.}
\label{fig:muontrk}
\end{figure}

Figure~\ref{fig:muontrk} shows the expected hit
residuals, impact parameter and angular resolution for muons from a sample of
quasi-elastic interactions, assuming triangular strips of 3~cm width and 1.5~cm thickness (close to the final design values). Hit resolutions of $\sim 3 \, \hbox{mm}$ and angular resolutions of $< 0.5^\circ$
are expected. The coordinate resolution is degraded to approximately 1.5~cm if rectangular strips are employed instead of triangular ones, since interpolation based on charge is no longer possible.

\subsubsection{Vertex reconstruction}
\label{sect:vertex}

\begin{figure}[pthb]
\centerline{\psfig{file=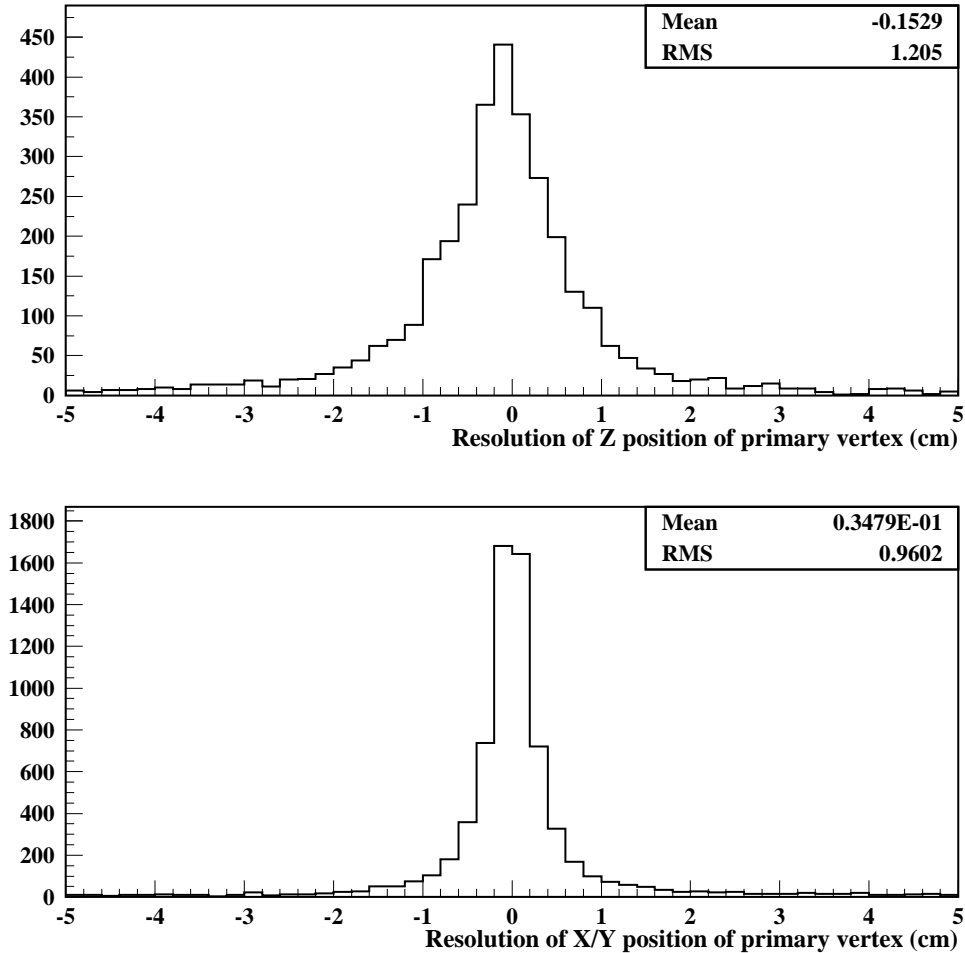,width=5in}}
\caption[Vertex resolution for charged-current quasi-elastic interactions]
{Reconstructed vertex resolution for two track charged current quasielastic events. Shown
are (top) the resolution in the longitudinal position of the vertex (Z) and (bottom) the
resolution of the transverse position of the vertex (X and Y).}
\label{fig:vertex}
\end{figure}

In this study, reconstructed tracks are associated to vertices using Monte Carlo truth information.  The vertex positions are then fit using a Kalman filter algorithm.  Track directions at the vertex are updated taking account of the constraint.  This is equivalent to a least squares fit, but mathematically more tractable since it does not involve inversion of large matrices and can be easily extended to a helical track model.  The primary vertex resolution for a sample of simulated quasi-elastic interactions with two visible tracks is shown in Figure~\ref{fig:vertex}. The vertex
postion can be measured to a precision of better than a centimeter.

\subsubsection{Particle identification}
\label{sect:PID}

Particle identification in \minerva\ will rely on measuring specific energy loss ($dE/dx$) as well as topology (hadron and electromagnetic showers, decay signatures).  

\paragraph {Electromagnetic showers}

Electromagnetic showers are easily identifiable by their diffuse track and characteristic $dE/dx$ profile in the fully-active central detector and energy deposition in the electromagnetic calorimeters.
Section~\ref{sect:oscBeamNue} describes a preliminary technique to separate electrons and photons, when the primary vertex is known, using distance to shower onset and shower length (Figure~\ref{fig:elgamdif}).

\paragraph{Muons}

Energetic muons can be identified by their penetration of material in the calorimeters and/or MINOS near detector. Muons with a momentum measurement in the magnetic field, or which stop inside the detector can be distinguished from protons and kaons by $dE/dx$.  In addition, the delayed $\mu \to e$ decay signature can be detected.

\paragraph{Hadrons}

Hadrons can be identified as such by their interactions in the inner detector and/or hadron calorimeters.  Hadrons which stop without interacting or have their momentum measured by the magnetic field can also be distinguished as $\pi$, $K$ or $p$ with good efficiency using $dE/dx$.

\paragraph{dE/dx analysis}

\begin{figure}[htbp]

\centerline{\epsfig{figure=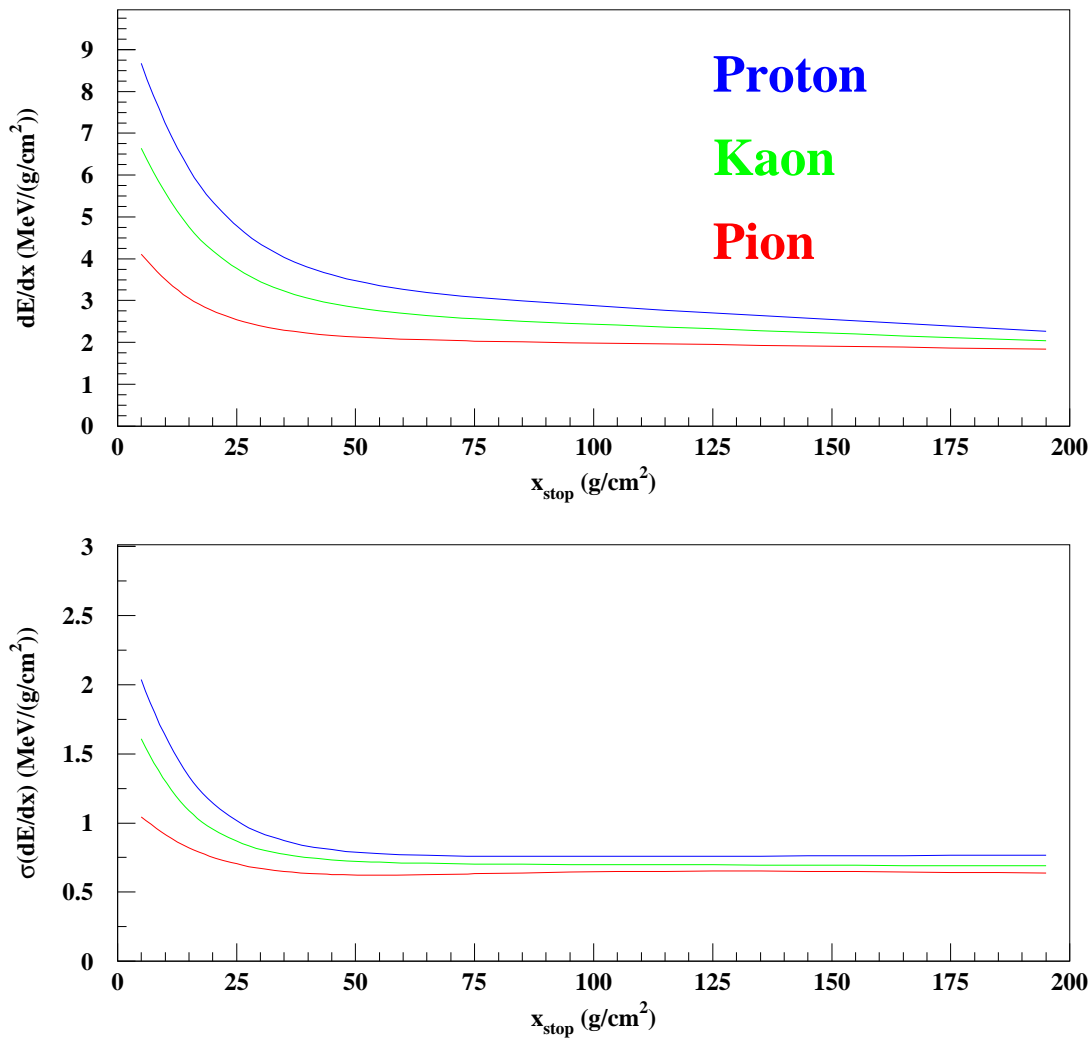,width=\textwidth}}
\caption[$dE/dx$ vs. range from stopping point for $\pi^\pm$, $K^\pm$ and protons]{The top figure shows the average specific energy loss $dE/dx$ for stopping $\pi^\pm$, kaons and protons, vs. range from the stopping point (in $\hbox{g/cm}^2$), for the simulated \minerva\ inner detector. The bottom figure shows the estimated standard deviation of the energy loss, which is used to form a $\chi^2$ estimator for particle identification.}
\label{fig:dedxCalib}
\end{figure}

\begin{figure}[htbp]

\centerline{\epsfig{figure=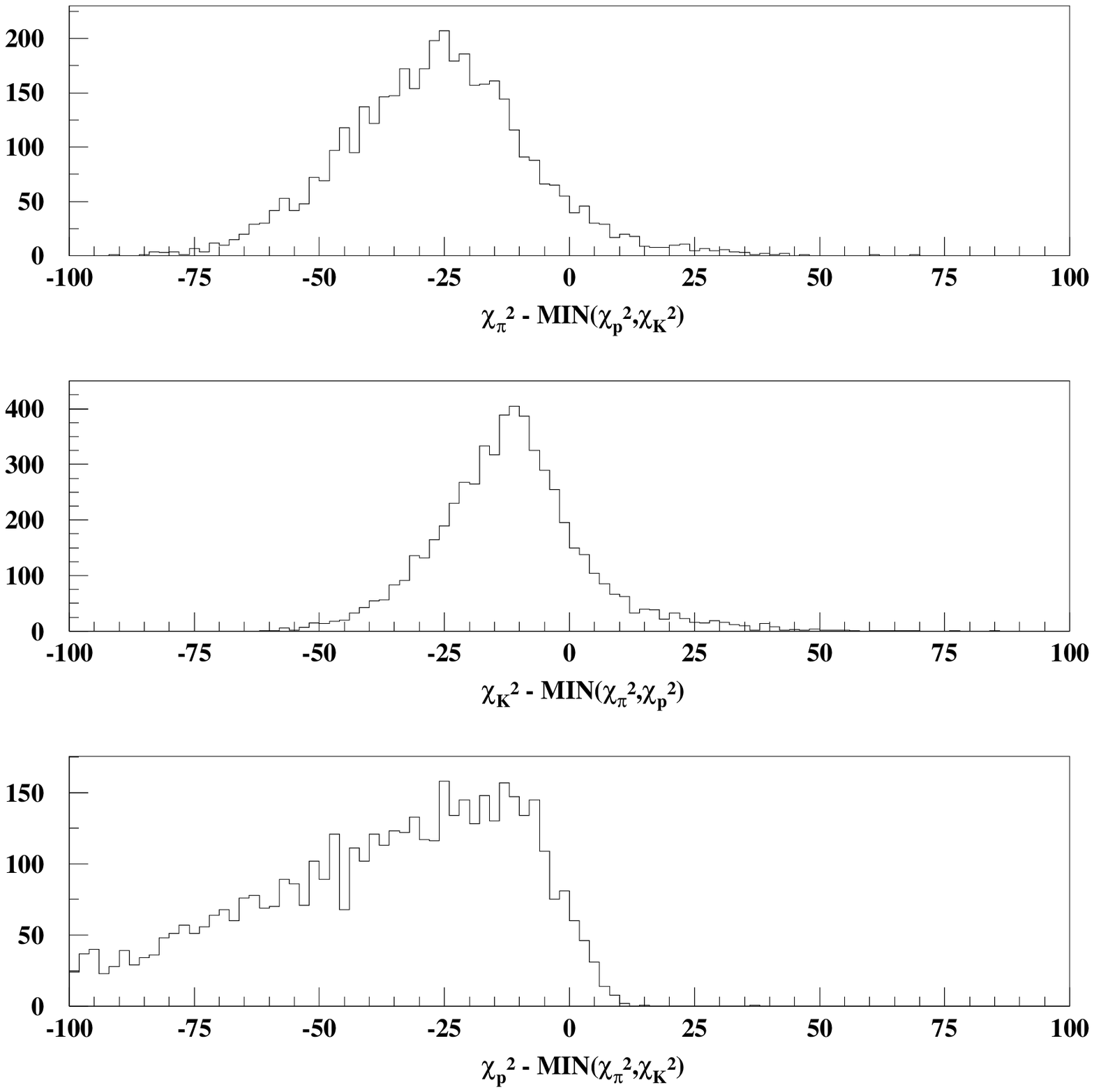,width=\textwidth}}
\caption[Particle identification performance for stopping tracks, using $dE/dx$.]{The three plots show the $\Delta\chi^2$ $dE/dx$ estimator for simulated and reconstructed charged pions(top), kaons(middle) and protons(bottom) stopping in the inner detector.  Tracks with $\Delta\chi^2 < 0$ are correctly identified.}
\label{fig:dedxPid}
\end{figure}

Specific energy loss ($dE/dx$) will be an important tool for particle identification in \minerva.  For tracks which stop in the inner detector, the charge deposited near the end of the track (corrected for sample length) can be compared with expected curves for, {\it e.g.,} the $\pi^\pm$, $K^\pm$ and proton hypotheses.  This technique does not require an independent momentum measurement, since the range ($x_{stop}$, in $\hbox{g/cm}^2$) from the stopping point to a given sampling point is closely correlated with the momentum at the sampling point.  The algorithm is calibrated by fitting the expected $dE/dx$ vs. $x_{stop}$, and the standard deviation of this quantity, $\sigma_{dE/dx}$, as a function of $x_{stop}$ for the three different particle types (see Figure~\ref{fig:dedxCalib}).  The measured $dE/dx$ for a track is compared to the expected value at each sample, to form $\chi^2$ estimators reflecting the goodness of fit to each of the three particle identification hypotheses:

$${\chi^2(\alpha) = \D \sum_{i=1}^{N_{sample}} \left[{\D \left(\frac{dE}{dx}\right)^{obs}_i - \left(\frac{dE_\alpha}{dx}\right)^{exp}_i \over\D \sigma_i^\alpha}\right]^2},$$

\noindent where the sum runs over all measured samples, and $\alpha = \{\pi,K,p\}$.  The hypothesis $\alpha$ with the minimum $\chi^2$ is assigned to the track.  The frequency of misidentification can be visualized most easily by plotting the difference $\Delta\chi^2$ between the correct $\chi^2$ (for the particle's true type) and the smallest of the two (incorrect) others (Figure~\ref{fig:dedxPid}).  With this na\"ive $dE/dx$ analysis, \minerva\ correctly identifies 85\% of stopping kaons, 90\% of stopping pions, and $>95\%$ of stopping protons.  A similar analysis can be applied to tracks with momenta measured in the magnetic regions of the detector.

\subsubsection{Energy reconstruction and containment}

The energy of muons from charged-current interactions will be measured using
range and/or curvature in the magnetized regions of the detector and the MINOS
spectrometer. For muons
stopping in the detector, the momentum resolution will
be $\frac{\Delta p}{p} \sim 5\%$. If the MINOS detector is
used, the momentum resolution will be 13\%\cite{MINOS_TRD_NDHALL}. Preliminary work on
hadronic energy reconstruction suggests that the energy of hadrons 
which rangeout in the detector will also be measured to a precision
of 5\% whereas the resolution for isolated showering hadrons 
will be $35\%/\sqrt{E(\hbox{GeV})}$. The resolution for hadronic showers
in deep-inelastic scattering will be approximately $55\%/\sqrt{E}$ 
assuming the ``ideal'' light collection and smearing effects described in section~\ref{sec:lightsim}. For
electromagnetic showers, the estimated energy resolution is $6\%/\sqrt{E}$.

\begin{figure}[thb]
\centerline{\psfig{file=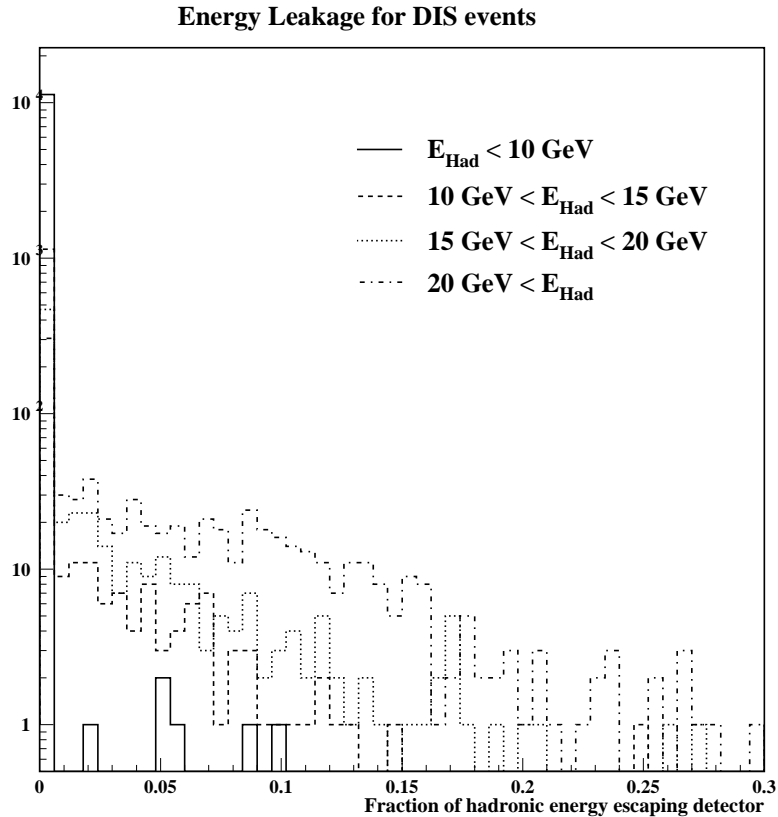,width=4in}}
\caption[Fraction of hadronic energy escaping detector] 
{Fraction of hadronic energy escaping the detector for deep-inelastic scattering
in the fully-active central region.}
\label{fig:hermeticity}
\end{figure}

\begin{figure}[thb]
\centerline{\psfig{file=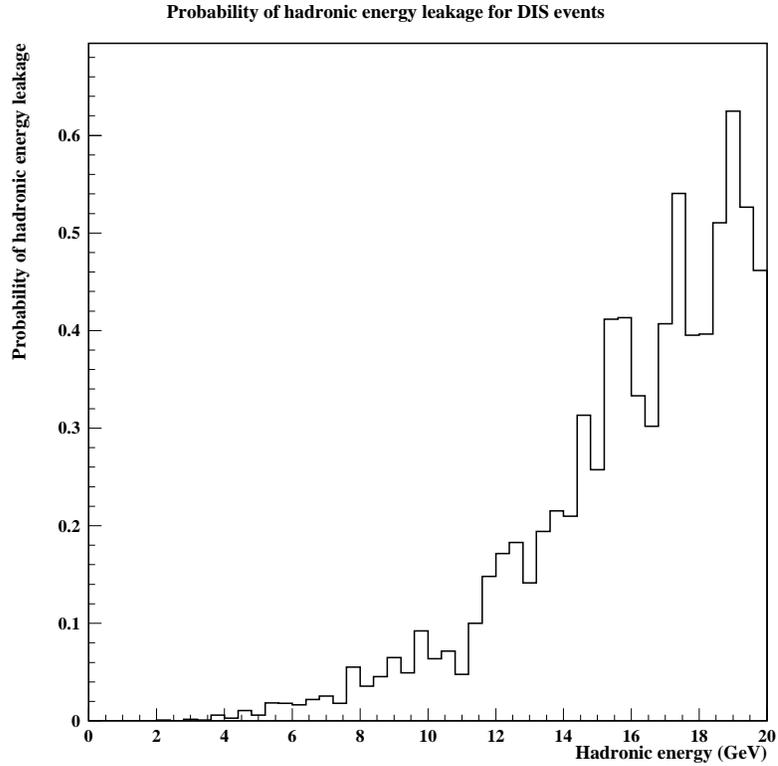,width=4in}}
\caption[Probability of visible hadronic energy leakage.]
{Probability that visible hadronic energy from a deep-inelastic event escapes undetected vs. total
hadronic energy.}
\label{fig:probleak}
\end{figure}

Containment of hadronic energy is a significant design consideration, as it assists
in meeting many of the experiment's physics goals.
Studies show that the visible hadronic component of 
quasi-elastic and resonant events in the fully-active central
region of the detector are completely contained, apart from secondary neutrinos and low-energy
neutrons. 
Figure~\ref{fig:hermeticity} shows the fraction of escaping visible hadronic energy
for deep-inelastic reactions in several hadronic energy ranges, and 
figure~\ref{fig:probleak} shows the probability that
a deep-inelastic event will leak visible energy as a function of the true hadronic energy. 
Only for hadronic energies greater than 8~GeV is
there any significant probability of leakage and only above
15~GeV is the average fraction of escaping energy 
greater than 10\%. The fraction of deep-inelastic interactions with hadronic energies
over 15~GeV in the low-energy, semi-medium or semi-high energy 
beams is $<1\%$, and so visible energy leakage should be insignificant. 
These estimates ignore downstream components beyond the forward hadron
calorimeter, such as a muon ranger and/or the MINOS detector, and are
therefore conservative.

\subsection{Event Categorisation}

Particle identification and event classification  
will play a central role in the analysis of data from \minerva\/. 
One possible method of event classification is use
of artificial neural 
network (ANN) techniques.

Event classification will be based on on 
topological characteristics as well as on particle ID.
Separation of CC from NC interactions will be based 
on muon identification. Detection of muon decays for low energy muons
stopping in the carbon gives the potential for accurate CC identification
even at high y$_{Bj}$.  
In each such class further event 
identification will be based on other particle ID, energy/momentum 
measurements and kinematics.  Neural networks are designed for such categorisation
and have been frequently used in the analysis of data from high energy physics experiments
(see, for example, the DONUT\cite{DONUT} experiment).

%% file: detector.tex
\section{Detector Design}
\label{sect:detector}

This section describes the basic elements of the \minerva\ detector,
including the arrangement of active elements and absorber,
photosensors and scintillator strip details, and the electronics.  A
summary of detector parameters along with an estimate of costs and
construction schedule are provided in Section~\ref{sect:costs}.

\input{detector-overview}

\input{photosensor}

\input{light_requirements}

\input{electronics}

\input{detector-parameters}

%% file: detector-overview.tex
\input{kevin-detector-tour}

\subsection{Overview of \minerva\ Detector Design}

For \minerva\ to meet its physics goals, detector must break new ground in the design of high-rate
neutrino experiments.  With final states as varied as high-multiplicity deep-inelastic reactions,
coherent single-$\pi^0$ production and quasi-elastic neutrino
scattering, the detector is a hybrid of a fully-active
fine-grained detector and a traditional calorimeter.

At the core of the \minerva\ design is a solid scintillator-strip detector,
similar in principle to the recently commissioned K2K
SciBar\cite{Hayato:2003}.  The plastic inner detector serves as the
primary fiducial volume, where the precise tracking, low density of
material and fine sampling ensures that some of the most difficult
measurements can be performed.  These include multiplicity counting in
deep-inelastic scattering, tracking of photons, detection of recoil protons in
low-$Q^2$ quasi-elastic events, and particle identification by $dE/dx$.

\begin{figure}[tbp]
\psfig{figure=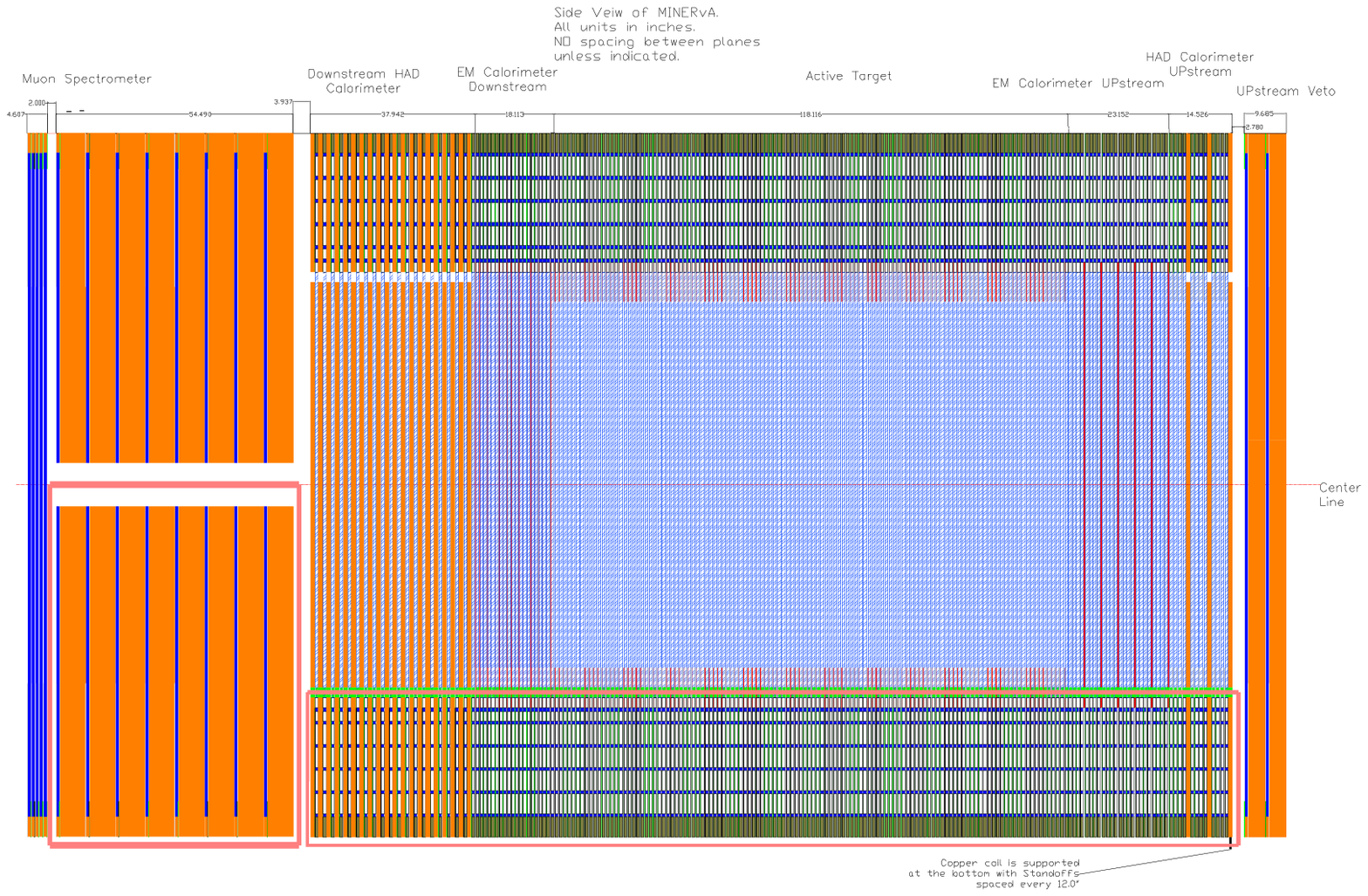,width=0.95\textheight,bbllx=60,bblly=20,bburx=522,bbury=325,angle=90}
\caption[Side view of \minerva\ detector]{A side view of the \minerva\ detector (landscape).  A schematic view of the same with labelled detectors is shown in Figure~\ref{fig:sideSchematic}.}
\label{fig:sideview}
\end{figure}
    
The scintillator detector cannot contain events due to its
low density and low $Z$, and therefore, the \minerva\ design surrounds
the scintillator fiducial volume with sampling detectors.  At the low
energies needed to study cross-sections of interest to neutrino-oscillation
studies, many of the events contain sideways-going and
backward-going particles, and therefore these sampling detectors extend to the
sides, and even to the back of the detector where they also serve as
high $A$ targets for studies of nuclear dependence in cross-sections.
Finally, it is important to contain or measure the final-state muon
in charged-current events, and for this purpose, the outer side
detector and downstream muon ranger of \minerva\ are magnetized toroids.
A side view of the complete \minerva\ design is shown in 
Figure~\ref{fig:sideview}.

\begin{figure}[tbp]
\begin{center}
\epsfxsize=\textwidth
\mbox{\epsffile{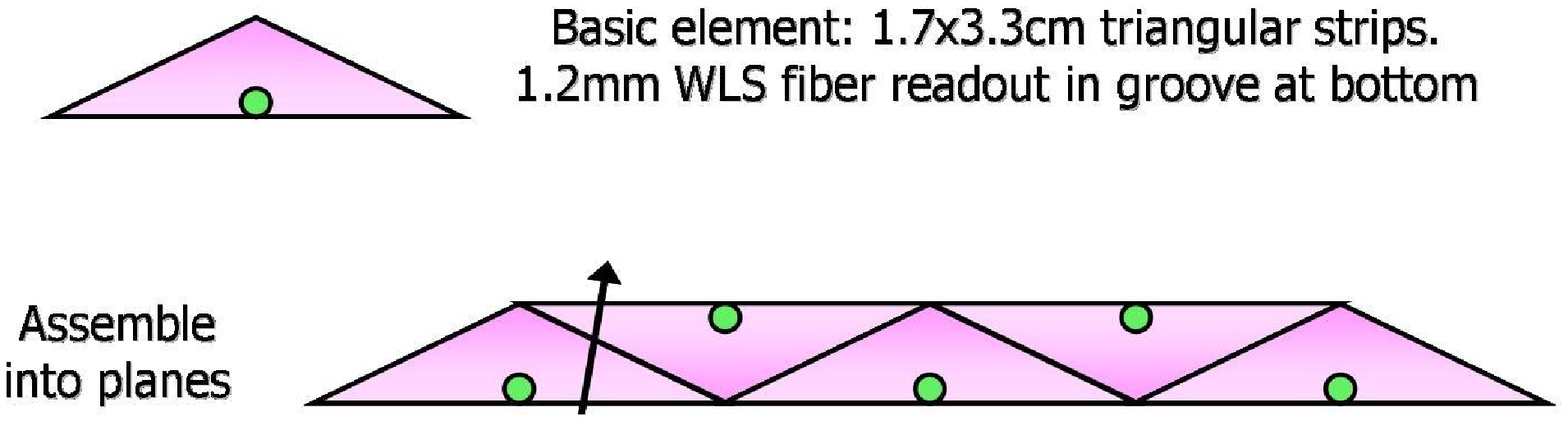}}\\
\end{center}
\caption{Assembly of scintillator strips into planes.}
\label{fig:scint-planes}
\end{figure}
    
\begin{figure}[tbp]
\begin{center}
\epsfxsize=\textwidth
\mbox{\epsffile{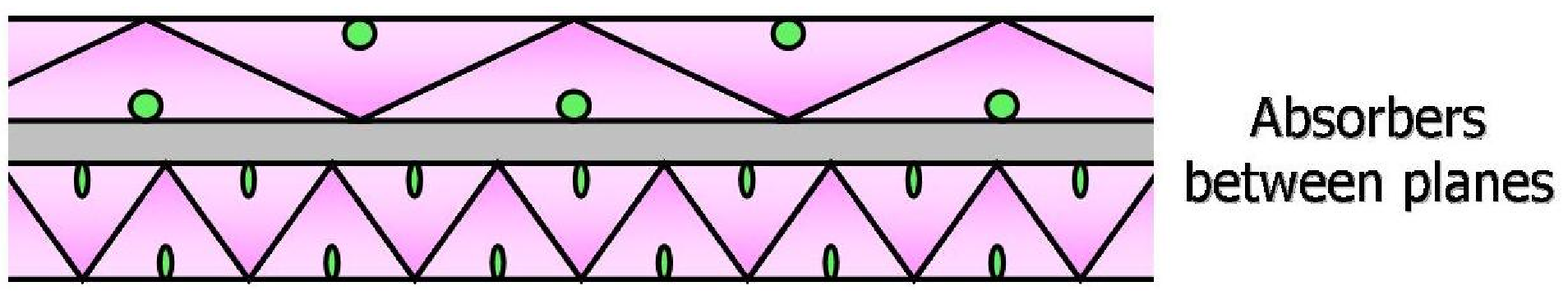}}
\vspace{2ex}\\
\epsfxsize=\textwidth
\mbox{\epsffile{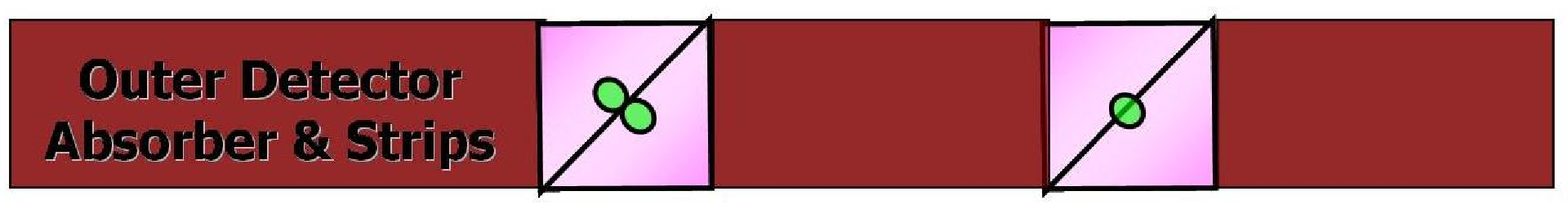}}\\
\end{center}
\caption[Planes and absorbers in inner and outer detectors]{Integration of planes with absorbers in calorimeters or
nuclear targets in the inner (above) and outer (below) detectors.}
\label{fig:planes-absorbers}
\end{figure}
    
The sensitive elements of \minerva\ are extruded
triangular scintillator strips, 1.7~cm height with a 3.3~cm base,
embedded with WLS fibers as detailed in
Section~\ref{sect:lightOutput}.  To improve coordinate resolution
while maintaining reasonably large strips, these elements are
triangular in shape and assembled into planes as shown in
Figure~\ref{fig:scint-planes}; this allows charge-sharing between neighboring strips
in a single plane to interpolate the coordinate position. Calorimetric detectors and nuclear targets
in the central region of the detector are constructed by inserting
absorber between adjacent planes as illustrated in
Figure~\ref{fig:planes-absorbers}.  In the outer detector (OD), strips
of steel absorber and scintillator are assembled in a picture frame
around the inner detector.  In the case of the triangle, the
scintillator strips are not the full size, but rather half (right)
triangles, 1.7~cm in height with a 1.65~cm base and are assembled in
doublets between steel absorber strips.

\begin{figure}[p]
\begin{center}
\mbox{
\epsfysize=0.8\textheight
\epsffile{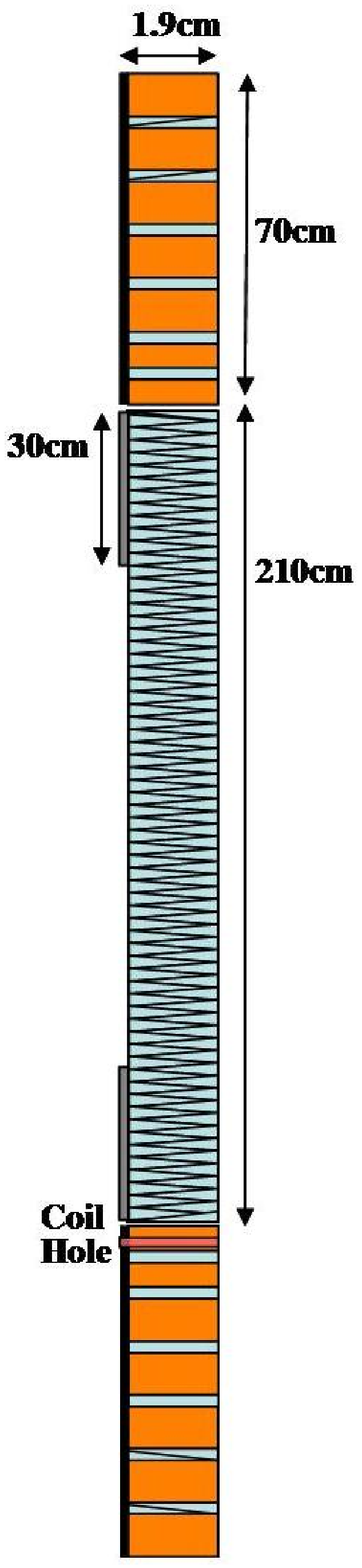}\hspace{0.2\textwidth}
\epsfysize=0.8\textheight
\epsffile{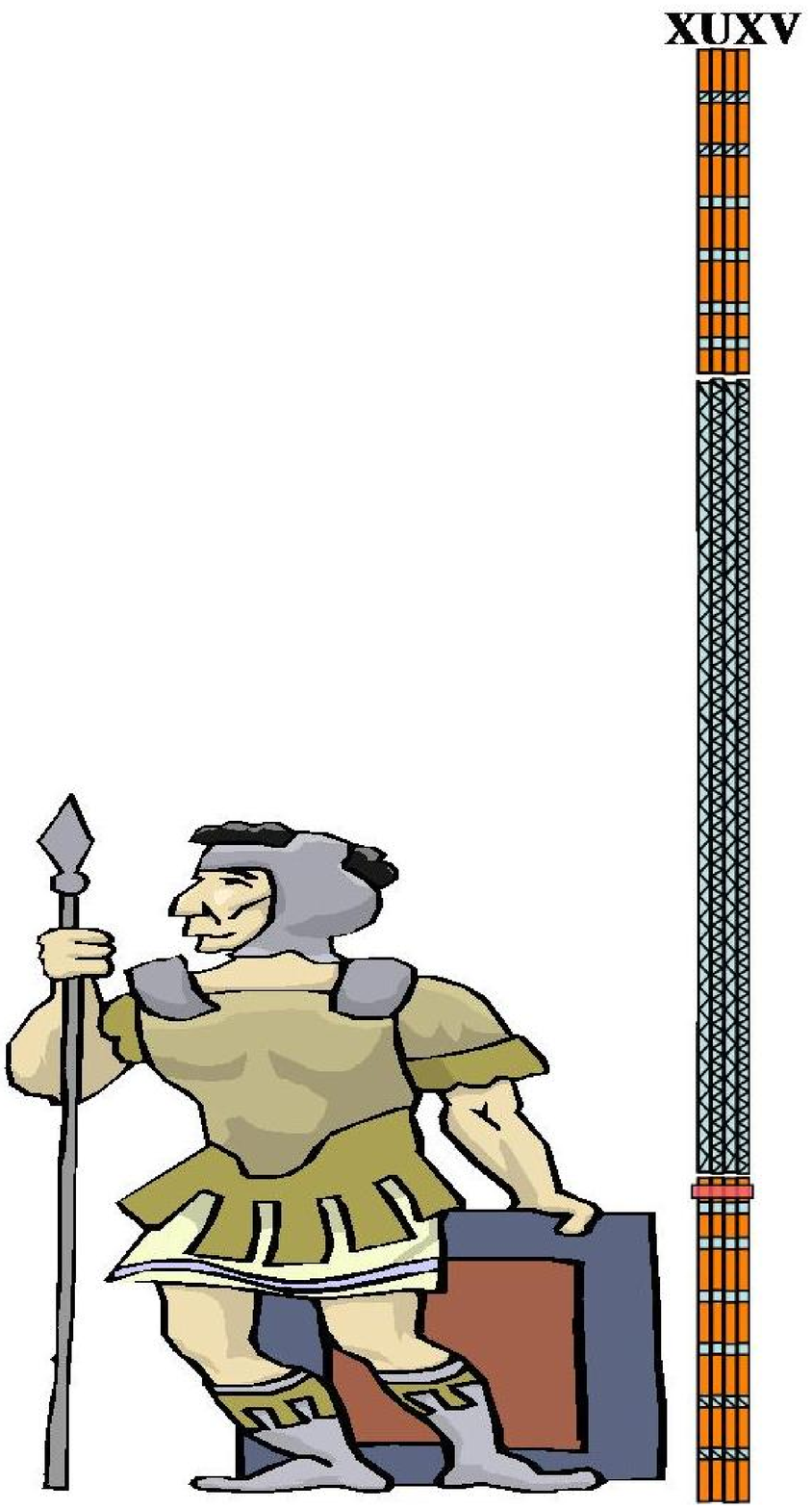}
}\\
\end{center}
\caption[Planes and modules in cross-section in the active target
region] {Plane assembly (left) and module assembly (right) in the
active target region for \minerva.  On each drawing, the scale is
exaggerated in the horizontal direction to show details.}
\label{fig:plane-module}
\end{figure}

\begin{figure}[tbp]
\begin{center}
\mbox{
\epsfxsize=0.6\textwidth
\epsfbox[20 20 574 441]{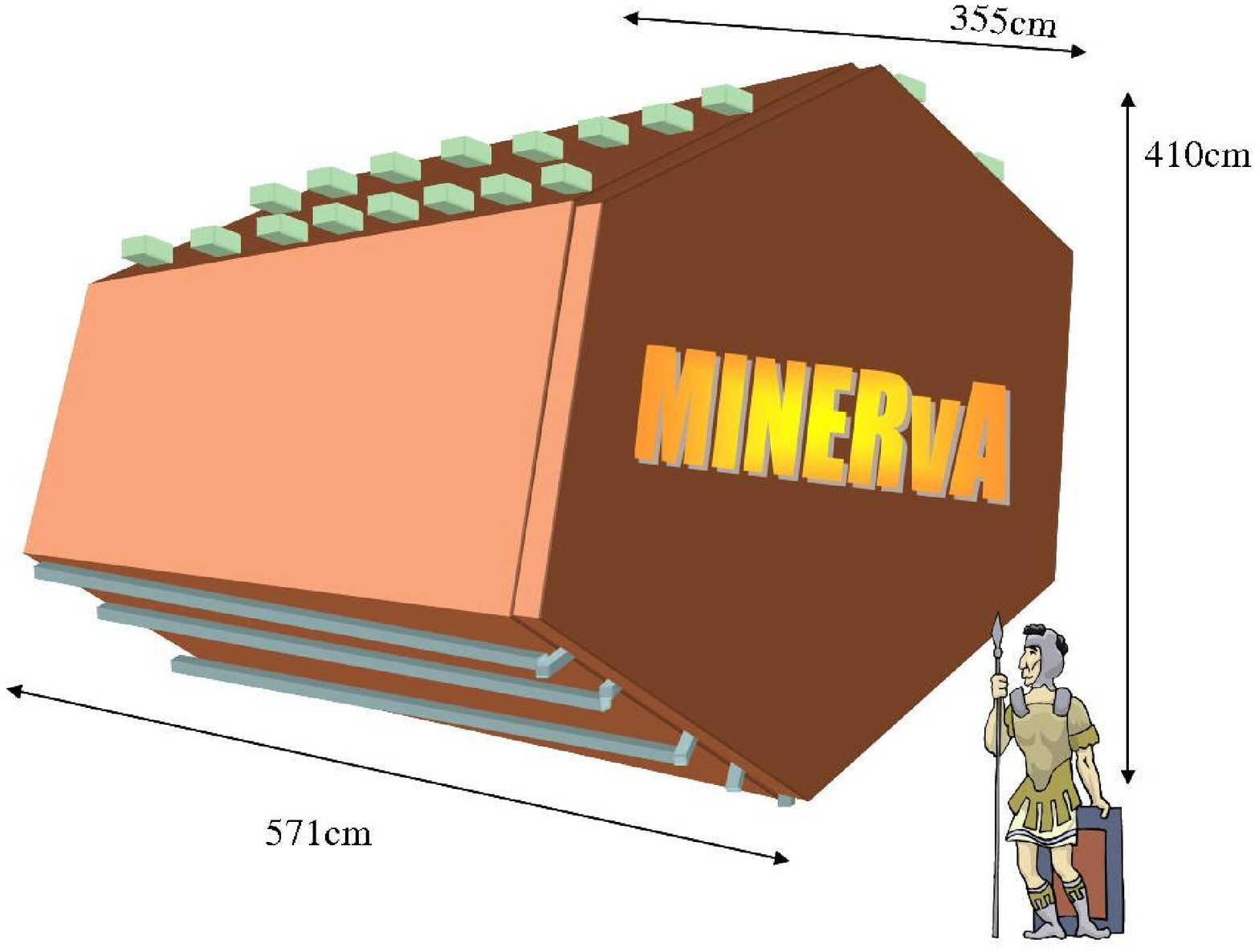}
}\\
\end{center}
\caption[Outline of \minerva\ detector]{Outline of \minerva\ detector
to illustrate shape and scale.  Note the locations of the PMT readout
boxes on top of the detector, coils on the bottom, and the support stands.}
\label{fig:minerva-cartoon}
\end{figure}

For construction and handling convenience, a single plane of \minerva,
shown in cross-section in Figure~\ref{fig:plane-module},
incorporates both the inner detector and OD ``picture frame'' as well
as an outer picture frame support structure.  Groups of four planes
(occasionally two planes only in the upstream veto and downstream muon
ranger components) are ganged together into modules, again as
illustrated in cross-section in Figure~\ref{fig:plane-module}.  There
are three distinct orientations of strips in the inner detector, muon
ranger and veto, separated by 60$^{\circ}$, and labelled X, U, V.  A
single module of \minerva\ has two X layers to seed two-dimensional
track reconstruction, and one each of the U and V layers to
reconstruct three-dimensional tracks.  The 60$^{\circ}$ offset makes
the hexagon a natural transverse cross-section for the detector, and
the size and shape of \minerva\ are illustrated in
Figure~\ref{fig:minerva-cartoon}.

\begin{figure}[p]
\begin{center}
\psfig{figure=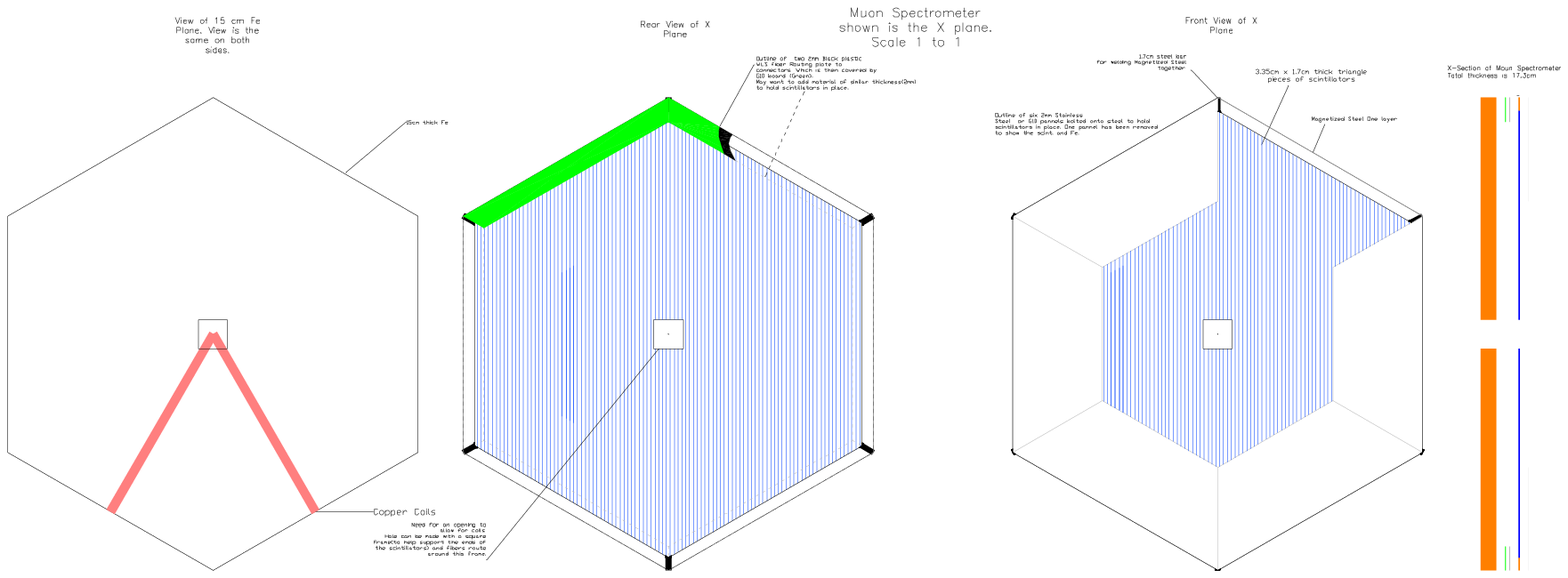,width=0.95\textheight,bbllx=20,bblly=90,bburx=560,bbury=305,angle=90}
\end{center}
\caption[Muon range/toroid and upstream veto plane design]{Muon range/toroid and upstream veto plane design (landscape).}
\label{fig:muon}
\end{figure}
    
Except for the upstream veto and downstream muon range (MR) detector,
the entire \minerva\ detector is segmented transversely into
an inner detector with planes of solid strips and an outer picture
frame magnetized toroid (OD).  In Figure~\ref{fig:sideview}, the
upsteam and downstream most detectors, the veto and muon range toroid,
respectively, are shown in Figure~\ref{fig:muon}.  As shown, the
scintillator strips extend the full length of the hexagon and range
between 205 and 400~cm in length.  The toroid steel/absorber is 15~cm
thick in the muon ranger and 10~cm thick in the veto; note also that
the final module in the muon ranger is constructed without
steel to ensure one final three dimensional spatial point free of
local multiple scattering.  The magnetic properties of the OD and the
MR detectors are discussed in Section~\ref{sect:toroids}.

\begin{figure}[p]
\begin{center}
\psfig{figure=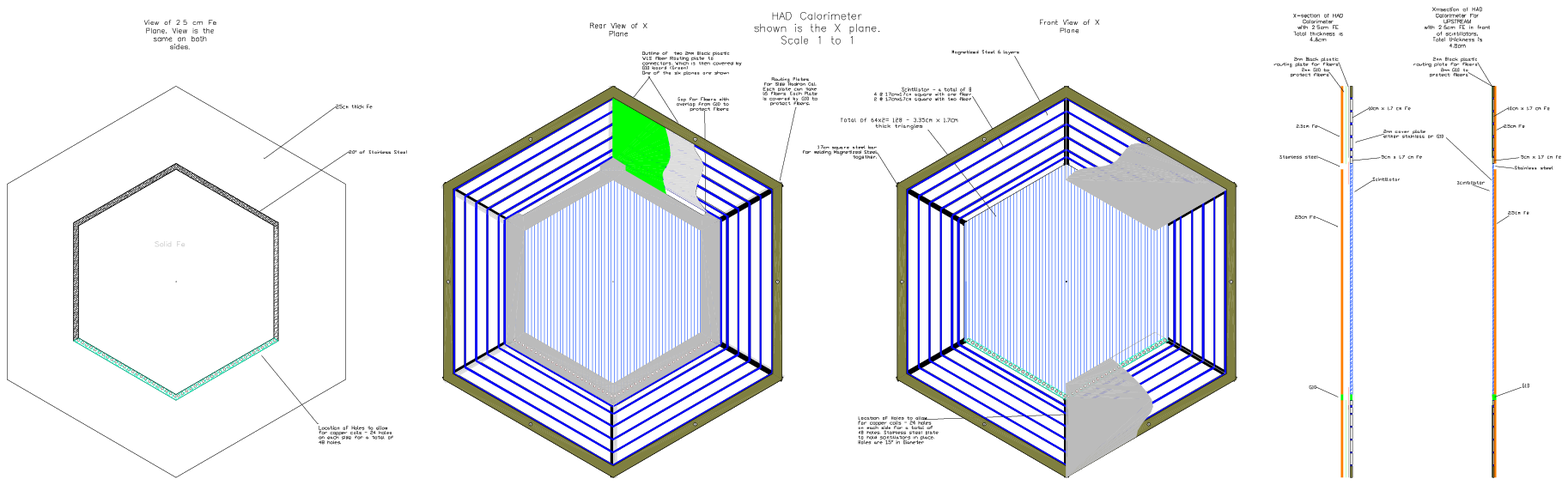,width=0.95\textheight,bbllx=20,bblly=190,bburx=560,bbury=370,angle=90}
\end{center}
\caption[The Hadronic Calorimeter plane design]{The Hadronic Calorimeter Plane Design (landscape).}
\label{fig:HCAL}
\end{figure}

Moving towards the center of the detector from each end, the next
detectors are the downstream and upstream hadronic calorimeters
(HCALs), shown in Figure~\ref{fig:HCAL}, with 2.5~cm absorbers, one
per plane downstream and one per module upstream.  This detector
is surrounded by the picture frames of absorber and scintillator
strips that make up the outer detector (OD).  Note that the
strips in the OD run only in one direction, in the bend plane of
the magnetic field.  Three-dimensional tracks must therefore be
matched from the inner detector and extrapolated outwards for an
energy measurement or muon momentum measurement.  A complication of
the design is illustrated by the fact that the inner detector strips, which
range in length from 120 to 240 cm, end inside the OD, and therefore
the WLS fibers from must be routed out to the detector edge through a
grooved plastic guide plate through the region of the OD.  Note also
the holes for the OD muon toroid coil in the lower region of the
detector.  Magnetic flux will be isolated in each region frame of the
OD, and will be prevented from leaking into the inner detector by a
guard ring of stainless steel as part of the HCAL absorber.

\begin{figure}[tbp]
\begin{center}
\psfig{figure=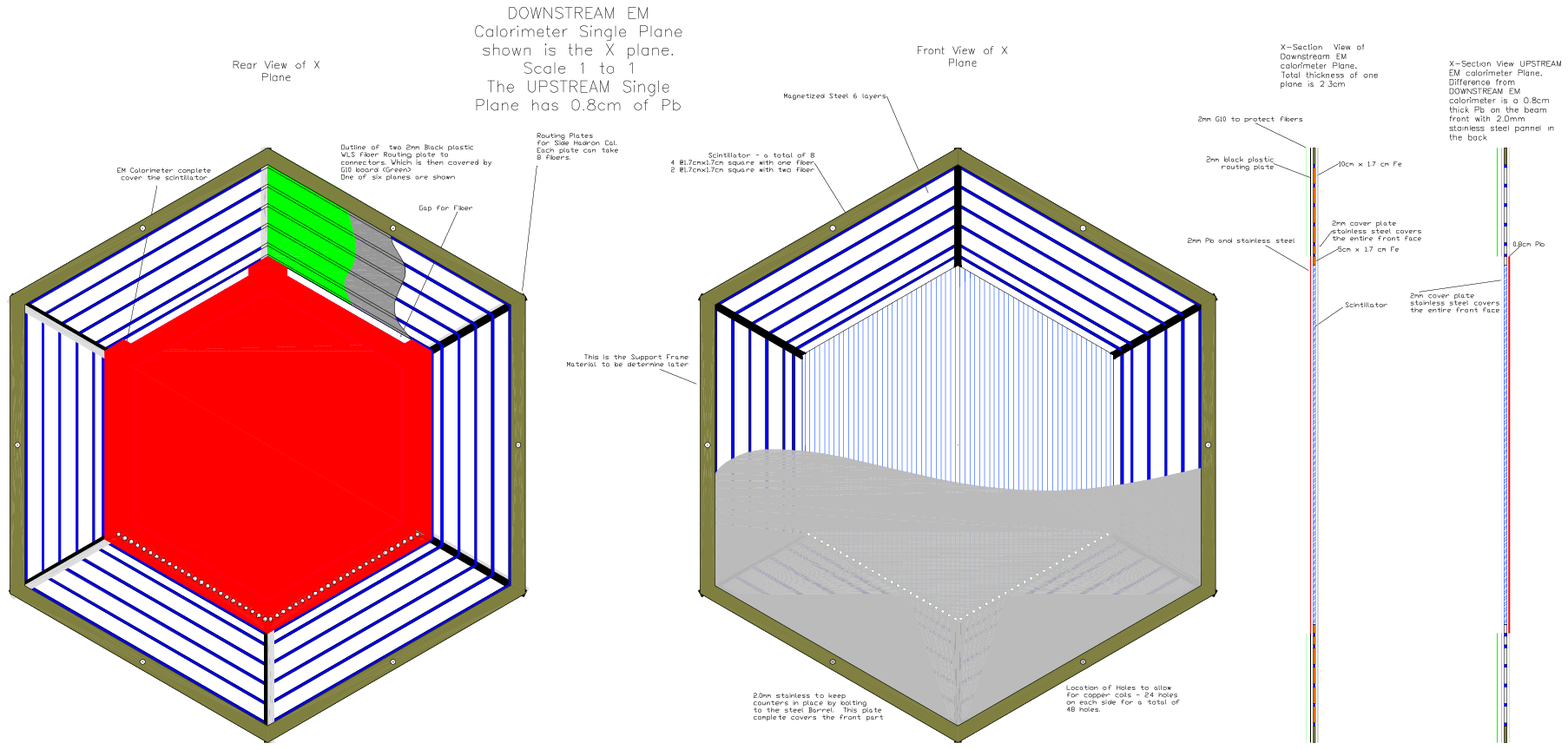,width=0.95\textheight,bbllx=20,bblly=130,bburx=560,bbury=415,angle=90}
\end{center}
\caption
[Electromagnetic calorimeter plane design]
{Electromagnetic calorimeter plane design (landscape).}
\label{fig:ECAL}
\end{figure}

Moving in again from upstream and downstream, the next detector module
elements are the electromagnetic calorimeters (ECALs), which have
0.2~cm Pb/Stainless absorbers downstream, one per plane, and 0.8~cm Pb
absorbers upstream, one per module.  Their design is shown in
Figure~\ref{fig:ECAL}.  Note that the absorber only overlaps the inner
detector and not the outer detector where it would represent a
negligible fraction of the absorber material.  The fine granularity of
the ECAL ensures excellent photon and electron energy resolution as
well as a direction measurement for each.

\begin{figure}[tbp]
\begin{center}
\psfig{figure=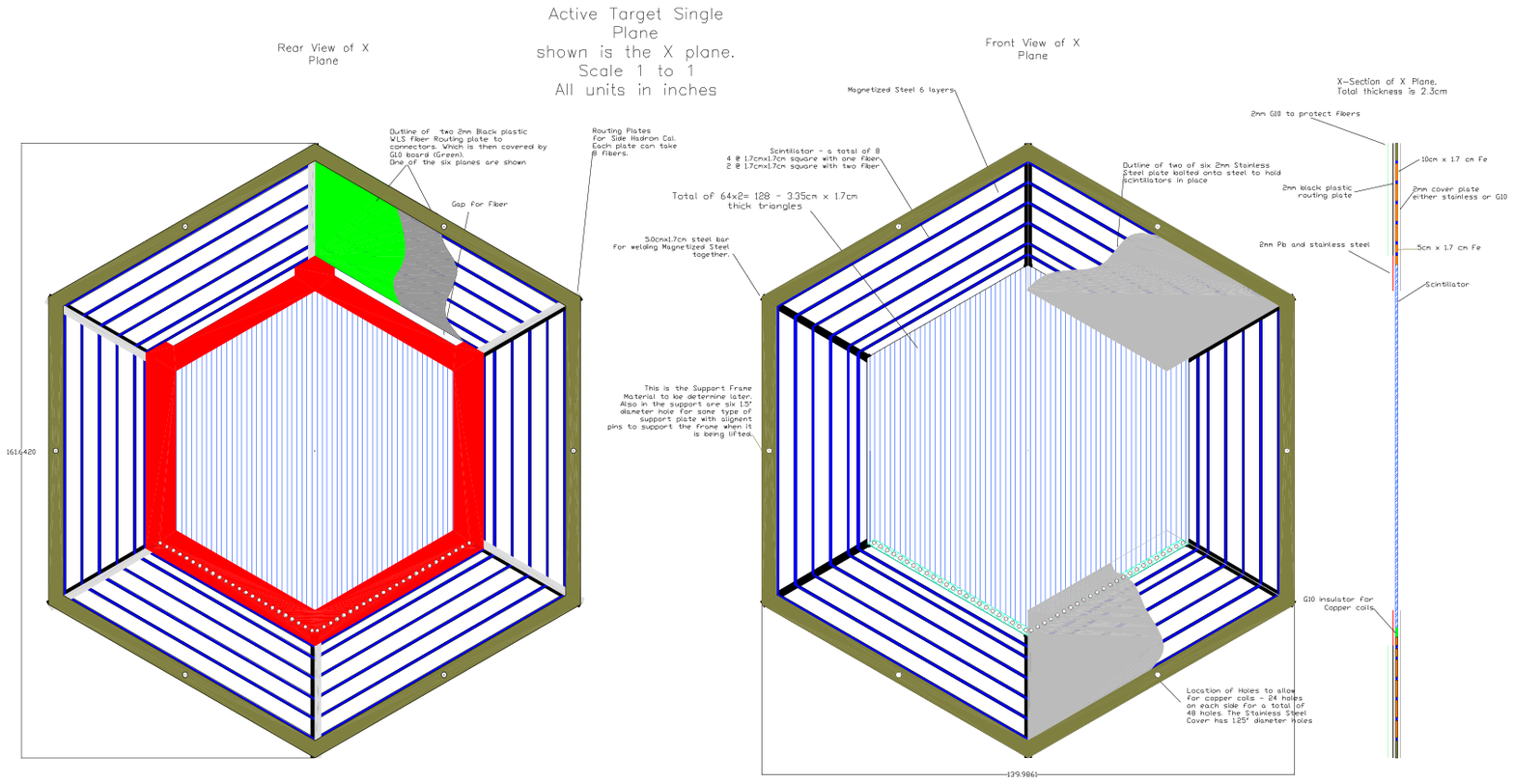,width=0.95\textheight,bbllx=20,bblly=75,bburx=560,bbury=327,angle=90}
\end{center}
\caption
[Active target plane design]
{Active target plane design (landscape).}
\label{fig:active}
\end{figure}

Finally, we reach the center of the detector, the fully-active inner
detector (ID), whose plastic core represents the fiducial volume for
most analyses in \minerva.  A plane of the active target is shown in
Figure~\ref{fig:active}.  In the center region, there is no absorber
at all; however, 30~cm from the edge of the ID, there are
lead/stainless absorbers identical in thickness to the downstream ECAL,
which act as a side electromagnetic calorimeter.  This part represents
the bulk of the detector in length, and the outer calorimeter
surrounding the fully active planes are the largest part of the
detector in mass.

Note that \minerva\ is, by design, entirely modular along the beam
direction.  Individual elements may be easily lengthened or shortened
by omitting modules from the design or adding new modules.  One
configuration that would be attractive is to forgo installation of the
muon ranger and perhaps a portion of the downstream HCAL in order to
move as close as possible to the front face of the MINOS near detector,
thus allowing MINOS to serve as a calorimeter and muon detector.  For
the purposes of cost and schedule, however, we proceed to make
estimates under the assumption that the full stand-alone detector will
be built.

\subsection{Muon Toroid Performance}
\label{sect:toroids}
The \minerva\ design calls for toroidal muon spectrometers in the
outer detector (OD) and downstream of the HCALs in the muon range (MR)
detector.  (Again, however, it
should be noted that if \minerva\ were situated immediately upstream
of MINOS, the downstream muon toroid may be omitted.)  This section
describes the momentum reconstruction and range capabilities of these
detectors for $\mu$ produced in the inner plastic fiducial volume.

The OD has a total of 50~cm of magnetized steel sampled by active planes
that are traversed by muons in a direction perpendicular to the beam.
It is magnetized by a 48~turn coil with 700~Amp current.  The average
magnitude of $H$ in the OD is therefore about 30~Gauss.  We plan to
use Armco specialty steel~\cite{MI-Armco} for the OD absorber which
would give a magnetic field of about $16$~kGauss.  For muons which
exit the side of the OD, the fractional momentum resolution measured
from the bend angle varies from 22\% to 30\% for muons with an angles
of 30$^\circ$ to 90$^\circ$ with respect to the beam.  In practice, of
course, the resolution will be better because of the loss of momentum
with $dE/dx$ in the OD.  The OD will run to focus muons forward with a
transverse momentum kick of 0.5~GeV (0.25~GeV) 30$^\circ$ (90$^\circ$)
angle.  Focusing will serve to lengthen the path length through
the OD and to direct the muons into a downstream muon range detector,
be it the \minerva\ MR or the MINOS near detector.

The downstream MR toroid has a total thickness of 1.2~m of
magnetized steel with a 48 turn coil and 1200~Amp current, resulting
again in an average field of 16~KGauss.  This yields a typical $p_T$
kick of 0.6~GeV and a momentum resolution of 20\% from the bend, which
is, again, improved by the muon's energy loss in passing through the
steel.

In summary, the \minerva\ detector has, on its own, excellent
acceptance and momentum resolution for muons.  This resolution can be
improved, especially for forward-going high-energy muons, by use of
the MINOS near detector as a downstream muon toroid.

%% file: kevin-detector-tour.tex
\begin{figure}[bhp]
\epsfxsize=\textwidth
\epsfbox{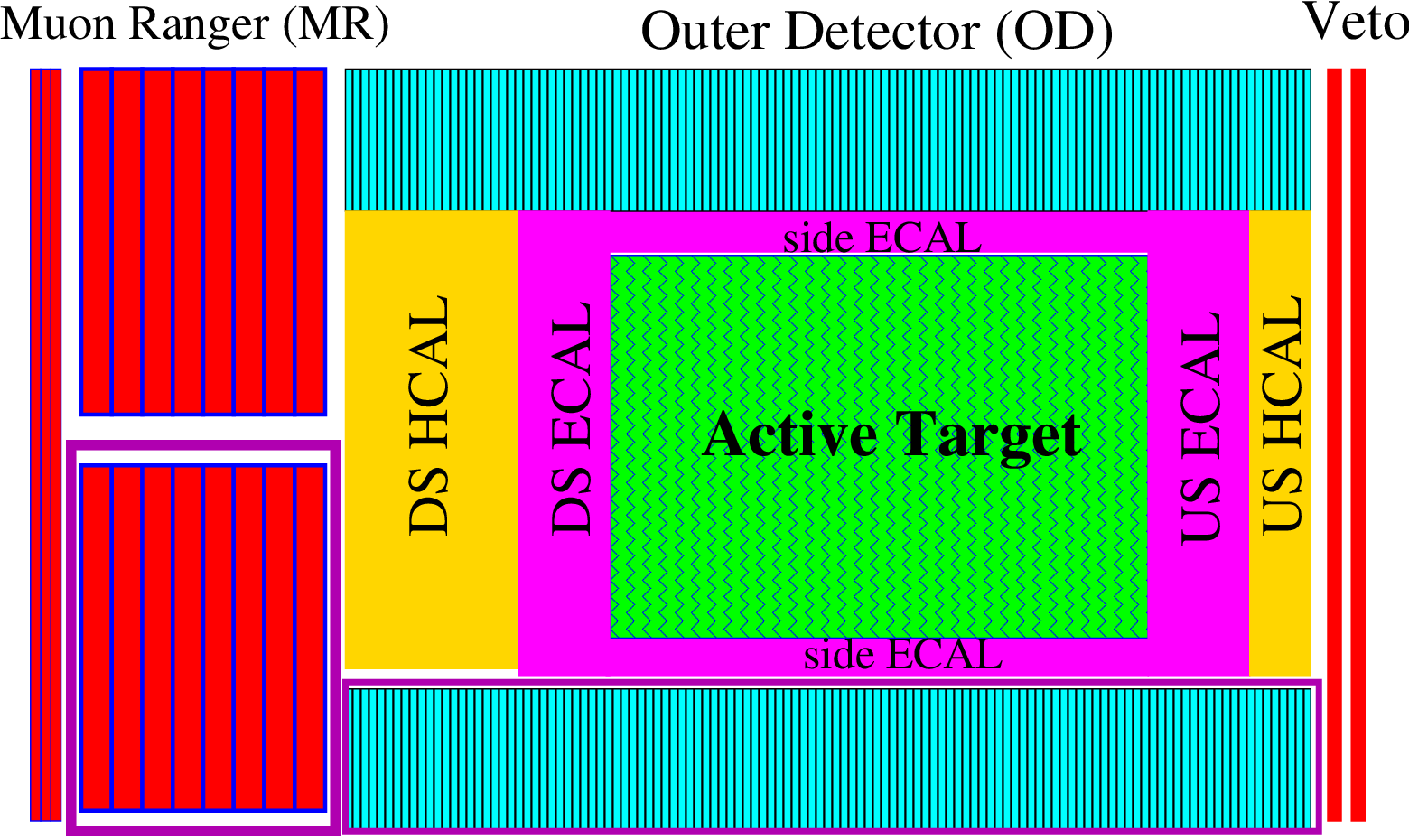}
\caption[Sub-detectors of the \minerva\ detector]{A schematic side
  view of the \minerva\ detector with sub-detectors labeled.  The
  neutrino beam enters from the right.}
\label{fig:sideSchematic}
\end{figure}
    
The \minerva\ detector is made up of a number of sub-detectors with
distinct functions in reconstructing neutrino interactions.  The
fiducial volume for most analyses is the inner ``Active Target'' shown
in Figure~\ref{fig:sideSchematic}, where all the material of the
detector is the scintillator strips themselves.  In other regions of
the detector, the strips are intermixed with absorbers.  For example,
the side, upstream (US) and downstream (DS) electromagnetic
calorimeters (ECALs) have lead foil absorbers.  Surrounding the ECALs
are the US and DS hadronic calorimeter (HCAL) where the absorbers are
steel plates.  On the side of the detector, it is the outer
detector (OD) that plays the role of the HCAL; however, note also that
the OD is a magnetized toroid which will focus and bend muons, thus
allowing a momentum measurement for muons which exit the detector.
Upstream of the detector is a veto of steel and scintillator strips to
shield \minerva\ from incoming soft particles produced upstream in the
hall.  Finally, the most downstream element, the muon range
detector/toroid (MR) gives \minerva\ the capability to fully
reconstruct even high energy muons without the use of the MINOS
near detector as an external muon spectrometer.  The presence or
absence of the MR in the final design will depend upon the location
chosen for \minerva.

%% file: photosensor.tex
\subsection{Photosensors for \minerva}

With an inexpensive active detector technology, the dominant
equipment costs for \minerva\ are photosensors and
their associated readout electronics.  The path through the
parameter space of available technologies is determined by
the answers to three questions.  First, is the light output of the
detector for a MIP signal sufficient to support a low quantum-efficiency
detector such as photomultiplier tubes (PMTs) or image
intensifier tubes (IITs)?  For \minerva\, there is
sufficient light to use a $1/6$ quantum efficiency photocathode with a
WLS fiber diameter of at least $1.2$~mm as demonstrated in
~\ref{sect:lightOutput}.  Second, is timing within the spill important
or can a technology that only integrates over a long spill, such as
IITs be used?  We concluded that timing within the spill, both to flag
overlapping events and measure time of flight and decay times at rest
was important for our physics goals.  Third, what level of technical
risk, R\&D time and cost is acceptable?  We concluded that to allow
\minerva\ to operate as early as possible in the NUMI beamline and
given the modest size of our collaboration and expected detector
costs, we should choose low technical risk over lengthy R\&D programs
designed to reduce those costs or improve performance.

In our design exercise, we considered four
technologies for photosensors: multi-anode photomultiplier tubes
(MAPMTs), IITs, avalanche photodiodes (APDs) and visible light photon
counters (VLPCs).  Ultimately, we chose to pursue a solution based on
MAPMTs which results in a sensor+electronics cost (including EDIA
and overhead but without contingency) of approximately \$40 per
channel, which breaks down approximately as \$15 per channel for the
sensor, \$15 for the electronics and \$10 for EDIA and testing.  To
defend this important decision, we discuss the alternative
technologies mentioned above.

Image intensifying tubes coupled to CCDs as a readout device are an
extremely appealing low cost solution for reading out bundles of
fibers, in part because the CCD itself is the final stage photosensor
and readout device.  This device is well-matched to the pulsed
structure of the neutrino beam with one readout corresponding to one
beam pulse.  Costs per channel are largely proportional to the total
photocathode surface required, which is set by the number of channels
and fiber diameter.  Cross-talk in adjacent channels is a non-trivial
issue, but can be addressed because of the high density of CCD
channels relative to fiber granularity, even with intermediate
spatially demagnifying stages.  We were driven to relatively expensive
CCD cameras because of the need maintain reasonable linearity.  Our
candidate system, based on Hamamatsu C8600 2-stage multi-channel
plate (MCP) intensifiers and C7190 bombardment CCDs, was approximately
\$15 per channel, including photosensor and CCD readout but not
including required demagnification optics.  Nevertheless, a complete
IIT/CCD system would still likely be half the cost of the chosen MAPMT
solution.  Our concerns about the system were the smaller effective
dynamic range, even with relatively costly IIT/CCD systems, and the
relatively low mean time to failure per device reported in other large
systems (4 years per two stages in the CHORUS experiment).  However,
the missing capability of timing within a single main injector spill
was enough for us to discard this otherwise promising
option.

Avalanche photodiodes (APDs) were also considered because
of their recent successful application in the CMS ECAL and their proposed use
in the NUMI off-axis far detector.  APDs are low gain ($\sim$100),
high quantum efficiency (85\% for Y11 WLS fibers) devices which offer
significant cost savings in the photodetector.  Complications of
operation include the need to cool the sensors below room temperature
to reduce noise, but this is a fairly easily solved problem as
cryogenic temperatures are not required.  The primary problem we
identified with APDs for \minerva\ was the need for
significant electronics R\&D to develop a relatively low-cost system
capable of controlling noise over the long NUMI spill.  For \minerva\
we set a requirement of keeping the photosensor and electronics 
noise well below $10$ delivered photon equivalents to maintain good
sensitivity to a MIP (typically $70$ photons in a doublet of
triangular scintillators) and a low rate of detector noise.   Over a 12
$\mu$sec gate (the NUMI spill plus $2\tau_\mu$) at -10$^{\circ}$C with an
operating gain of $100$ (optimal), the signal from $10$ photons is
$850$ electrons and the noise on the best existing candidate
electronics, the MASDA chip, is $900$ electrons.  To achieve the
better signal to noise that is the goal of the proposed NuMI off-axis
R\&D program requires
design of a new ASIC, which would imply at least a one-year development project.
In short, although the APD is a potentially
promising technology, we were not convinced it could be in production
on the timescale required for \minerva.

The final option we considered was the VLPC.  These have the advantage
of successful past deployment and electronics design in the D0 fiber
tracker and preshower detectors.  However, the costs for just the
VLPCs themselves, even with optimistic assumptions about the outcome
of future R\&D, are expected to exceed \$50 per channel, and are thus
significantly more expensive than the MAPMT solution.  Given that the
low quantum-efficiency solution gives sufficient resolution, it is
difficult therefore to justify VLPCs.

The MAPMT we have tentatively selected as our default photosensor is
the Hamamatsu R7600U-00-M64.  These are an incremental design
improvement from the R5900-00-M64 MAPMTs used in the MINOS near
detector, and we expect much of the experience gained by the MINOS
collaboration with these detectors to be applicable.  In particular,
we have confidence in costing the testing, housing for and optical
connectors to the PMTs because of our ability to scale costs from the
MINOS experience.

Having chosen MAPMTs, a low quantum-efficiency device with good
timing, low noise and a large dynamic range, the following two
sections address the issues of the photoelectron yield from the strips
married to the MAPMTs and the electronics to readout these MAPMTs,
respectively.

%% file: light_requirements.tex
\subsection{Scintillator Strips}
\label{sect:lightOutput}

The MINOS experiment has successfully demonstrated that co-extruded
solid scintillator with embedded wavelength shifting fibers and PMT
readout produces adequate light for MIP tracking and can be
manufactured with excellent quality control and uniformity in an
industrial setting.  The performance characteristics of the MINOS
scintillator modules produced at the three `module factories' are now
well known, both through measurements taken with radioactive sources
post-fabrication at the factories and through measurements of cosmic
rays at Soudan.  We intend to use the same technology for the active
elements of \minerva.  

The basic active element in \minerva\ is a co-extruded
triangular scintillator strip with a wavelength-shifting fiber glued
into a groove.  Like MINOS, the scintillator strips are polystyrene
(Dow 663) doped with PPO (1\% by weight) and POPOP (0.03\% by
weight), co-extruded with a reflective coating of TiO$_2$ loaded
polystyrene\cite{Adamson:2002mj}.  The strip cross-sections have
width 3.35 cm and height 1.7~cm.  Strip lengths vary throughout the
detector and range from 1.4~meters to 2.2~meters in the inner tracking
detector to 4~meters for the veto and muon ranger sections.  The WLS fiber
(Kurrary Y11, 175ppm dopant) is 1.2~mm in diameter, glued into an
extruded groove and covered with aluminized mylar tape in the same
fashion as MINOS.  The WLS fibers are brought to optical connectors at
the edge of the modules, and clear optical cables bring the light to a
PMT box.  Single-ended readout is used, and the far strip/fiber ends
are mirrored.

Physics simulation studies indicate that for a triangular extrusion,
average~\footnote{Note that this is an not only an average over
photostatistical fluctuations, but also an average over all locations
for normally incident tracks to enter the strip.  The average light
through the full thickness of scintillator in a plane, a doublet of
triangles, is twice this average.}  light levels above $3.9$
photo-electrons(``PE'')/MeV of $dE/dx$ for a minimum-ionizing particle (MIP)
are required in the
inner detector in order to obtain good particle identification as
described in Section~\ref{sect:PID}.  Coordinate resolution, vertex
finding, and track pointing are also affected by light levels, but to
a lesser extent.  For this design we have targeted an average light
level of $7.8$ PE/MIP on average through the strips.  This allows for
losses expected to be 25\% in the clear fiber and connectors and
possible effects from the degradation of the scintillator over time,
the latter of which was measured to be as large as $20\%$ over $10$
years for MINOS\cite{Border:vk}.

\begin{figure}[tb]
\center
\epsfxsize=0.65\textwidth\leavevmode
\epsffile{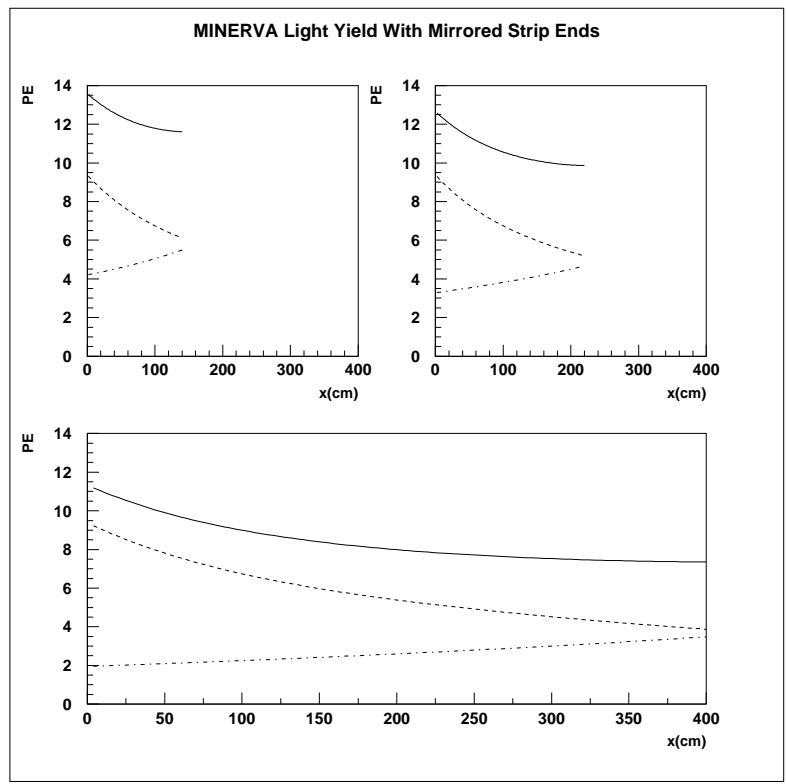}
\caption[Light yield vs. distance along strip]
{Light yield vs. distance along strip for \minerva\ scintillator
strips with one-ended readout and a mirrored end.  Dot-dashed line 
is light collected from reflections off the mirrored end, dashed line 
is light travelling directly to the readout end; solid line is 
the sum. The top plots correspond to the shortest and longest strips
used in the fully-active inner detector, and the bottom plot is for
the longest strips in the veto and muon ranger.}
\label{fig:lmirrors_pe}
\end{figure}

The overall light levels from 3 lengths of strips, as calculated using
the photon transport Monte Carlo described in
Section~\ref{sec:lightsim}, are shown in Figure~\ref{fig:lmirrors_pe}.
Here we have assumed a 90\% reflectivity from the mirror end of the
fiber, and in all cases a 1~meter WLS `pigtail' from the end of the
near end of the strip to the PMT face.  Clear fiber lengths and
connectors are not included.  In the MINOS near detector, the far
strip end was not mirrored; here we assume the strip ends are mirrored
with 100\% reflectivity.  Because the light produced in the
scintillator is generally collected within a few~cm of the MIP
crossing location, this approximation only affects the calculation of
collection efficiency at the very far end of the strip.  Shown are the
light levels predicted for three strip lengths.  In each plot, the lowest
curve corresponds to light collected from reflections off the mirrored
end, the middle line corresponds to light travelling directly from the
MIP to the readout end, and the upper line is the sum.  As the figure
shows, the light level in the inner tracking detector, with a maximum
length of 2.2~m, exceeds the design requirement of 7.8~PE/MIP over the
entire length by about 25\%.  In the longer strips (only used in the
downstream muon range detector and upstream veto counters), the light
falls slightly short of this target at the far ends of the strips;
however, because these detectors are not used for particle
identification by $dE/dx$, this is still acceptable.   

%% file: electronics.tex
\subsection{Electronics}
\label{sect:electronics}

\begin{table}[bp]
\begin{center}
\begin{tabular}{|l|l|l|}
\noalign{\vspace{-8pt}} \hline
Parameter                & Value                & Comments\\ \hline
Active Spill Width       & 12$\mu$sec          & Spill plus $2\tau_\mu$  \\  
Repetition Time          & $>$ 1.9~sec        &         \\ 
Number of Channels       & 37478           &         \\ 
Occupancy per Spill      & 0.02            & LE beam, 2.5E13 POT/spill \\
Front-end noise RMS      & $<$ 1 PE            &         \\
Photodetector gain variation & 4.5 dB & extremes of pixel-to-pixel variation \\
Minimum Saturation       & 500 PE    &  proton range-out or DIS event \\ 
Maximum Guaranteed Charge/PE  & 50 fC & lowest possible charge at highest gain \\
Time Resolution          &  3ns            & Identify backwards tracks by TOF   \\
                         &                 & Identify decay-at-rest $K^\pm$ \\  \hline

\end{tabular}
\end{center}
\caption{Electronics design requirements and parameters for \minerva}
\label{tab:requirements}
\end{table}

The requirements for the \minerva\ electronics are summarized in
Table~\ref{tab:requirements}.  To minimize technical risks, we studied
a number of existing solutions, including those used for the MINOS
design.  Major components of the electronics system include the
front-end boards, the PMT and electronics housing and the slow
controls and readout systems.

\subsubsection{Front-end boards}

For the front-end digitization and timing, the best performing system
with the least required R\&D is a scheme based on D0 TRiP ASIC.  The
TRiP chip is a redesign of the readout ASIC for the D0 fiber tracker
and preshower originally motivated by the need to run at 132~ns bunch
crossings in the TeVatron.  A production run of one version of TRiP
with 7000 produced chips has been completed; however, since the
TeVatron run plans do not now call for 132~ns operation, these chips
will not see their original use.  These existing chips, however, could
be recycled into use in \minerva.

\begin{figure}[tbp]
\begin{center}
\epsfxsize=\textwidth
\mbox{\epsffile{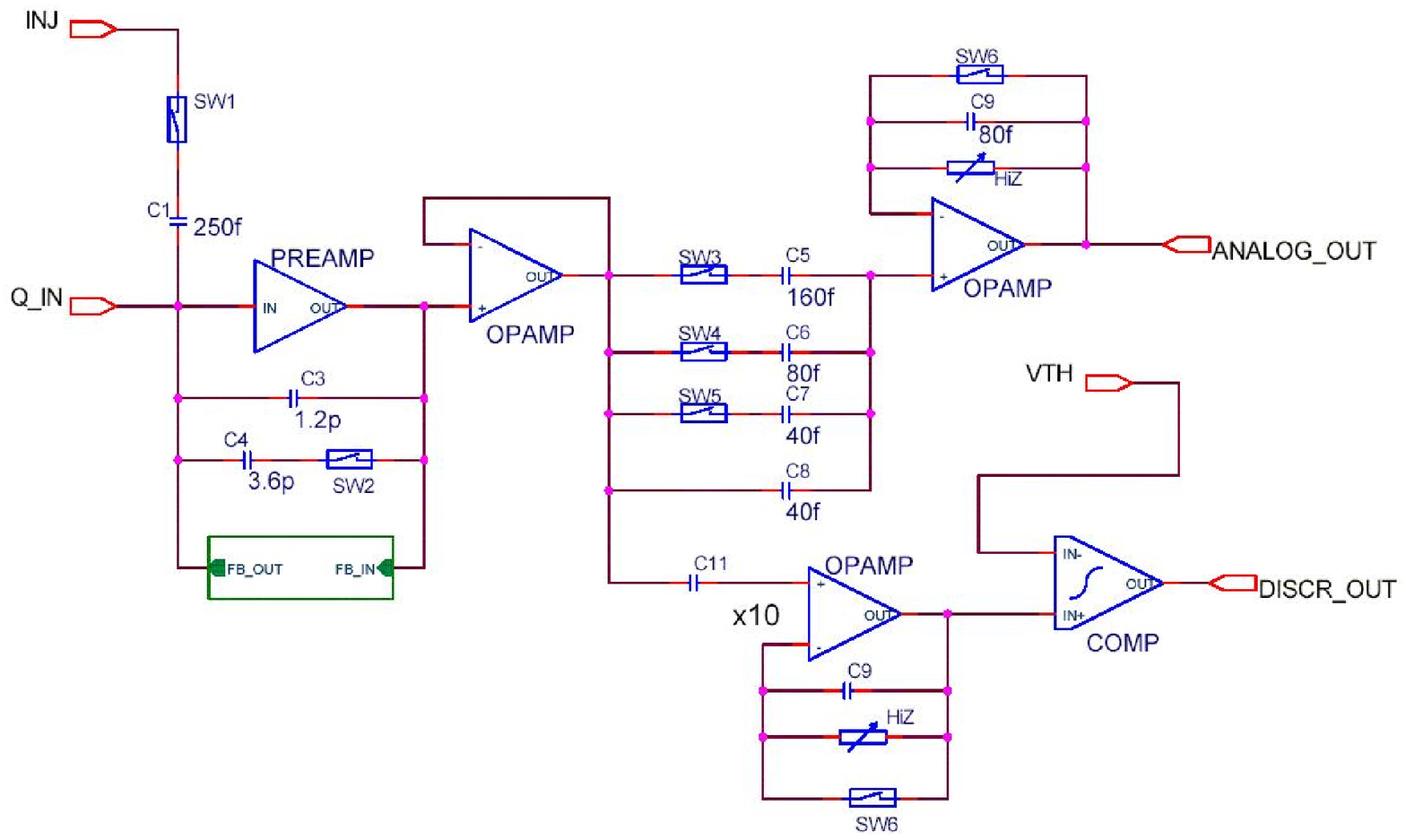}}\\
\end{center}
\caption{ A simplified schematic of the front end electronics of the TRiP chip}
\label{front_end}
\end{figure}
   
\begin{figure}[tbp]
\begin{center}
\epsfxsize=0.5\textwidth
\mbox{\epsffile{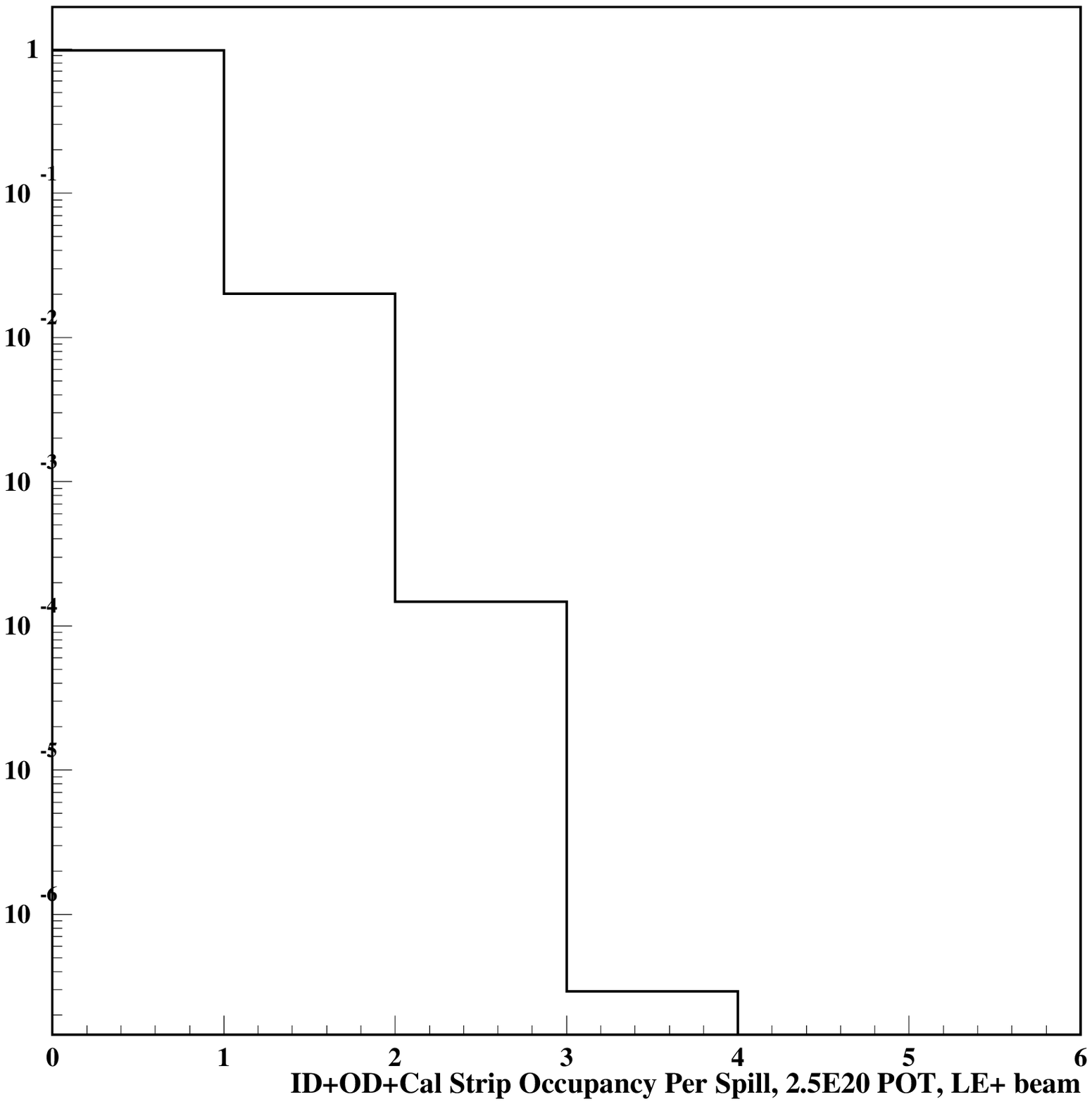}}\\
\end{center}
\caption[\minerva\ strip occupancy]
{The mean number of neutrino interactions in which a strip is 
hit per 2.5E13 POT spill, NUMI LE neutrino beam}
\label{fig:occupancy}
\end{figure}
   
TRiP was designed by Abder Mekkaoui of the Fermilab ASIC group and
successfully met the specs in its first submission and has undergone
extensive testing by D0~\cite{Rubinov:TRiP}.  Its analog readout is
based on the SVX4 chip design, and each TRiP chip supports 32 channels
for digitization, but only half that number of channels for timing.  A
simplified schematic of the TRiP ASIC is shown in
Figure~\ref{front_end}.  The preamp gain is controlled by SW2 and has
two settings which differ by a factor of four.  The gain of the second
amplifier stage is controlled by SW3-SW5.  We will set the chip to the
lowest gain setting for the preamp and largest integration
capacitor. This gives a linear range with a maximum charge readout of
5 pC. The ``ANALOG\_OUT'' goes into a analog pipeline, which is
identical to the one used on the SVX4 chip and 48 cells deep.  The
SVX4 chip can read out four of these 48 buffers, and although the TRiP
chip was also designed to read out four buffers, it can empirically
only read out one buffer.  It is not known why only one buffer can be
read out; however, this is not an issue in 
\minerva\, as shown by the per channel per spill occupancy
illustrated in Figure~\ref{fig:occupancy}.
To gain dynamic range, \minerva\ will increase the input
range of the electronics by using a passive divider to divide charge
among two TRiP channels with a ratio of a factor of 10.  This ``high
range'' channel, then, will give a equivalent total readout charge of
50 pC.  Each TRiP channel will be digitized by a 12 bit ADC.

\begin{figure}[tbp]
 \centerline{\psfig{figure=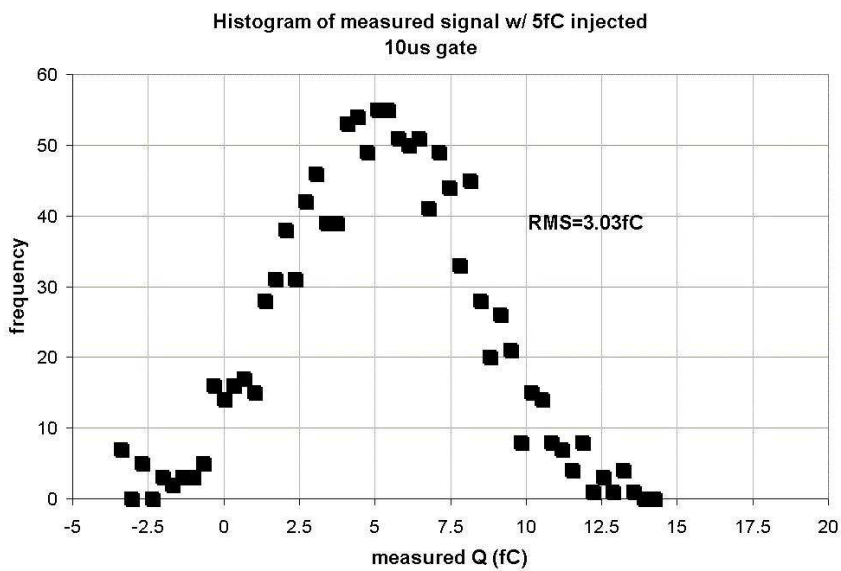,height=3.0in,angle=0}}
\caption{Response of the TRiP chip to 5~fC injected with a 10~$\mu$s gate.}
\label{noise_10mu_36pf}
\end{figure}

In \minerva\ the integration time for the ADC will be 10--12~$\mu$s,
much less than the hold time for the charge in the capacitor of
100~$\mu$s. The TRiP chip has been tested explicitly with a 10~$\mu$s gate,
and Figure~\ref{noise_10mu_36pf} shows pedestal RMS of 3~fC for the
estimated \minerva\ input capacitance of 36~pF.
%
The \minerva\ design requires no saturation below 500 photoelectron (PE)
and RMS noise well below 1 PE.  Matching this to the 5~pC charge
limit, the highest gain anodes in a tube would be set at 100~fC/PE and
therefore the lowest gain anodes would be run at 33 fC/PE.  Since the 
RMS of the noise is about 3.0~fC, this will put
a single photoelectron approximately a factor of 10 above the pedestal RMS,
well within our design spec.  The maximum PMT gain for the lowest gain
anode will be 50~fC/PE, safely within the desired parameters above.

\begin{figure}[tbp]
\begin{center}
\mbox{\epsfxsize=0.46\textwidth
\epsffile[60 0 530 520]{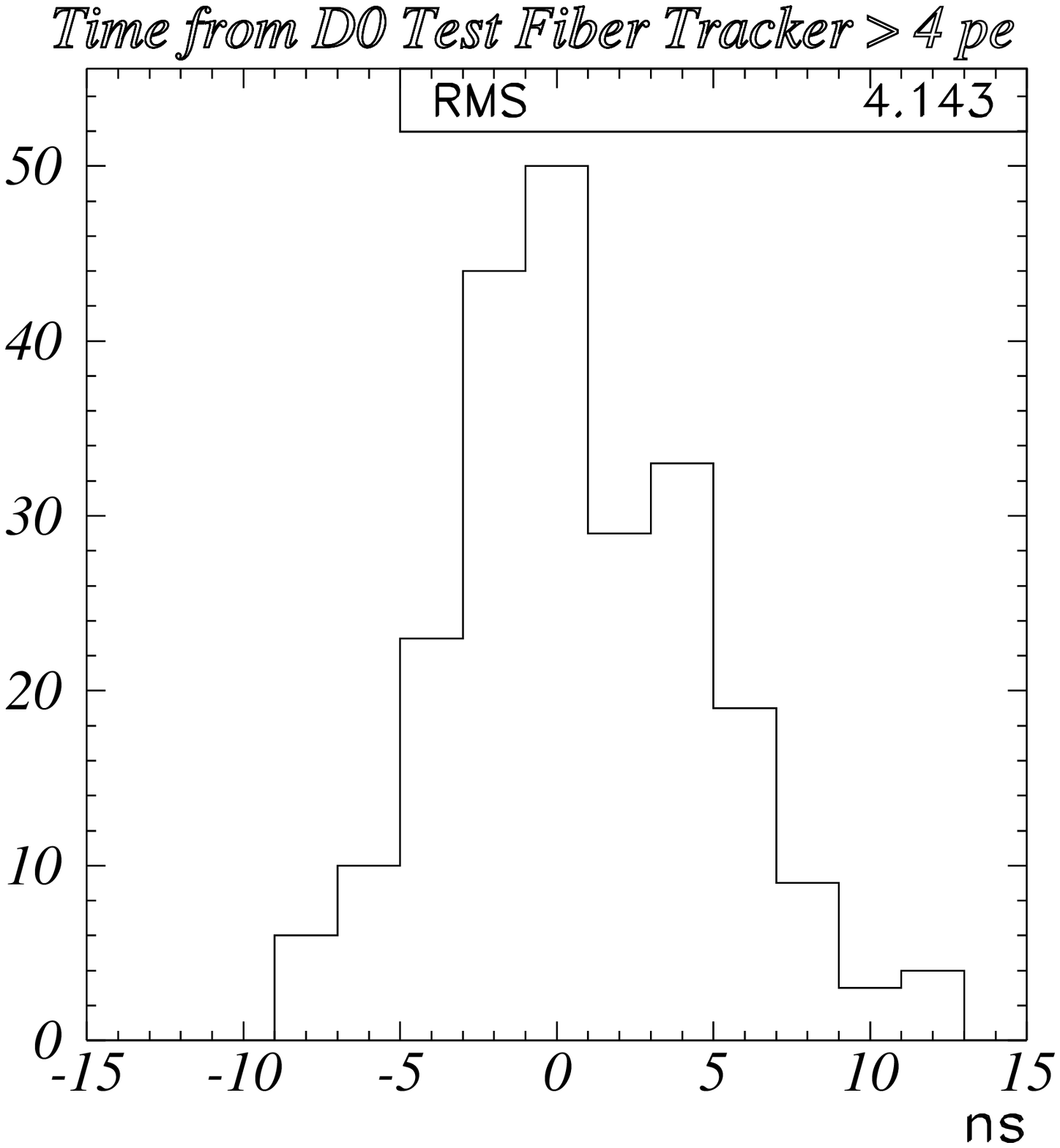}
\hspace{0.05\textwidth}\epsfxsize=0.43\textwidth
\epsffile[60 0 520 530]{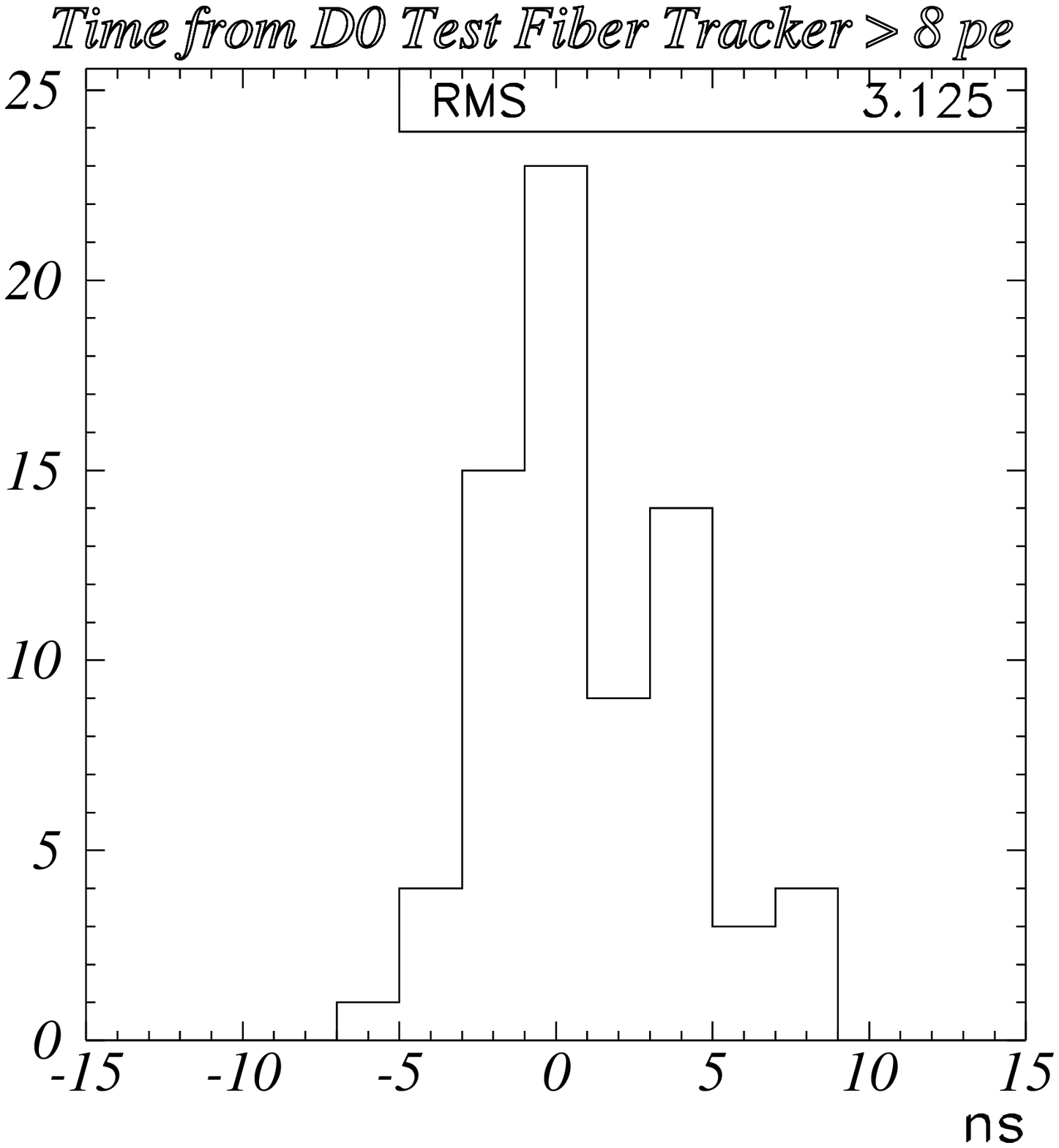}}\\
\caption[Timing resolution of the TRiP chip]{Results of the D0 test of timing of TRiP using the Test Fiber 
Tracker with signals $>$ 4pe and signals $>$ 8pe}
\label{timing_hist}
\end{center}
\end{figure}

Only one of every two input channels to the TRiP chip have a latched
discriminator output (latch) which can be used for timing information.
Hence, only the lower range channels will feed the latch whose output
will then go into an FPGA.  The internal clock of the FPGA can be used
to get timing with a granularity of 5~ns, and with a delay line scheme
this can be improved to below 3~ns.  The reset for the latch is only
15~ns, so inside the spill the latch will be in the ready state by
default.  When the signal exceeds a threshold of 1.5~PE, the latch
will fire.  After storing the time, the latch is reset, incurring
minimal deadtime.  Figure~\ref{timing_hist} shows result of the D0
timing test of the TRiP chip using their fiber tracker and VLPCs.
They get a timing RMS of 3ns for signals with $\ge$8~PE.  Y-11 (the
waveshifter in the \minerva\ fibers) has an equivalent decay time to
3HF (the dye used in D0's fiber tracker), and hence in a doublet of
triangles in the scintillator, we can reasonable expect similar timing
resolution.

Note in the \minerva\ scheme, there is no trigger.  Two charges are
read from all channels along with all latched times at the end of each
spill.

Although individual parts of this system have been tested by D0, the
system described above has not been tested.  Ray Yarema's group at
FNAL has begun layout of a board for a vertical slice test of this
system, and we expect a proof of principle from this test by early
summer 2004.

\subsubsection{PMT boxes, readout electronics and controls}

The MAPMTs will be mounted directly on the front-end boards to
reduce input capacitance.  Therefore, the front-end board and PMT need
to reside in a single light-tight box with optical cables from the
detector as input.  In our preliminary design, each PMT box will
have a single PMT and front-end electronics for its
64 channels, along with a Cockcroft-Walton HV generator.

In addition to the optical input cable, each PMT box has three
electrical cables.  The slow-control cable and the low-voltage cable
will travel from box-to-box in a daisy chain.  The digitial readout
LDVS cables will arranged in 16-box 'Token Rings', and will connect to
a VME card at the first and last box in the ring.  In addition to
reading out the data after each spill, the token ring will supply the
timing synchronization signal.  The low-voltage cable will likely run
at an intermediate DC voltage and step down at each box to minimize
the role of resisitive losses in the chain.  The slow-control cable
will be a MIL 1553b bus which is in wide use at FNAL.  An existing VME
card (the 1553 controller) will drive the slow control system.
Table~\ref{num_parts} shows the number of parts needed
for the complete electronics system.

To minimize the length of the clear fiber cables from the WLS
fibers to the PMT boxes, we plan to mount the PMT boxes directly on
the upper parts of the \minerva\ detector, on the two highest sides of
the hexagon to avoid conflict with the coils or side clearance of the
detector.  This will require magnetic shielding of the MAPMTs.

The DAQ requirements for this system are trivial as the data rate is 
expected to be under $100$~kByte/second.  A VME-resident PVIC interface 
in each of the four VME crates will be readout with a single Linux PC.
Data will be buffered in a local RAID system and transferred over the network
to FCC for permanant storage. 

\begin{table}[bp]
\begin{center}
\begin{tabular}{|l|l|l|}
\noalign{\vspace{-8pt}} \hline
Component             & Number      & Comments\\ \hline
Channel               & 37478       & WLS Fibers   \\
PMT boxes             & 587         & includes 90 empty M-64 anodes   \\
Readout Token Rings   & 37          & 16 PMTs/ring  \\
VME Readout Cards     & 10          & 4 rings/card  \\
VME Slow Control Cards& 20          & 30 PMTs/card \\
VME Crates            & 4           & \\
VME PVIC Interface    & 4           & one per crate  \\
DAQ PCs with PVIC,    & 1           & data rate is 120kB/spill\\
\hspace*{3ex}RAID system         &             &  \\
\hline
\end{tabular}
\end{center}
\caption{Parts count for \minerva\ Electronics Design}
\label{num_parts}
\end{table}

\subsubsection{Whither the TRiP chip?}

D0 is likely to redesign the TRiP chip as part of the fiber tracker
front end upgrade.  As part of that upgrade, features desirable for
MINERvA, including individual channel discriminator threshold and
front end buffer gain could be added.  A major goal for this
submission, independent of \minerva, would be to get the multi-buffer
readout mode, which would be a useful safety valve if rates in
portions of the detector were higher than we now predict.  The
\minerva\ readout board would be able to use either chip without any
modification to the board; hence, development of the front-end board
and a new TRiP submission could occur in parallel minimizing the
scheduling impact.  This upgrade is not essential for \minerva, but
could provide enhanced performance.  

We also note that if the new submission fails, DO might need to use
the existing chips.  If the yield is 90\%, as was found in a sampling
of 100 chips, we would have about 2460 chips after satisfying the D0
requirements, which is enough for 39000 channels.  Hence, in the
``worst-case'' scenario of both D0 needing the existing TRiP chips and
also a lower yield than the sampling to date would suggest, we might
have to make more TRiP chips with the existing design and masks.

%% file: detector-parameters.tex
\subsection{Parameters of the \minerva\ Detector} 
\label{sect:detparm}

\minerva\ combines
the fine granularity of an electronic bubble chamber with the final
state analyzing power of more traditional (but very fine grained)
sampling calorimeter and muon magnetic spectrometers.  To maintain the
segmentation required to identify each final state particle in a low
energy neutrino interaction and to accurately track final state
photons for $\pi^0$ reconstruction, the number of channels in
\minerva\ must be large.  To contain the produced final state
particles, the mass of \minerva\ must be large.  We attempt to break
down the contributions to mass and channels by sub-detector in this
section.

\begin{table}[bp]
\begin{center}
\begin{tabular}{|l|l|l|}
\noalign{\vspace{-8pt}} \hline
Sub-Detector & Channels in Inner Detector & Channels in Outer Detector
\\ \hline
Active Target and Side ECAL & 15360 & 5760 \\
US ECAL (Pb Target) & 3072 & 1152 \\
US HCAL (Fe Target) & 1536 & 576 \\
DS ECAL & 2560 & 960 \\
DS HCAL & 2560 & 960 \\
Veto & 426 & n/a \\
MR/Toroid & 2556 & n/a \\ \hline
{\bf Totals} & 28070 & 8408 \\ \hline
\end{tabular}
\end{center}
\caption{Channel count by sub-detector}
\label{tab:channels}
\end{table}

\begin{table}[bp]
\begin{center}
\begin{tabular}{|l|l|l|}
\noalign{\vspace{-8pt}} \hline
Sub-Detector & Mass (metric tons) \\
Active Target & 6.1 \\
Side ECAL & 8.5 \\
US ECAL (Pb Target) & 3.5 \\
US HCAL (Fe Target) & 7.0 \\
OD Framing the Target Regions & 126.5 \\ 
DS ECAL  & 19.8  \\ 
DS HCAL & 26.4 \\
Veto & 15.1 \\
MR/Toroid & 90.8 \\ \hline
{\bf Total} & 302.1 \\ \hline
\end{tabular}
\end{center}
\caption{Mass by sub-detector}
\label{tab:mass}
\end{table}

Table~\ref{tab:channels} lists the total number of channels by
sub-detector.  Predictably, it is the granuarlity required in the
plastic, Pb and Fe targets that dominates the channel count, with the
downstream calorimeters, side calorimeters, the muon and the veto
systems contributing 19\%, 17\%, 7\% and 1\% of the channels,
respectively.
As shown in Table~\ref{tab:mass}, the situation is very different with
the mass apportionment among the detectors where the OD and MR
dominate the mass.

The scintillator and optical system system of \minerva, though it
pales in comparison to long baseline neutrino experiments like MINOS,
is impressive on the scale of the CDF Plug calorimeter or CMS HCAL.
\minerva\ will use  19.2 metric tons of extruded polystyrene
scintillator, 93~km of wavelength-shifting fiber and another 46~km of
clear fiber in optical cables between the detector and the 587 M-64
multi-anode photomultipliers.

%% file: costSchedule.tex
\section{Cost and Schedule}
\label{sect:costs}

This section describes the cost and schedule associated with the
construction of the \minerva\ detector.  

Given the relatively modest cost of the detector, we plan to largely
fund the construction of the \minerva\ detector from a combination of
university program research funds and special program funds
indepedendent of the Fermilab budget.

Portions of the project that would be, by necessity, managed and
funded by Fermilab would include site outfitting and utilities (e.g.,
magnet and quiet power, cooling), crucial safety items for the NUMI
hall that must be designed at Fermilab (e.g., low voltage distribution
to the electronics, the magnet coils), and installation costs
associated with bringing modules to the NUMI near hall.  At the time
of the submission of this proposal, we do not have complete
evaluations of these costs.  As discussed in
Section~\ref{sect:utilities}, these costs have {\em not} been
estimated.  We are encouraged, however, to note that the utilities
requirements of \minerva\ appear to be within the capacity of the NUMI
near hall, and do not appear to require major infrastructure upgrades.
We expect to update this document with a good estimate of these costs
by the time of oral presentation to the PAC on December 12, 2003.

\input{cost-description}

\input{schedule}

%% file: cost-description.tex
\subsection{Description and Summary of Costs}

The cost of \minerva\ is dominated by three major categories of
expenses: external materials purchases, craft durable items and labor
to assemble the active elements and absorber into modules.  Each of
these has its own appropriate costing methodology.

For the large external equipment costs, the MAPMTs, the clear and WLS
fiber and the metal plate to construct absorbers, we have contacted
our preferred vendors directly to obtain quotes.  For the
photosensors, we have shown that the stock specifications of the
R7600U-00-M64 PMTs are adequate for our application.  Similar
phototubes are in wide use throughout the lab, and have performed
reliably.  Hamamatsu has provided a quote on our quantity with a three
month delivery time, and we are investigating cost savings that can be
realized through a more efficient custom packaging suitable for our
Cockcroft-Walton supplies.  It is worth noting that another
manufacturer, Burle Technologies, manufactures a product which would
likely meet our specifications with better channel gain uniformity,
the Planacon 85011-501, and we plan to pursue this possibility as
well.  The clear and WLS fiber vendor, Kurary, again is a vendor with
a long history at Fermilab, and they have provided similar fiber to
MINOS, CMS, CDF, etc.  We have also secured a quote on our quantities
independent of the concentration of WLS dopant should we chose to
reoptimze the dopant for our strip lengths.  Finally, the costs of the
absorber were provided by suppliers who have established relationships
with the Rutgers Physics Department machine shop, and variations here
would likely result only from movement in the bulk prices of the
relevant materials.  The machining costs have been estimated through
the Rutgers shop which is ready to perform the work as needed.
Because of the relative certainly of these costs, we
allow relatively low contingencies for these items, ranging from
20\% (MAPMT and fibers) to 30\% (absorbers).  We have not yet included
F\&A costs on most of our equipment purchases since we anticipate most
of these purchases will be made through University fully-costed shops
which will try to negotiate low F\&A costs on these large, bulk purchases.
  
The second category of costs come from the craft items which must be
constructed to assemble the detector, including the front-end
electronics and associated auxiliary systems, parts for the
PMT/front-end housing and the extruded scintillator strips.  Here the
strategy was to identify similar components from the construction of
the MINOS far or near detectors, or from the CMS HCAL or CDF Plug
construction, and to attempt to scale the actual project costs.  For
example, the clear fiber cable costs were scaled from the CMS HCAL
project which was of similar scope with similar fiber and connector
scheme.  These costs already include actual labor and EDIA costs.
Following this approach is very useful because one learns interesting
and relevant facts about hidden cost drivers.  For example, the
front-end boards require approximately 5~kWatts of power over the
entire system, and surprisingly, scaling from the MINOS far
front-end costs, one calculates a low voltage power supply system cost
in excess of \$135,000, excluding cables.  The reason turned out to be
the special fire protection requirements imposed by underground operation of
power supplies, so at least in this case, we found a cost that would
otherwise have been missed.  For these projects, we assigned
contingencies between 40--50\% of total sub-project cost,
based on our scaled estimates.   In the case of the electronics, there
is an additional contingency cost of \$70,000 for the case where we have to
re-submit the TRiP ASIC due to unexpected demand for these chips from
D0.

\begin{table}[b]
\begin{center}
\begin{tabular}{|l|r|r|r|r|}
\noalign{\vspace{-8pt}} \hline
& \multicolumn{4}{|c|}{ Cost (kUSD)}\\
 & M\&S  & SWF & EDIA & Contingency  \\ 
Sub-Project & (no F\&A) & (w/ F\&A) & (w/ F\&A) & (\%) \\ \hline
Extruded Scintillator & 151 & 12 & 30 & 78 (40\%) \\
Fiber and Glue & 262 & n/a & n/a & 52 (20\%) \\
~~WLS Fiber Prep.\ & 50 & 104 & 16 & 85 (50\%) \\
~~Optical Cables & 77 & 162 & 11 & 100 (40\%) \\
Absorbers & 310 & 67 & 32 & 122 (30\%) \\
Module Assembly & 11 & 473 & 53 & 268 (50\%) \\
Photosensors & 772 & n/a & 25 & 159 (20\%) \\
~~MAPMT Testing & 6 & 45 &n/a & 26 (50\%) \\
PMT Box and Optics & 278 & 95 & 51 & 212 (50\%) \\
Electronics and DAQ & 628 & 33 & 206 & 435 (50\%)\\ \hline
Totals & 2545 & 990 & 423 & 1537 (39\%)   \\ \hline
\end{tabular}
\end{center}
\caption[Summary of \minerva\ Detector Costs]
{Summary of \minerva\ detector costs in exclusive sub-project categories}
\label{tab:costs}
\end{table}

The final item is technical labor for component assembly and testing
for procedures that have not yet been prototyped.  These have been
estimated based on assembly models from the CMS HCAL and MINOS far
detector projects, and are generally more uncertain than other
estimates because of differences in construction between CMS HCAL,
MINOS and what we proposed in \minerva.  We have assigned
contingencies of 50\% for these projects.  Labor and EDIA costs which
dominate here are based on FY2005 projected costs for technician and
engineering staff on the CMS HCAL project at the University of Rochester.

A summary of the costs is shown in Table~\ref{tab:costs}.  The total
project construction cost is estimated to be \$3.96M, excluding the
installation and hall utilities costs.  Our calculated contingency,
\$1.54M, is 39\% of the total cost.  As previously noted, the M\&S costs
do not include F\&A.

A brief summary of what is included in each sub-project category follows.
\begin{description}
\item[Extruded Scintillator:] prototype and production extrusion dies;
purchase of plastics; extrusion in the Lab 5 facility; Q/C and
monitoring.
\item[Fiber and Glue:] WLS and clear fiber (1.2mm, Y11 0-400 ppm,
J-type, S-35), BC-600 epoxy.
\item[WLS Fiber Prep.:] Design and construct gluing assembly; cut
fibers, mirror one end and glue into scintillator, 
prepare fiber pigtail for connector.
\item[Optical Cables:] Purchase connectors and test equipment; 
EDIA for fiber termination procedure and cable layout; 
bundle fibers into conduit; insert WLS and
clear fibers into two pairs of connectors; polish ends; test for
transmission and light tight.
\item[Absorbers:] ECAL: purchase and machine Pb and stainless sheets, epoxy,
stainless to Pb, ship to assembly site; OD: order strips pre-cut from
vendor and ship to assembly site; HCAL: order partial plates pre-cut
from vendor, weld and ship to assembly site; Coil: purchase Cu AWG4
wire for coil and fabricate bus bars; Plastic fiber router plate for
OD: purchase polypropylene sheets, program and route groves on CNC
router and ship to assembly site.
\item[Module Assembly:] prototype procedures; laminate sub-planes of ID strips; connect
inner OD frame; connect stainless stop to frame and attach to
strongback; construct OD in layers; attach plastic routing plates; lay
in ID strips and route fiber; add stainless retainer; layer in additional planes;
join OD at outside layer; attach WLS bundle connector;
prepare for delivery to experimental hall
\item[Photosensors:] purchase
R7600U-00-M64 PMTs.
\item[MAMPT Testing:] design and assemble test stand; test for specs.
\item[PMT Box and Optics:] design and prototype PMT box; design and
assemble testing station; construct
box; add connectors; assemble internal optical system; mount front-end
board and PMT socket; add PMT; add fiber bundle from connector to
cookie and attach; attach internal cables to board; light tight and
Q/C.
\item[Electronics and DAQ:] prototype front-end TRiP design; design
front-end board; design VME data board; purchase VME crates,
controllers, PVIC interfaces and DAQ PC; TRiP checkout; produce,
 assemble and checkout front-end, VME data and slow controls boards; 
purchase LV system; purchase LDVS, slow controls and LV power cables.
\end{description}

%% file: schedule.tex
\subsection{Schedule}

The \minerva\ collaboration has not yet produced a resource-loaded
schedule for the experiment capable of reliably predicting the
schedule.  We plan to present such a schedule at the PAC meeting on
December 12, and to update this section when it is available.

This having been said, the schedule driving elements for the
experiment to be ready to be installed are three: construction of the
front-end electronics, assembly of the detector modules, and 
construction of the PMT boxes.  We will discuss each of these in turn.  With
the possible exception of the PMT boxes, we have high confidence that
the result of the resource-loaded schedule will be to produce detector
modules ready to install in the NUMI near hall approximately two years from
the project start, assuming that the bulk of the project funds for
M\&S items can be expended at the front end of the project.

Construction of custom electronics with an ASIC would usually be an
overriding concern in such an aggressive schedule.  However, as
Section~\ref{sect:electronics} explains, both the ASIC and the bulk of
the front-end designs are being recycled from the D0 fiber tracker
upgrade to 132~ns bunch crossings.  Design for a vertical slice test
of a prototype front-end system has already begun, and we are
confident that we can demonstrate success of this prototype 
front-end by 2004.  The VME data board will require only minor
modifications from existing designs, and the slow controls board is a
stock design which will require no modification.  Even with 
the earliest project start date of summer 2004, we would have a
completed design of all boards by the end of 2004, and be finishing
production in the middle of 2005.

The assembly of the modules is a very complicated task because of the
large number of channels in the detector, the complicated routing of
fibers in the detector and the need to reduce support mass in the
inner region of the detector.  Furthermore, assembly of modules cannot
begin until scintillator production, WLS fiber installation and  
absorber production are well underway.  We have developed an assembly
procedure and manpower assessment based on that procedure and the
University of Rochester CMS HCAL experience that suggests that seven
technician-years would be required to assemble the 53 \minerva\
modules.  It is aggressive, but perhaps plausible to attempt to fit
those seven technician years into twelve or fifteen months, after
completing a prototyping Q/C development phase of six months.  Of the
prerequisites for beginning module construction, it is most likely the
start of significant scintillator production  will most severely limit
our ability to prototype and construct modules.  We would expect to be
able to begin scintillator production approximately four months after
the project start, and therefore we conclude module production could
be complete 22 -- 25 months after project start.  Installation of the
modules in the near hall could proceed in parallel with the completion
of the last modules.

Finally, we are working to develop a complete model of the
construction of the PMT box and associated optical components.  With
over 550 boxes to construct and assembly of complicated optical
connectors and a fiber bundle in a tight space, the quality control
concerns are very non-trivial.  We have sufficient experience within
the collaboration from the design and construction of the MINOS ``MUX
boxes'' and the CMS HCAL to address this problem and expect to have a
confident assessment of the schedule and schedule risks associated
with the PMT box by the time of the PAC presentation.

%% file: appendixCryogenic.tex
\section{Cryogenic LH$_2$ and LD$_2$ Targets}

Some of the nuclear and fundamental particle physics described in this document may be
dramatically improved by the inclusion of cryogenic liquid Hydrogen and/or
Deuterium targets. The data from such targets would allow a detailed
comparison with Jefferson lab experiments that are currently using Liquid
Hydrogen and Deuterium as targets in electron and photon scattering
experiments. A comparison with these experiments in a similar momentum
transfer range with the high precision neutrino cross section
measurements made possible by \minerva\ would provide for unprecedented
studies of nucleon structure, particularly at large $x$ where it has
heretofore been very difficult. It is, for example, clear that
the substantial uncertainties on parton distribution
functions at large $x$, which are dominated by nuclear
corrections and uncertainties involved in
the flavor decomposition, would be removed.

Those measurements described in this proposal which may become limited by resolution
would benefit greatly from the inclusion of a cryogenic target
system. The interpretation of resonance data, for example, 
would no longer be complicated by uncertainties in 
nuclear binding and on shell extrapolation.
This would allow direct comparison with the Sato and Lee
pion cloud predictions without additional model systematics. 
The availability of a clean nucleon target would remove the 
complexity of the nuclear potential in heavy targets allowing the
underlying physics of strange particle production, for instance, to
be probed without interference and quasielastic studies would be greatly
helped by the lack of intra-nuclear proton scattering. Furthur, comparison
of the data from the liquid Hydrogen/Deuterium target with data
gathered from interactions in the \minerva\ nuclear targets would be of
great use in the determination of nuclear effects in neutrino interactions and
would help to limit even furthur systematic effects in the oscillation experiments.

The cryogenic target itself would be small and compact. It could be installed
upstream of the detector proper and would only require that the
veto array be moved to cover it. There exists the possibility that
the cryogenic target could be converted from a passive to an active 
target with the inclusion of CCD cameras to view the interactions;
however, even considered only as a passive target, a
high-statistics sample of neutrino interactions on liquid Hydrogen or Deuterium
would be of a great benefit in the understanding of neutrino interactions in this relatively
complicated few-GeV region.

%% file: appendixOffAxis.tex
\section{Off-Axis Running}

\begin{figure}[bhp]
\centerline{\epsfig{file=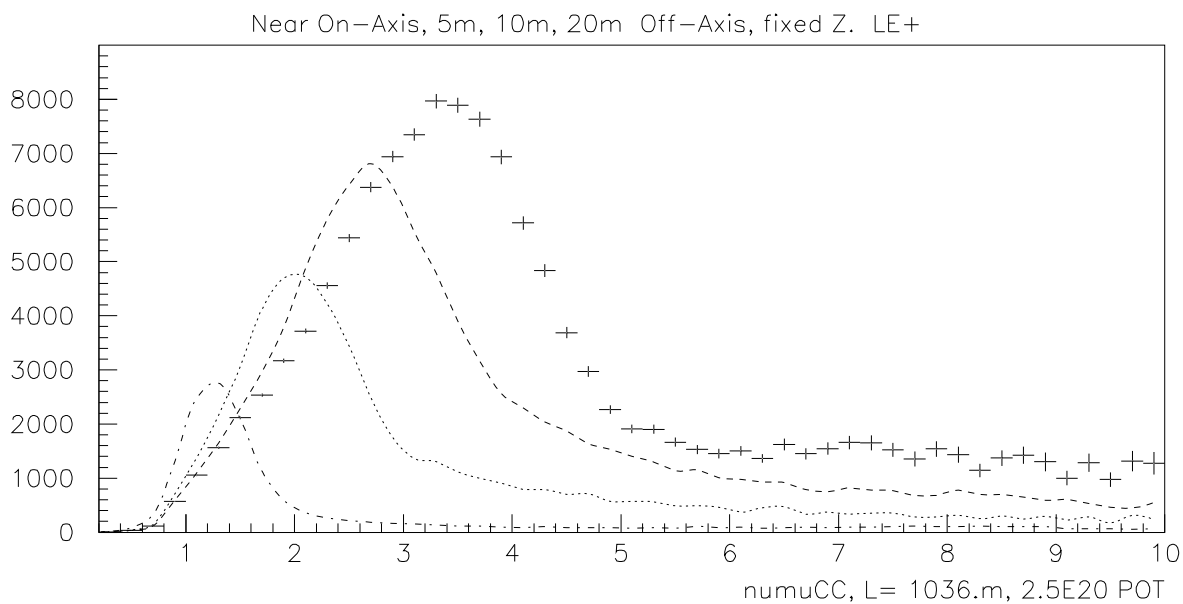, width=\textwidth,bbllx=10,bblly=280,bburx=564,bbury=564}}
\caption[Off-axis energy distributions]{Distribution of off-axis 
events that are available at different distances off-axis at the distance
from the target of the MINOS near detector in the LE beam.  In order of the peak from left to right, the curves represent event rates 20~m off-axis, 10~m off-axis, 5~m off-axis and on-axis, for comparison.} 
\label{fig:oaevts}
\end{figure}

When the \minerva\ detector is located on the beamline axis of the NuMI 
beamline, it is exposed to a broad band of neutrino energies with a peak 
energy dependent on the momenta of the pions being focused by the horns.  
Because there are high energy mesons that travel through the holes of 
both horns, there is also a long tail of neutrino events with 
energies well above the peak energy.   
However, as the detector moves off the axis, the 
peak neutrino beam energy spectrum decreases and becomes much more narrowly 
distributed in energy, and the highest energy mesons are no longer pointing 
at the detector, essentially removing the ``high energy tail'', as shown in 
Figure~\ref{fig:oaevts}. Note that at 10mrad away from the beamline axis the
$\nu_\mu$ event rate is peaked at 2GeV.
This is solely a result of the 2-body kinematics
of the $\pi\to \nu_\mu \mu$ decay.  Although the event rates are highest 
when the \minerva\ detector is on axis, the energy of the incoming neutrino 
is not known {\em a priori} and so by moving the \minerva\ detector off axis 
for a given running period the experiment can make measurements of cross 
sections in a more narrowband beam. This is particularly useful for neutral current
measurements, where the total incoming neutrino energy cannot be reconstructed
because of the loss of the outgoing neutrino. Due to the intensity of the NuMI beamline, \minerva\ can collect appreciable statistics
for precision measurements of low energy neutral current processes when running for short periods  
of time off the beam axis.

The NuMI underground complex itself was excavated primarily 
by a 21.5 foot diameter tunnel
boring machine (TBM), and because of this there are large sections of 
excavated regions underground which will be unused once the MINOS near
detector is in place.  These regions are located anywhere from 0 to 20mrad
off the NuMI beamline axis, and could house a future off axis near detector.
The Off Axis experiment is likely to place a near detector in these 
drifts to be able to predict the $\nu_e$ appearance backgrounds at a far 
off axis detector.  

\begin{figure}[htb]
\centerline{\epsfig{file=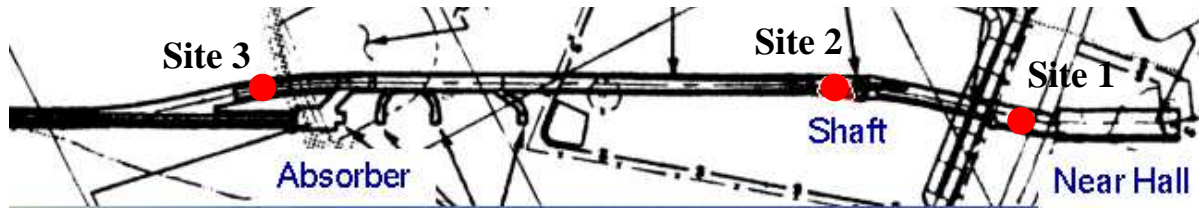, width=\textwidth}}
\caption[Possible sites for off-axis running]{Possible sites
for off axis running in the NuMI Underground area}
\label{fig:oasites} 
\end{figure}

Figure \ref{fig:oasites} shows three possible locations for the 
\minerva\ detector to be placed for off axis running.  We discuss
from downstream to upstream, the advantages and disadvantages of 
each location, keeping in mind that none of these locations would 
require any additional excavation.

\begin{figure}[htb]
\centerline{\epsfig{file=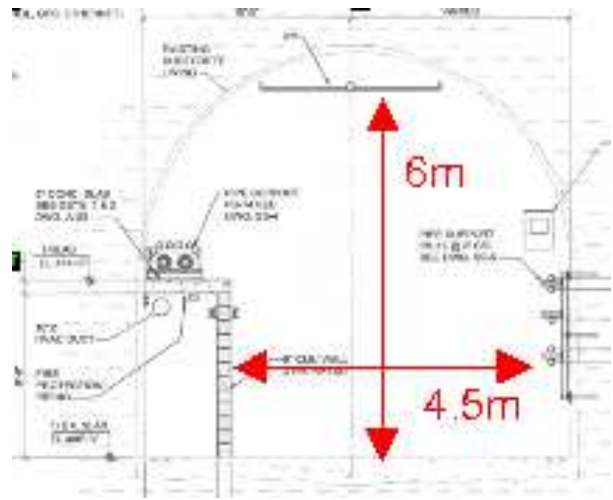, width=.5\textwidth}}
\caption[Off-axis drift cross-section]{Off-axis drift cross-section
for site 1.} 
\label{fig:oadriftxsec}
\end{figure}

The most downstream site, site 1 in figure \ref{fig:oasites}, is 
the easiest site to use, as it is the closest to the MINOS near 
hall.  It is just downstream of the shaft, and the floor is flat 
between the base of the MINOS shaft and the MINOS near hall.  
This location views off axis beams anywhere from 5 to 10mrad 
off the NuMI Axis.  The drift itself has an access tunnel on the east 
side for emergency personnel egress, and some cable
tray and utilities on the west side, but there is 
a region in the middle of the drift which measures 4.5m wide by
6m tall which is currently vacant, as shown in figure \ref{fig:oadriftxsec}.  
The neutrino energy spectrum 
in this hall would be anywhere from 1.5 to 3 GeV, depending on the location, 
since at the downstream end of the access tunnel the tunnel is nearly on 
axis, and at the upstream end of the tunnel, near the shaft, the tunnel is 
about 10m off axis.
%
%

Moving upstream, the next site is just upstream of the MINOS shaft
(called site 2 in figure \ref{fig:oasites}).
This location is also relatively easy to get to since it is near 
the MINOS shaft and therefore close to utilities, but the floor in 
this region has a 9\% slope.  The available cross section in this area
is also wide, at least as wide as the area for site 1, but some space
must be left for access to points upstream, since there is no longer an 
independent egress tunnel as with site 1.   
Here the mean neutrino energy is about 1.5GeV.  
Finally, the most remote site is located in a drift that was created
when the tunnel boring machine had to change angles between excavating 
for the decay volume itself and excavating for the access tunnel to the 
downstream areas.  This site is substantially closer to the decay pipe, 
located about 725m from the NuMI target.  This site has the widest cross 
cross section available since it is located in a ''dead end'', and would 
allow off axis distances from 5 to 15m off axis, which correspond to between 
8 and 20mrad off axis angles.  To run in this location 
the detector would have to be 
moved up the 9\% sloped floor about 200m from the base of the MINOS shaft, 
and then re-assembled in this new hall, which also has a floor with a 9\% slope. 
There are always several meters of earth between the NuMI hadron 
absorber and the detector, but neutron radiation issues would 
be worse here than in the other two sites.  However, this site's weakness
is also its strength, in that a detector in this location would have the
least amount of interference with the MINOS experiment during construction. 
The closer location would also provide a significantly higher event rate.  
The downstream portion of this site has utilities for the NuMI absorber, 
but there are about 10 meters of cleared space upstream of those utilities 
where the \minerva\ detector could be placed. 
%
%


%% file: summary.tex
\section{Updates to the Physics Case \\ 
(submitted to the Fermilab PAC 30 March 2004) }

\subsection{Introduction}

As detailed in the \minerva\ proposal, the imminent completion of the
NUMI beamline offers the particle and nuclear physics communities a
new opportunity to study neutrino interactions in an environment
unprecedented in granularity of detectors and event
rate.  The construction of \minerva, with its fully active core target,
would allow a wide variety of measurements in neutrino interaction
physics, which both support future and current neutrino oscillation
efforts, and are interesting for their own sake.

Since the proposal was submitted to the FNAL PAC at the December 2003
meeting, the \minerva\ collaboration has continued to make progress in
understanding the ultimate capabilities of the experiment.  This
addendum serves to document these continuing studies for the benefit
of the PAC as we seek approval of the experiment.

This document is broken down by physics topic, and each
section contains a brief summary of the physics goals, 
the status of our understanding at the time of the proposal, and then
documents improved understanding, since the proposal, in more detail.
The topics considered in this addendum are:
\begin{itemize}
  \item Quasi-Elastic Cross-Sections and Form Factors
  \item Coherent Pion Production
  \item Physics Opportunities in the Resonance Production Region
  \item Nuclear Effects in Neutrino Scattering
  \item The Impact of \minerva\ on Oscillation Experiments
\end{itemize}

This document is intended to be read as a supplement to the updated
\minerva\ proposal in the main body of this document.  
This document, along with
other documentation of the status of the \minerva\ experiment, 
is available from the \minerva\ collaboration web page,
\url{http://www.pas.rochester.edu/minerva/}

%% file: qe_analysis.tex
\subsection{Quasi-Elastic Scattering}

At the lowest neutrino energies relevant for future long baseline
efforts, it is the quasi-elastic scattering that dominates the charged
current interaction rate.  As outlined in the \minerva\
proposal\footnote{see the discussion in Chapter 6 of the main
proposal}, there are some interesting physics issues to
be addressed in quasi-elastic neutrino scattering.  The first of these is
understanding the impact of nuclear effects on the quasi-elastic
kinematics at low $Q^2$ which dominates the signal rate for
oscillation experiments.  The second is understanding the axial
form-factor of the proton at high $Q^2$ which will contribute to the
blossoming body of new measurements on high $Q^2$ nucleon form-factors
where some surprises have already been seen in charged-lepton
scattering (see main proposal above).  
Since the proposal, we have made significant progress in
simulating our expected analysis of quasi-elastics, focusing on the
important issues of maintaining high efficiency at low $Q^2$ and low
backgrounds at high $Q^2$.

In $\nu n\to\mu^- p$, the outgoing proton carries a kinetic energy
that is approximately $Q^2/2M_N$.  So for low $Q^2$,  the challenge is
identifying events with a very soft recoil proton; for high $Q^2$,
this proton is high energy and may interact in the detector, making
particle identification more challenging.  The main strategies of the
current analysis are:
\begin{itemize}
  \item At low $Q^2$, accept quasi-elastic candidates with a single
      (muon) track, and discriminate from background by requiring low
      activity in the remainder of the detector
  \item At high $Q^2$, reconstruct both the proton and the muon, and
  require kinematic consistency with $x=1$ and $p_T=0$
\end{itemize}
Simple cuts deriving from these ideas allow for reasonable efficiency
with good purity, even at high $Q^2$.

\subsubsection{Details of Quasi-elastic reconstruction}

The analysis uses the NEUGEN generation and the hit level \minerva\
simulation and tracking package in order to simulate signal selection
and background processes.  

The initial event identification proceeds by requiring one or two
tracks in the active target. One of these tracks must be long range
($400$~g/cm$^2$) and is the putative muon.  If a second track forms a
vertex with this track, it is assumed to be the proton.  No
other tracks are allowed to be connected with this event vertex.  The
muon track momentum is reconstructed with a fractional uncertainty of
between $10$--$20\%$.

\begin{figure}[tp]
\begin{center}
\epsfxsize=\textwidth
\mbox{\epsffile[0 0 566 530]{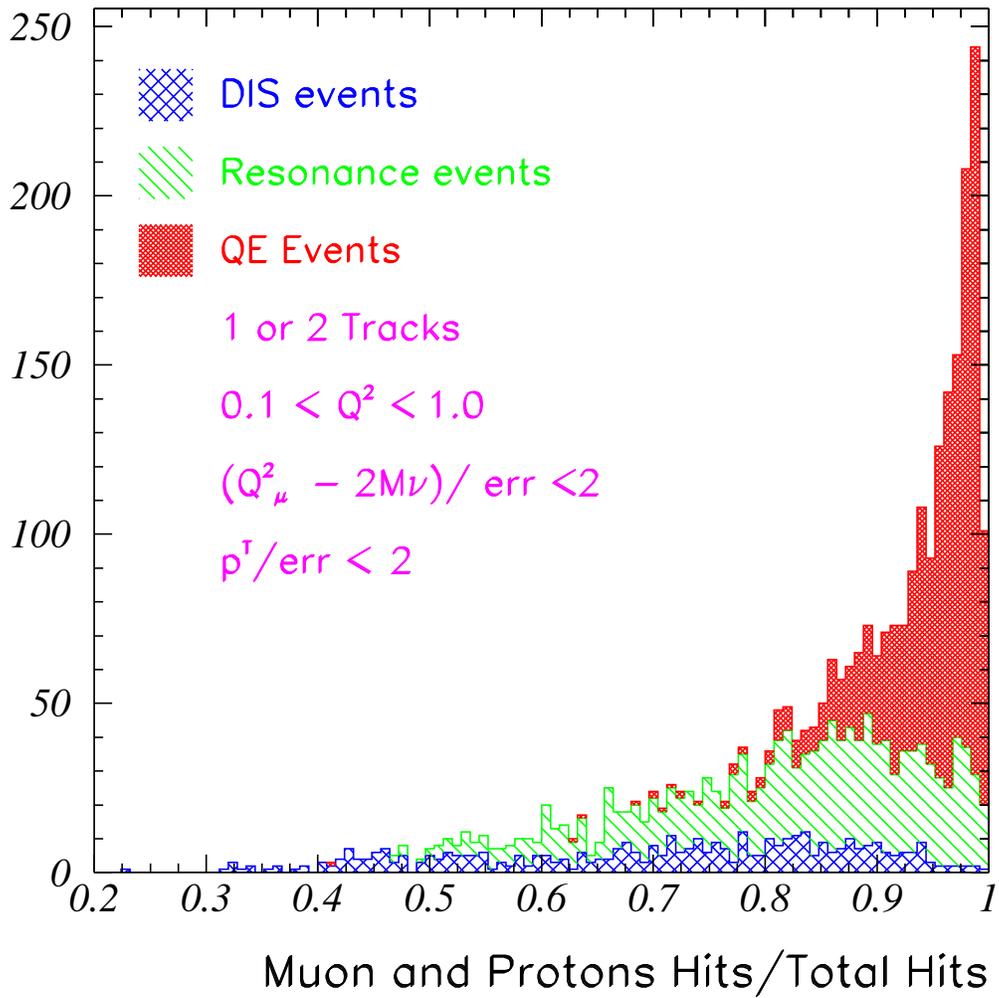}}\\
\end{center}
\caption[Fraction of Hits Associated with the Muon and Proton in
  Quasi-Elastic Candidates]{The fraction of hits associated with
the muon and the proton tracks in quasi-elastic candidates.
The events for the plot may have have one or two vertex tracks, 
  pass additional kinematic requirement and are required to have
$0.1~GeV^2 < Q^2 < 1~GeV^2 $.}
\label{hits_q2l0}
\end{figure}
In the low $Q^2$ case, the proton track (if found) would be
effectively required to lose energy by range since only a limited
amount of detector activity not associated with the primary tracks is
allowed by the event selection.  We attempt to recover some of the
lost efficiency at higher $Q^2$ due to this cut by allowing hits on
tracks near the proton track to be associated with the proton track
itself.  Figure~\ref{hits_q2l0} shows the fraction of hits not
associated with the lepton or proton in the quasi-elastic events and
in expected background processes.  For higher $Q^2$ events a similar
requirement could in principle be applied, but it is not particularly
effective nor efficient.

 The energy of the proton for the high $Q^2$ case (where the
proton almost always interacts) is reconstructed calorimetrically with
an expected fractional energy resolution that is well parameterized by
$35\%/\sqrt{E_{\textstyle proton}}$.

Although muons are identified by requiring a single track with a long
range in the detector, no attempt is made in the analysis to improve
particle identification by requiring $dE/dx$ consistent with the muon
or proton tracks.  This requirement is expected to be particularly
effective for protons of ${\cal O}(1)$~GeV momentum\footnote{see
Section 15.5.5 of the main proposal}, and such a
requirement can be optimized to improve the analysis in the future.
In addition, it may be possible to improve the efficiency by allowing
a lower range muon with a $dE/dx$ requirement without sacrificing
purity.

If a quasi-elastic interaction is assumed, one can reconstruct the event
kinematics from only the momentum and direction of the final state
$\mu$.  Neglecting the binding energy of the final state proton, 
$$
E_\nu^{\textstyle QE} = \frac{M_N E_\mu - \frac{m_\mu^2}{2}}
                             {M_N - E_\mu+ p_\mu\cos\theta_\mu}.
$$ If a proton track is required and its angle and energy are
measured, one can additionally require consistency with the
quasi-elastic hypothesis.  Two constraints are possible, one on the
$x$ of the reconstructed interaction and one on the $p_T$ of the
observed final state.  

\begin{figure}[tp]
\begin{center}
\epsfxsize=\textwidth
\mbox{\epsffile[0 0 566 534]{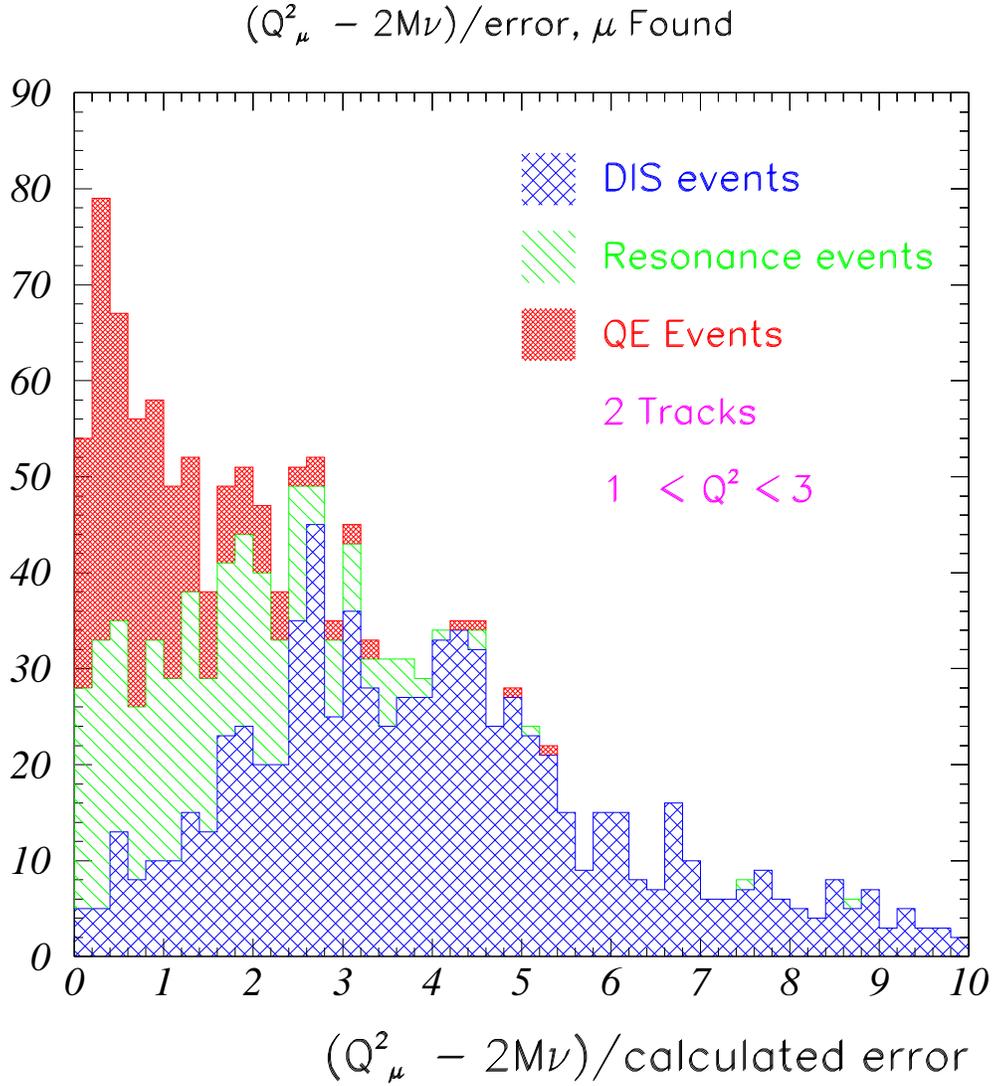}}
\end{center}
\caption[$Q^2$ Difference Significance for Quasi-Elastic
  Reconstruction]
{The significance of the 
difference between $Q^2$ from the quasi-elastic hypothesis and $Q^2$
  from the final state energy}
\label{q2dif_q2l1_q2hi3}
\end{figure}
If the interaction is truly quasi-elastic, then
$x=1$, and therefore $Q^2=2M_N\nu$ where 
$\nu$ = $E_{had}$ - $M_{nucleon}$, and 
$E_{had}$ is the energy of the hadronic final state.  
In this analysis, we test
this by comparing $Q^2$ reconstructed from the lepton kinematics under
the quasi-elastic hypothesis to $2M_N\nu$ and forming
$(Q^2_{\mu}-2M_N\nu)/{\textstyle error}$ where the
dominant part of the calculated error for this term comes from the
smearing of hadronic final state energy.
Figure~\ref{q2dif_q2l1_q2hi3} shows this $Q^2$ difference significance
for two track quasi-elastic candidates with observed $1~GeV^2 <
Q^2 < 3~GeV^2$, for quasi-elastic,
resonance and DIS events.   Note that this cut can be applied without
identifying a proton track if the visible energy, less the muon energy, is
assumed to be $\nu$.
 
The $Q^2$ significance ($x$) cut does not use information on the
proton direction, and so we impose a second kinematic cut on the $p^T$
of the final state relative to the incoming neutrino direction.  This
selection requires that a proton track is identified and we cut on the
significance of the difference from $p_T=0$.  We impose a cut of
$p^T/{\textstyle error}<$2 except for $Q^2>3~GeV^2$, for which the cut is
3.  Note also that if we impose a $p^T$ cut first, the $Q^2$ difference
cut still improves the result, i.e. both cuts are needed.

In summary, the selection requirements for quasi-elastic candidates
are:
\begin{itemize}
  \item One or two tracks for $Q^2<1~GeV^2$ and two tracks for $Q^2>1~GeV^2$.
  \item One track must have $400$~g/cm$^2$ range (muon).
  \item $(Q^2_{\mu}-2M\nu)/({\textstyle error}) < 2 $.
  \item $p_T/({\textstyle error}) < 2$ for $Q^2<3~GeV^2$ and
  $p_T/({\textstyle error}) < 3$ for  $Q^2>3~GeV^2$.  
  \item Hit fraction associated with muon and proton $> 0.9$, for 
       $Q^2 < 0.5~GeV^2$, or $>$ 0.85, for $0.5~GeV^2 < Q^2 < 1.0~GeV^2$. 
\end{itemize}

\subsubsection{Results}

\begin{table}
\begin{center}
\begin{tabular}{|c|cc|cc|cc|cc|cc|}
\noalign{\vspace{-8pt}} \hline
  &  \multicolumn{2}{|c|}{$\mu$} &   \multicolumn{2}{|c|}{\qdif}      & \multicolumn{2}{|c|}{p$_T$ /err}& \multicolumn{2}{|c|}{Hits}      \\ 
$Q^2$ bin  & Effic& Purity & Effic & Purity & Effic & Purity & Effic &Purity \\ \hline   
 0.1-0.5   & 0.926& 0.246 & 0.918& 0.442 & 0.866& 0.559 & 0.775& 0.842 \\
 0.5 - 1   & 0.775& 0.199 & 0.765& 0.410 & 0.624& 0.486 & 0.528& 0.685 \\
1 - 2      & 0.600& 0.199 & 0.541& 0.416 & 0.397& 0.555 & 0.338& 0.598 \\   
2 - 3      & 0.456& 0.146 & 0.400& 0.375 & 0.344& 0.554 & 0.278& 0.676 \\
3 - 10     & 0.689& 0.123 & 0.600& 0.310 & 0.467& 0.420 & 0.311& 0.700 \\ \hline
\end{tabular}
\end{center}
\caption[Efficiency and Purity in $Q^2$ Bins for Quasi-Elastic
  Candidates]{Efficiency and purity in $Q^2$ bins for quasi-elastic
  candidates}
\label{eff_purity}
\end{table}

\begin{figure}[tp]
\begin{center}
\epsfxsize=4.0in
\mbox{\epsffile{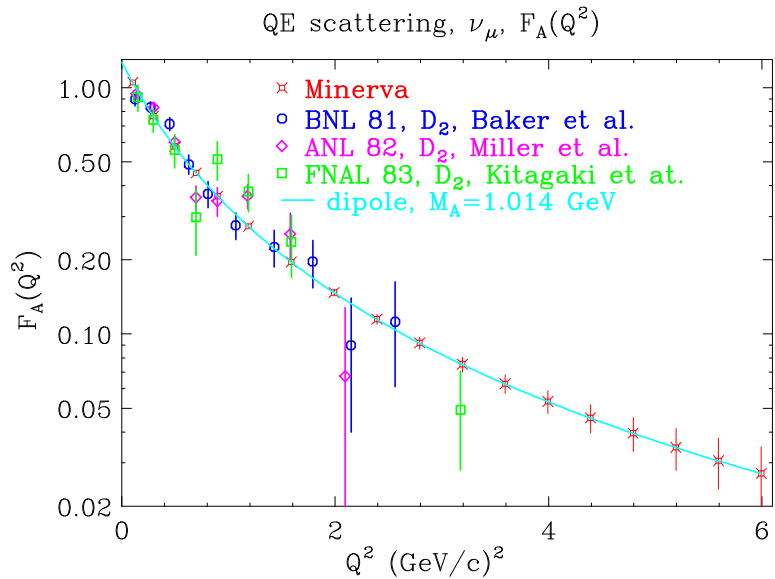}}\\
\end{center}
\caption[Extraction of $F_A$ in \minerva]{Estimation of $F_{A}$ from a
sample of Monte Carlo neutrino quasi-elastic events recorded in the
 \minerva\ active carbon target.  Here, a pure dipole form for $F_{A}$ is
assumed, with $M_{A} = 1.014~\hbox{GeV}/c^2$.
The simulated sample and error bars correspond to four years of NuMI running.
Also shown is $F_A$ extracted
from deuterium bubble chamber experiments
using the $d\sigma/dq^2$ from the papers of FNAL 1983~\cite{Kitagaki_83}
BNL 1981~\cite{Baker:1981su}, and ANL 1982~\cite{Miller_82}}
\label{log:ps2}
\end{figure}

\begin{figure}[tp]
\begin{center}
\epsfxsize=4.0in
\mbox{\epsffile{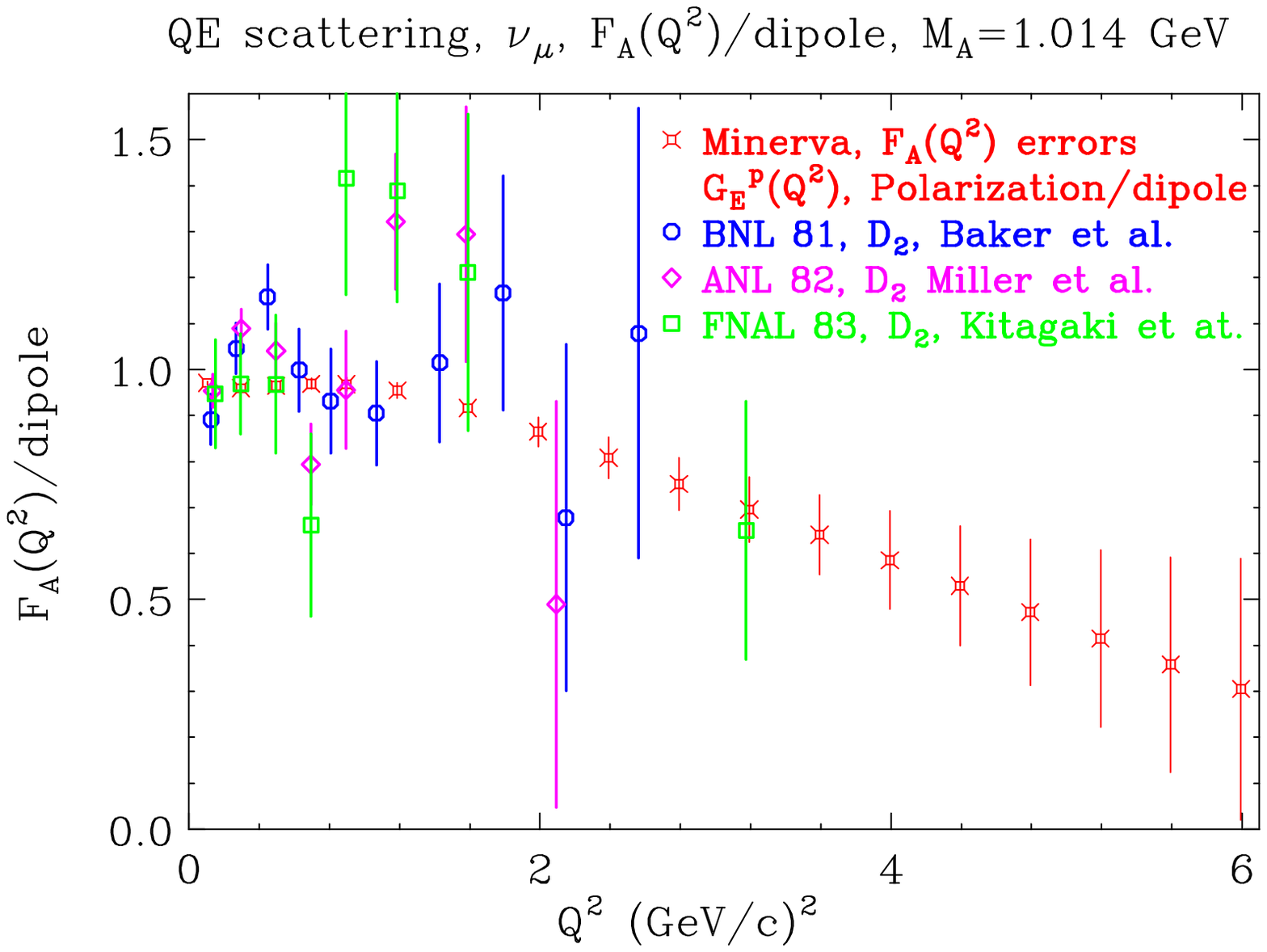}}\\
\epsfxsize=4.0in
\mbox{\epsffile{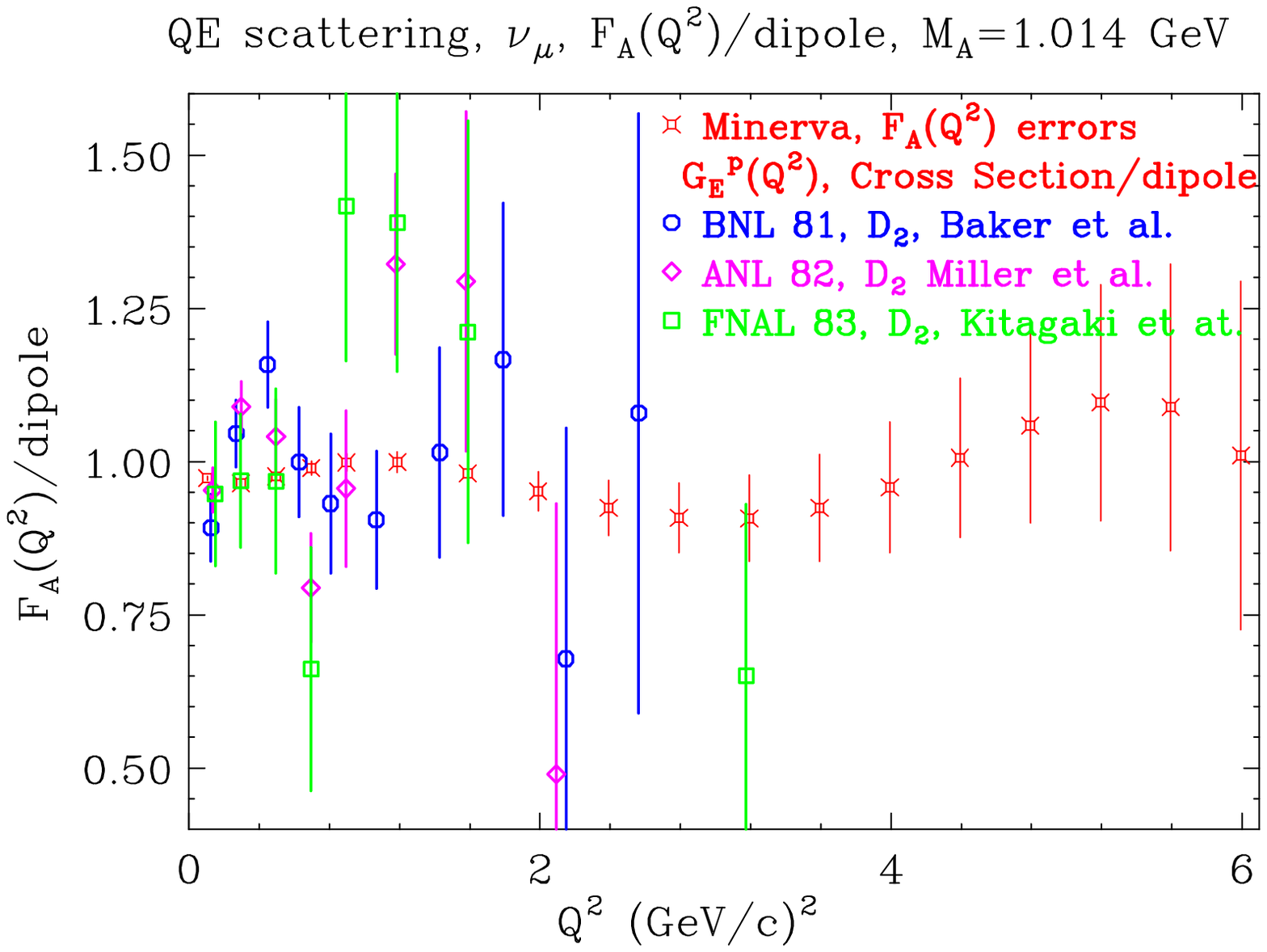}}\\
\end{center}
\caption[\minerva\ sensitivity to $F_A/F_A(\hbox{Dipole})$]{
Extracted ratio $F_{A}$/$F_{A}$(Dipole) from \minerva\ active
target (4 NuMI-years) under two
scenarios for the true $F_{A}$.
{\bf (Top)} $F_{A}$/$F_{A}$(Dipole) is  $G_E^p$/$G_E^p$(dipole) from
polarization transfer measurements. 
{\bf (Bottom)} $F_{A}$/$F_{A}$(Dipole) is  
$G_E^p$/$G_E^p$(dipole) from Rosenbluth separation technique.
Also shown is $F_A$ extracted
from deuterium bubble chamber experiments
using the $d\sigma/dq^2$ from the papers of FNAL 1983~\cite{Kitagaki_83}
BNL 1981~\cite{Baker:1981su}, and ANL 1982~\cite{Miller_82} }
\label{f_a_polar:ps2}
\end{figure}

Table~\ref{eff_purity} shows the efficiency and purity of the 
quasi-elastic sample for different $Q^2$ bins after each cut.
Using the calculated efficiency and purity, we have updated the
uncertainties on $F_A$ derived in the main 
proposal which did not include efficiency or background
effects.  

Figure~\ref{log:ps2} shows the extracted values and errors on
$F_{A}$ in bins of $Q^{2}$ from a sample of simulated quasi-elastic
interactions in the \minerva\ active carbon target, for a four-year
exposure in the NuMI beam.  Figure~\ref{f_a_polar:ps2} shows these
results as a ratio of $F_{A}$/$F_{A}$(Dipole), demonstrating
\minerva's ability to distinguish between different models of $F_A$.
Note that resolution effects are still not included in this extraction
of $F_A$; however, the typical $Q^2$ resolution for quasi-elastic
events at high $Q^2$ is $\stackrel{<}{\sim}0.2$~GeV$^2$ which is
smaller than the bin size.

%% file: coherent_addendum.tex
\subsection{Coherent Neutrino-Nucleus Scattering}

With its high statistics, fine segmentation, excellent tracking, good
particle ID, and range of nuclear targets, the \minerva\ experiment
will be able to obtain data samples of coherent
interactions several orders of magnitude larger than those published
from previous experiments.  The study of neutrino coherent scattering
is an excellent tool for exploring the `hadronic' nature of the weak
current, particularly the axial component.

The capabilities of \minerva\ for coherent scattering are compelling
for a number of reasons:
\begin{enumerate}
\item
Coherent $\pi$ production is largely a neutrino-specific process, as the
coupling is dominated by axial current.  In the Rein-Seghal model
\cite{Rein:1983pf}, for instance, the coupling is entirely axial.
Therefore there is no electron-scattering analogue to constrain
models for this process.
\item
Existing measurements in the energy range of the NuMI beam are quite poor.
There is only one experiment (SKAT) with data on charged current coherent 
production below $E_\nu$ of 10 GeV, and it only had 59 events.     
\item
The charged current is an extremely clean measurement in \minerva\
because distinct muon and pion tracks will be clearly visible and will
allow precise measurement of the interaction point and kinematic
quantities.
\item
With nuclear targets spanning A=12 to A=207, \minerva\ will be 
the first experiment to measure the A-dependence of the coherent 
cross section. 
\item
An improved knowledge of coherent cross sections will be crucial for
 future oscillation experiments, as the neutral current reaction
 producing a single \piz\ is one of the main backgrounds for
 subdominant $\numu \rightarrow \nue$ mixing searches.  
\item The \piz\
 reconstruction capabilities of \minerva\ make it possible to
 statistically separate coherent from non-coherent \piz\ production
 to directly determine this background and to check the expected relationship
 to the charged-current process. 
\end{enumerate} 


\minerva\ selects coherent events in both the charged and neutral
current by relying on the distinct kinematics of the events.  Because
the coherence condition requires that the nucleus remain intact,
the process is tagged by low-energy transfer ($|t|$) to the nuclear
system which is reconstructed by 
\begin{equation}
-|t| = -(q - p_\pi)^2 =
(\Sigma_i (E_i - p_i^{||}))^2 - (\Sigma_i(p_i^\perp))^2.
\end{equation}
Candidate events are generally selected as coherent by requiring a low final
state multiplicity and by requiring kinematics consistent with low
$|t|$.

\begin{table}
\begin{center}
\begin{tabular}{|l|r|r|}
\hline
Cut                    &   Signal Sample & Background Sample \\
\hline
                       &        5000     &             10000 \\
2 Charged Tracks       &        3856     &              3693 \\
Track Identification   &        3124     &              3360 \\
$\pi^o$/neutron Energy &        3124     &              1744 \\
Track Separation       &        2420     &               500 \\
x$<$0.2                  &        2223     &               100 \\
t$<$0.2                  &        2223     &                19 \\
$p_\pi<600$ MeV        &        1721     &                12 \\
\hline
\end{tabular}
\label{tab:cohcuts2}
\end{center}
\caption[Purity and Efficiency in the Coherent Charged Current Analysis]{Analysis cuts to isolate a sample of coherent interactions.  
The cuts are described in the text.}
\end{table}

The \minerva\ event selection is described in detail in the
proposal\footnote{Chapter 8 of the main
proposal}.  The efficiency and purity of the selection
in the charged current case is summarized in Table~\ref{tab:cohcuts2},
and the expected uncertainties in the final cross-section as a
function of energy is shown in Figure~\ref{fig:cohminerva}.

\subsubsection{A-Dependence of the Cross Section}

\begin{figure}[tp]
\center
\epsfxsize=90mm\leavevmode
\epsffile{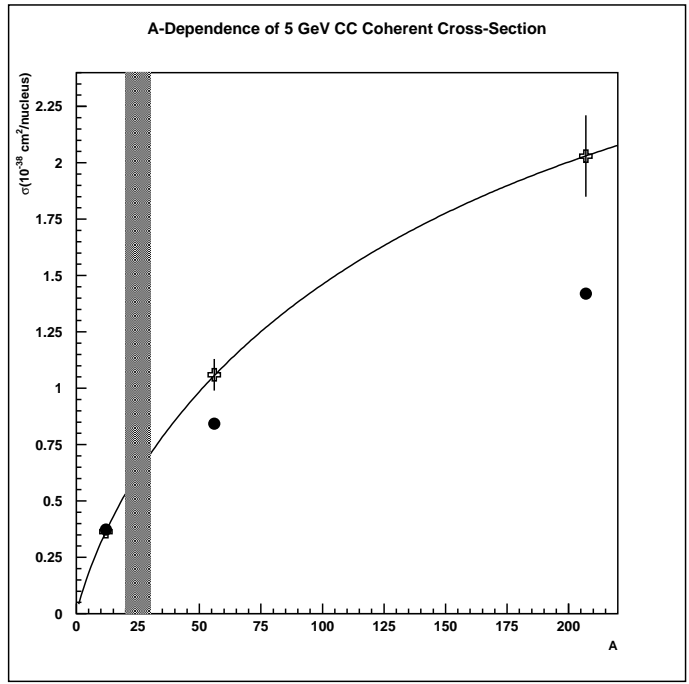}
\caption[\minerva\ Measurement of Nuclear Dependence of Coherent Cross-Sections]{Coherent cross-sections for 5 GeV neutrinos 
as a function of atomic number.  Solid 
curve is the prediction of Rein-Seghal, solid circles are the predictions
of Paschos-Kartavtsev for carbon, iron and lead.  Crosses indicate the 
expected measurement errors for minerva assuming the measured cross 
section is that of Rein-Seghal.   The shaded band indicates the region 
explored in previous experiments, primarily from measurements on 
aluminum (A=27), neon (A=20), and freon ($<$A$>$=30), all performed 
at higher energies than are relevant for \minerva\ or future
oscillation experiments.}
\label{fig:coh_adep}
\end{figure}

As noted above, there are no existing measurements of the coherent
cross-sections on light nuclei (e.g., $H$, $C$, $O$).
These measurements are important for planned $\nu_e$
appearance experiments. Current
predictions must rely on extrapolation in A in addition to extrapolation to
lower energies.  

Figure \ref{fig:coh_adep} shows the predicted A-dependence of the
charged current coherent cross section from Rein-Seghal and Paschos-Kartavtsev
\cite{Rein:1983pf,Paschos:2003hs} models, and the expected
measurement errors from \minerva.  As this analysis indicates, the
high statistics and large dynamic range in A of the \minerva\
experiment will make possible detailed examinations of the coherently
produced meson-nucleus interaction.  Although K2K and MiniBooNE are
expected to significantly improve the knowledge of coherent $\pi$
production on $CH_n$ at very low energies, only \minerva\ will be able
to reach this level of precision and to perform this systematic
examination of the A dependence of the cross-section.

%% file: nuclupdate.tex
\subsection {Measuring Nuclear Effects with the \minerva\ Detector}

As indicated in our proposal, to study nuclear effects in \minerva, carbon,
iron and lead targets will be installed upstream of the pure
scintillator active detector.  The currently preferred configuration
involves a total of 6 planes, with each plane divided transversely
into C, Fe and Pb wedges.  As one proceeds from upstream to downstream, the
C, Fe and Pb targets exchange (rotate) positions.  As always, a
scintillator module of four views (x,u,x,v) separates each of the
planes.  The total mass is over 1~ton of Fe and Pb and somewhat over
0.5 ton of C.  Since the pure scintillator active detector essentially
acts as an additional 3-ton carbon target (CH), the pure graphite (C)
target is mainly to check for consistency.  For the standard four-year
run described in the proposal, \minerva\ would collect 1 M events on
Fe and Pb, 600~k events on C as well as 2.8~M events on the
scintillator within the fiducial volume.  In this section we give more 
experimental background than was given in the proposal\footnote{see
  Chapter 12 of the main proposal}, as well as describe
more specifically the analysis technique that will be used to measure 
nuclear effects.

\minerva 's goals in measuring nuclear effects can be summarized as
follows:

\begin{itemize}

\item measure final-state multiplicities, and hence absorption
probabilities, as a function of A with initial $\nu$;

\item measure the visible hadron energy distribution as a function of
target nucleus to determine the relative energy loss due to final state
interactions (FSI);

\item investigate if the correction factors for observed multiplicity
and hadron energy are a function of the muon kinematics for a more
directed application of nuclear effect corrections.

\item measure $\sigma(x_{Bj})$ for each nuclear target to compare
$x_{Bj}$-dependent nuclear effects measured with both $\nu$ and
charged lepton beams.

\item If sufficient $\nubar$ running is available, 
measure the nuclear effects on
F$_2$(x,Q$^2$) and xF$_3$(x,Q$^2$) to determine whether sea and
valence quarks are affected differently by the nuclear environment.

\end{itemize}

\subsubsection {Pion Absorption Effects in Neutrino Interactions}  

Interactions of few GeV neutrinos in nuclei easily produce
resonances which decay to pions.  Any attempt to reconstruct
the incident neutrino energy based on the total observed 
energy must take into account the interactions of the pions
in both the interaction nucleus and the detector.  Current 
neutrino interaction Monte Carlos (such as INTRANUKE~\cite{intranuke}) handle
pion interactions crudely and have generally not yet incorporated
the latest knowledge of pion interactions.  We would like to
summarize here the current knowledge of pion interactions and
discuss plans for using \minerva\ to better account for pion
interactions for neutrino-nucleus interactions.

Our concern here is mainly with pions in the energy range of
100 to 500 MeV, where the interaction cross sections are the 
highest.  In this range the pion-nucleon cross section is
dominated by the very strong $\Delta(1232)$ resonance. The
$\Delta$ is a fairly broad (about 100 MeV) resonance, and
the pion-nucleon cross section reflects this, with a peak
near 200 MeV pion energy which drops quickly above and below this.
The pion nucleus cross section exhibits a similar behavior,
with a less pronounced drop-off at higher energy.  The charged pion
nucleus cross section has four important components in the
intermediate energy range: elastic scattering (nucleus left
in ground state), inelastic scattering (nucleus left in excited
state or nucleon knocked out), true absorption (no pion in
the final state), and single charge exchange (neutral pion in
the final state).  

\begin{figure}[tp]
\centerline{
\epsfxsize=0.75\textwidth
\epsffile{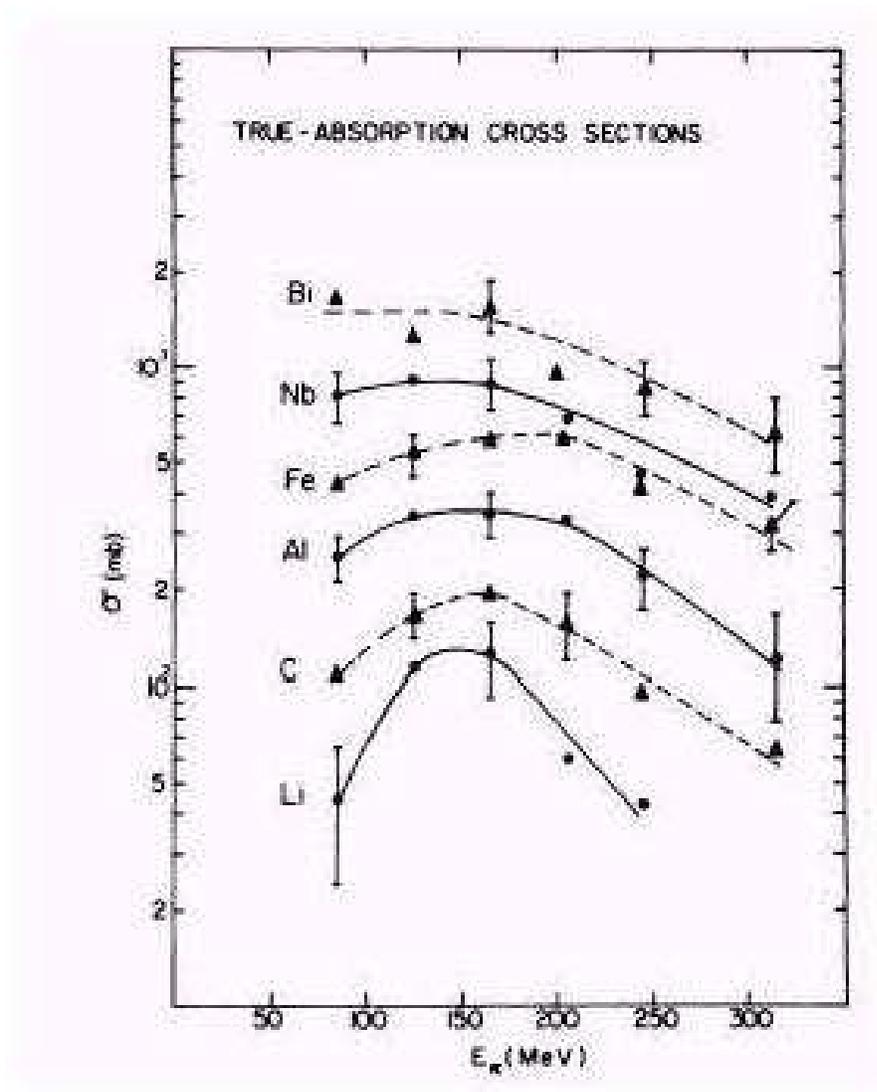}}
\caption{The absorption cross-sections for various nuclei as a
function of pion energy.}
\label{piabs}
\end{figure}

Neutrino detectors are mainly iron (absorber), oxygen (water) and
carbon (scintillator). The total pi-carbon cross section is 600 mb,
with elastic and inelastic cross sections about 200 mb each, and
absorption about 160 mb.  The total pi-iron cross section is about
1700 mb, with elastic and absorption about 600 mb each, and inelastic
about 400 mb.  Cross sections for positive and negative pions are
nearly the same because nuclei contain about the same number of
protons and neutrons.  These very large cross sections mean that
essentially all pions will undergo some nuclear reaction, many within
the interaction nucleus, and in most circumstances nearly all will be
absorbed rather than stopping or exiting the detector.  The absorption
probability within the interaction nucleus is order 30\% while the
absorption probabilities in the detector are about 1\%/cm in
scintillator and 4\%/cm in iron.  Figure~\ref{piabs} \cite{ashery}
shows absorption cross sections for various nuclei as a function of
pion energy.

In elastic and most inelastic reactions the scattered pion will
not, because of its light mass, lose much energy.  However, absorbed
pions will lose all of their kinetic and mass energy.  Most of that
energy will go into nucleons.  We want to discuss here
what happens to that energy.

Pion absorption cannot occur on a single nucleon due to energy
and momentum conservation.  The simplest absorption mechanism
is on two nucleons.  Because
absorption appears to proceed mainly through the $N-\Delta$
intermediate states, an isospin zero (np) pair is the primary
candidate.  Such an absorption for a positive pion would give
two energetic protons whose kinetic energy nearly equaled the
total pion energy.  However, early studies of pion absorption
found that was not the most probable mechanism.

In the 1990's two large solid angle detectors, the LAMPF BGO Ball
and the PSI LADS detector, were built to study pion absorption.
Both detectors had large solid angles (both more than 90\% of
the full solid angle) and low proton thresholds (about 20 MeV
for each).  The LADS detector also had reasonable neutron 
detection efficiency and energy measurement.  The somewhat surprising
result from both detectors was that pion absorption was dominated
by three body absorption \cite{ingram}.  For positive pions, the absorption
on a $pnn$ triplet (leading to a $ppn$ final state) was the most
common.  This was observed even in $^4$He.  The absorption in
heavier nuclei also appears to proceed mainly through a 
three-body mechanism, although increased initial state interactions
(pion re-scattering) and final state interactions (nucleon
re-scattering) result in four to five nucleons
being emitted.  Typically the final state contains more neutrons
than protons. The absorption process, which is still not
well understood theoretically, largely fills the available 
phase space thus giving a wide range of nucleon energies with
little angular dependence.

\begin{figure}[tp]
\centerline{
\epsfxsize=0.9\textwidth
\epsffile{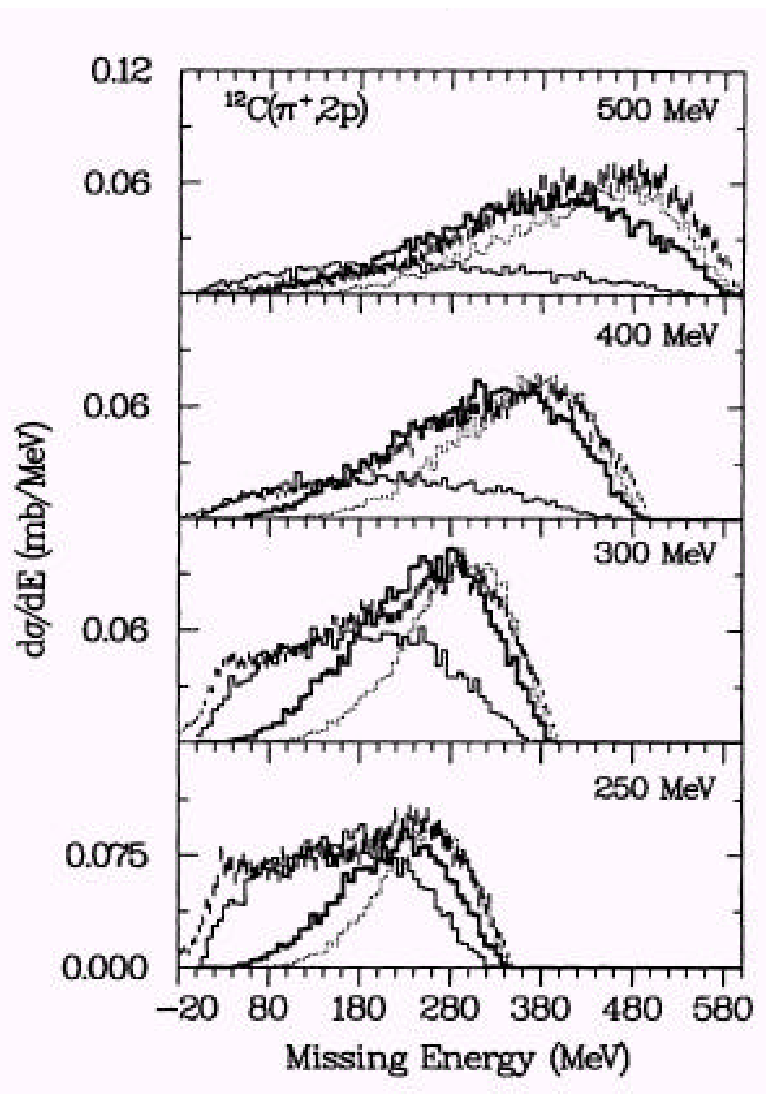}}
\caption{The missing energy (total pion energy minus the total proton
kinetic energy) for the absorption of 250-500 MeV pions on carbon.}
\label{deC}
\end{figure}

\begin{figure}[tp]
\centerline{
\epsfxsize=0.65\textwidth
\epsffile{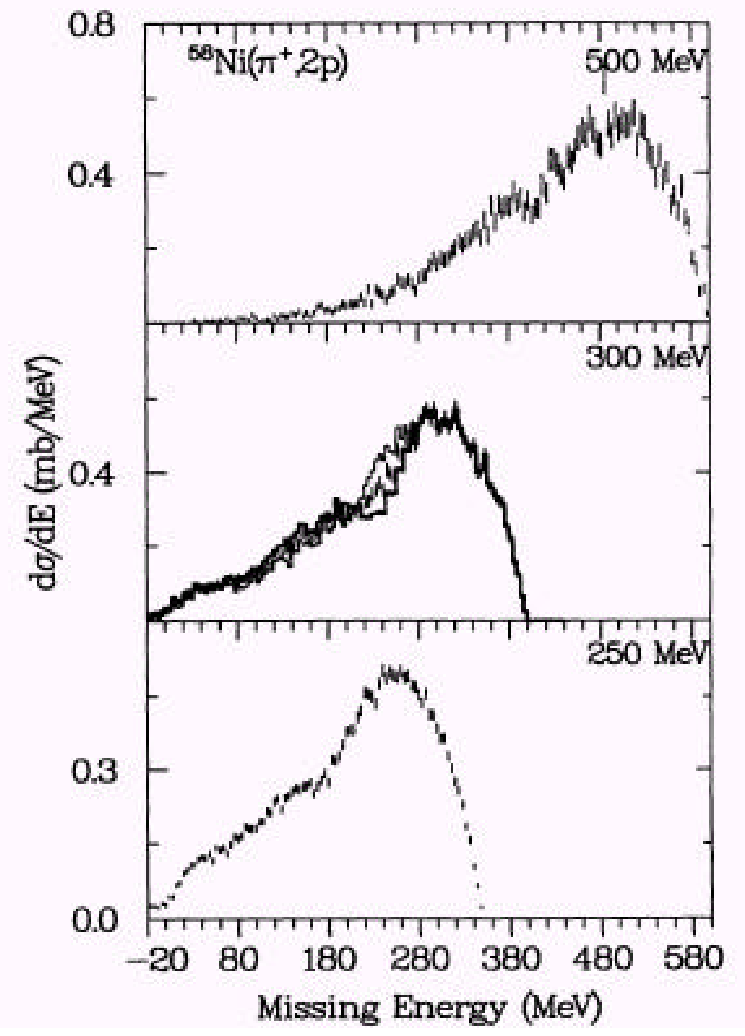}}
\caption{The missing energy (total pion energy minus the total proton
kinetic energy) for the absorption of 250-500 MeV pions on nickel.}
\label{denickel}
\end{figure}

Because much of the energy is in neutrons, the observed energy is well
below the total pion energy.  Figure~\ref{deC} and
Figure~\ref{denickel} \cite{jones} show missing energy (total pion
energy minus the total proton kinetic energy) for absorption of
250-500 MeV positive pions on $^{12}$C and $^{58}$Ni.  As can be seen,
even in carbon more than half the energy is lost to unobserved
particles, a fraction which increases with pion energy and with A.

The situation is of course worse for negative pions.  Charge
symmetry would indicate that the primary absorption should be
on a $ppn$ triplet leading to a $pnn$ final state.  In this case,
most of the pion energy would be in neutrons, and hence not
directly observed.  However, if the interaction vertex and
one proton energy is known, and the angles of the outgoing
neutrons are known, the total energy of the three nucleons
can be estimated.  Monte Carlo studies with realistic absorption
models will be needed to determine the accuracies of such estimates.

Although neutral pions escaping the nucleus will decay, usually to two
photons, the mean distance traveled before decay is a few nanometers,
much greater than the size of the nucleus.  Thus the absorption of
neutral pions in the interaction nucleus must also be taken into
account in any study of resonance production.

We have begun
studies with INTRANUKE to determine the sensitivity to 
the probability of pion absorption in the interaction 
nucleus.  We are currently in the process of modifying Monte Carlo routines
to treat pion absorption more realistically. Because \minerva\ will also
have good neutron detection capability, we expect to be able to
substantially improve the determination of the incident neutrino
energy.

We also wish to note that there are essentially no measurements of
pion absorption above 500 MeV.  The fine spatial resolution and full
solid angle detection capability of \minerva\ will allow us to study
these interactions, especially in carbon.

\subsubsection {Nuclear Transparency in Neutrino Interactions}  
  
\begin{figure}[tp]
\center
\epsfxsize=90mm\leavevmode
\epsffile{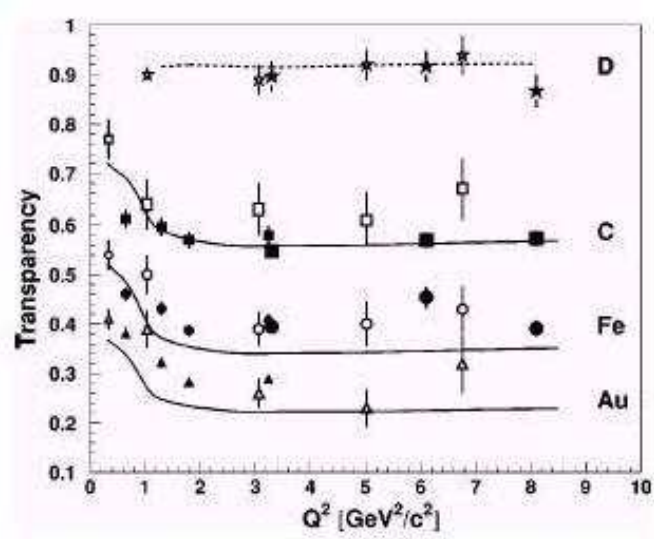}
\caption{Probability for the outgoing proton to escape the
nucleus as a function of $Q^2$.}
\label{fig:transp}
\end{figure}

A second nuclear interaction process which affects the observed final
state energy is the final state interaction of a nucleon in the
struck nucleus.  An outgoing nucleon has a substantial
probability of interacting in the nucleus. These probabilities have
been measured, most recently at JLab, with some precision.  The
experiments used $(e,e'p)$ coincidence reactions.  The cross section
for finding the scattered electron in the quasi-elastic peak was
compared to the cross section for finding the coincident proton.  A
summary of the results are shown in Fig. \ref{fig:transp}.  

In contrast to pion absorption, there is little available information
on what happens to the scattered nucleon.  Of course, most either
scatter from a single nucleon quasi-elastically or produce a pion (for
protons above 600 MeV).  Improving Monte Carlo routines to model this
interaction should allow us to better estimate the total final state
energy.  As with pion absorption, the good resolution, neutron
detection capability, and full solid angle coverage of \minerva\
should allow us to experimentally determine the actual final states
and constrain the Monte Carlo routines.

\subsubsection {Proposed Experimental Analysis}

The NEUGEN Monte Carlo has been used to study the sensitivity of the
\minerva\ experiment to nuclear effects.  The nuclear effects in the
NEUGEN Monte Carlo are controlled by the INTRANUKE processor. This
processor incorporates a probability for pion absorption based on
earlier eletroproduction absorption studies and lower-statistics
Ne/$H_2$ neutrino bubble chamber data.  The observed phenomena of
hadron formation length, which increases the transparency, is
incorporated as well.  The particular model used for pion absorption,
which is currently being improved and updated, assumes that the
absorption process eliminates a pion and the resulting nucleons are
themselves either absorbed in the nucleus or are too low in energy to
be observed in the detector.

\begin{figure}[tp]
\centerline{ \epsfxsize=0.5\textwidth 
\epsffile[40 160 550 650]{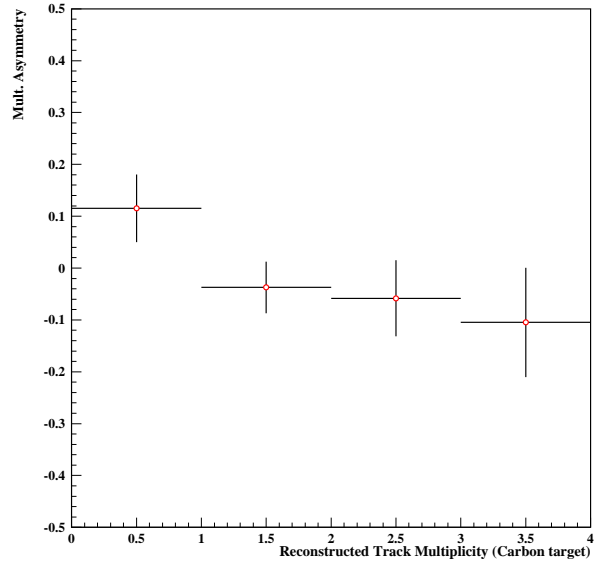}}
\caption{The fractional change in multiplicity distributions between the two values assumed for pion absorption on carbon described in the text.}
\label{M_C}
\end{figure}

\begin{figure}[tp]
\centerline{
\epsfxsize=0.5\textwidth
\epsffile[40 160 550 650]{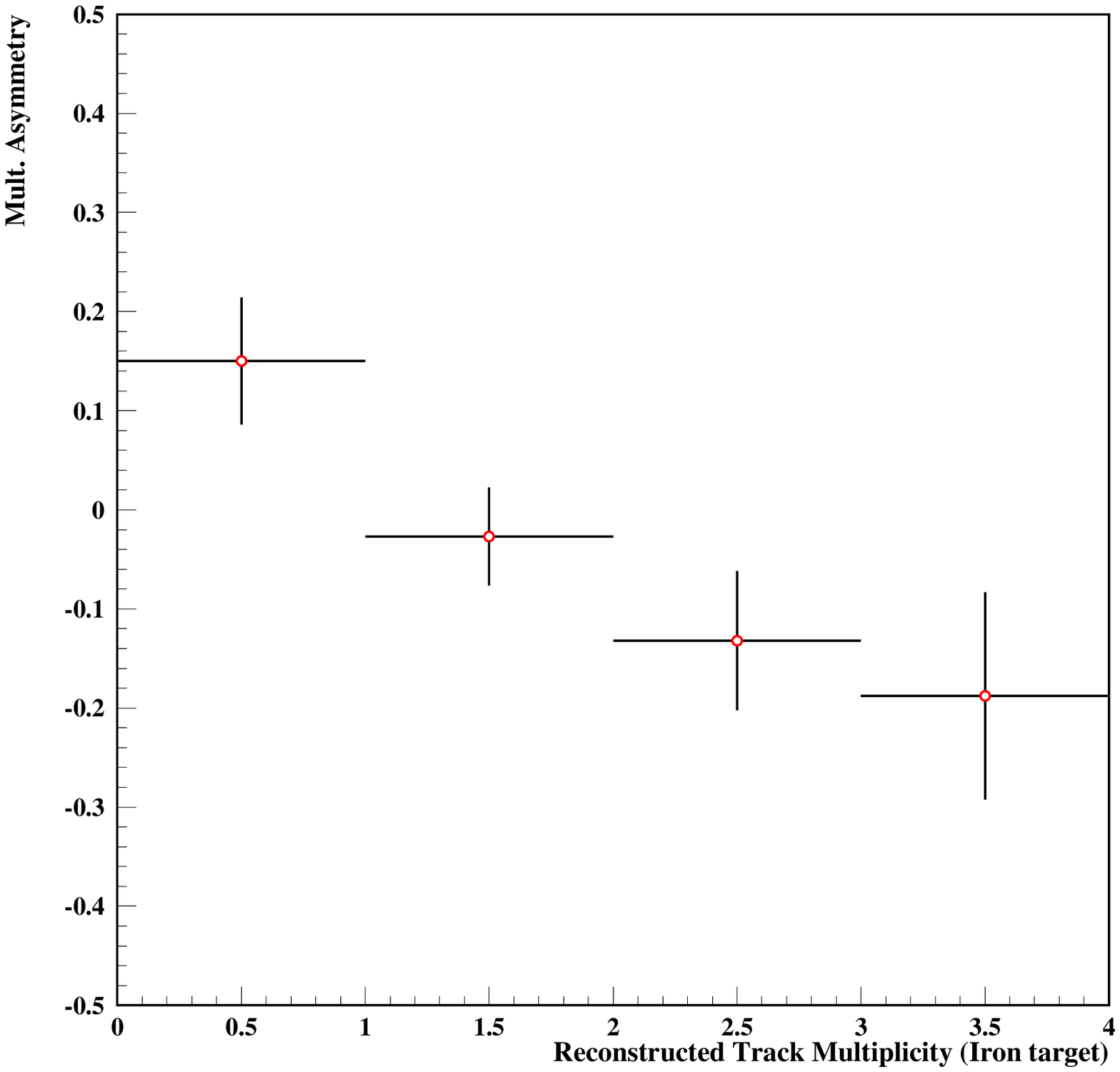}}
\caption{The fractional change in multiplicity distributions between the two values assumed for pion absorption on iron discussed in the text.}
\label{M_Fe}
\end{figure}

\begin{figure}[tp]
\centerline{
\epsfxsize=0.5\textwidth
\epsffile[40 160 550 650]{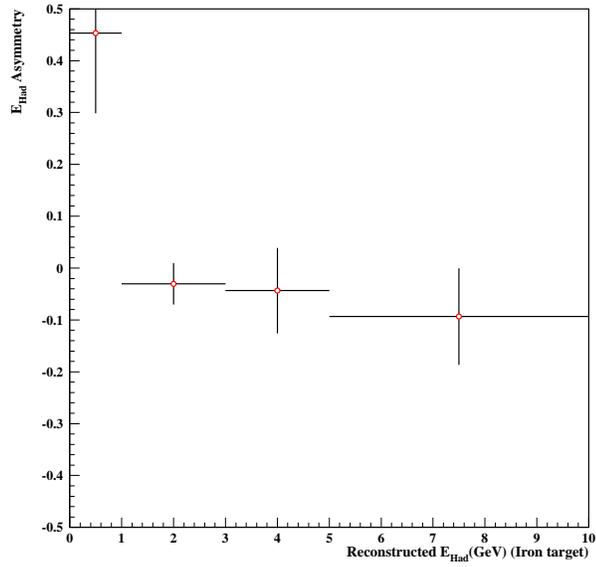}}
\caption{The fractional change in the visible hadron energy distributions between the two values of pion absorption on iron discussed in the text.}
\label{E_Fe}
\end{figure}

\begin{figure}[tp]
\centerline{
\epsfxsize=0.5\textwidth
\epsffile[40 160 550 650]{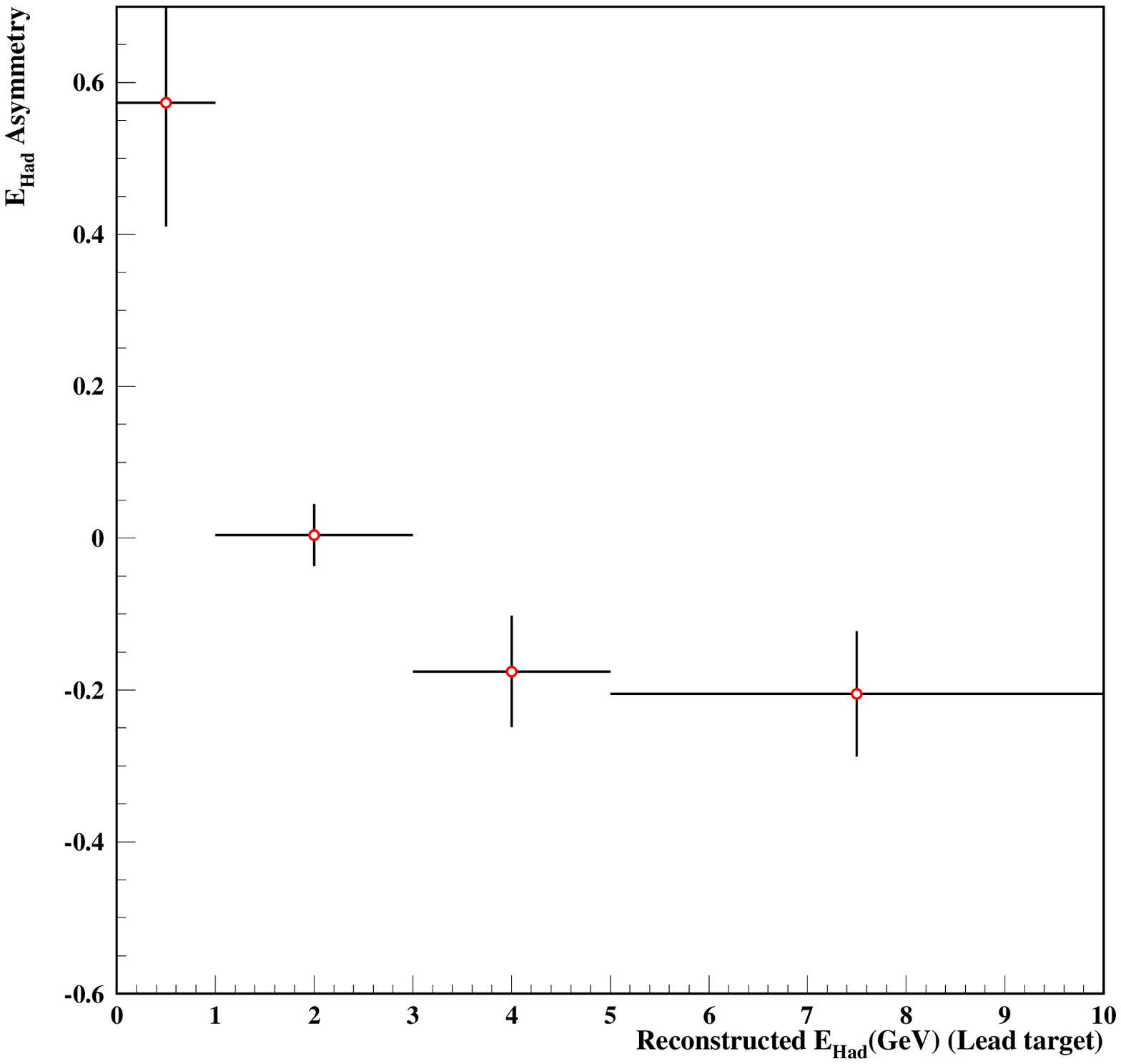}}
\caption{The fractional change in the visible hadron energy distributions between the two extremes in pion absorption on lead discussed in the text.}
\label{E_Pb}
\end{figure}

To determine the sensitivity of \minerva\ measurements to the
predictions of this model, the assumed probability for pion absorption
in INTRANUKE has been increased by three standard deviations 
and then decreased by the
same amount.  The multiplicity and a very crude estimate of the
visible hadron energy have been examined under these extreme
conditions. In the next series of figures the predicted asymmetry in
the multiplicity and visible hadron energy are shown.  Currently we
have only a very small sample of 2500 generated events available for
this analysis.  The asymmetry is defined as the percentage change
under these extreme assumptions.  That is, the bin contents at plus three standard deviations
minus the bin contents at minus three standard 
divided by bin contents at minus three standard deviations.  
Figure~\ref{M_C} shows the predicted change in the multiplicity
distributions for carbon, while Figure~\ref{M_Fe} shows the same
distribution for iron.  Both figures are consistent with the model  
that the decrease
in observed multiplicity is caused by an increase in absorption
probability and therefore the effect should be stronger in Fe than in C.  The final
determination of the visible hadron energy will be an involved process
for this experiment.  For now, we are using the most primitive
estimate of this quantity, namely an uncorrected version derived from 
the total light output of the hadron shower.  
In the data analysis this will be refined for example, through
measurements of stopping/decaying particles.  With this crude
estimate, the change in hadron energy for iron is shown in
Figure~\ref{E_Fe} and for lead and Figure~\ref{E_Pb}.  There is a dramatic change, even with these low simulation statistics, in
the lowest energy bin.  \minerva\ will collect several times these statistics and should be capable of measuring this effect at even higher hadron energy.  

Since the incoming neutrino energy is not {\it a priori} known, the
measured {\bf muon} kinematics will be tested as a basis to compare
characteristics of the visible hadron shower across nuclear targets
and to determine whether the nuclear-effects correction-factor is a
function of the observed muon kinematics.  The muon is relatively free
from nuclear dependent effects and serves well as an A-independent
normalization.  For example, the quantity:
\begin{equation}
Q^\prime = 4 E_{\mu} sin^2({\theta}/2) 
\end{equation}
is representative of the square of the 4-momentum transfer to the
nucleon or quark, weighted (inversely) by $E_{\nu}$.  This quantity then
reflects the energy-momentum transferred to the hadronic vertex.  The
distribution of events in this quantity are peaked toward low $Q^\prime$
with half the events below $Q^\prime$ = 1.0.  The Monte Carlo statistics
are still too low to make any conclusions on this study at this point,
however the trend is encouraging and the study will continue with
increased statistics.

%% file: res_addendum.tex
\subsection{Resonance Production at \minerva}

Resonance production is an important issue both for future oscillation
experiments, where resonant final states comprise much of both the
signal and backgrounds, and for {\minerva} itself where the comparison
of resonance production in electron and neutrino scattering provides
an important motivation for our physics program in the low energy beam
at \numi.  For the latter studies, a joint effort has been launched
between the neutrino and electron communities to better utilize the
available electron scattering data for improved neutrino cross section
modeling, considering nuclear effects as well as production
mechanisms. This collaboration was recently approved by the Jefferson
Lab Program Advisory Committee to measure low $Q^2$, separated
structure functions in the resonance region on a variety of targets of
interest to the neutrino community (JLab Hall C Experiment E04-001).
Many details of the \minerva\ program in resonance production are
given in the proposal\footnote{see chapters 7, 10 and 13 of the
main proposal}.

Broadly, the resonance production measurements will focus
on two areas.  Fundamentally, {\minerva} will measure pion production cross
sections on a variety of nuclear targets nuclear targets, carbon and
heavier.  Secondly, the understanding of nuclear effects gained from
these measurements will allow us to ``extrapolate'' the measurements
to the free nucleon.  Comparing these measurements to electron
scattering will allow a better understanding of the axial component of
neutrino resonance production.  As nuclear effects will be important in
all the measurements, we review below approaches to understanding
the nuclear effects.  This is followed by
a discussion of ongoing studies of
event and particle identification techniques relevant to single pion
production.

\subsubsection{Nuclear Corrections}

The interaction of neutrinos in nuclei produce secondary particles
which propagate though the nucleus. Among them are protons produced in
quasi-elastic scattering and pions produced in the resonance
region. These effects are usually accounted for in Monte Carlo programs,
such as the {\minerva} simulation, and in an analytic method described
here.

The propagation is viewed as a two step process:
\begin{itemize}
\item A proton (QE) or a pion is produced on a bound proton or
   neutron, corrected for Pauli blocking and Fermi motion.
\item The produced particle propagates through the nucleus performing
   a random walk and it may exchange its charge in each interaction.
\end{itemize}
The assumptions allow a general description of the problem based
on charge symmetry and isospin conservation.  The main property is
a factorization of production and subsequent propagation.  For the
produced pions for instance we can write,
\begin{equation}
\begin{array}{c} N_f(\pi^+) \\ N_f(\pi^0)\\N_f(\pi^-)\end{array}
= M \left ( \begin{array}{c} N_i(\pi^+) \\ N_i(\pi^0)\\N_i(\pi^-)\end{array}
\right )
\end{equation}
where the subcript $i$ denotes the number of pions produced initially at
the neutrino interaction and, similarly, the subscript $f$ denotes the
number of pions emerging from the nucleus. The matrix $M$ has a special
form, which follows from isospin symmetry.
\begin{equation}
\frac{M}{A} = \left ( \begin{array}{ccc}
  1-c-d & d    & c \\
  d     & 1-2d & d \\
  c     & d    & 1-c-d
\end{array}
\right )
\end{equation}
The parameter $A$ contains absorption of the pion and the Pauli
blocking in the pion interactions. The parameters c and d involve the
effects of charge exchange in the re-scattering. The formalism can be
used in two ways:
\begin{itemize}
\item To observe the yields in a specific nucleus and compare them to
  production of pions on free protons and neutrons.  Simple algebra
  relates the parameters to the observations.  Thus determining the
  parameters in electroproduction allows one to use them in charge and
  neutral current neutrino reactions.
\item To calculate $A$, $c$ and $d$ theoretically by solving the random walk
  problem. The numbers produced by
  the two methods are compatible with each other.
\end{itemize}
More results are expected from the present experiments and {\minerva}
will contribute on this topic. The method is needed to interpret the
data and to determine oscillation parameters.

\begin{figure}[tp]
\begin{center}
\epsfxsize=0.7\textwidth\epsfbox[50 80 510 390]{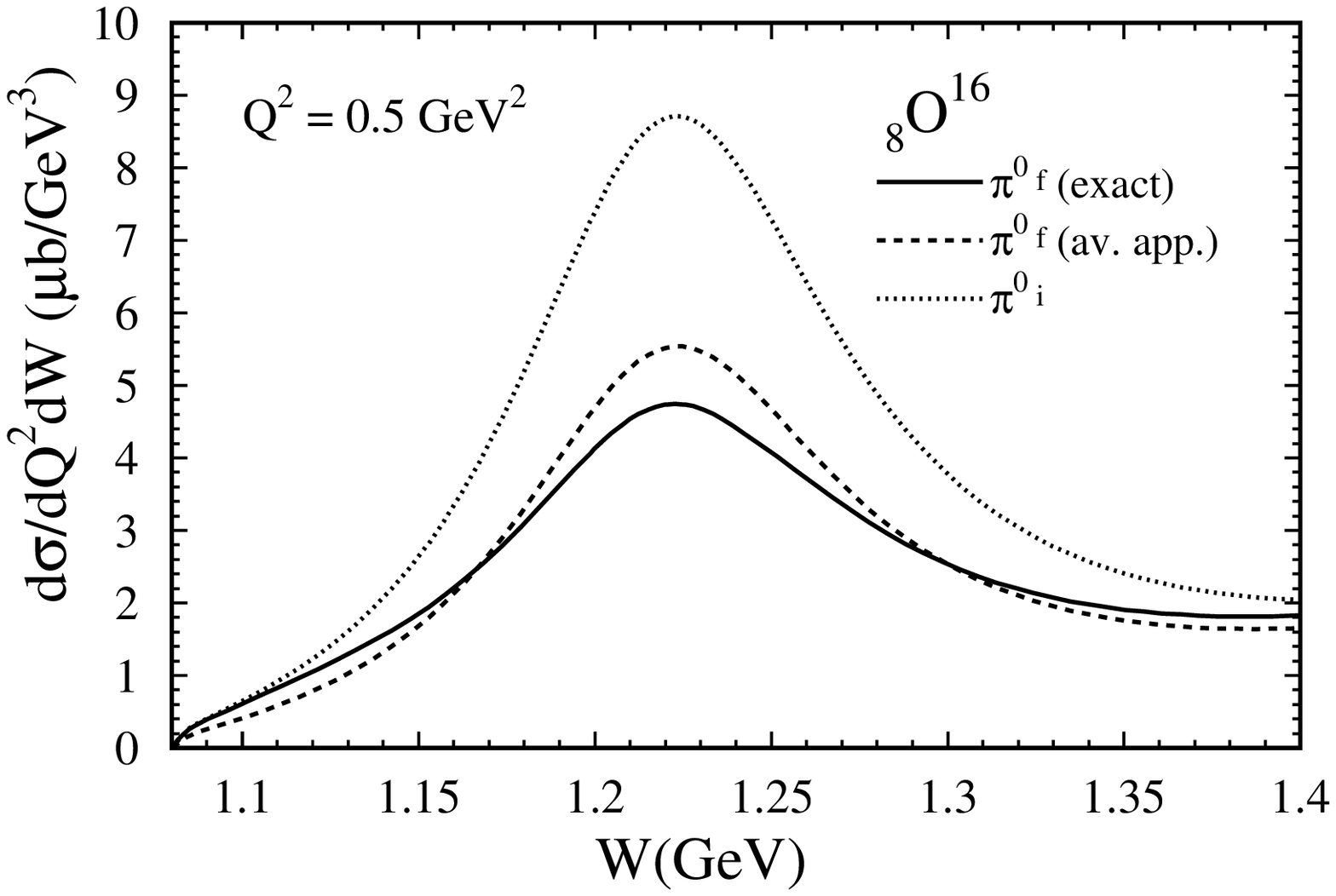}
\epsfxsize=0.7\textwidth\epsfbox[50 80 510 390]{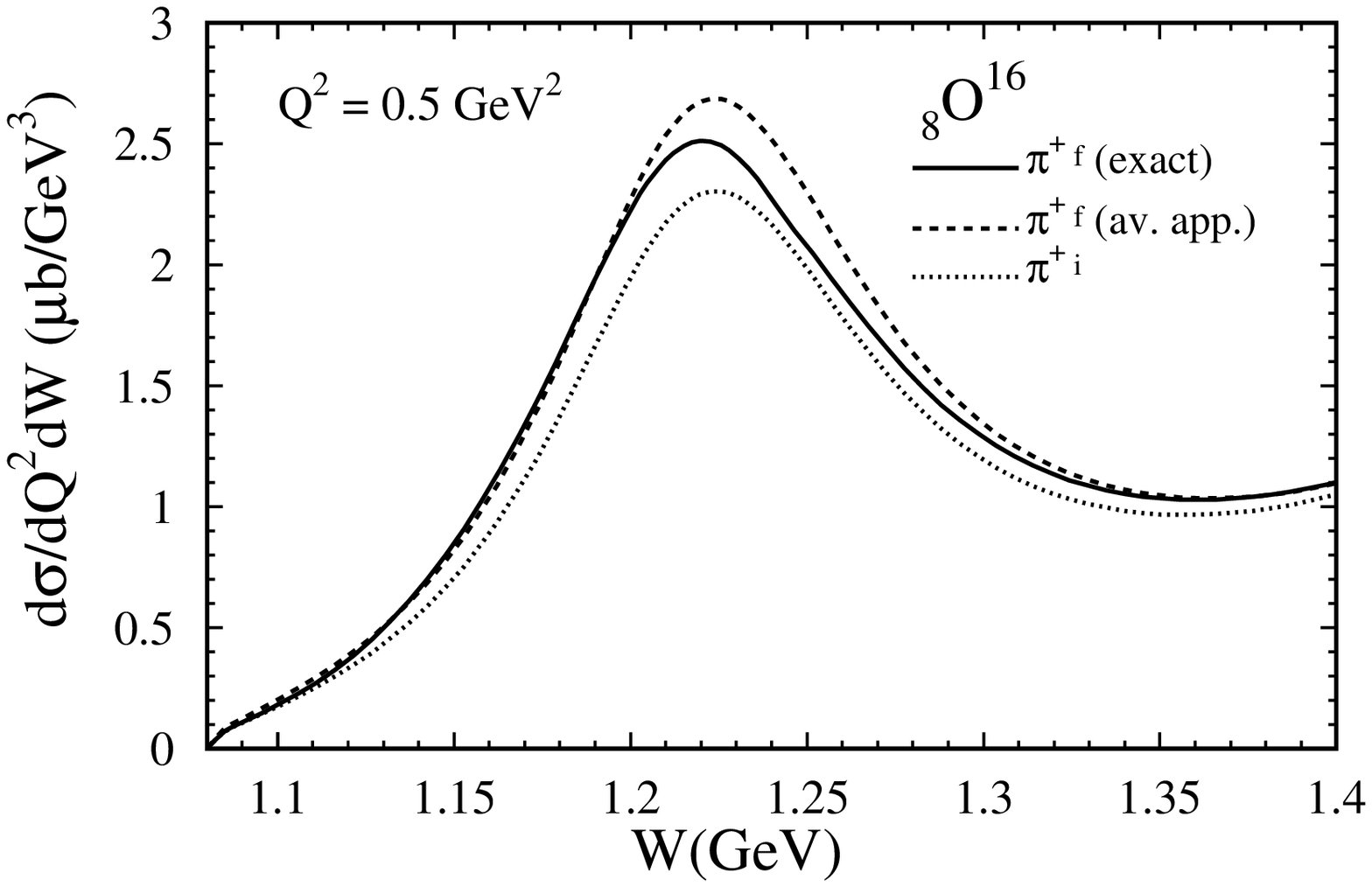}
\end{center}
\caption[Nuclear effects in electroproduction of pions]{Dotted lines show
electroproduction cross sections for {\piz} and {\pip} on an
${}^{16}{\rm O}$ target, with no final state interactions.  The solid curves
show these cross sections after the nuclear effects of pion
re-scattering and absorption, according to the model of Paschos
et.al.\cite{}  The final state interactions greatly attenuate the
{\piz} yield.  However, charge exchange from the larger {\piz}
production cross section serves to roughly balance the losses for the
{\pip} channel, keeping the the {\pip} yield about the same before and
after pion reinteractions.  Except for the overall scale of the cross
section, these same results are predicted for neutrino neutral current
reactions.}
\label{addendumresonant:invmassdist}
\end{figure}

As an illustration, we show in
Figure~\ref{addendumresonant:invmassdist}, the yields of pions in
electroproduction and neutrino experiments on an ${}^{16}{\rm O}$
target.
The curves in the upper graph show the yields for {\piz}s.  The dotted
curve shows the production on free isoscalar target,i.e. $1/2(n+p)$. The
solid curve shows the reduction produced on an oxygen target,
where the change is substantial.  The lower graph shows yields of {\pip}s
where the nuclear corrections are much smaller.

\begin{figure}[tp]
\epsfxsize=0.8\textwidth\epsfbox[25 180 515 515]{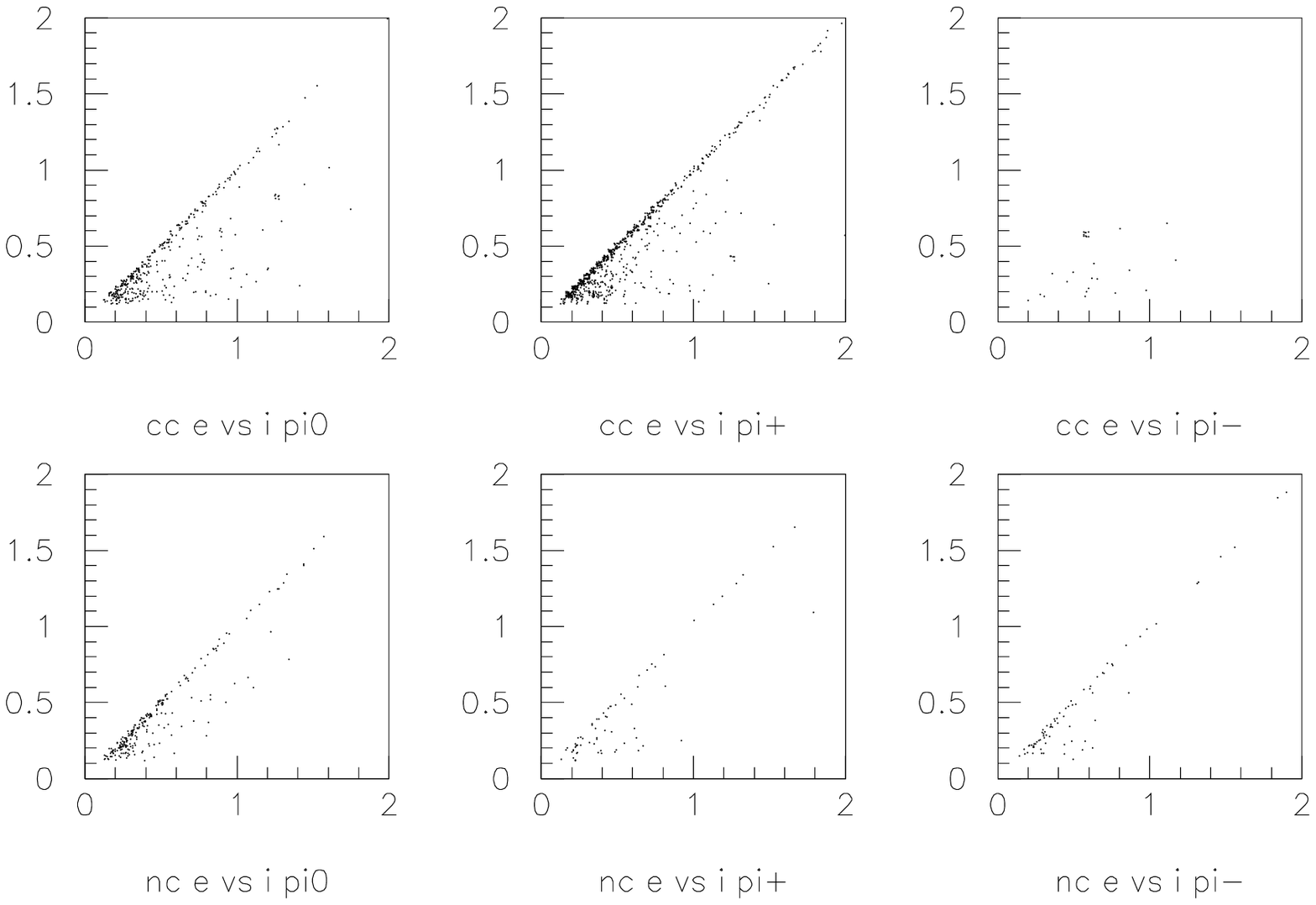}
\caption[Correlation of pion energies before and after reinteraction
in NEUGEN]{
Correlation of pion energies before and after intranuclear
re-scattering as implement in NEUGEN for the {\minerva} simulation.
Events are generated within the central plastic region (carbon and
hydrogen scatterers) of the
detector.  Each plot shows on the vertical axis, the pion energy after
it exits the nucleus versus the energy at the point that it is
produced inside the nucleus.  The first row of figures if for CC
interactions while the second if for NC scattering.  The left column
is neutral pions, while the middle and right columns are for positive
and negative pions.  Pions that do not re-scatter in the nucleus fall
along a diagonal line, while re-scattered pions fall below the line.
These correlations indicate that while re-scattering is important,
pions exiting the nucleus will still carry information about their
initial state.  The top right figure shows the correlation for
negative pions arising from CC scattering.  Production of
single negative pions is forbidden due charge conservation, so
negative pions will only be created by charge exchange from neutral
pions.  
}
\label{addendumresonant:rescatcorrelation}
\end{figure}

The {\minerva} monte carlo simulation, which uses the NEUGEN neutrino
event generator also includes a type of random-walk
simulation of pion re-interactions.  These pion re-scatterings are
implemented using the INTRANUKE \cite{intranuke} pion cascade model.  With this model,
the correlation of pion momenta before and after final state
interactions can be studied.  For ${}^{12}{\rm C}$,
(Figure~\ref{addendumresonant:rescatcorrelation}), while some pions do
scatter and loose momentum, many retain their original momentum.  This
correlation suggests that pion production data on carbon will be able
to constrain free nucleon resonance production cross sections.  While
these correlations are not observable, {\minerva} will be able to test
the pion re-scattering models by studying the $A$ dependence of pion
yields and distributions.

\subsubsection{Event Identification and Particle ID}

In electron scattering, resonance excitation spectra can be measured by
detecting just the electron.  However, in neutrino scattering, $W$ and
$Q^2$ must be reconstructed using energies of the final state hadrons.
Since the proposal, we have begun a program of studying resonance
reconstruction in the {\minerva} detector using using our hit-level Monte
Carlo simulation and the NEUGEN  neutrino event generator.  
The techniques initially being studied
are topological cuts to identify overall reaction type, and particle
identification of individual tracks.

We have performed a first
analysis of single pion production detection efficiencies using simple
topological cuts.  The event selection here is very simple; tracks in
the fully active target pointing to the vertex are counted and are
identified as charged or neutral by the distance between the vertex
and the first hit of a track.  Tracks with a first hit less than $5$ cm
from the vertex are labeled as charged, while tracks further from the vertex are
labeled as neutral (photons from $\pi^0$ decay or neutrons). 

\begin{figure}[tp]
\epsfxsize=0.85\textwidth\epsfbox{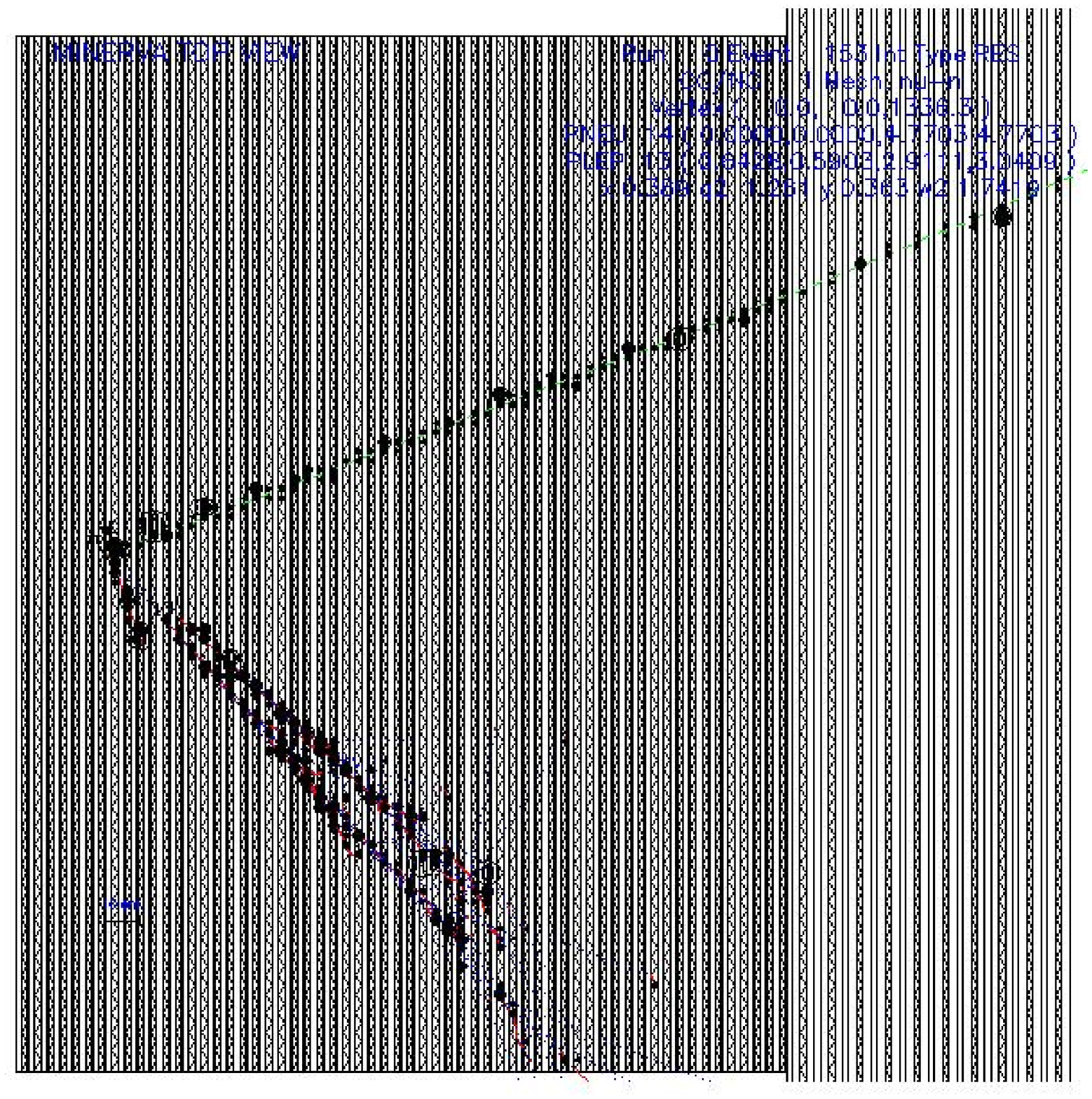}
\caption[Charged Current Resonance event ($\nu N\to p \pi^0$ final state]{
 Event display example of charged current event with a muon, proton and neutral
 pion.  This event passes a simple topological selection of $\pi^0$
 events.  The proton has hits close to the vertex, while the $\pi^0$
 photons convert well away from the vertex. }
\label{addendumresonant:resppi0}
\end{figure}

With these simple track multiplicity and vertex cuts, for example,
$\mu^- p \pi^{\pm}$ events are identified with over a 90\% efficiency
nearly independent of pion energy.  Background processes comprise less
than 2\% of events passing this two charged track cut even without
particle identification cuts.  Events with neutral pions are also
identified with good efficiency with these cuts.  For example, a
topological cut designed to isolate a sample of neutral pions
accompanied by zero or one nucleons and zero or one muons, such as the
$\mu^-p\pi^0$ event illustrated in Figure~\ref{addendumresonant:resppi0},
results in identification of a sample containing 75\% of single
neutral pion resonance events with only 10\% background.  This is in
the absence of particle identification requirements on the $\pi^0$,
such as invariant mass reconstruction techniques that are described in
the \minerva\ proposal\footnote{see Sections 8.3 and 13.6.2 of the
main proposal}.

\begin{figure}[tp]
\epsfxsize=\textwidth\epsfbox[25 180 515 515]{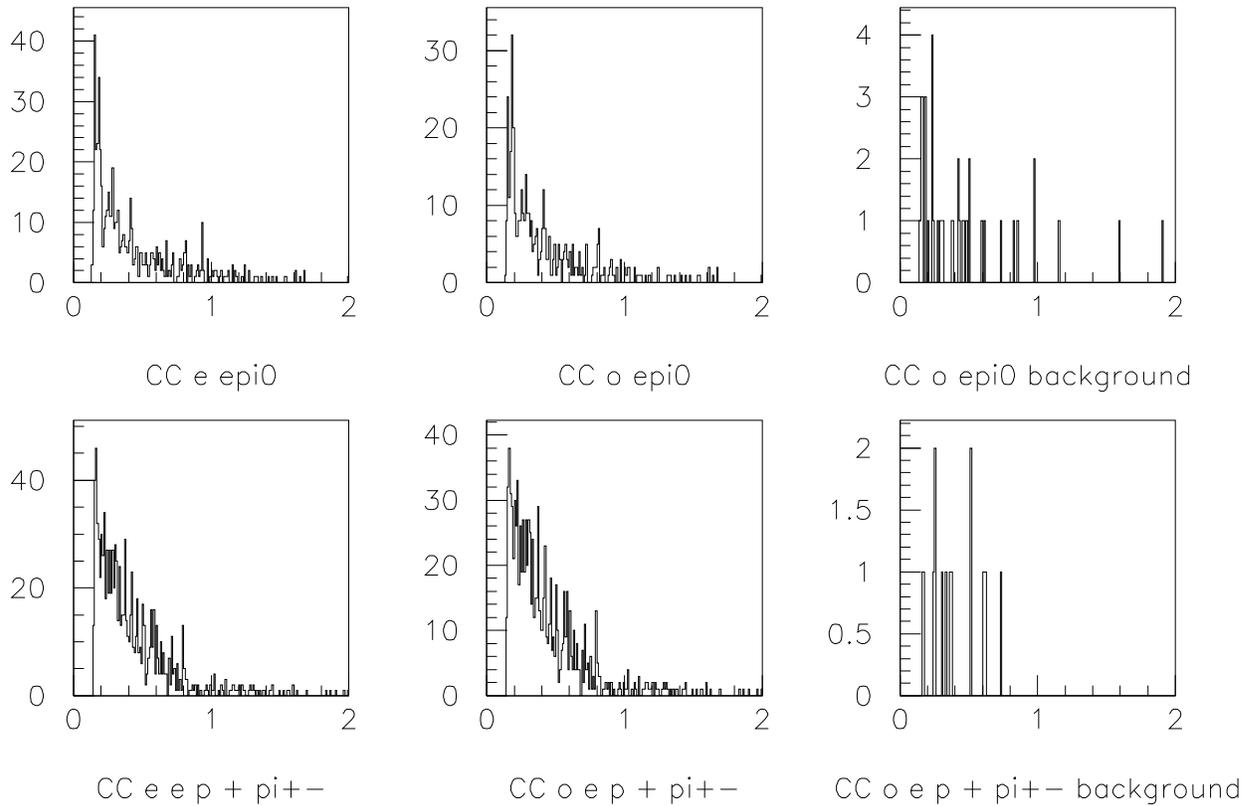}
\caption[Distributions of identified pion events]{The top row shows
  pion energy distributions for events that contain a single $\pi^0$ and
  zero or one nucleons, while the bottom row is of events with a
  single charged pion (positive or negative) and a single proton.  The
  left column shows the distributions using known/truth particle
  identification from the monte carlo. The second and third column are
  events that have been identified using the simple topological cuts
  discussed in the text.  The middle column is events that have been
  correctly identified, while the right column is events of other
  types that have been misidentified as the respective single pion
  event types.  The simple cuts correctly identify about 75\% of the
  $\pi^0$ events, admitting about a 10\% back from other types of events.
  The charged pion efficiency is about 95\% with less than 2\%
  contamination of background events.  Within the statistics of this
  sample, the
  efficiencies of event type identification are independent of pion energy.
}
\label{addendumresonant:ident}
\end{figure}

For charged pions, this topological analysis depends on detecting a
nucleon with the pion, which biases against pions with low momentum
nucleons.  Since the full tracking will likely be able to distinguish
pions by $dE/dx$, particularly for low energy pions, it should be
possible to produce inclusive energy and angle spectra of single pions
that can be compared directly against pion re-scattering models such as
Paschos, et.al., described above,  or the pion nuclear cascade models
in neutrino event
generators.  The shapes of initial and re-scattered pion spectra
predicted by NEUGEN are shown in Figure \ref{addendumresonant:ident}.
Additional capabilities in particle identification,
such as identifying production of pions in ``forbidden charge''
states, e.g., $\mu^-p \pi-$, 
will provide a check of pion reinteraction models as these can not
arise from single pion production on a nucleon through the charged
current reaction.  Understanding these reinteractions is ultimately
very important for oscillation analysis as it is these reinteractions
that limit the ability to transfer knowledge of resonance production 
from charged lepton data to neutrino scattering.

In addition to identification of resonance production events by simple
vertex topology, particle identification will be required to
differentiate the proton and charged pion in a $p\pi^{\pm}$ final
state, and possibly identify pion charge.
As described in the proposal, 
particle identification in {\minerva} will rely on measuring discriminating 
particle characteristics, such as specific energy loss ($dE/dx$) as well as 
topology (hadron and electromagnetic showers, decay signatures).  
As noted above, photons 
will appear as showers disconnected from the vertex, and {\piz} will
appear as
two photons pointing back to the vertex, with invariant mass 
consistent with the {\piz} mass. 

\begin{figure}
\begin{center}
\epsfxsize=0.7\textwidth
\epsfbox{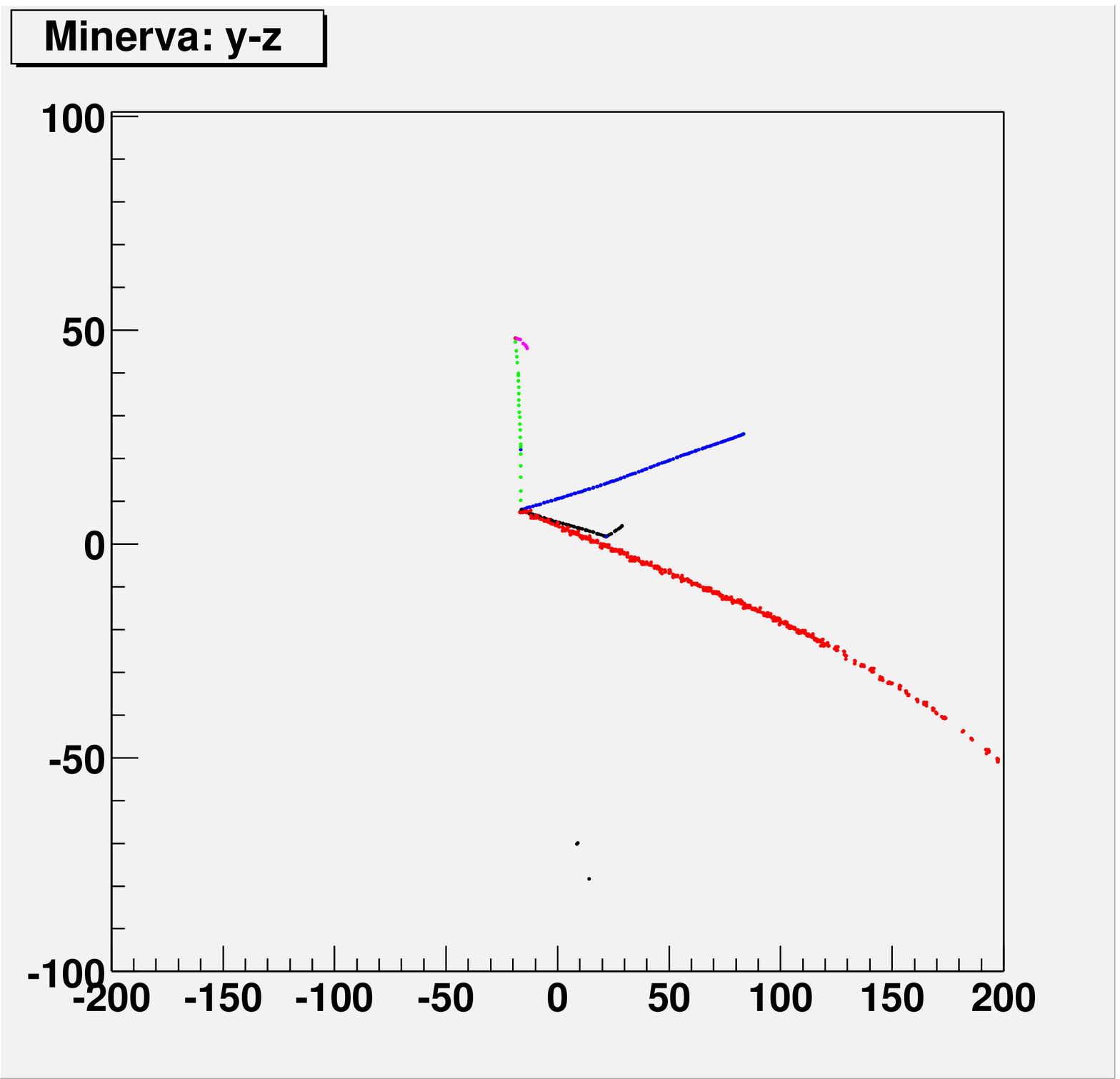}
\end{center}
\caption{A $\pi^+$ Decay Chain in \minerva}
\label{tz:fig1}
\end{figure}
Of particular consideration are low energy particles that stop in the
target area without entering the magnetic field. We have already
demonstrated in the proposal our ability to separate $\pi^{\pm}$ from
$K^{\pm}$ and protons.  However, we would like to be able to
distinguish $\pi^+$ from $\pi^-$ in the inner tracker alone.  Indeed,
there are physics processes visible in the highly segmented low density target
can allow such separation.  Shown in Figure~\ref{tz:fig1} is the
display of a neutrino interaction in the target area before GEANT
digitization.  A muon (red), proton (blue), $\pi^+$ (green) that
decays in a $\pi\to\mu\to e$ sequence (green-magenta, the muon is not
seen).  However, a $\pi^-$ (black) scatters before stopping without
visible interaction or may charge exchange with nuclei in the target
to produce a neutral pion.

\begin{figure}
\begin{center}
\epsfxsize=0.45\textwidth
\epsfbox{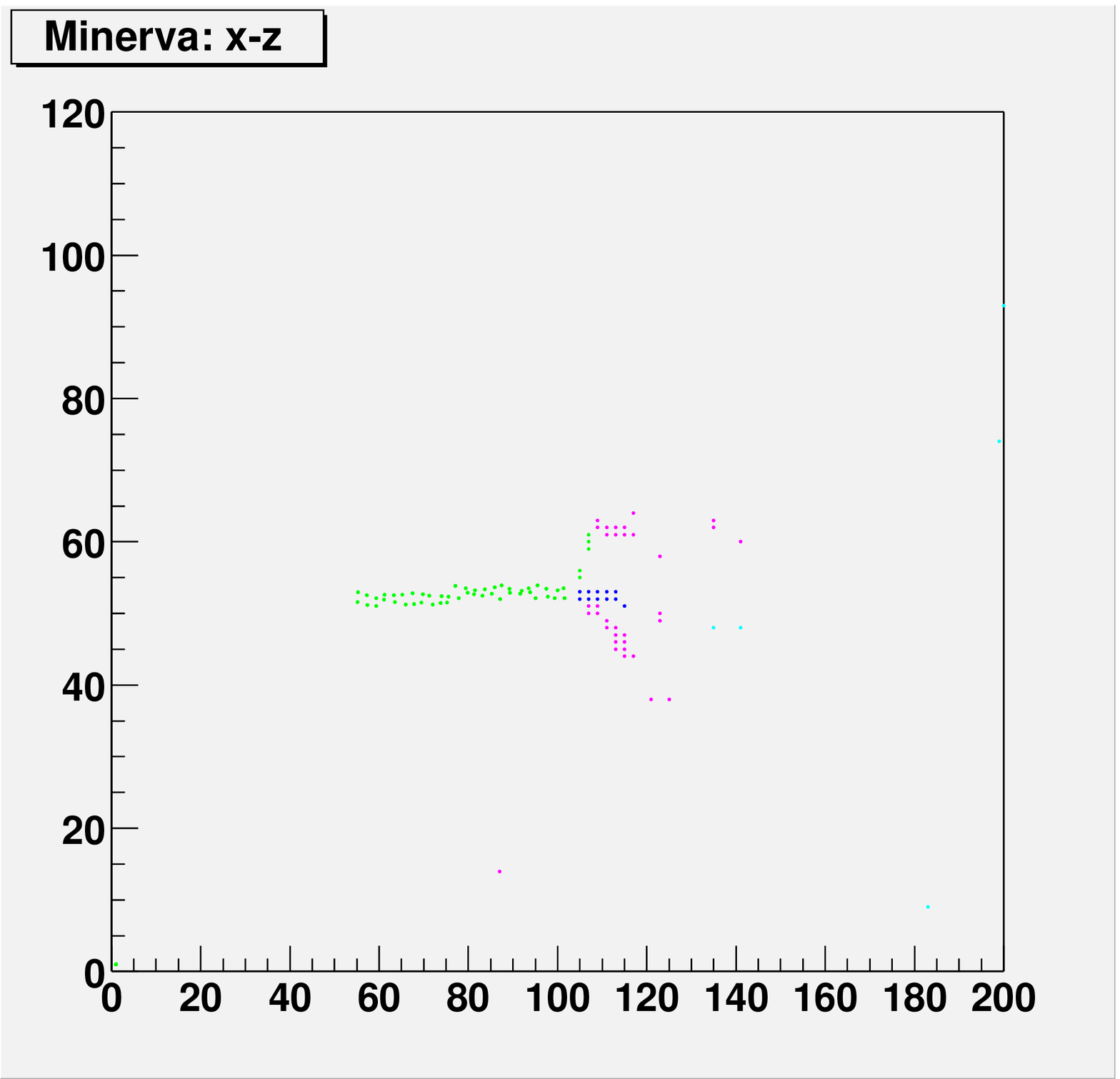}
\epsfxsize=0.45\textwidth
\epsfbox{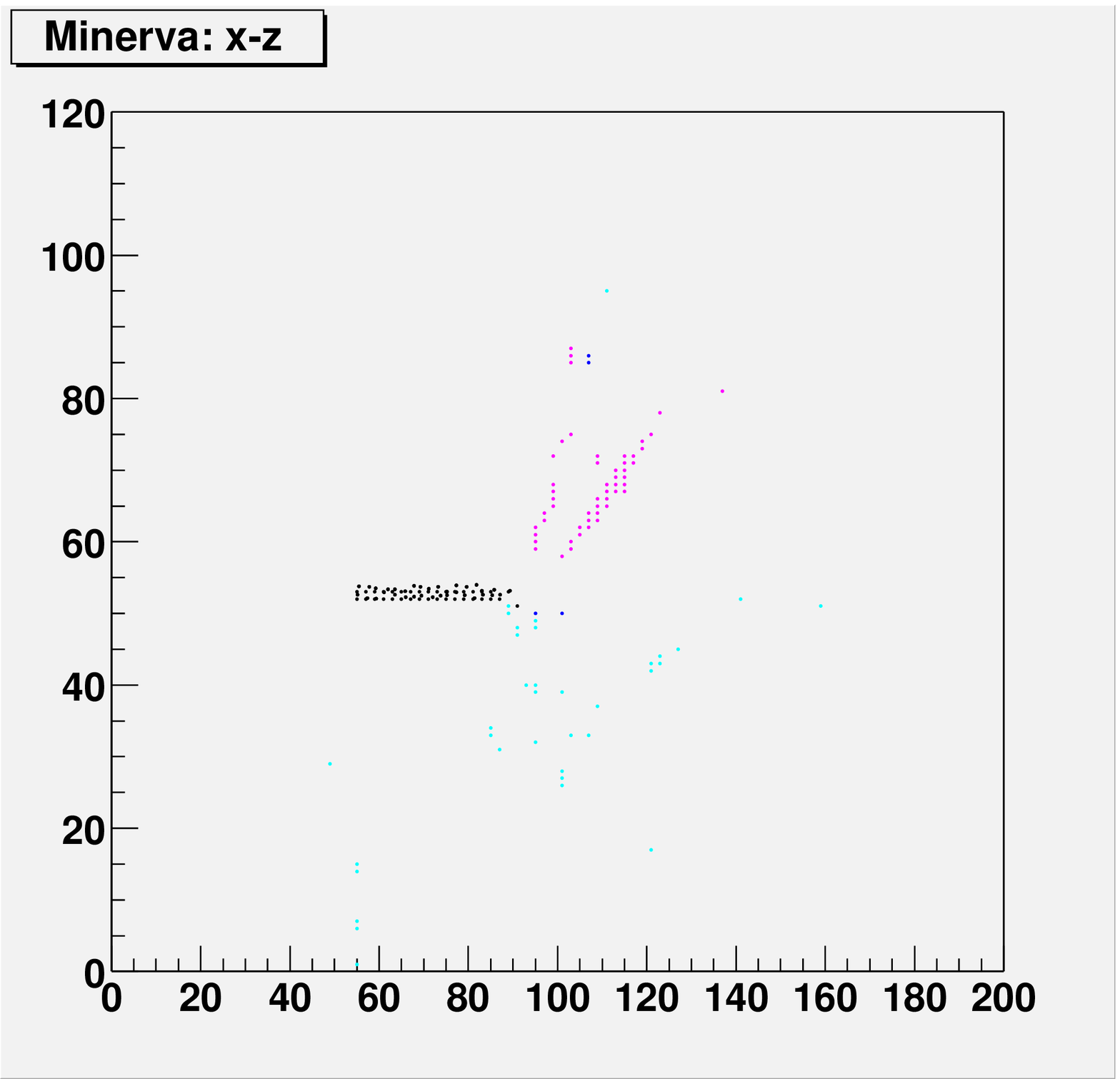}
\end{center}
\caption{$\pi^+$ and $\pi^-$ Interactions in \minerva}
\label{tz:fig23}
\end{figure}
\begin{figure}
\begin{center}
\psfig{file=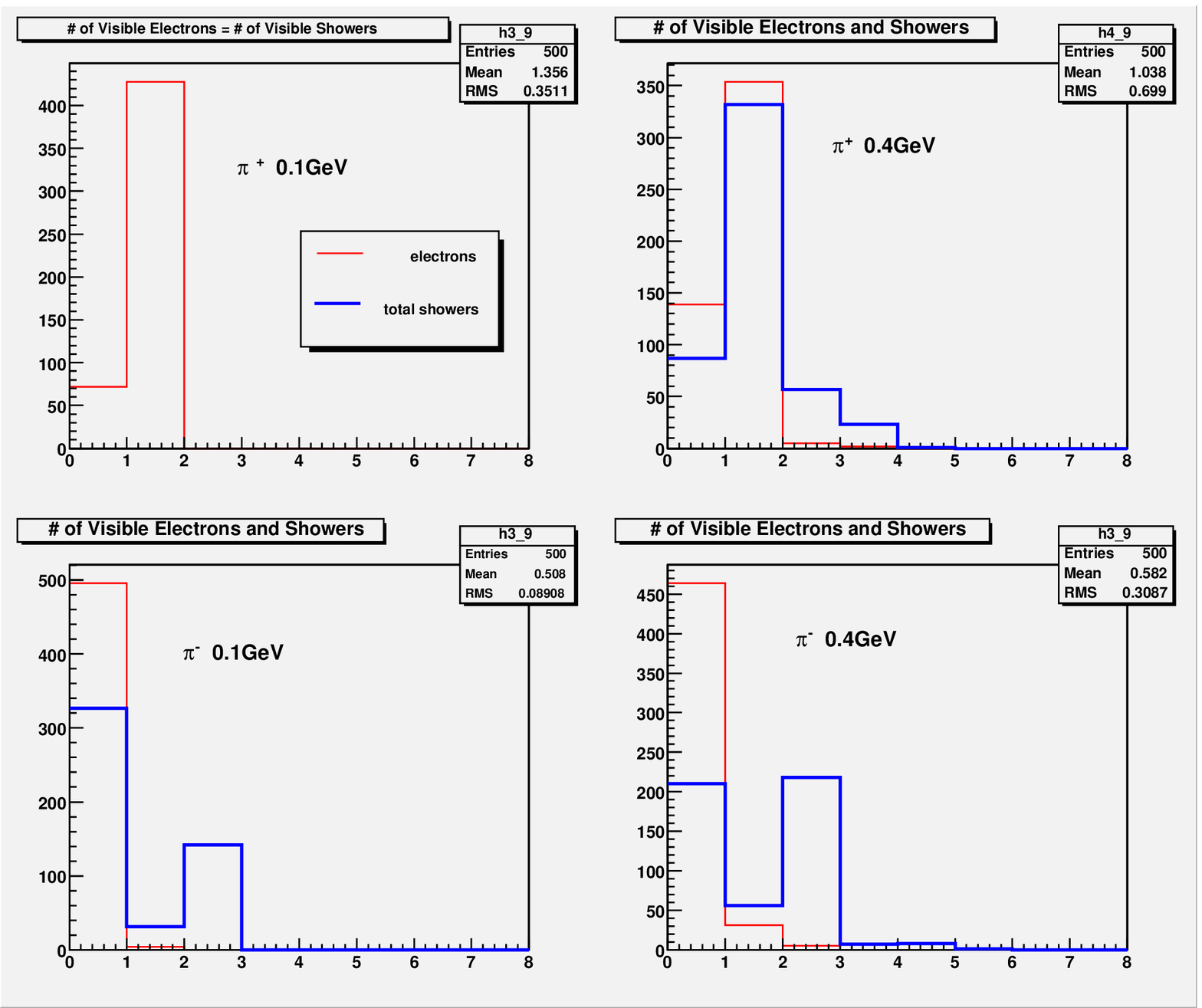,width=4.5in,height=3.0in}
\psfig{file=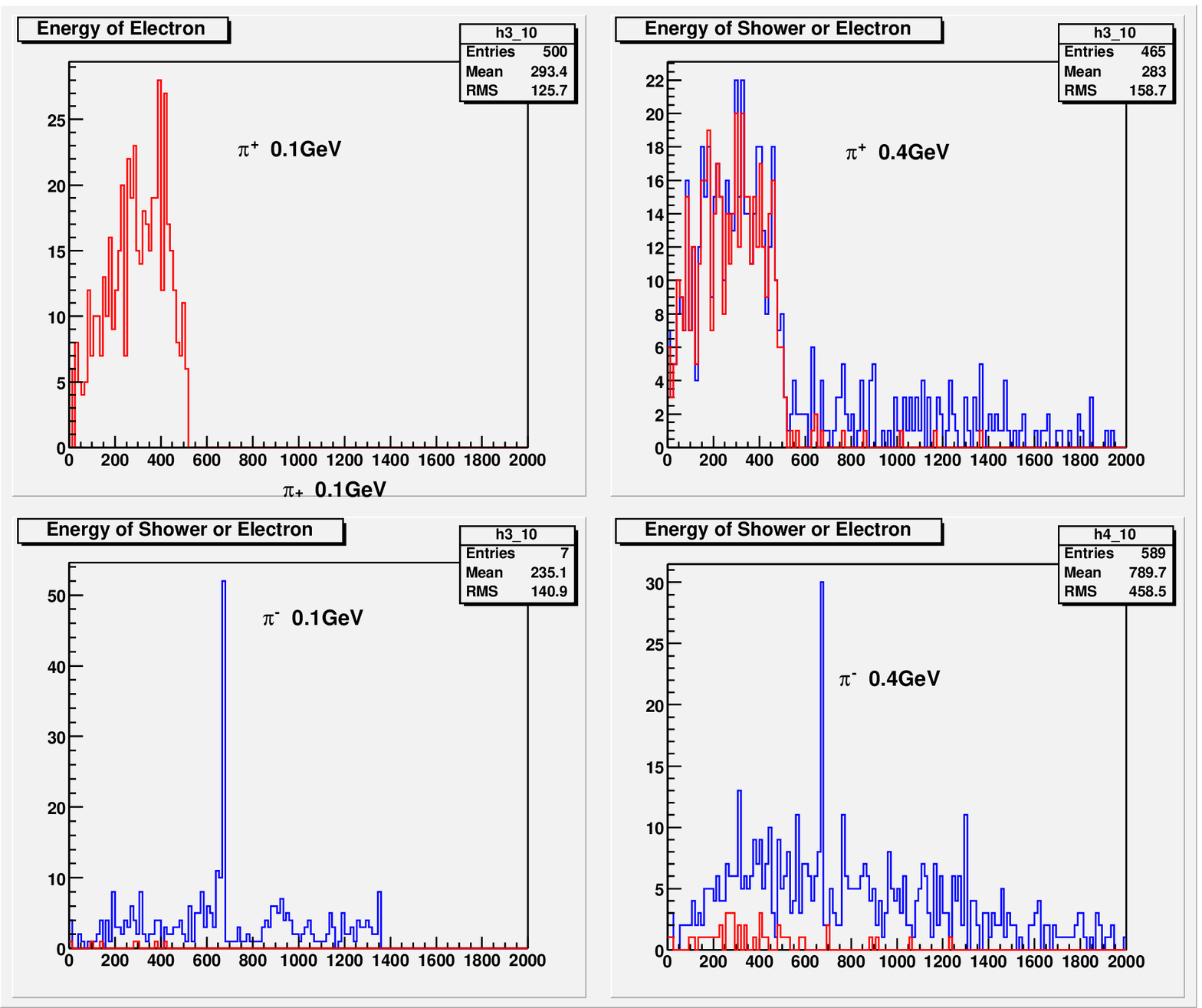,width=4.5in,height=3.0in}
\end{center}
\caption{Electron shower multiplicity and energies for $\pi^+$ and $\pi^-$ Interactions in \minerva}
\label{tz:fig45}
\end{figure}
In order to study the $\pi^+$/$\pi^-$ differences we generated 
pion interactions at 0.1, 0.4, 0.8, and 1 GeV/c in our target using
the \minerva\ simulation.  As a case study, the two interactions in
Figure~\ref{tz:fig23}, show a 1 GeV/c $\pi^+$ 
in \minerva.  The final $\pi^+$ 
decays ($\pi\to\mu\to e$) with the muon not seen, but with a clear 
positron (magenta). A proton (blue) is also seen coming out of 
the vertex as well as two gammas (magenta) pointing to the vertex.  
The $\pi\to\mu\to e$ sequence is characteristic to all stopping $\pi^+$.
In the second interaction, we see a $\pi^-$ interaction with charge
exchange $\pi^-p\to\pi^0 n$ as seen in the central detector (after 
GEANT digitization).  The  characteristic $\pi^0$ (two showers magenta) 
accompanies low energy $\pi^-$.
These differences in final
state can be observed by studying the electron shower multiplicity and
energy associated with charged $\pi$ tracks as shown in
Figure~\ref{tz:fig45}.  We conclude that although these issues need
still more study that the \minerva\ inner tracker
is capable of statistical $\pi^{\pm}$  separation
even without a magnetic field.

%% file: MINERVA-osc.tex
\subsection{\minerva\ and Oscillation Measurements }

     Examples of oscillation measurements that will benefit from 
\minerva\ are the $\Delta m^2$ determination by
MINOS, and the $\nu_\mu \to \nu_e$ oscillation probability measurement by
the NuMI Off-axis $\nu_e$ Appearance (\nova) and the T2K
experiments.  In chapter 13 of the proposal we presented an estimate
for how much uncertainties in nuclear effects would contribute to
uncertainty in $\Delta m^2$ in MINOS; here we update that estimate
with an improved treatment of nuclear effects, and  
consider two consequences of these effects separately.  After
discussing the $\Delta m^2$ measurement we then describe how \nova\
can utilize \minerva\ measurements of cross sections to minimize the
systematic error on its oscillation probability measurement.  
In the proposal we
described the challenge of measuring the $\nu_\mu \to \nu_e$ oscillation
probability due to the uncertainty in backgrounds that must be
rejected at a high level, and improvements that  \minerva\ can 
provide to
neutrino event generators that will used to predict those backgrounds.
We present here a new analysis (which is also part of the \nova\
proposal, chapter 9) which shows quantitatively how uncertainties in
cross sections translate to uncertainties in background
predictions.  We also show that information
\minerva\ provides will be important regardless of the size of
oscillations that \nova\ observes. If \nova\ sees no evidence for
$\nu_e$ appearance then \minerva's most important contribution
will be to improve the determination of the
background at the far
detector.  However, if \nova\ does see a signal, then the reduction of the
cross section uncertainties plays an
even larger role in helping \nova\ achieve the best precision on its
measurement of the $\nu_\mu \to \nu_e$ oscillation probability.

\subsubsection{Nuclear Effects and a $\Delta m^2$ measurement} 

     The key to a precise measurement of $\Delta m^2$ is the ability
to measure the oscillation probability as a function of neutrino
energy.  Although MINOS has undergone an extensive program to determine
the response of its near and far detectors
to specific charged particles,
it cannot measure the likelihood with which those particles are
produced in a neutrino interaction.  At these low neutrino energies,
there are two effects that become important, and therefore
contribute significantly to the uncertainty in a
measurement of $\Delta m^2$.  One effect, which is independent of the
target nucleus, results from the rest masses of the secondary particles
which contribute an important fraction to the reconstructed neutrino energy.
A measurement of final state particle multiplicities and species as a function
of hadron energy, which cannot be measured in MINOS, 
is therefore important for 
accurate reconstruction of the neutrino energy spectra at both
MINOS detectors. Secondly, as shown in the
proposal, the multiplicity distribution is a function of target
nucleus, since secondary particles can either scatter in the nucleus,
or be completely absorbed.  Either process results in a change in the
visible hadron energy for an event, again contributing to the uncertainty
in the measurement of the neutrino energy spectrum.

\begin{figure}[tp]
\epsfxsize=\textwidth
\epsfbox{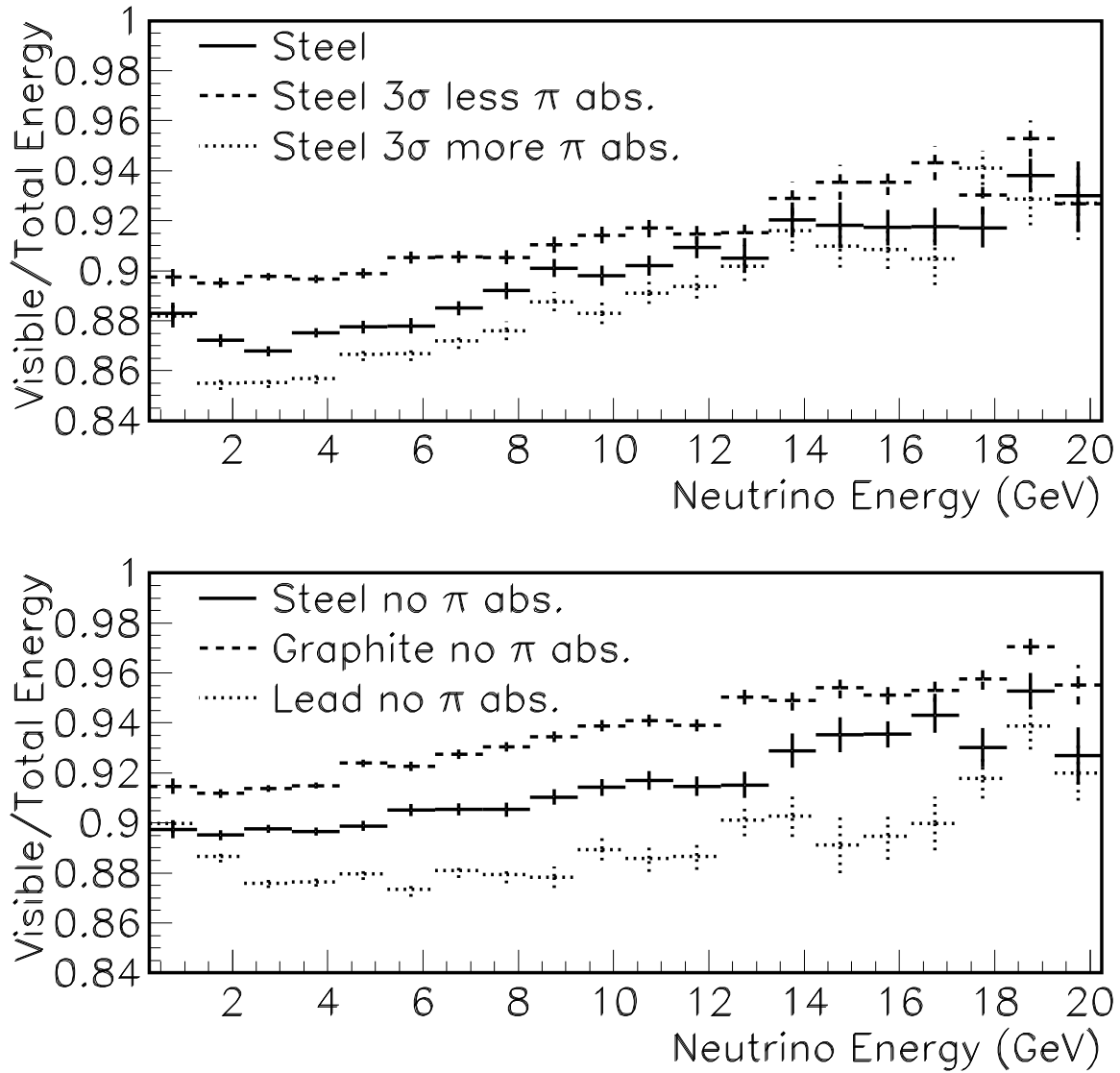}
\caption[Ratio of Visible to True Neutrino Energy versus Neutrino
  Energy with Nuclear Effects]
{Ratio of visible to true neutrino energy for several different 
models of nuclear effects.  The top plot shows the ratio for steel (solid) 
with the nominal pion absorption, as well as the same ratio for the pion
absorption cross section varied by plus and minus three standard deviations.  
The bottom plot shows the differences in the average of this ratio for three 
different target nuclei, where the absorption effects are turned off to 
see more clearly the effects of pion rescattering.}
\label{fig:oscplot_a}
\end{figure}
\begin{figure}[tp]
\epsfxsize=\textwidth
\epsfbox{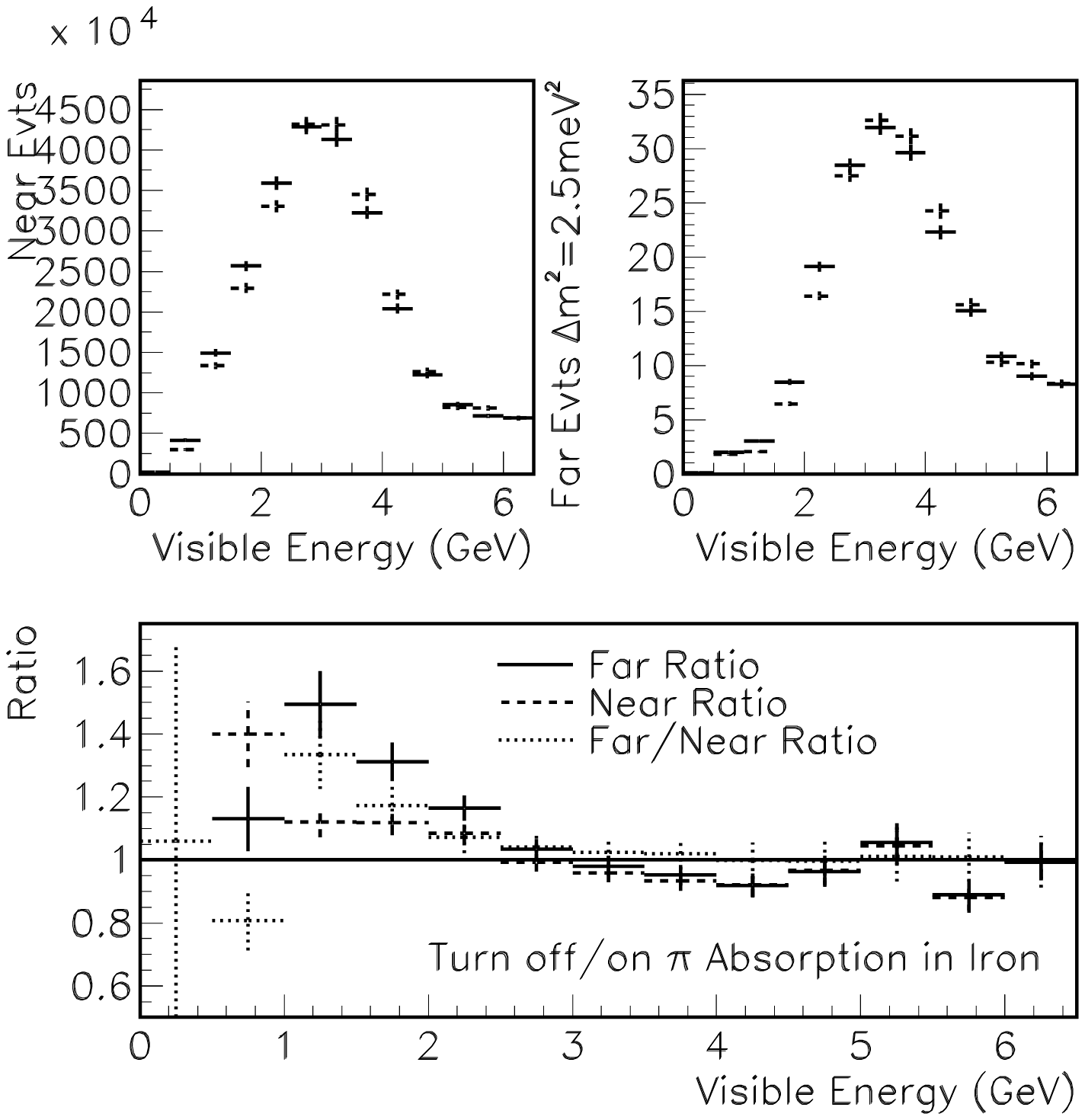}
\caption[Near and Far Neutrino Visible Energy Spectra for Different
  Assumed Pion Absorption Effects]{{\bf (Top left)} visible neutrino
  energy spectrum for the near detector for $\nu_\mu$ CC events, for
  nominal and three sigma high pion absorption in iron.  {\bf (Top
  right)} visible neutrino energy spectrum for the far detector for
  $\nu_\mu$ CC events, again for nominal and three sigma high pion
  absorption.  $\Delta m^2$ in this case is assumed to be $2.5\times
  10^{-3}eV^2$.  {\bf (Bottom)} Ratio of changed divided by nominal
  pion absorption model, for the far (solid) and near (dashed) energy
  spectra, as well as the ratio of far over near.  Note that the
  effect of pion absorption cancels somewhat in the ratio of far over
  near event spectra, but not completely.}
\label{fig:oscplot_b}
\end{figure}

Figure \ref{fig:oscplot_a} shows the changes in the ratio of visible
to total neutrino energy for uncertainties in 
absorption (top) and scattering (bottom)
separately.  In the proposal these two effects were not shown
separately. Furthermore, nuclear effects were assumed to be the same
for carbon as for steel, and a flat 20\% uncertainty on the  
nuclear plus rest mass effect was assumed. The visible energy is defined
as the sum of the kinetic energies of all the charged final state
particles, plus the total energy for the neutral pions, (since it is
assumed they deposit all their energy in the form of electromagnetic
showers).  In the top plot, for a steel target, the
parameter in the event generator that describes pion absorption is
varied by three standard deviations. In the bottom plot pion absorption
is turned off, and the visible energy to total energy ratio
is compared for steel, carbon, and lead.  Figure
\ref{fig:oscplot_b} shows the effect of changing the pion absorption
on the visible neutrino energy spectra for both near
and far detectors (with $\Delta m^2$ is $2.5\times 10^{-3}eV^2$). 
The effect is also shown (Figure \ref{fig:oscplot_b} bottom)
as a ratio for both the near and far
spectra as well as the effect of the change on the far/near ratio.
It is clear that although the visible energy difference produces the
same kind of
effect in near and far spectra, it only partially cancels because of the
different underlying neutrino spectra.

\begin{figure}[tp]
\epsfxsize=0.85\textwidth
\epsfbox{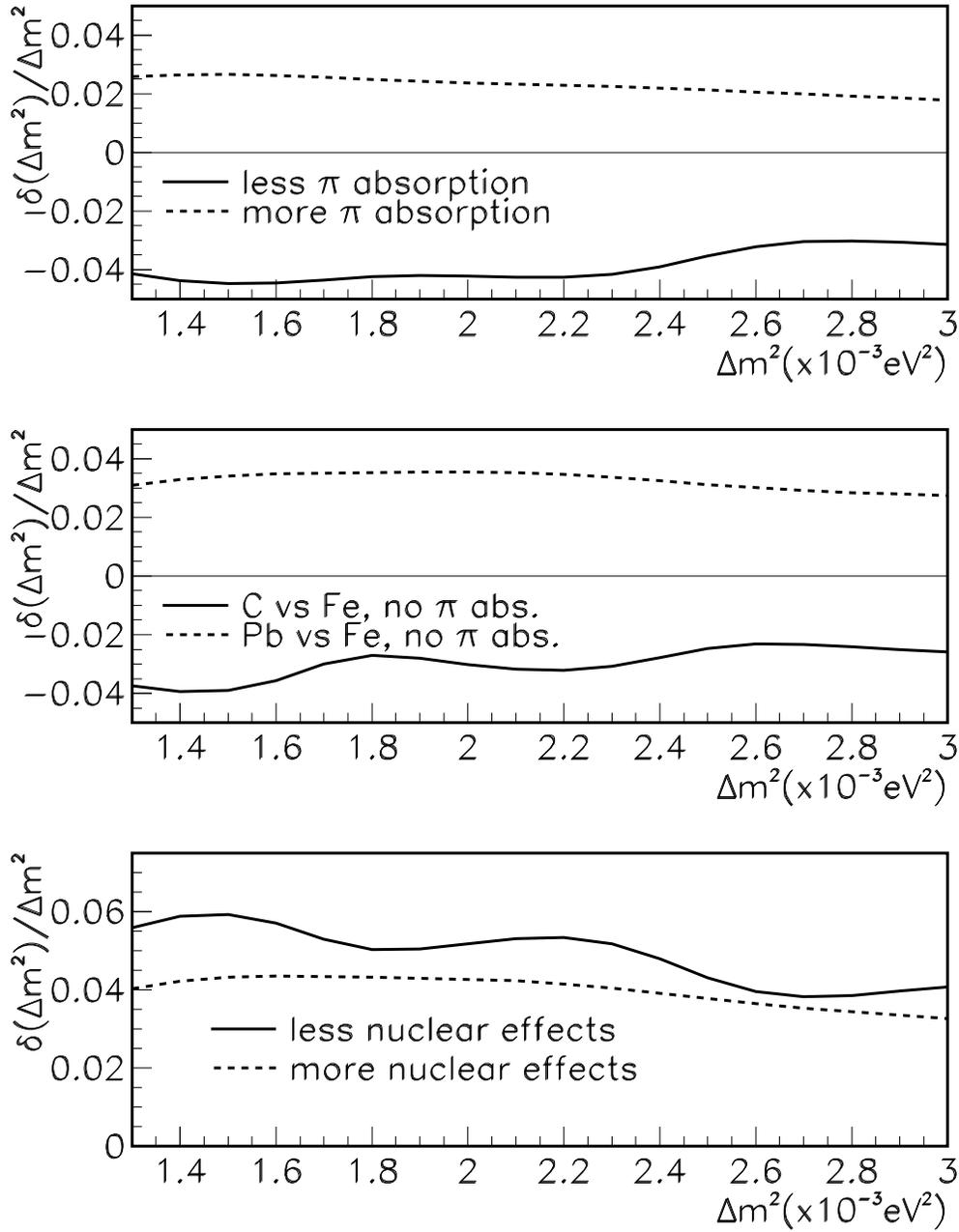}
\caption[Change in Measured $\Delta m^2$ with Different Models for
  Nuclear Effects]{{\bf (Top)} Increase (decrease) in the measured $\Delta m^2$
due to a decrease (increase) in the pion absorption cross section on
Iron.  Middle plot: Increase (decrease) in the measured $\Delta m^2$
due to assuming the target had the nuclear effects of Carbon (Lead)
compared to Steel, when pion absorption is turned off.  {\bf (Bottom)}
The errors due to increases (decreases) in $\Delta m^2$, added in
quadrature, as a function of $\Delta m^2$.}
\label{fig:oscplot_c}
\end{figure}

Evaluating the appropriate uncertainty in the size of nuclear effects
in neutrino scattering is non-trivial, because the only precise
data on differences between nuclei 
come from charged lepton scattering, and one has
to use theoretical models to translate the effects from the charged
leptons to the neutral leptons.  The ``three standard deviations'' 
in pion scattering were measured in a bubble chamber experiment, not  
on steel, therefore, a theoretical extrapolation must be
made for the target in this case. If we naively take the
differences described above as ``one standard deviation'' for pion
absorption, and the differences between steel and carbon or lead
as ``one standard deviation'' for pion rescattering, we can
determine how this systematic error
would compare to the MINOS statistical error.  The
top plot in Figure \ref{fig:oscplot_c} shows how raising and lowering
pion absorption would lower and raise, respectively, the measured
$\Delta m^2$.  The central plot shows how assuming nuclear effects for
lead or carbon, compared to steel, would again lower or raise,
respectively, the measured $\Delta m^2$.  The bottom plot in Figure
\ref{fig:oscplot_c} shows the two sets of errors added in quadrature.

\begin{figure}[tp]
\epsfxsize=\textwidth
\epsfbox{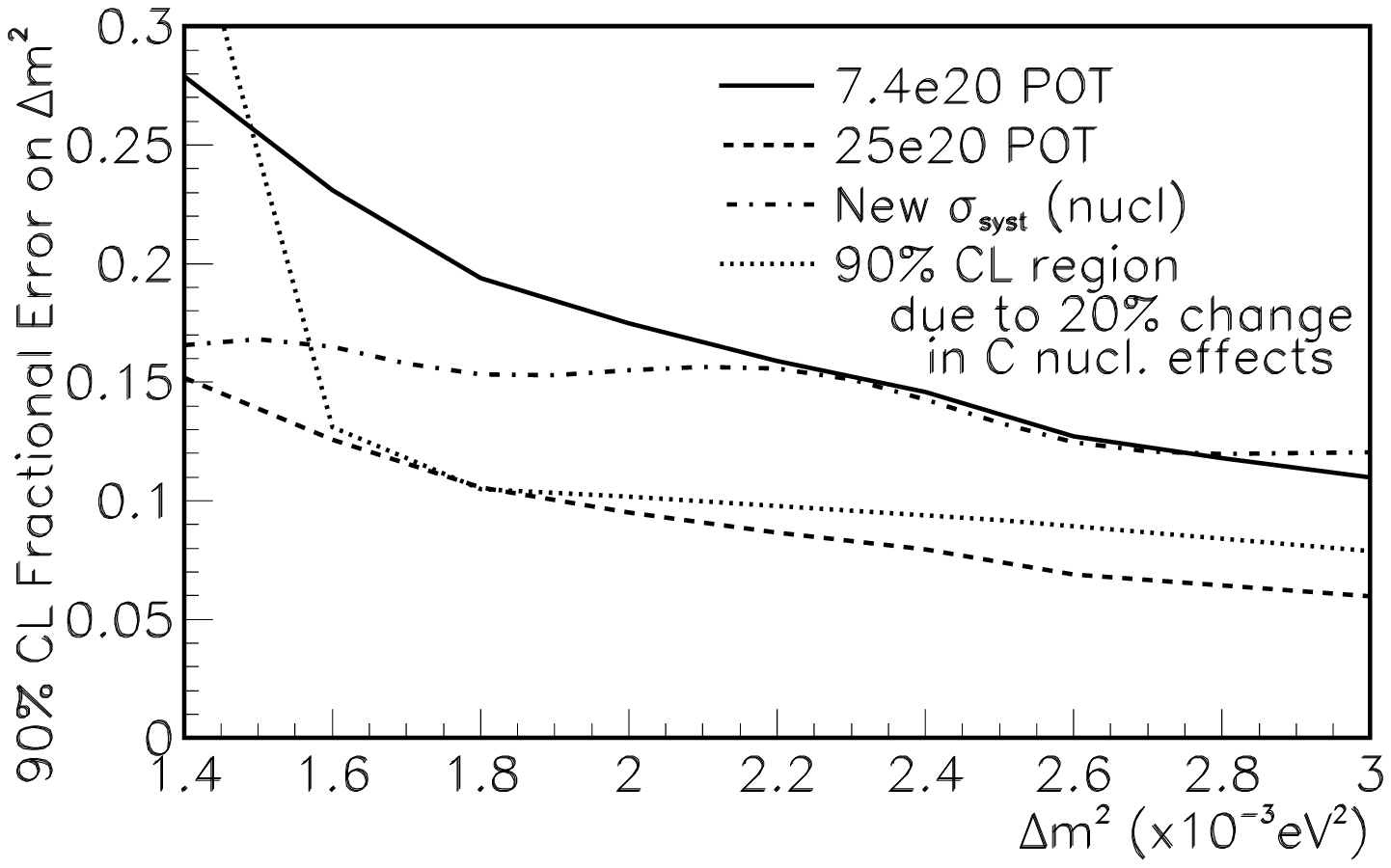}
\caption[Systematic error on $\Delta m^2$ from Nuclear Effects]
{Fractional size of the 90\% confidence level region at
$\sin^2 2\theta_{23}=1$ for the changes in nuclear effects described
earlier, assuming the uncertainties for nuclear effects in neutrinos
are three times the uncertainties coming from the charged lepton
measurements.}
\label{fig:oscplot_d}
\end{figure}

If these errors are appropriate, then they are comparable with the
statistical error expected by MINOS for $7.6\times 10^{20}$ protons on
target, as shown in figure \ref{fig:oscplot_d}.  Also shown on this
plot is the previous uncertainty that was shown in the proposal, which
was evaluated assuming the nuclear effects were known to 20\%,
and which assumed the size of nuclear effects on carbon
were the same as those on steel.  As described in the section on
nuclear effects in this addendum, \minerva\ will be able to
measure these effects in neutrino scattering directly and precisely, 
for several different targets.

\subsubsection{Measurements of $\nu_\mu \to \nu_e$ oscillation Probability} 

In order to understand the full importance of \minerva\ cross section
measurements for the \nova\ experiment, it is helpful to revisit how
experiments will determine the $\nu_\mu \to \nu_e$ oscillation
probabilities.  The number of events in the far
detector can be described as

\begin{equation} 
N_{\textstyle far}  =  \phi_\mu P(\nu_\mu \to \nu_e) \sigma_e
\epsilon_e M_{\textstyle far} + B_f  
\label{eqn:prob1}
\end{equation}

Where $\phi_\mu$ is the muon neutrino flux at the far detector, $P$ is
the oscillation probability, $\sigma_e$
and $\epsilon_e$ are the electron neutrino cross section and
efficiency, respectively, and $M_{far}$ is the far detector mass.  The
backgrounds at the far detector, $B_{far}$, can be expressed as

\begin{equation}
B_{far} = \Sigma_{i=e,\mu} \phi_i P(\nu_i \to \nu_i)\sigma_i\epsilon_i M_{far} 
\label{eqn:prob2} 
\end{equation}       

Where the notation is the same as equation \ref{eqn:prob1}, but
$\epsilon_i$ is the efficiency for a neutrino of type $i$ to be
misreconstructed as an electron neutrino.  Backgrounds come from
both muon and electron neutrinos, and from several different
neutrino interaction channels.  Both equation
\ref{eqn:prob1} and \ref{eqn:prob2} must be summed over neutrino
channels, as well as integrals over neutrino energy.

The error on the oscillation probability, in this simplified notation,
can then be expressed as

\begin{equation}
\frac{\delta P}{P} = 
\frac{N_{\textstyle far}+(\delta B_{\textstyle far})^2}{(\phi_\mu
  \sigma_e\epsilon_e M_{\textstyle far})^2} +
 (N_{\textstyle far}-B_{\textstyle far})\left[ (\frac{d\phi}{\phi})^2
    + 
(\frac{\delta \sigma}{\sigma})^2 + 
(\frac{\delta \epsilon}{\epsilon})^2 \right] 
\label{eqn:proberr1} 
\end{equation}

The two terms in equation \ref{eqn:proberr1} suggest two regimes: in
the case where the number of events in the far detector is comparable
to the background prediction, the error on the probability is
dominated by a combination of statistics and the uncertainty on the
background prediction. The background prediction uncertainty is 
dominated by the uncertainty in
the background process cross sections and efficiencies.  In the
other extreme, where the number of events is dominated by the signal
events, the uncertainty on the probability comes from the statistics, and
the uncertainties on the signal channel cross sections and efficiencies.

\begin{table} 
\begin{tabular}{lccccc} 
Process & Statistics & QE & RES & COH & DIS \\ 
$\delta \sigma/\sigma$ &   & 20\% & 40\% & 100\% & 20\% \\ 
Signal $\nu_e$ & 175 ($\sin^2 2\theta_{13}=0.1)$ & 55\% & 35\% & n/I & 10\% \\ 
NC & 15.4 & 0 & 50\% & 20\% & 30\% \\ 
$\nu_\mu CC $ & 3.6 & 0 & 65\% & n/I & 35\% \\ 
Beam$\nu_e$ & 19.1 & 50\% & 40\% & n/I & 10\% \\ 
\end{tabular}
\caption[Signal and Background Processes at \nova]
{List of the signal and background processes than can
contribute events in the \nova\ far detector, for a 50kton detector
located 12km from the NuMI axis, 820km from Fermilab, assuming a
$\Delta m^2$ of $2.5\times 10^{-3}eV^2$.  Also given are the cross
section uncertainties on those processes before \minerva\ runs.}
\label{tab:nova1}
\end{table} 

For the \nova\ detector simulation running at an off axis location
12km from the NuMI beamline and 820km from the source at Fermilab, the
signal and background statistics for the nominal 5 year run are given
in table \ref{tab:nova1}.  As has been described in more detail in
both the \nova\ and \minerva\ proposals, the three categories of
backgrounds are $\nu_e$'s originally produced in the neutrino beam,
the neutral current events with energetic neutral pions which can fake
electrons, and $\nu_\mu$ charged current events, where the final state
muon is low energy and the event contains a high energy neutral pion.
Also given in Table \ref{tab:nova1} are the fractions that each
neutrino interaction process contributes to the events of that type
that pass all cuts, as well as the cross section uncertainty on that
process.

Without a near detector, the errors from cross sections,
for the case that there are no $\nu_\mu$ oscillations, are 16\%, which
is equivalent to the statistical error for that case.  For the
case of mixing at the level indicated in the table, the statistical
error on the probability would be 8\%, while
the errors from cross section uncertainties alone would be 31\%.

In the case where there is a near detector that is identical to the
far detector, one can try to cancel out these uncertainties.  Consider
first the prediction of the background events.  The events in an
identical near detector that pass the same analysis cuts as those made
at the far detector can be described as

 \begin{equation}
N_{near} = \Sigma_{i=e,\mu} \phi_i \sigma_i\epsilon_i M_{near} 
\label{eqn:neardet} 
\end{equation}       

And then one can use the simulation to predict the number of
backgrounds at the far detector by the following equation:

\begin{eqnarray}
B_{\textstyle far} &=&  N_{\textstyle near} \frac{M_{\textstyle
    far}}{M_{\textstyle near}} R
\label{eqn:neardet2} 
\end{eqnarray}       
where,
\begin{eqnarray}
R &=& \frac{\Sigma_{i=e,\mu} \phi_{i,{\textstyle far}} \sigma_i\epsilon_i}
{\Sigma_{i=e,\mu} \phi_{i,{\textstyle near}} \sigma_i\epsilon_i}
\end{eqnarray}       

\begin{figure}[tp]
\epsfxsize=\textwidth
\epsfbox{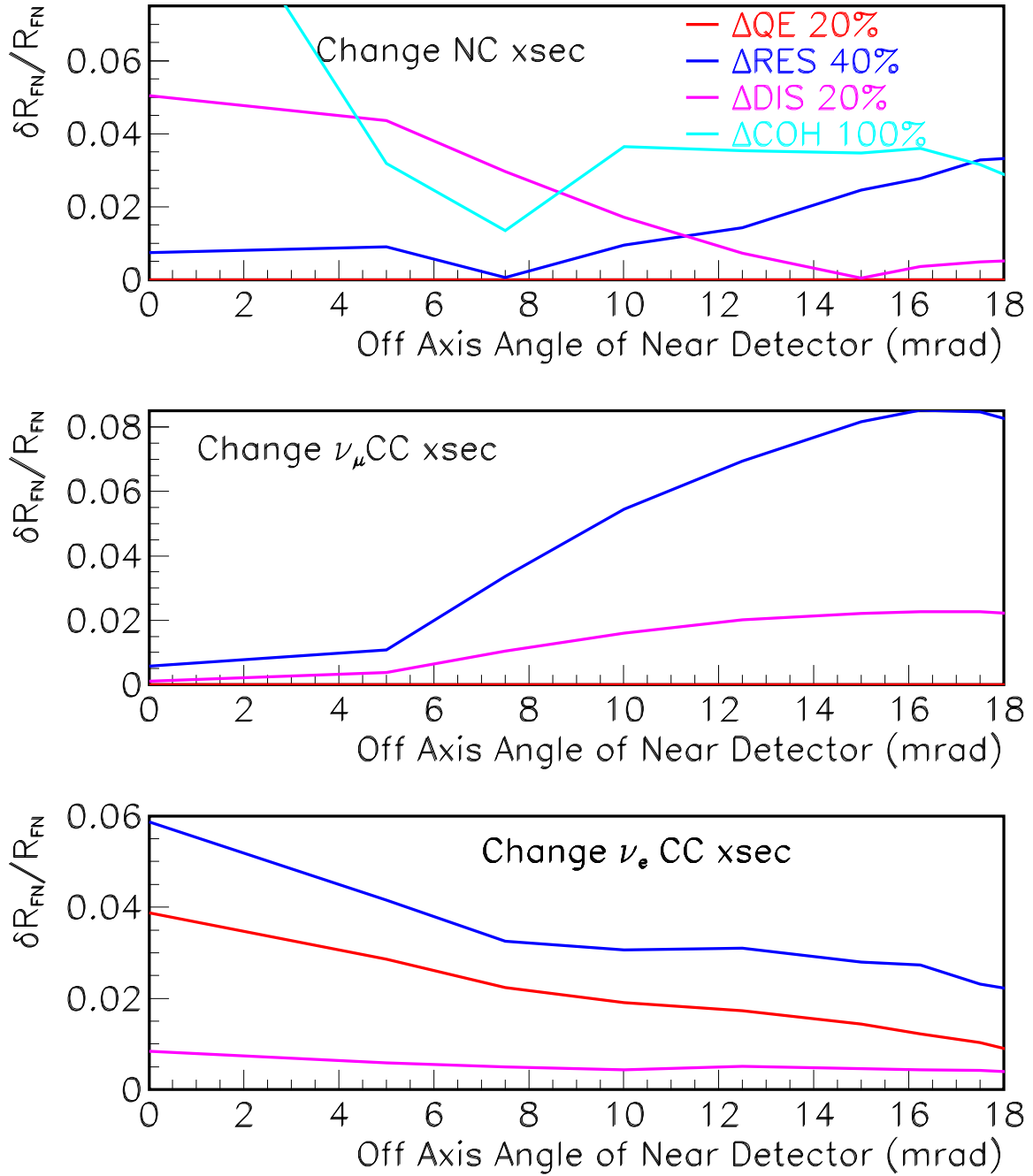}
\caption[Uncertainty in \nova\ Background Prediction from Ambiguity of
  Background Source as a function of Near detector position]
{The fractional change in the far detector background
prediction coming from an identical near detector, as a function of
near detector off axis angle, for a far detector located 12km off the
NuMI axis and 820km from Fermilab.  The top plot shows the fractional
change when the neutral current cross sections are varied by their
uncertainties, the middle plot shows the fractional change when the
$\nu_\mu$ charged current cross sections are varied by their
uncertainties, and the bottom plot shows the fractional change when
the $\nu_e$ charged current cross sections are varied by their
uncertainties.}
\label{fig:raterr}
\end{figure}
For different near detector off axis angles, there are different
fractions of background events that pass all cuts, but at no value
of the angle is
the mix of backgrounds the same as that in the far detector.
Figure \ref{fig:raterr} shows the
fractional change in the variable $R$ defined above due to the cross
section errors listed in table \ref{tab:nova1}.  This translates
directly into an uncertainty on the far detector background prediction
due to current cross section uncertainties.

\begin{figure}[tp]
\epsfxsize=\textwidth
\epsfbox{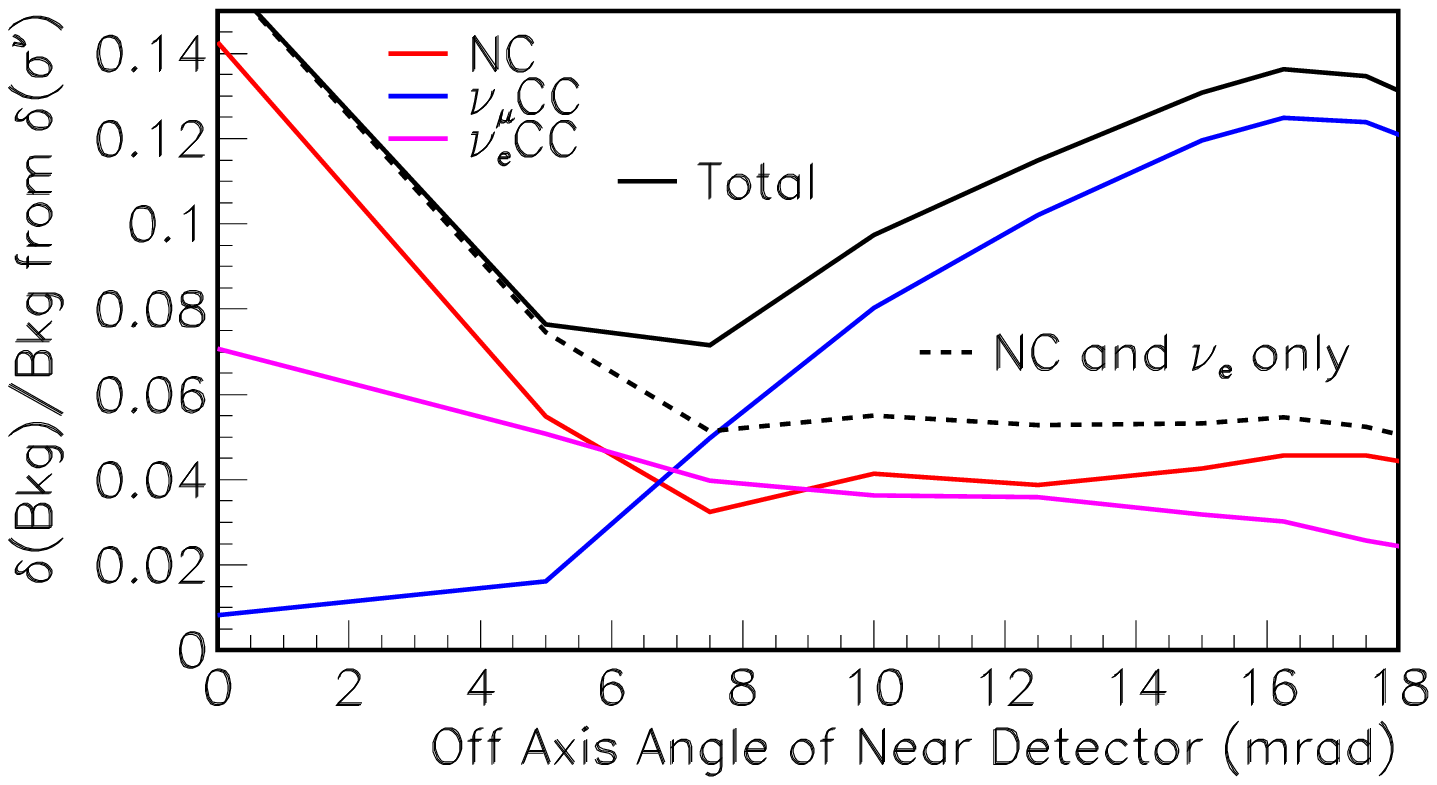}
\caption[Total Uncertainty in \nova\ Background Prediction as a 
function of Near detector position]
{The fractional error in the background prediction at the far detector 
from uncertainties in each process (Quasi-elastic, resonance, deep inelastic
scattering, and neutral current coherent $\pi^0$ production), 
added in quadrature for each source (NC, $\nu_\mu CC$, beam $\nu_e$), plotted
as a function of near detector off axis angle.}
\label{fig:toterr}
\end{figure}
Note that for low off axis angles the systematic error due to the
$\nu_\mu$ charged current uncertainties is minimum, yet the error due
to the neutral current and electron neutrinos is maximum there.
Figure \ref{fig:toterr} shows the above errors added in quadrature, as
a function of near detector off axis angle.  Note that at best the
cross section uncertainties can be reduced from 16\% with no near
detector to about 10\% with an identical near detector, but given that
this is only one of several systematic errors in the \nova\
experiment, it is an unacceptably large fraction of the total error.
The \minerva\ experiment can significantly reduce the errors due to
the charged current processes, because although it is on the NuMI axis
where the neutrino flux is different from the off-axis
fluxes, for the charged current processes the final state energy is
close to the neutrino energy, and the flux prediction from the hadron
production combined with NuMI horn B-field measurements means that
charged current cross section channels will be measurable at the 5\%
level overall.  So with the presence of \minerva\ the error due to
cross section uncertainties on the background at the far detector can
be reduced from 10\% to better than 5\%.  Furthermore, by measurements
of charged current coherent pion production on axis one can infer the
neutral current coherent pion process off axis using theoretical
models constrained by the charged current process.

For the case where the number of signal events is well above the
number of background events, the challenge to keep the uncertainties 
due to cross section
errors low is even harder, since in that case the composition of
events near to far is even more different than it is in the case of no
signal events.  Furthermore, because the total number of events is
higher, the improved statistical precision will required an
even more precise far detector prediction.

As a final note, it should be pointed out that the only cross section
errors considered here are uncertainties in the overall levels
of each of the processes.  There is an additional uncertainty in the energy 
dependence of these processes, which will again contribute uncertainties
in the far detector predictions because of the differing spectra.  

\subsubsection{Summary} 

This section has quantified how \minerva's cross section
measurements will have important consequences on both the current and
future generations of neutrino experiments with the NUMI beamline.
These improvements to MINOS and \nova\ measurements are important
regardless of the ultimate value of $\Delta m^2$ that MINOS measures,
and regardless of the size of the $\nu_\mu \to \nu_e$ oscillation
probability that \nova\ ultimately measures.

%% file: minerva_proposal_E938.bbl
\clearpage


\begin{thebibliography}{999}
%
%

\bibitem{EOI} The EOIs from the groups that combined to form the \minerva
collaboration, the NuMI "On-axis" and NuMI "Off-axis" groups, can be found on
the \minerva\ home page at: \url{http://www.pas.rochester.edu/minerva/}.

\bibitem{Vovenko} 
A.~Vovenko, ''Total cross sections and structure functions for neutrino
interactions in  3-30 GeV energy range and near plans for 1-5 GeV energy
range'', submitted  to the Proceedings of NuInt01 : Proceedings of The First
International Workshop on Neutrino-Nucleus Interactions in the Few GeV Region,
pg 116. (2002).

\bibitem{Gallagher} 
A recently-formed working group coordinated by Hugh Gallagher, Costas
Andreopoulos and Sam Zeller is attempting to gather all existing neutrino
scattering data for the Durham and PDG data bases.  

\bibitem{Rein:1980wg}
D.~Rein and L.~M.~Sehgal,
Annals Phys. {\bf 133}, 79 (1981).

\bibitem{LeeSato}
T.~S.~Lee and T.~Sato
Few Body Syst. Suppl. {\bf 15}, 183 (2003).

\bibitem{Paschos}
E.~A.~Paschos, L.~Pasquali and J.~Y.~Yu,
Nucl. Phys. B {\bf 588}, 263 (2000) and E.~A.~Paschos, J.~Y.~Yu and M.~Sakuda
[arXiv:hep-ph/0308130].

\bibitem{Wood}
S.~Wood, ''Resonance Region to DIS, Quark-Hadron Duality'', submitted
to the Proceedings of NuInt01 : The First International Workshop on
Neutrino-Nucleus Interactions in the Few GeV Region, (2002).

\bibitem{Bodek2002}
A.~Bodek and U.~K.~Yang,
[arXiv:hep-ex/0203009.]
 	
\bibitem{Shrock} R.~Shrock, Phys. Rev. {\bf D12}, 2049 (1975); A.~A.~Amer, Phys. Rev.
{\bf D18}, 2290 (1978); H.K.~Dewan, Phys. Rev. {\bf D24}, 2369 (1981).
 
\bibitem{solomey}
N.~Solomey,
``Physics prospects with an intense neutrino experiment,''
[arXiv:hep-ex/0006021] and ``Neutrino Interactions-Strange Particle 
Production", contribution to the Workshop on New Initiatives for the NuMI
Neutrino Beam, Fermilab, 2002.

\bibitem{MIPP} P-907: Proposal to Measure Particle Production in the Meson Area
Using Main Injector Primary and Secondary Beams, May 2000
\begin{verbatim}(http://ppd.fnal.gov/experiments/e907/Proposal/E907_Propsal.html)\end{verbatim}

\bibitem{numiinst} NuMI Technical Design Handbook
\begin{verbatim}(http://www-numi.fnal.gov/numiwork/tdh/tdh_index.html)\end{verbatim}


\bibitem{MacFarlane:1984ax}
D.~MacFarlane {\em et~al.},
\newblock Z. Phys. {\bf C26}, 1 (1984).

\bibitem{Berge:1987zw}
J.~P.~Berge {\em et~al.},
\newblock Z. Phys. {\bf C35}, 443 (1987).

\bibitem{Morfin:1981kg}
J.~G.~Morfin {\em et~al.},
\newblock Phys. Lett. {\bf B104}, 235 (1981).

\bibitem{Colley:1979rt}
D.~C.~Colley {\em et~al.},
\newblock Zeit. Phys. {\bf C2}, 187 (1979).

\bibitem{Mukhin:1979bd}
A.~I.~Mukhin {\em et~al.},
\newblock Sov. J. Nucl. Phys. {\bf 30}, 528 (1979).

\bibitem{Baranov:1979sx}
D.~S.~Baranov {\em et~al.},
\newblock Phys. Lett. {\bf B81}, 255 (1979).

\bibitem{Baltay:1980pr}
C.~Baltay {\em et~al.},
\newblock Phys. Rev. Lett. {\bf 44}, 916 (1980).

\bibitem{Barish:1979pj}
S.~J.~Barish {\em et~al.},
\newblock Phys. Rev. {\bf D19}, 2521 (1979).

\bibitem{Baker:1982ty}
N.~J.~Baker {\em et~al.},
\newblock Phys. Rev. {\bf D25}, 617 (1982).

\bibitem{Ciampolillo:1979wp}
S.~Ciampolillo {\em et~al.},
\newblock Phys. Lett. {\bf B84}, 281 (1979).

\bibitem{Barish:1977qk}
S.~J.~Barish {\em et~al.},
\newblock Phys. Rev. {\bf D16}, 3103 (1977).

\bibitem{Baker:1981su}
N.~J.~Baker {\em et~al.},
\newblock Phys. Rev. {\bf D23}, 2499 (1981).

\bibitem{Miller_82} 
K.L.~Miller {\em et al.}, 
\newblock Phys. Rev. {\bf D26} (1982) 537.

\bibitem{Kluttig:1977gb}
H.~Kluttig, J.~G. Morfin, and W.~Van~Doninck,
\newblock Phys. Lett. {\bf B71}, 446 (1977).

\bibitem{Erriquez:1978yc}
O.~Erriquez {\em et~al.},
\newblock Phys. Lett. {\bf B73}, 350 (1978).

\bibitem{Krenz:1978sw}
W.~Krenz {\em et~al.},
\newblock Nucl. Phys. {\bf B135}, 45 (1978).

\bibitem{Pohl:1979ya}
M.~Pohl {\em et~al.},
\newblock Phys. Lett. {\bf B82}, 461 (1979).

\bibitem{Lee:1977wr}
W.-Y. Lee {\em et~al.},
\newblock Phys. Rev. Lett. {\bf 38}, 202 (1977).

\bibitem{Barish:1974fe}
S.~J.~Barish {\em et~al.},
\newblock Phys. Rev. Lett. {\bf 33}, 448 (1974).

\bibitem{Derrick:1980nr}
M.~Derrick {\em et~al.},
\newblock Phys. Lett. {\bf B92}, 363 (1980).

\bibitem{Derrick:1981xw}
M.~Derrick {\em et~al.},
\newblock Phys. Rev. {\bf D23}, 569 (1981).

\bibitem{Baker:1981pj}
N.~J.~Baker {\em et~al.},
\newblock Phys. Rev. {\bf D23}, 2495 (1981).

\bibitem{ANLpi1}
J.~Campbell {\em et~al.},
\newblock Phys. Rev. Lett. {\bf 30}, 335 (1973).

\bibitem{BNL1pi}
T.~Kitagaki {\em et~al.},
\newblock Phys. Rev. {\bf D34}, 2554 (1986).

\bibitem{FNALpi}
J.~Bell {\em et~al.},
\newblock Phys. Rev. Lett. {\bf 41}, 1008 (1978).

\bibitem{BEBC1pi}
P.~Allen {\em et~al.},
\newblock Nucl. Phys. {\bf B264}, 221 (1986).

%
%
\bibitem{Fukada_98} Y.~Fukada {\em et~al.}, Phys. Rev. Lett. {\bf 81} 1562 (1998).

\bibitem{budd} H.~Budd, A.~Bodek and J.~Arrington, [arXiv:hep-ex/0308005];
A.~Bodek, H.~Budd  and J.~Arrington, [arXiv:hep-ex/0309024]
(to be published in Proceedings of CIPANP2003, New York City, NY 2003.)

\bibitem{Lle_72} C.H.~Llewellyn Smith, Phys. Rep. {\bf 3C} (1972).



\bibitem{Bernard_01}V.~Bernard, L.~Elouadrhiri, U.G.~Meissner,
 J.  Phys. {\bf G28} (2002).

\bibitem{JRA_03} J.~Arrington, nucl-ex/0305009.



\bibitem{paschos} M.~Paschos,
\newblock (private communication).



\bibitem{halla}
M.~K.~Jones {\em et~al.}, Phys. Rev. Lett {\bf 84}, 1398 (2000);
O.~Gayou {\em et~al.}, Phys. Rev. Lett {\bf 88}, 092301 (2002).

\bibitem{Krutov_02} A.~F.~Krutov, V.~E.~Troitsky, Eur. Phys. J. {\bf A16}, 285 (2003).


\bibitem{Kitagaki_83} T.~Kitagaki {\em et~al.}, Phys. Rev. {\bf D26}, 436 (1983).
  







\bibitem{pdg} Particle Data Group, Eur. Phys. J {\bf C15}, 1 (2000).

\bibitem{Mann_73} W.A.~Mann {\em et~al.}, Phys. Rev. Lett. {\bf 31}, 844 (1973).

\bibitem{Brunner_90} J.~Brunner {\em et~al.}, Z. Phys. {\bf C45}, 551 (1990).

\bibitem{Pohl_79} M.~Pohl {\em et~al.}, Lett. Nuovo Cimento {\bf 26}, 332 (1979).

\bibitem{Auerbach_02} L.B.~Auerbach {\em et~al.},  Phys. Rev. {\bf C66}, 015501 (2002).

\bibitem{Belikov_85} S.V.~Belikov {\em et~al.}, Z. Phys. {\bf A320}, 625 (1985).

\bibitem{Bonetti_77} S.~Bonetti {\em et~al.}, Nuovo Cimento {\bf 38}, 260 (1977).

\bibitem{Armenise_79} N.~Armenise {\em et~al.}, Nucl. Phys. {\bf B152}, 365 (1979).

\bibitem{Zeller_03} G.~Zeller (private communication).

\bibitem{Smith_72} R.~A.~Smith and E.~J.~Moniz, Nucl. Phys. {\bf B43}, 605 (1972) ;
 E.~J.~Moniz  {\em et~al.}, Phys. Rev. Lett. {\bf 26}, 445 (1971); 
E.~J.~ Moniz, Phys. Rev. {\bf 184}, 1154 (1969).


\bibitem{Casper_02} D.~Casper, Nucl. Phys. Proc. Suppl. {\bf 112}, 161 (2002). 

\bibitem{Tsushima_03} K.~Tsushima, Hungchong Kim, K.~Saito, 
                     nucl-th[0307013].


\bibitem{belgium} Ghent Theory group in Belgium,
\newblock Jan Ryckebusch (jan@inwpent5.UGent.be). 



\bibitem{e94110} JLab hydrogen experiment 94-110, C.E.~Keppel spokesperson.
\begin{verbatim}(http://www.jlab.org/exp_prog/proposals/94/PR94-110.pdf)\end{verbatim}
\bibitem{e02109} JLab deuterium  experiment 02-109, C.E.~Keppel, M.E.~Christy,
spokespersons.
\begin{verbatim}(http://www.jlab.org/exp_prog/proposals/02/PR02-109.ps)\end{verbatim}
\bibitem{e99118} JLab experiment 99-118 on the nuclear dependence
of R at low $Q^2$,  A.~Brull,
C.E.~Keppel spokespersons.


\bibitem{brooks} W.~K.~Brooks, NUINT02 proceedings.

\bibitem{nstarclas} M.~Ripani, Nucl.~Phys. {\bf A699}, 270 (2002),
V.~D.~Burkert, Proceedings of the 9th International Conference on the
Structure of Baryons, Baryons 2002, p. 29.
 

\bibitem{wood02}
S.~A.~Wood,
\newblock To be published in {\em Proceedings of the Second Workshop on Neutrino-Nucleus Interactions in the Few-GeV Region (NUINT02),} Irvine, California (2002).
%

\bibitem{satolee} T.~Sato and T.-S.~Lee, Phys.~Rev. {\bf C54}, 2660
 (1996);  T.~Sato and T.-S.~Lee, Phys.~Rev.{\bf C63}, 055201 (2001).

\bibitem{satounolee} T.~Sato, D.~Uno and T.-S.~Lee, Phys.~Rev. {\bf C67},
  065201 (2003).

\bibitem{shreinervonhippel}  P.~A.~Schreiner and F.~Von Hippel,
  Nucl.~Phys. {\bf B58}, 333 (1973).

\bibitem{reinsehgal} D.~Rein and L.~Seghal, Nucl.~Phys. {\bf B223}, 29
  (1983); D.~Rein, Z.~Phys. {\bf C35}, 43 (1987).



\bibitem{rcar} C.E. Carlson, Phys. Rev. {\bf D 34}, 2704 (1986).
\bibitem{rsto} P. Stoler, Phys. Reports, {\bf 226 3}, 103-171 (1993).
\bibitem{rstlr} P. Stoler, Phys. Rev. Lett. {\bf 66}, 1003 (1991).
\bibitem{rgast} M. Gari and N.G. Stephanis, Phys. Lett. {\bf B175}, 462 (1986).
\bibitem{rgakr} M. Gari and W. Krupelmann, Phys. Lett. {\bf B173}, 10 (1986).
\bibitem{rrmdav} R.M. Davidson {\em et~al.}, Phys. Rev. {\bf D43}, 71 (1991).
\bibitem{rbwf} E.P. Wigner, Proc. Cambridge Philosophical Soc. {\bf 47}, 790 (1951).
\bibitem{rral} J.P. Ralston and B. Pire, Univ. of Kansas preprint 5/92.
\bibitem{rfarr} G.R. Farrar and D.R. Jackson, Phys. Rev. Lett. {\bf 35}, 21 (1975).
\bibitem{rcarl} C.E. Carlson and J.L. Poor, Phys. Rev. {\bf D 38}, 2758 (1988).
\bibitem{rble} G.P. Lepage and S.J. Brodsky, Phys. Rev. {\bf D 22}, 2180 (1980).
\bibitem{rvzak} A.I. Vainshtein and A.V. Zakharov, Phys. Lett. {\bf B72}, 368 (1978).
\bibitem{becchi} C. Becchi and G. Morpurgo, Phys. Lett. {\bf 17}, 352 (1969).
\bibitem{HallC} V.~V.~Frolov {\em et al.}, Phys. Rev. Lett. {\bf 82}, 45 (1999).
\bibitem{ct1}
S.J.~Brodsky and A.H.~Mueller, Phys. Lett. {\bf B 206}, 685 (1988).
\bibitem{ne18}
N.C.R. Makins {\em et al.}, Phys. Rev. Lett. {\bf 72}, 1986 (1994);
T.G. O'Neill {\em et al.}, Phys. Lett. {\bf 351}, 87 (1995).
\bibitem{ct4} K.~Garrow {\em et al.}, Phys. Rev. C {\bf 66}, 044613 (2002).
\bibitem{BNL1} A.S. Carroll {\em et al}, Phys. Rev. Lett. {\bf 61}, 1698 (1988).
\bibitem{BNL2} Y. Mardor {\em et al}, Phys. Rev. Lett. {\bf 81}, 5085 (1998);
A. Leksanov {\em et al}, Phys. Rev. Lett. {\bf 87}, 212301 (2001).
\bibitem{dutta} D. Dutta, R. Ent, K. Garrow, {\em et al}, Jefferson Lab Hall
C Experiment E-01-107, approved with A- priority and expected to run in 2004.
\bibitem{dave}
D.~Gaskell,
\newblock To be published in {\em Proceedings of the Second Workshop on Neutrino-Nucleus Interactions in the Few-GeV Region (NUINT02),} Irvine, California (2002).
%
%
%
\bibitem{Kopeliovich:1993ym}
B.~Z. Kopeliovich and P.~Marage,
\newblock Int. J. Mod. Phys. {\bf A8}, 1513 (1993).

\bibitem{Rein:1983pf}
D.~Rein and L.~M. Sehgal,
\newblock Nucl. Phys. {\bf B223}, 29 (1983).

\bibitem{Belkov:1987hn}
A.~A. Belkov and B.~Z. Kopeliovich,
\newblock Sov. J. Nucl. Phys. {\bf 46}, 499 (1987).

\bibitem{Piketty:1970sq}
C.~A. Piketty and L.~Stodolsky,
\newblock Nucl. Phys. {\bf B15}, 571 (1970).

\bibitem{Paschos:2003hs}
E.~A. Paschos and A.~V. Kartavtsev,
\newblock (2003), [arXiv:hep-ph/0309148].

\bibitem{Faissner:1983ng}
H.~Faissner {\em et~al.},
\newblock Phys. Lett. {\bf B125}, 230 (1983).

\bibitem{Isiksal:1984vh}
E.~Isiksal, D.~Rein, and J.~G. Morfin,
\newblock Phys. Rev. Lett. {\bf 52}, 1096 (1984).

\bibitem{Bergsma:1985qy}
CHARM, F.~Bergsma {\em et~al.},
\newblock Phys. Lett. {\bf B157}, 469 (1985).

\bibitem{Vilain:1993sf}
CHARM-II, P.~Vilain {\em et~al.},
\newblock Phys. Lett. {\bf B313}, 267 (1993).

\bibitem{Allport:1989cq}
BEBC WA59, P.~P. Allport {\em et~al.},
\newblock Z. Phys. {\bf C43}, 523 (1989).

\bibitem{Marage:1986cy}
BEBC WA59, P.~Marage {\em et~al.},
\newblock Z. Phys. {\bf C31}, 191 (1986).

\bibitem{Grabosch:1986mt}
SKAT, H.~J. Grabosch {\em et~al.},
\newblock Zeit. Phys. {\bf C31}, 203 (1986).

\bibitem{Baltay:1986cv}
C.~Baltay {\em et~al.},
\newblock Phys. Rev. Lett. {\bf 57}, 2629 (1986).

\bibitem{Ammosov:1987kr}
V.~V. Ammosov {\em et~al.},
\newblock Yad. Fiz. {\bf 45}, 1662 (1987).

\bibitem{Willocq:1993fv}
E632, S.~Willocq {\em et~al.},
\newblock Phys. Rev. {\bf D47}, 2661 (1993).

\bibitem{Paschos-K}
E.~Paschos and A.~Kartavtsev (private communication).

\bibitem{Zeller:Nuint02}
G.~Zeller,
\newblock To be published in {\em Proceedings of the Second Workshop on Neutrino-Nucleus Interactions in the Few-GeV Region (NUINT02),} Irvine, California (2002).
%
%
\bibitem{Deden} H. Deden {\em et~al.}, Phys. Lett. {\bf 58B}, 361 (1975).

\bibitem{Erriques} O. Erriques {\em et~al.}, 
 Phys. Lett. {\bf B70}, 385 (1977); 
 Nucl. Phys. {\bf B140}, 123 (1978).

\bibitem{Barish_74} S.J.~Barish {\em et~al.}, Phys. Rev. Lett.
 {\bf 33}, 1446 (1974)

\bibitem{Baker_81Strange} N.J. Baker {\em et~al.}, 
 Phys. Rev. {\bf D24}, 2779 (1981).
 
\bibitem{Cazzoli}E.~G. Cazzoli {\em et~al.},  
 Phys. Rev. Lett. {\bf 34}, 1125 (1975).

\bibitem{nomad} P.~Astier {\em et~al.}, Nucl. Phys. {\bf B611}, 3-39 (2001).


\bibitem{Amer} A.~A. Amer, Phys. Rev. {\bf D18}, 2290 (1978). 

\bibitem{Dewan} H.~K. Dewan, Phys. Rev. {\bf D24}, 2369 (1981).

\bibitem{Mann_86} W.A.~Mann {\it et al.}, Phys. Rev. D {\bf 34}, 2545 (1986).

\bibitem{NextGenWC} K. Nakamura in proceedings of 
 {\em  Neutrinos and Implications for Physics 
 Beyond the Standard Model}, SUNY at Stony Brook, October 2002, pp. 307-317.
 {\em Physics Potential and Feasibility of UNO},
 (http://nngroup.physics.sunysb.edu/uno/), report SBHEP-01-03 (2000).

\bibitem{Mine} SuperKamiokande Collaboration, S. Mine, 
 presented at {\em Neutrino Interactions at Few GeV Energies NUINT02},
 (http://nuint.ps.uci.edu/) UC Irvine, California, December 2002; see
 also, S. Mine, Nucl. Phys. B (Proc. Suppl.) {\bf 112}, 154 (2002).

\bibitem{NEUGEN} H. Gallagher, Nucl. Phys. B (Proc. Suppl.) 
 {\bf 112}, 188 (2002).

\bibitem{Day}For the cross section excitation of Equation (\ref{eq:35}), measurements at ANL 
(Table 4.5 in D.L.~Day, Ph.D. thesis, 
University of Kansas, 1977) agree with those 
reported by BNL~\cite{Baker:1981su} to within large errors, 
however the $\sigma$(E$_{\nu}$) nominal
values are higher by factors 1.5 to 2.0.



\bibitem{lach} J. Lach and P. Zenczykowski, Int. J. Mod. Phys. {\bf
A10}, 3817 (1995).

\bibitem{singer} P. Singer, Nucl. Phys. B (Proc. Suppl.) {\bf 50}, 202
(1996).

\bibitem{ktev}  A. Alavi-Harati {\em et~al.}, Phys. Rev. {\bf 86}, 3239 (2001).

\bibitem{schmidth} S.A. Schmidth, Ph.D. Thesis, Univ. Mainz, D-55099 Mainz
(November 2001).

\bibitem{zen} P. Zenczykowski, [arXiv:hep-ph/0205311] (October 2002).

\bibitem{gilman} F.J. Gilman and M.B. Wise, Phys. Rev. {\bf D19}, 976
(1979).

\bibitem{beta_ktev} A. Alavi-Harati {\em et~al.}, Phys. Rev. {\bf 87} 132001 (2001).

\bibitem{ratclif} P.G. Ratcliffe, Phys. Rev. {\bf D59}, 014038 (1999).

\bibitem{cabibbo} N. Cabibbo {\em et~al.}, Semileptonic Hyperon Decay and CKM Unitarity,
[arXiv:hep-ph/0307214] (July 2003).

\bibitem{beta_e871} N. Solomey, Recent Rare $\Omega$ Decay Results from the 
HyperCP Experiment, talk at DPF 2003, Philadelphia.

\bibitem{pais} A. Pais and S.B. Treiman, Phys. Rev. {\bf 178}, 2365 (1969).

\bibitem{singh} Y. Singh, Phys. Rev. {\bf 161}, 1497 (1967).

\bibitem{pquark} T. Nakano {\em et~al.}, [arXiv:hep-ex/0301020]; V.V. Barmin {\em et~al.}, [arXiv:hep-ex0304040]; 
S. Stepanyan [arXiv:hep-ex/0307018].

\bibitem{jaffe} R. Jaffe and F. Wilczek, Di-quarks and Exotic Spectroscopy, [arXiv:hep-ph/0307341] (July 2003).


\bibitem{saghai}B. Saghai, Nuclear Physics {\bf A639}, 217-226 (1998).

\bibitem{nutev} M. Goncharov {\em et~al.}, Phys. Rev. {\bf D64}, 112006 (2001).


\bibitem{Shifman} M. Shifman, Handbook of QCD, Volume 3, 1451,
World Scientific (2001).


\bibitem{MRS} 
A.D. Martin, R.G. Roberts, and W.J. Stirling, Phys. Rev. 
{\bf D50} (1994) 6734.

\bibitem{barbieri}
R. Barbieri, J. Ellis, M.K. Gaillard, G. Ross, Nucl. Phys. 
{\bf B117}, 50 (1976).

\bibitem{Keppel}
C.~E.~Keppel,
``Quark hadron duality studies at Jefferson Lab: An overview of new and existing results,''

{\em Prepared for Exclusive Processes at High Momentum Transfer, Newport News, Virginia, 15-18 May 2002}.

\bibitem{Dasu}
L.W. Whitlow, E.M. Riordan, S. Dasu, S. Rock and A. Bodek,
Phys. Lett. B {\bf 282}, 475 (1992).

\bibitem{ioana1} 
I.~Niculescu {\em et~al.},
Phys. Rev. Lett.  {\bf 85}, 1186 (2000).

\bibitem{g1hermes}
A.~Fantoni  [HERMES Collaboration],
Eur. Phys. J. {\bf A17}, 385 (2003).

\bibitem{ioana2}
I.~Niculescu {\em et~al.},
Phys. Rev. Lett. {\bf 85}, 1182 (2000).


\bibitem{lattice}
D.~Dolgov {\em et~al.}  [LHPC collaboration],
Phys. Rev. {\bf D66}, 034506 (2002)
[arXiv:hep-lat/0201021].

\bibitem{DGP}
A. De Rujula, H. Georgi, and H.D. Politzer, Ann. Phys. {\bf 103} 315 (1977).

\bibitem{SPIN0}
R.~P.~Feynman,
{\em Photon Hadron Interactions}
(Benjamin, Reading, Massachusetts, 1972);
%
F.~E.~Close,
Phys. Lett. {\bf B43}, 422 (1973);
Nucl. Phys. {\bf B80}, 269 (1973);
%
R.~D.~Carlitz,
Phys. Lett. {\bf B58}, 345 (1975);
%
N.~Isgur,
Phys. Rev. {\bf D59}, 034013 (1999).

\bibitem{GLOBAL}
A.~D.~Martin, R.~G.~Roberts, W.~J.~Stirling and R.~S.~Thorne,
Eur. Phys. J. {\bf C14}, 133 (2000);
%
H.~L.~Lai {\em et~al.},
Eur. Phys. J. {\bf C12}, 375 (2000);
%
M.~Gl\"uck, E.~Reya and A.~Vogt,
Eur. Phys. J. {\bf C5}, 461 (1998).

\bibitem{FJ}
G.~R.~Farrar and D.~R.~Jackson,
Phys. Rev. Lett. {\bf 35}, 1416 (1975).

\bibitem{KUHL}
S.~Kuhlmann {\em et~al.},
Phys. Lett. {\bf B476}, 291 (2000).

\bibitem{WallyTony}
W.~Melnitchouk and A.~W.~Thomas
Phys. Lett. {\bf B377}, 11 (1996).

\bibitem{Yang:1998zb}
U.~K.~Yang and A.~Bodek,
Phys. Rev. Lett.  {\bf 82}, 2467 (1999).
[arXiv:hep-ph/9809480]

\bibitem{Lai:1994bb}
H.~L.~Lai {\em et~al.},
Phys. Rev. {\bf D51}, 4763 (1995)
[arXiv:hep-ph/9410404].

\bibitem{Bodek-Yang}
U.~K.~Yang and A.~Bodek,
Eur. Phys. J. {\bf C13}, 241 (2000)
[arXiv:hep-ex/9908058].

\bibitem{Gargamelletwist}
H.~Deden {\em et~al.},
Nucl. Phys. {\bf B85}, 269 (1975).

\bibitem{Varvell:1987qu}
K.~Varvell {\em et~al.},
Z. Phys. {\bf C36}, 1 (1987).

\bibitem{webb}
J.~Webb, Thesis "Measurement of Continuum Dimuon Production in 
800-GeV/c Proton-Necleon Collisions" (New Mexico State Univ., Sept 2002), 
unpublished.

\bibitem{cteq6}
J.~Pumplin, D.~R.~Stump, J.~Huston, H.~L.~Lai, P.~Nadolsky and W.~K.~Tung,
JHEP {\bf 0207}, 012 (2002)
[arXiv:hep-ph/0201195].

\bibitem{BONUS}
Jefferson Lab Experiment E03-012 Proposal (approved).

%
%
\bibitem{e03110} A. Bodek, C. Keppel, et al, E93-110 Collaboration
Proposal to Jefferson Lab Hall C (2003).

\bibitem{dualnuc}
J.~Arrington, R.~Ent, C.~E.~Keppel, J.~Mammei and I.~Niculescu,
arXiv:nucl-ex/0307012 (submitted to {\it Phys. Rev. Lett.}).


\bibitem{Ji97} X. Ji, Phys. Rev. Lett. {\bf 78}, 610 (1997).
\bibitem{Ji97b} X. Ji, Phys. Rev. {\bf D55}, 7114 (1997).
\bibitem{Rady96} A. V. Radyushkin, Phys. Lett. {\bf B380}, 417 (1996).
\bibitem{Rady96b} A. V. Radyushkin, Phys. Lett. {\bf B385}, 333 (1996).
\bibitem{Collins} J.C. Collins, L. Frankfurt, and M. Strikman,
Phys. Rev. {\bf D56}, 2982 (1997).
\bibitem{Rady02}  A. V. Radyushkin, Nucl. Phys. {\bf A711}, 99 (2002).
\bibitem{Vander} M. Vanderhaeghen, Nucl. Phys. {\bf A711}, 109 (2002).
\bibitem{Diehl} M. Diehl, hep-ph/0307382 (2003).
\bibitem{Lehm01} B.~Lehmann-Dronke and A.~Schafer,
Phys. Lett. {\bf B521}, 55 (2001).

%
%

\bibitem{garrow02} K.~Garrow {\em et~al.}, Phys. Rev. {\bf C66}, 044613 (2002).

\bibitem{ashery} D. Ashery {\em et~al.}, Phys. Rev. {\bf C23}, 2173 (1981).

\bibitem{rans} M.K.~Jones {\em et~al.}, Phys. Rev. {\bf C48}, 2800 (1993);
R.D.~Ransome {\em et~al.}, Phys. Rev. {\bf C46}, 273 (1992); R.D. Ransome {\em et~al.},
Phys. Rev. {\bf C45}, R509 (1992).

\bibitem{lads} D. Rowntree {\em et~al.}, Phys. Rev. {\bf C60}, 054610 (1999);
B.~Kotlinksi {\em et~al.}, Eur. Phys. J. {\bf A9}, 537 (2000).

\bibitem{Frankfurt:2002kd}
L.~Frankfurt, V.~Guzey, M.~McDermott and M.~Strikman,
JHEP {\bf 0202}, 027 (2002), [arXiv:hep-ph/0201230].

\bibitem{Kulagin}
S.~A.~Kulagin, [arXiv:hep-ph/9812532].

\bibitem{Eskola}
K.~J.~Eskola, V.~J.~Kolhinen, P.~V.~Ruuskanen and C.~A.~Salgado,
Nucl. Phys. {\bf A661}, 645 (1999), [arXiv:hep-ph/9906484].

\bibitem{Kumano}
S.~Kumano, [arXiv:hep-ph/0109152].

\bibitem{Benvenuti:1994bb}
A.~C.~Benvenuti {\em et~al.}, 
Z. Phys. {\bf C63}, 29 (1994).

\bibitem{Vakili:1999qt}
M.~Vakili {\em et~al.},
Phys. Rev. {\bf D61}, 052003 (2000), [arXiv:hep-ex/9905052].

\cite{Zeller:2001hh}
\bibitem{Zeller:2001hh}
G.~P.~Zeller {\em et~al.},
Phys. Rev. Lett. {\bf 88}, 091802 (2002), [arXiv:hep-ex/0110059] and references therein.

\bibitem{Kovalenko:2002xe}
S.~Kovalenko, I.~Schmidt and J.~J.~Yang, [arXiv:hep-ph/0207158].

%
%
\bibitem{k2kosc}M.H.Ahn {\em et al}, Phys. Rev. Lett. {\bf 90} 041801 (2003).

\bibitem{McGregor:ds}
G.~McGregor  [MiniBooNE Collaboration],
AIP Conf.\ Proc.\  {\bf 655}, 58 (2003).

\bibitem{kamland} K.~Eguchi {\em et al}, Phys. Rev. Lett. {\bf 90}, 021802 (2003).

\bibitem{MSW}
L.~Wolfenstein, Phys. Rev. {\bf D 17} (1978) 2369; Phys. Rev. {\bf D 20} (1979) 2634; \\
S.~P.~Mikheyev and A.~Yu.~Smirnov, Sov. J. Nucl. Phys. {\bf 42}
(1986) 913.

\bibitem{minos5yr} ``Proposal for a Five Year Run Plan for MINOS'', 
The MINOS Collaboration, NuMI-930, May 2003, 
submitted to the Fermilab Directorate, 
see \verb+http://hep.caltech.edu/~michael/minos/fiveyear.ps.+ 

\bibitem{minosnue} M.Diwan, M.Messier, B.Viren, L.Wai, 
``A Study of $\nu_\mu \to \nu_e$ Sensitivity in MINOS'' 
NUMI-L-714, February 2001.

\bibitem{numioa} D. Ayres {\em et al}, \verb+hep-ex/0210005+ and \verb+http://www-off-axis.fnal.gov.+

%
%
\bibitem{MINOS_TRD_NDHALL} MINOS Collaboration, ``MINOS Technical Design Report``
,NuMI-NOTE-GEN-0337 (1998).

\bibitem{MARS} N. V. Mokhov, ``The MARS Monte Carlo'', FERMILAB FN-628 (1995);
N. V. Mokhov and O. E. Krivosheev, ``MARS Code Status'', FERMILAB-Conf-00/181 (2000);
http://www-ap.fnal.gov/MARS/.


\bibitem{GNuMI} M.~Messier (private communication)

\bibitem{GEANT} GEANT Manual, CERN Program Library Long Writeup W5013.

\bibitem{beamTypes} M. Kostin et~al.,
``Proposal for Continuously-Variable Beam Energy'',
NuMI-B-783.

\bibitem{ref:Mars_nuInteractions} Nikolai Mokhov and Andreas Van Ginneken,
``Neutrino Radiation at Muon Colliders and Storage Rings'',
FERMILAB-Conf-00/065.

%
%

\bibitem{Benhar:2003ka}
O.~Benhar,
[arXiv:nucl-th/0307061].

\bibitem{Ruddick}
K.~Ruddick (private communication).

\bibitem{Mualem}
L.~Mualem (private communication).

\bibitem{Fruhwirth}
R.~Fruhwirth,
\newblock Nucl.~Inst.~Meth. {\bf A262}, 444 (1987).

\bibitem{DONUT}
K.~Kodama {\em et~al.}, 
\newblock Nucl. Phys. Proc. Suppl {\bf 98}, 43-47 (2001)

\bibitem{Hayato:2003} Y. Hayato, \newblock
To be published in {\em Proceedings of the Second Workshop on
Neutrino-Nucleus Interactions in the Few-GeV Region (NUINT02),}
Irvine, California (2002).

\bibitem{MI-Armco} J.F.~Ostiguy, FNAL-MI-0127.

\bibitem{Adamson:2002mj}
MINOS, P.~Adamson {\em et~al.},
\newblock IEEE Trans. Nucl. Sci. {\bf 49}, 861 (2002).

\bibitem{Border:vk}
P.~Border {\em et~al.},
Nucl. Instrum. Meth. {\bf A463}, 194 (2001).

\bibitem{Rubinov:TRiP} P.~Rubinov, FNAL-TM-2226. P.~Rubinov, FNAL-TM-2227. 
P.~Rubinov, FNAL-TM-2228.

\bibitem{ingram} for a review of pion absorption results see
C.H.Q. Ingram, Nucl. Phys. {\bf A684}, 122c (2001).

\bibitem{jones} M.K. Jones et al., Phys. Rev. C {\bf 48}, 2800 (1993).

\bibitem{intranuke}
R.~Merenyi {\em et al.}, Phys.~Rev.~D {\bf 45}, 743 (1992).



\end{thebibliography}
